\documentclass[a4paper,12pt]{article}
\pdfoutput=1
\usepackage{epsfig,amsmath,amsfonts,amsthm}
\usepackage[figuresright]{rotating}
\usepackage[left=28mm,right=20mm,top=30mm,bottom=20mm]{geometry}
\usepackage{subfig}
\usepackage{graphicx}
\usepackage{rotating}
\usepackage{booktabs}
\usepackage{moreverb}

\DeclareGraphicsExtensions{.pdf,.png,.jpg}
\usepackage{natbib}
\numberwithin{equation}{section}

\def\e{\hbox{E}}

\def\var{\hbox{Var}}

\def\min{\hbox{min}}

\usepackage[nokeyprefix]{refstyle}
\usepackage{varioref}
\usepackage{xr}
\usepackage{hyperref}

\begin{document}

\title{Simulation study of  $Q$ statistic with constant weights for testing and estimation of  heterogeneity of  standardized mean differences in meta-analysis}

\author{Ilyas Bakbergenuly, David C. Hoaglin,  and Elena Kulinskaya}

\date{\today}

\maketitle

\begin{abstract}
Cochran's $Q$ statistic is routinely used for testing heterogeneity in meta-analysis. Its expected value  is also used for estimation of between-study variance $\tau^2$. Cochran's $Q$, or $Q_{IV}$, uses estimated inverse-variance weights which makes approximating its distribution rather complicated.

As an alternative,  we are investigating a new $Q$ statistic,  $Q_F$, whose constant weights use only the studies' effective sample sizes. For standardized mean difference as the measure of effect, we study, by simulation, approximations to distributions of $Q_{IV}$ and $Q_F$, as the basis for tests of heterogeneity and for new point and interval estimators of the between-study variance $\tau^2$. These include new DerSimonian-Kacker (2007)-type moment estimators based on the first moment of $Q_F$, and novel median-unbiased estimators of $\tau^2$.
\end{abstract}

\section{Moment estimation of $Q$}
We consider the following random-effects model (REM): For Study $i$ ($i = 1, \ldots, K$),  the estimate of the effect is $\hat\theta_i \sim G(\theta_i, v_i^2)$, where the effect-measure-specific distribution $G$ has mean $\theta_i$ and variance $v_i^2$, and $\theta_i \sim N(\theta, \tau^2)$.
The $\hat\theta_i$ are unbiased estimates of the true conditional effects $\theta_i$, and the $v_i^2 = \var(\hat\theta_i | \theta_i)$ are the corresponding true conditional variances.

Cochran's $Q$ statistic is a weighted sum of squared deviations of the estimated effects $\hat\theta_i$ from their weighted mean $\bar\theta_w = \sum w_i\hat\theta_i/\sum w_i$:
\begin{equation} \label{Q}
Q=\sum w_i(\hat\theta_i-\bar\theta_w)^2.\end{equation}
In \cite{cochran1954combination} $w_i$ is the reciprocal of the \textit{estimated} variance of $\hat{\theta}_i$. In meta-analysis those $w_i$ come from the fixed-effect model. We denote this standard version of $Q$ by $Q_{IV}$.  In our simulations, we examine a version of $Q$, discussed by \cite{dersimonian2007random}, in which the $w_i$ are arbitrary positive constants. We denote this version by $Q_{F}$. 

Define $W = \sum w_i$,  $q_i = w_i / W$, and $\Theta_i = \hat\theta_i - \theta$.  In this notation, and expanding $\bar\theta_w$, Equation (\ref{Q}) can be written  as
\begin{equation} \label{Q1}
Q = W \left[ \sum q_i (1 - q_i) \Theta_i^2 - \sum_{i \not = j} q_i q_j \Theta_i \Theta_j \right].
\end{equation}

Under the above REM, it is straightforward to obtain the first moment of $Q$ as
\begin{equation} \label{M1Q}
\e(Q) = W \left[ \sum q_i (1 - q_i) \var(\Theta_i) \right] = W \left[ \sum q_i (1 - q_i) (\e(v_i^2) + \tau^2) \right].
\end{equation}
Rearranging the terms gives the moment-based estimator of $\tau^2$
\begin{equation} \label{tau_DSK}
\hat\tau^2_{M} = \frac{Q / W - \sum q_i (1 - q_i) \e(v_i^2)} {\sum q_i (1 - q_i)}.
\end{equation}
This equation is similar to Equation (6) in \cite{dersimonian2007random}; they use the estimate $s_i^2$ instead of $\e(v_i^2)$.  We study this estimator of $\tau^2$, which uses the conditional estimated variances, $\hat v_i^2$, in the sample-size-based weights, and denote it by $\hat\tau^2_{SSC}$ or simply SSC. We also study the corresponding estimator that uses the unconditional estimated variances, $\hat\e(v_i^2) $, which we denote by SSU.

\section{Approximations for the distribution of $Q$}
We also study approximations for the distributions of $Q$ statistics. For SMD, $Q_F$ is a quadratic form in $t$ variates.  The  \cite{Farebrother1984} algorithm for the exact distribution of a quadratic form in normal variables  may provide a satisfactory approximation, especially for larger sample sizes. To apply it, we plug in estimated variances. We investigate the quality of that approximation, which we denote by F SW, and the two-moment approximation (M2 SW) based on the gamma distribution.

The null distribution of $Q_{IV}$ is usually approximated by the chi-square distribution with $K - 1$ degrees of freedom.  For mean difference and standardized mean difference, however, this approximation is not accurate for small sample sizes (\cite{Viechtbauer-2007a}). For SMD \cite{kulinskaya2011moments} provided an improved approximation to the null distribution of $Q_{IV}$ based on a chi-square distribution with degrees of freedom equal to the estimate of the corrected first moment; we denote this approximation by KDB. \cite{biggerstaff2008exact} used the Farebrother approximation as the ``exact'' distribution of $Q_{IV}$. We denote this approximation by BJ.
When $\tau^2 = 0$, the BJ approximation to the distribution of $Q_{IV}$ reduces to the $\chi^2_{K - 1}$  distribution. For comparison, our simulations include these three approximations.

\section{Other point and interval estimators of $\tau^2$}

Distributions of the $Q$ statistics depend on the $\theta_i$ and the $v_i^2$ and on the between-study variance $\tau^2$. We denote such a distribution by $F(\cdot | \theta_i, v_i^2,\;\tau^2)$. In practice, an approximation uses estimated parameters, and we calculate its value at the observed value of $Q$, obtaining the approximate upper-tail p-value
$p(Q | \tau^2) = 1 - F(Q | \hat\theta_i, \hat v_i^2,\;\tau^2)$.

A confidence interval for $\tau^2$ at confidence level $1 - \alpha$ can be obtained as $$\{\tau^2: p(Q | \tau^2) \in [\alpha/2, 1 - \alpha/2]\}.$$
Similarly, a point estimator of $\tau^2 $ can be found as $$\hat\tau^2_{med} = \min(0,\; \{\tau^2: F(Q|\tau^2)=0.5\})$$

Our simulations for SMD use the Farebrother approximation to the distribution of $Q$ in two versions. One uses the conditional variances of the $\hat\theta_i$ (i.e., $\hat v_i^2$), and the other uses the unconditional variances (i.e.,  $\hat \e(v_i^2)$).  The corresponding point estimators of $\tau^2$, based on the median, are denoted by SMC and SMU, respectively.  The corresponding intervals are denoted by FPC and FPU. The P in their names reminds us that they are profile intervals.

For comparison our simulations include four point estimators that use inverse-variance weights: DerSimonian-Laird (DL), restricted maximum likelihood (REML), Mandel-Paule (MP), and an estimator (KDB) based on the work of \cite{kulinskaya2011testing}, which uses an improved first moment of $Q$.

We also include three interval estimators: the Q-profile (QP) interval, the profile-likelihood (PL) interval, and the KDB interval, which is based on the chi-square distribution with the corrected first moment.

\section{Study-level estimation of standardized mean difference} \label{sec:EffectSMD}

Each of the $K$ studies consists of two arms, treatment (T) and control (C), with sample sizes $n_{iT}$ and $n_{iC}$. The total sample size in Study $i$ is $n_i = n_{iT} + n_{iC}$, and the ratio of the control sample size to the total is $f_i = n_{iC}/n_{i}$. We define the effective sample size in Study $i$ as $\tilde{n}_i = n_{iC} n_{iT} / n_i$.

The subject-level data in each arm are assumed to be normally distributed with means $\mu_{iT}$ and $\mu_{iC}$ and equal variances $\sigma_{i}^2$. The sample means are $\bar{x}_{ij}$, and the sample variances are $s^2_{ij}$, for $i = 1,\ldots, K$ and $j = C$ or $T$.

The standardized mean difference effect measure is
$$\delta_{i} = \frac{\mu_{iT} - \mu_{iC}} {\sigma_{i}}.$$
The unbiased estimator of $\delta_i$ is Hedges's $g$:
\begin{equation} \label{eq:delta}
{g}_i = J(m_i) \frac{\bar{x}_{iT} - \bar{x}_{iC}} {s_{i}},
\end{equation}
where the standard deviation, $\sigma_i$, is estimated by the square root of the pooled sample variance $s_i^2$, $m_{i} = n_{iT} + n_{iC} - 2$, and the factor
$J(m) = {\Gamma \left( \frac{m} {2} \right) } / {\sqrt{\frac{m}{2}}\Gamma \left( \frac{m - 1} {2} \right) }$
corrects for bias.

For the variance of $g_i$ we use the unbiased estimator
\begin{equation} \label{eq:g_var}
v_{i}^2 = \frac{n_{iT} + n_{iC}} {n_{iT } n_{iC}} + \left(1 - \frac{(m_{i} - 2)} {m_{i}J(m_{i})^2} \right) g^2_{i},
\end{equation}
derived by \cite{hedges1983random}. The literature contains several other estimators of the variance of $g_i$ and its biased counterpart, $d_i$. \cite{LinAloe-2021} provide a comprehensive assessment.

The sample SMD ${g}_i$ has a scaled non-central $t$-distribution with non-centrality parameter
$\gamma_i = \tilde{n}_{i}^{1/2} \delta_i$ \citep{HO-1985}:
\begin{equation} \label{eq:g_dist}
\sqrt{\tilde{n}_i} J(m_i)^{-1} g_i \sim t_{m_i} (\tilde{n}_{i}^{1/2} \delta_i).
\end{equation}

\section{Sketch of the simulations}

The design of the simulations follows that described in \cite{BHK2018SMD}. Briefly,
we vary five parameters: the overall true SMD ($\delta$), the between-studies variance ($\tau^2$), the number of studies ($K$), the studies' total sample size (both equal, $n$, and unequal, $\bar{n}$), and the proportion of observations in the Control arm ($f$). Table~\ref{tab:altdataSMD} lists the values of each parameter. We use a total of $10,000$ repetitions for each combination of parameters.

We generate the true effect sizes $\delta_{i}$ from a normal distribution:  $\delta_{i} \sim N(\delta, \tau^2)$. We generate the values of Hedges's estimator ${g}_{i}$ directly from the appropriately scaled non-central $t$-distribution with non-centrality parameter $\tilde{n}_{i}^{1/2} \delta_i$:
\begin{equation}
\sqrt{\tilde{n}_i} J(m_i)^{-1} g_i \sim t_{m_i} (\tilde{n}_{i}^{1/2} \delta_i).
\end{equation}

\begin{table}[ht]
	\caption{\label{tab:altdataSMD} \emph{Values of parameters in the simulations for SMD}}
	\begin{footnotesize}
		\begin{center}
			\begin{tabular}
				{|l|l|l|}
				\hline
				Parameter & Equal study sizes & Unequal study sizes \\
                                & &\\	\hline
				$K$ (number of studies) & 5, 10, 30 & 5, 10, 30 \\
				$n$ or $\bar{n}$  (average (individual) study size ---  & 20, 40, 100, 250,640, 1000 & 30 (12,16,18,20,84), \\
				total of the two arms) & 30, 50, 60, 70 & 60 (24,32,36,40,168), \\
				For  $K = 10$ and $K = 30$, the same set & & 100 (64,72,76,80,208), \\
				of unequal study sizes is used & &160 (124,132,136,140,268) \\
twice or six times, respectively. & & \\
				$f$ (proportion of each study in the control arm) & 1/2, 3/4 &1/2, 3/4 \\
				$\delta$ (true value of the SMD) & 0, 0.2, 0.5, 1, 2 & 0, 0.2, 0.5, 1, 2 \\
				$\tau^{2}$ (variance of random effects) & 0, 0.5, 1, 1.5, 2, 2.5 & 0, 0.5, 1, 1.5, 2, 2.5 \\
				\hline
			\end{tabular}
		\end{center}
	\end{footnotesize}
\end{table}

\section{Summary}

In estimating $\tau^2$, our moment estimator based on $Q_F$ with effective-sample-size weights is almost unbiased, the Mandel-Paule estimator has some negative bias in some situations, and the DerSimonian-Laird and restricted-maximum-likelihood estimators have considerable negative bias.

All 95\% interval estimators have coverage that is too high when $\tau^2 = 0$, but otherwise the Q-profile interval performs very well.

An approximation based on an algorithm of Farebrother follows both the null and the alternative distributions of $Q_F$ reasonably well, whereas the usual chi-square approximation for the null distribution of $Q_{IV}$ and Biggerstaff-Jackson approximation to its alternative distribution are poor.

The test for heterogeneity based on $Q_F$ and F SW has error rates that somewhat exceed the nominal 5\%, and the test based on $Q_{IV}$ and the chi-square approximation has error rates that are noticeably too low.

When $\tau^2 > 0$, F SW provides robust empirical levels at all values of $\tau^2$ and $\delta$. Those levels, however, are somewhat higher than the nominal .05 (for larger $K$). M2 SW has below-nominal levels, which decrease further for larger $\delta$ but do not depend on $\tau^2$. The levels of the BJ approximation are even lower, and they decrease further as $\tau^2$ increases.

\section{Detailed results}
The appendices contain plots of the results of the simulations:
\begin{itemize}
	\item Appendix A. Plots for bias and coverage of estimators of $\tau^2$
	\item Appendix B. Plots of error in approximations for the distribution of $Q$
	\item Appendix C. Empirical level of the test for heterogeneity ($\tau^2$ = 0 versus $\tau^2 > 0$) based on approximations for the distribution of $Q$, plotted vs sample size
	\item Appendix D: Empirical level of the test for heterogeneity ($\tau^2 \leq \tau_{0}^2$ versus $\tau^2 > \tau_{0}^2$) based on approximations for the distribution of $Q$, plotted vs $\tau_{0}^2$
	\item Appendix E. Power of the test for heterogeneity ($\tau^2 = 0$ versus $\tau^2 > 0$) based on approximations for the distribution of $Q$
\end{itemize}

\bibliography{Qfixed_SMD.bib}

\clearpage

\section*{Appendices}
\begin{itemize}
	\item Appendix A. Plots for bias and coverage of estimators of $\tau^2$
	\item Appendix B. Plots of error in approximations for the distribution of $Q$
	\item Appendix C. Empirical level of the test for heterogeneity ($\tau^2$ = 0 versus $\tau^2 > 0$) based on approximations for the distribution of $Q$, plotted vs sample size
	\item Appendix D: Empirical level of the test for heterogeneity ($\tau^2 \leq \tau_{0}^2$ versus $\tau^2 > \tau_{0}^2$) based on approximations for the distribution of $Q$, plotted vs $\tau_{0}^2$
	\item Appendix E. Power of the test for heterogeneity ($\tau^2 = 0$ versus $\tau^2 > 0$) based on approximations for the distribution of $Q$
\end{itemize}

\clearpage
\setcounter{section}{0}

\renewcommand{\thesection}{A.\arabic{section}}
\section*{A. Bias and coverage of estimators of $\tau^2$}
\section{Bias of estimators of $\tau^2$}
Each figure corresponds to a value of $\delta$, a value of $f$, and either equal or unequal sample sizes. \\
For each combination of a value of $n$ or $\bar{n}$ and a value of $K$, a panel plots bias of estimators of $\tau^2$ versus $\tau^2$. \\
The point estimators of $\tau^2$ are
\begin{itemize}
	\item SSC (constant effective-sample-size weights and conditional variances)
	\item SSU (constant effective-sample-size weights and unconditional variances)
	\item SMC (the median $\tau^2$ value in Farebrother approximation with conditional moments)
	\item SMU (the median $\tau^2$ value in Farebrother approximation with unconditional moments)
	\item DL (standard DerSimonian-Laird: inverse-variance weights)
	\item REML (restricted maximum likelihood)
	\item MP (Mandel-Paule)
	\item KDB (inverse-variance method based on corrected first moment of null distribution of $Q$)
\end{itemize}

\clearpage
\renewcommand{\thefigure}{A1.\arabic{figure}}
\setcounter{figure}{0}

\begin{figure}[t]
	\centering
	\includegraphics[scale=0.33]{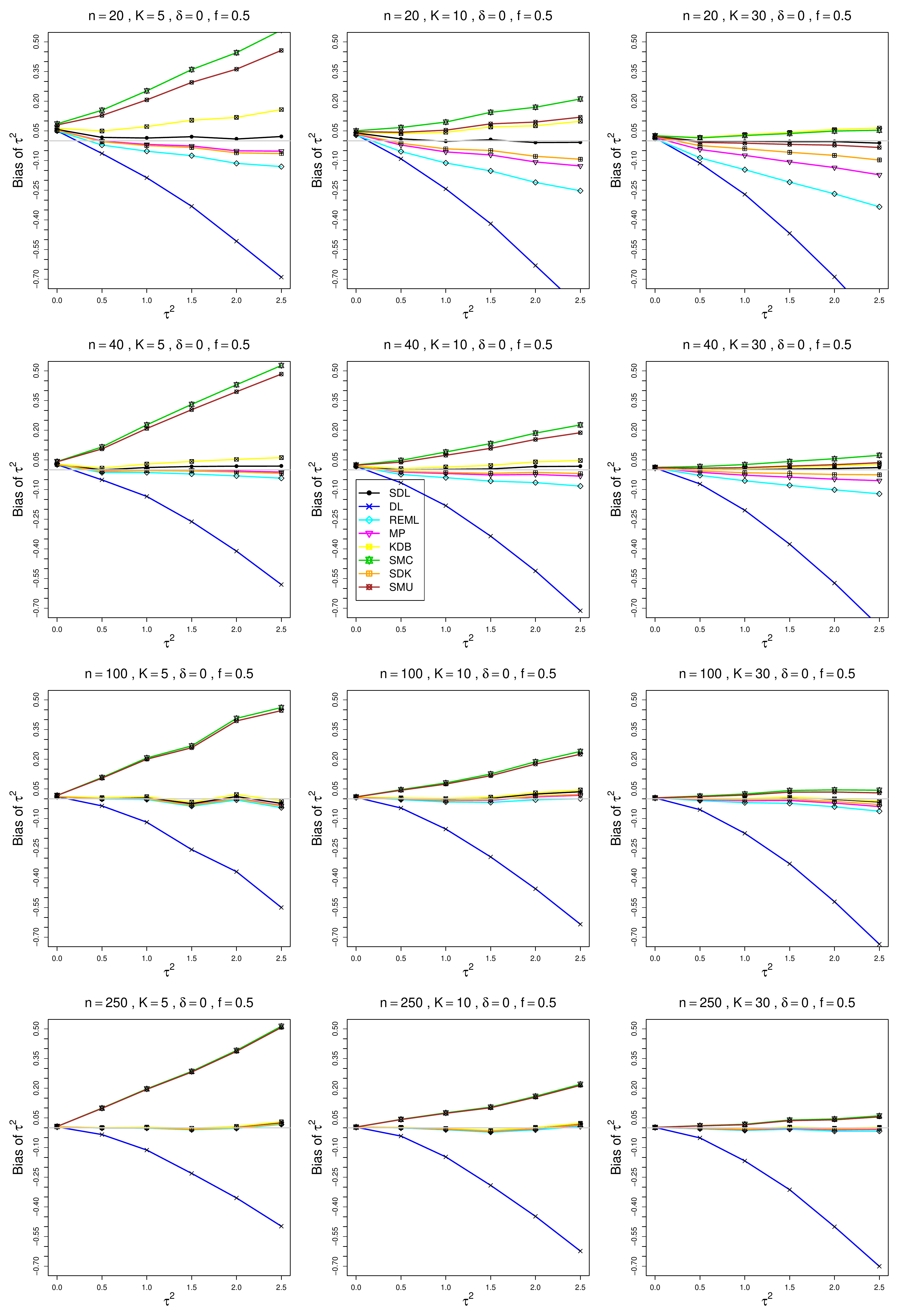}
	\caption{Bias of estimators of the between-studies variance $\tau^2$ for $\delta = 0$, $f = .5$, and equal sample sizes
		\label{PlotBiasOfTau2delta0andq05_SMD} }
\end{figure}

\begin{figure}[t]
	\centering
	\includegraphics[scale=0.33]{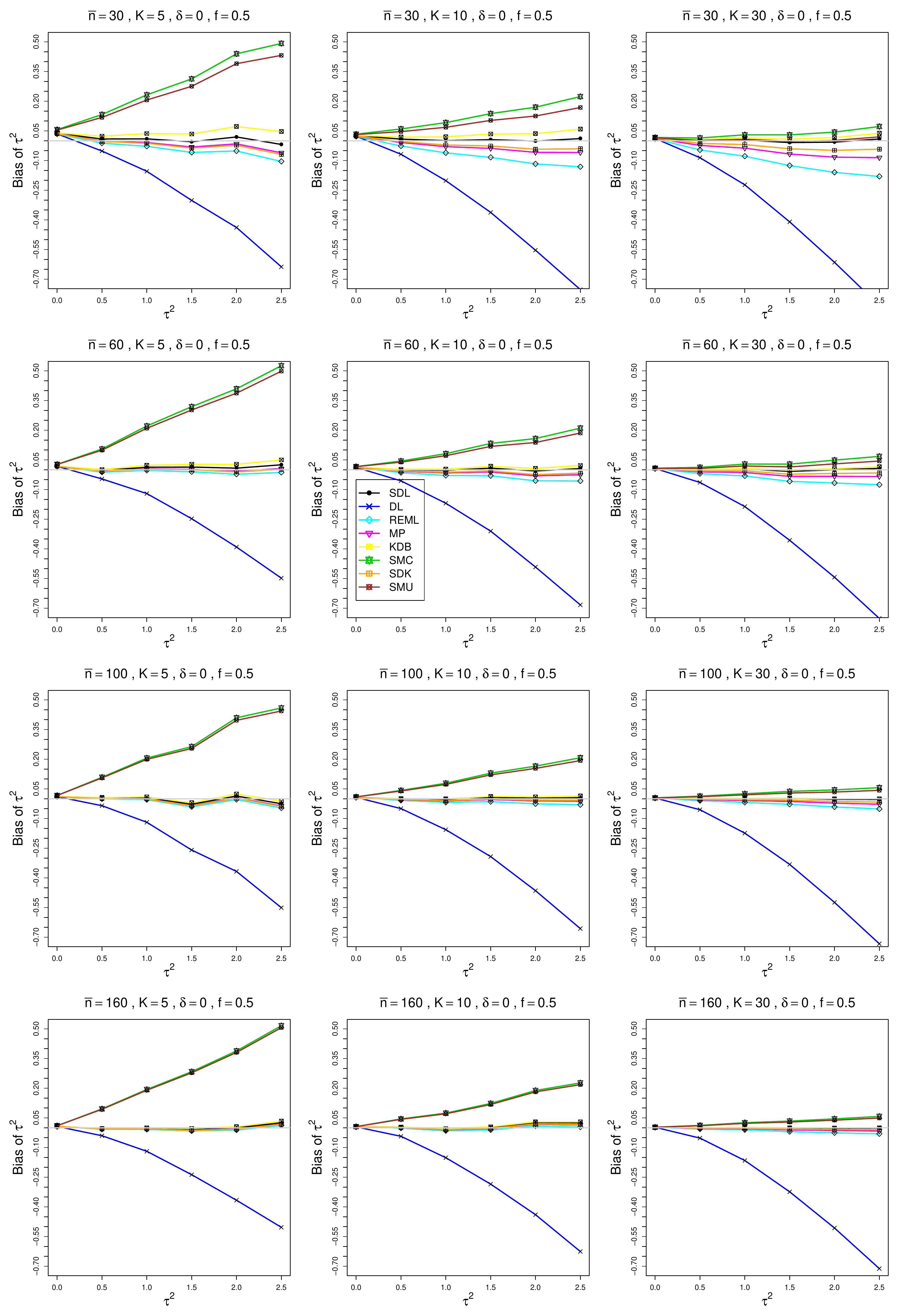}
	\caption{Bias of estimators of the between-studies variance $\tau^2 $ for $\delta = 0$, $f = .5$, and unequal sample sizes
		\label{PlotBiasOfTau2delta0andq05_SMD_unequal}}
\end{figure}

\begin{figure}[t]
	\centering
	\includegraphics[scale=0.33]{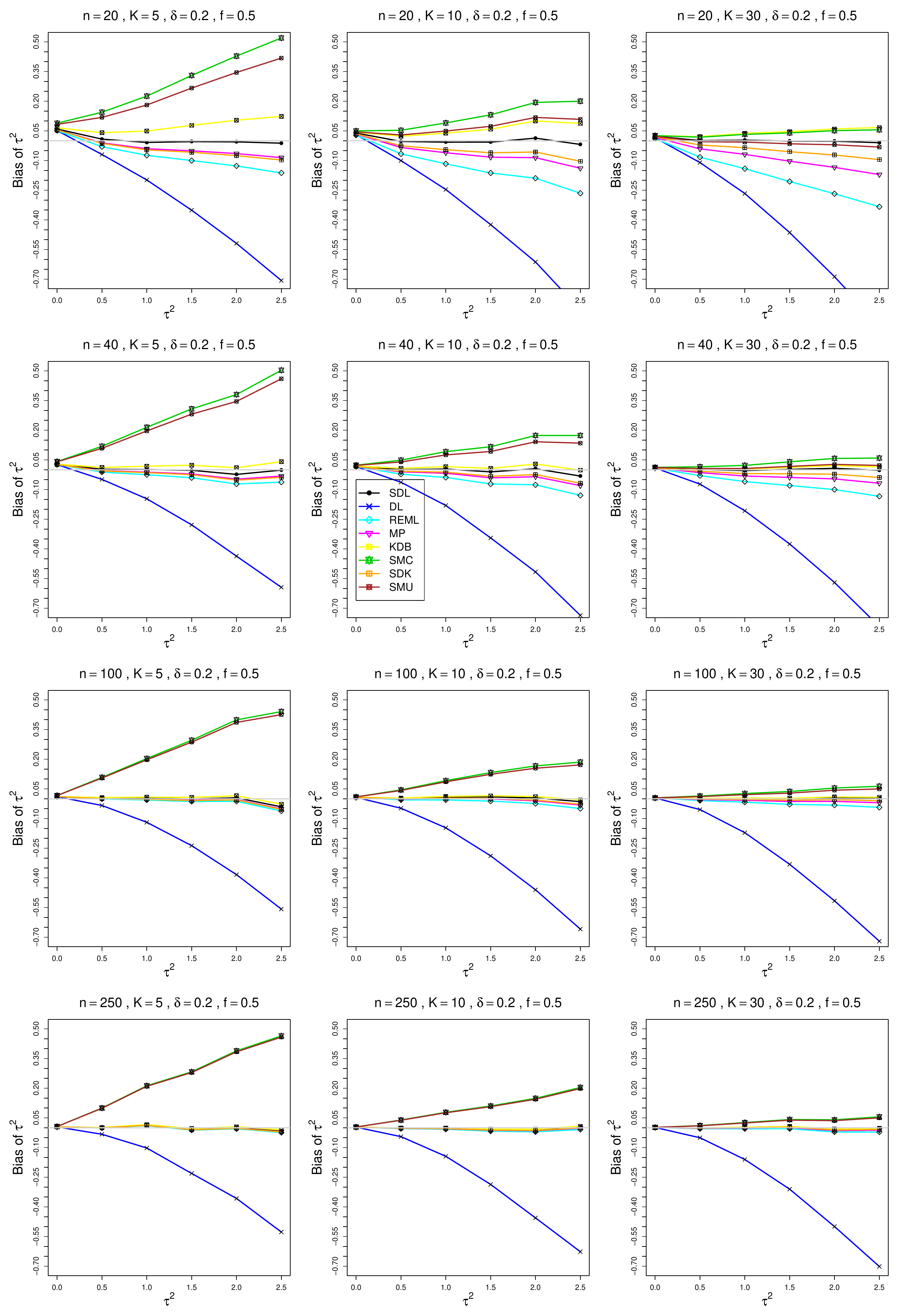}
	\caption{Bias of estimators of the between-studies variance $\tau^2$ for $\delta = 0.2$, $f = .5$, and equal sample sizes
		\label{PlotBiasOfTau2delta02andq05_SMD} }
\end{figure}

\begin{figure}[t]
	\centering
	\includegraphics[scale=0.33]{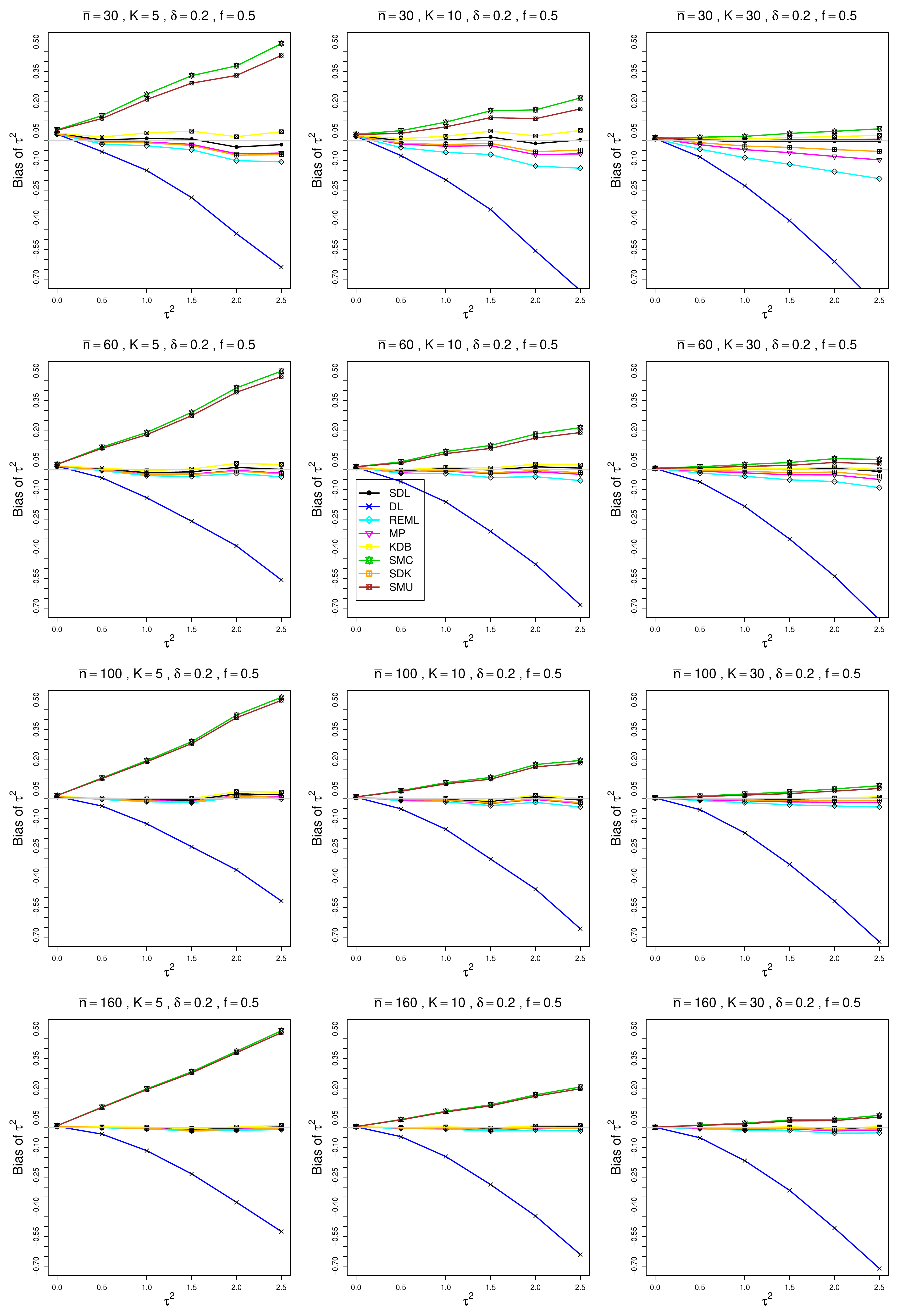}
	\caption{Bias of estimators of the between-studies variance $\tau^2$ for $\delta = 0.2$, $f = .5$, and unequal sample sizes
		\label{PlotBiasOfTau2delta02andq05_SMD_unequal} }
\end{figure}

\begin{figure}[t]
	\centering
	\includegraphics[scale=0.33]{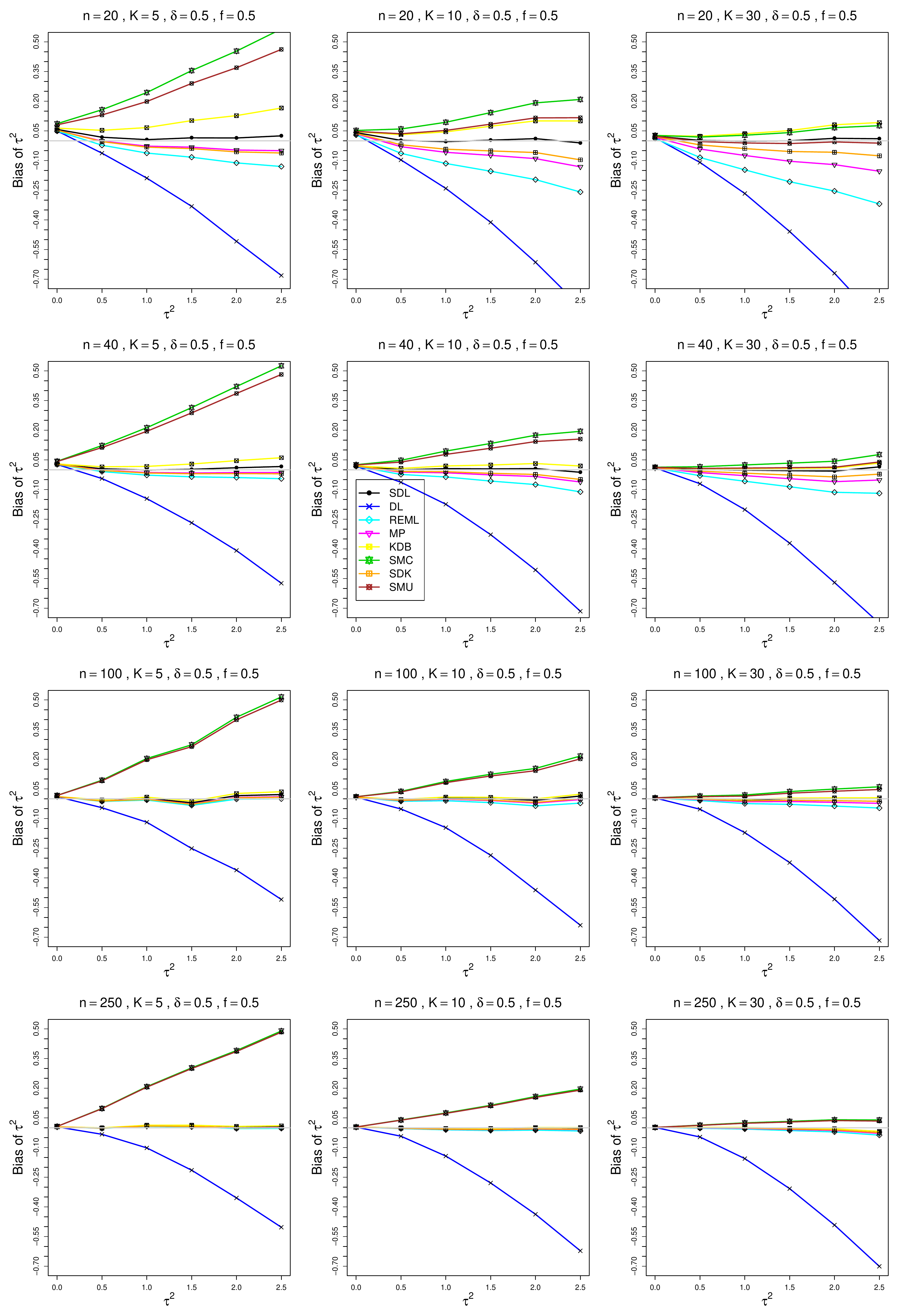}
	\caption{Bias of estimators of the between-studies variance $\tau^2 $ for $\delta = 0.5$, $f = .5$, and equal sample sizes
		\label{PlotBiasOfTau2delta05andq05_SMD}}
\end{figure}

\begin{figure}[t]
	\centering
	\includegraphics[scale=0.33]{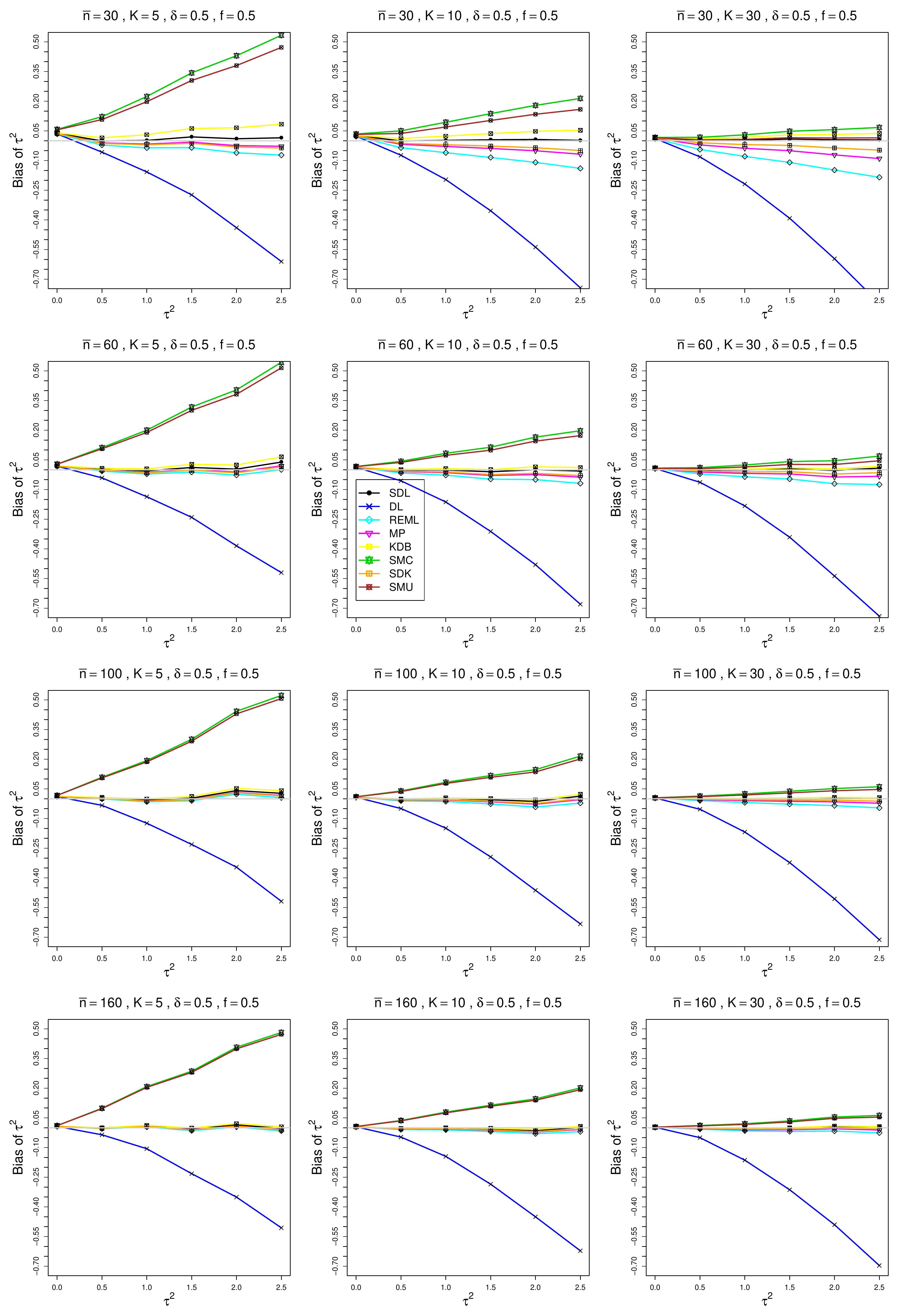}
	\caption{Bias of estimators of the between-studies variance $\tau^2$ for $\delta= 0.5$, $f = .5$, and unequal sample sizes
		\label{PlotBiasOfTau2delta05andq05_SMD_unequal} }
\end{figure}

\begin{figure}[t]
	\centering
	\includegraphics[scale=0.33]{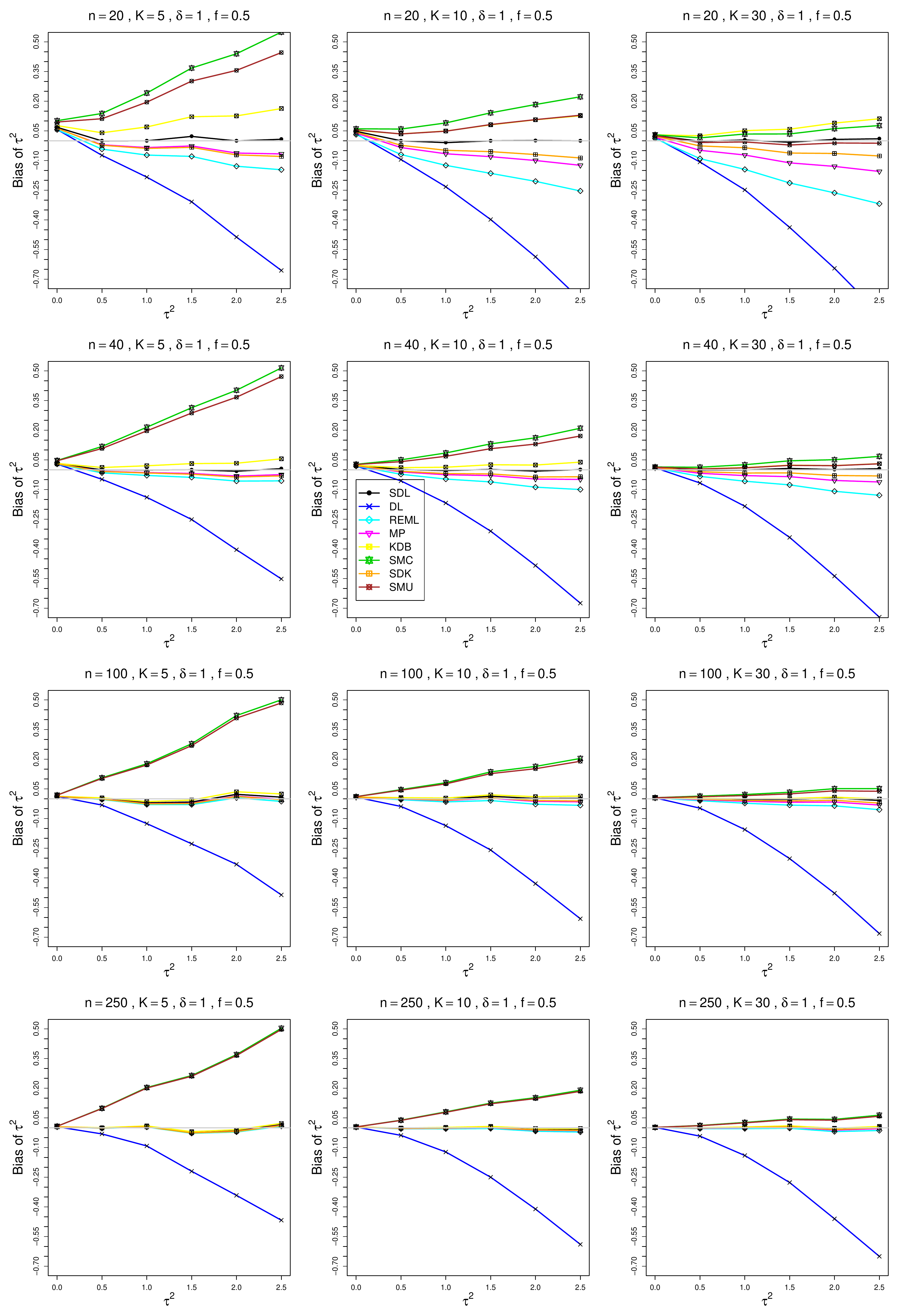}
	\caption{Bias of estimators of the between-studies variance $\tau^2$ for $\delta = 1$, $f = .5$, and equal sample sizes
		\label{PlotBiasOfTau2delta1andq05_SMD}}
\end{figure}

\begin{figure}[t]
	\centering
	\includegraphics[scale=0.33]{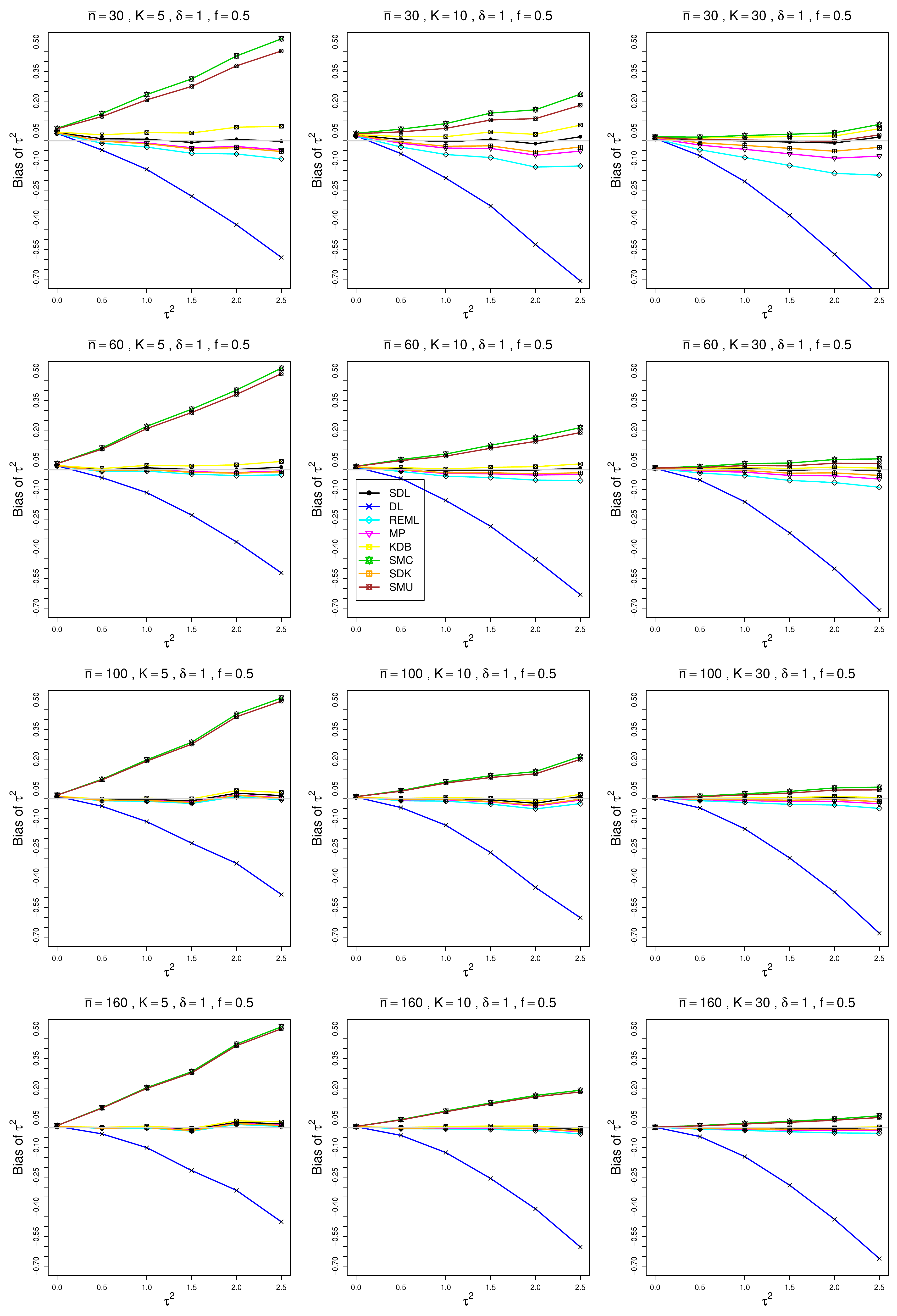}
	\caption{Bias of estimators of the between-studies variance $\tau^2$ for $\delta = 1$, $f = .5$, and unequal sample sizes
		\label{PlotBiasOfTau2delta1andq05_SMD_unequal}}
\end{figure}

\begin{figure}[t]
	\centering
	\includegraphics[scale=0.33]{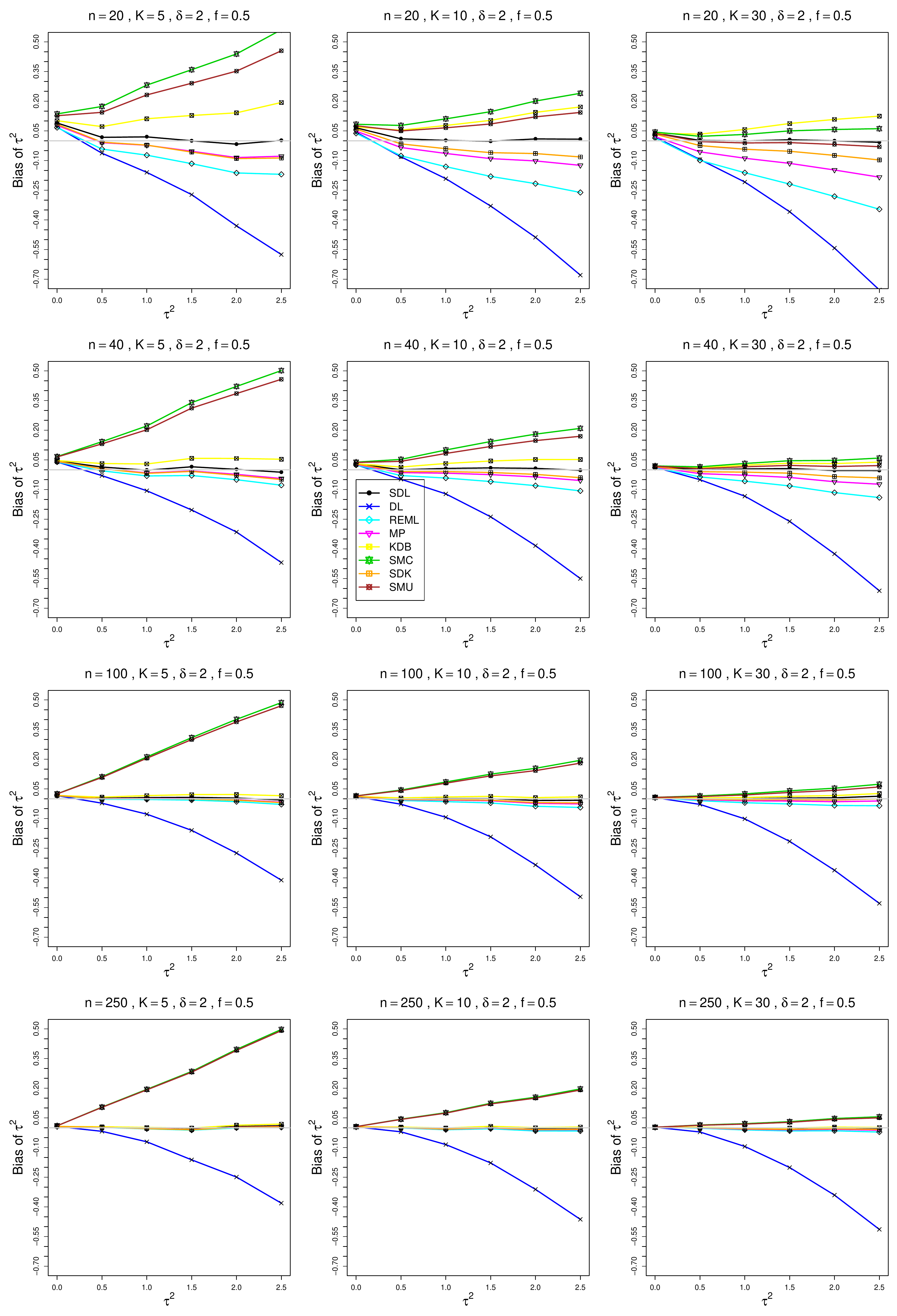}
	\caption{Bias of estimators of the between-studies variance $\tau^2 $ for $\delta = 2$, $f = .5$, and equal sample sizes
		\label{PlotBiasOfTau2delta2andq05_SMD}}
\end{figure}

\begin{figure}[t]
	\centering
	\includegraphics[scale=0.33]{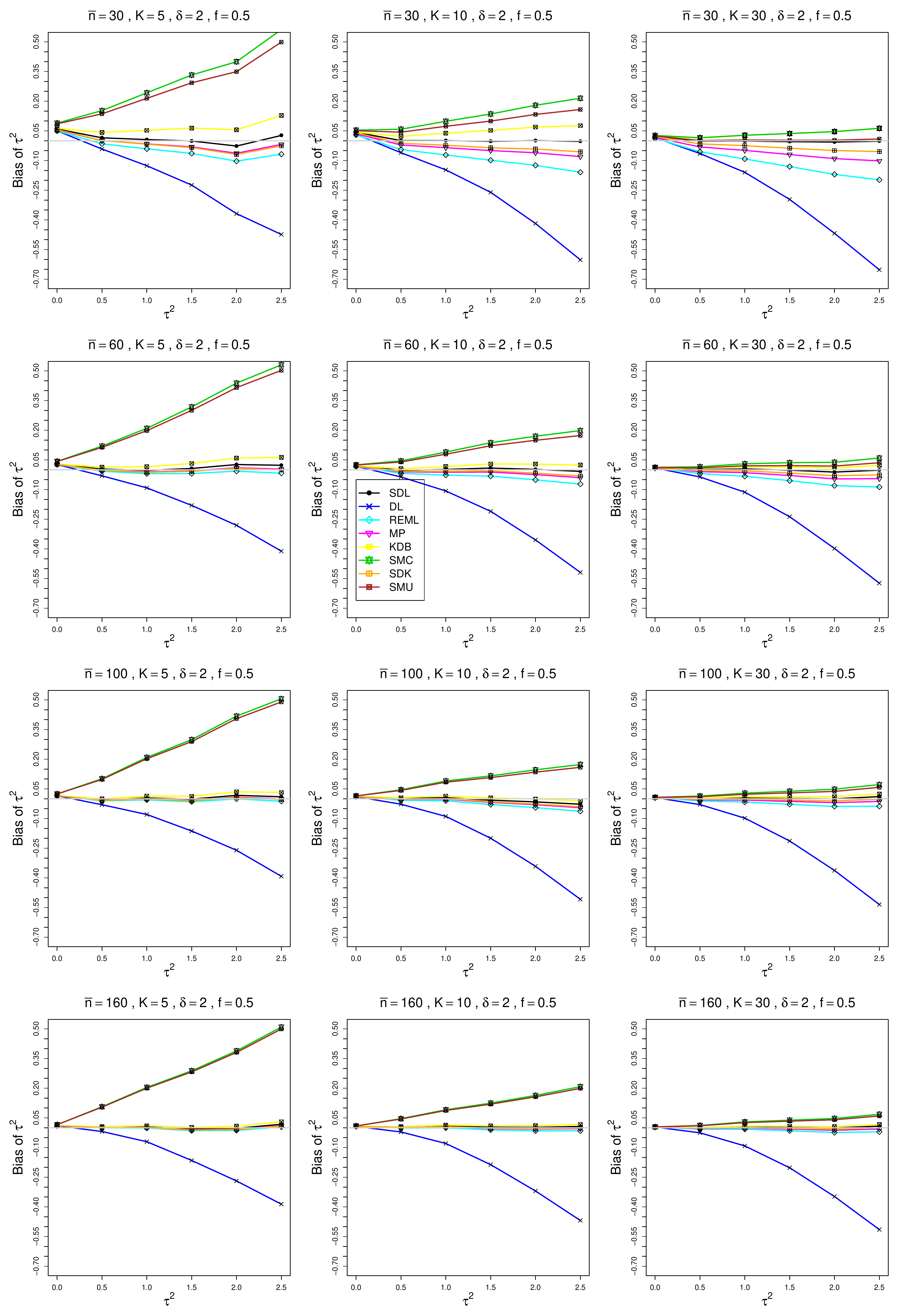}
	\caption{Bias of estimators of the between-studies variance $\tau^2$ for $\delta = 2$, $f = .5$, and unequal sample sizes
		\label{PlotBiasOfTau2delta2andq05_SMD_unequal}}
\end{figure}

%%%%%%%%%%%%%%%%%%%%%%%%%%%%%%%%%%%%%%%%%%%%%%%%%%%%%%%%%%%%%%%%%%%%%%%%%%%%%%%%%%%%%%%%%%%%%%%%%%%%%%%%%%%%%%%%%%%%%%%%%%%%%%
%q=0.75

\clearpage

\begin{figure}[t]
	\centering
	\includegraphics[scale=0.33]{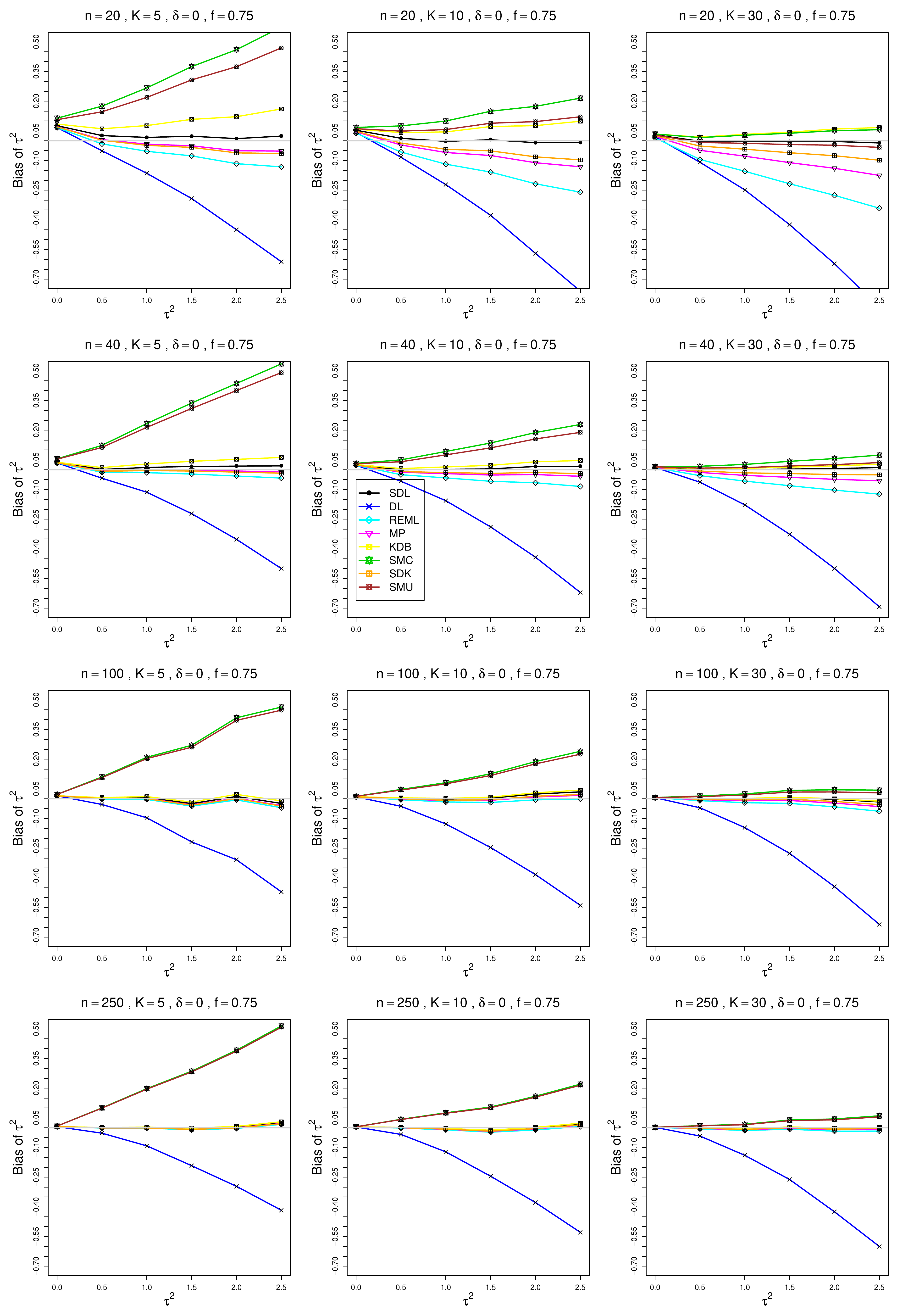}
	\caption{Bias of estimators of the between-studies variance $\tau^2$ for $\delta = 0$, $f = .75$, and equal sample sizes
		\label{PlotBiasOfTau2delta0andq075_SMD}}
\end{figure}

\begin{figure}[t]
	\centering
	\includegraphics[scale=0.33]{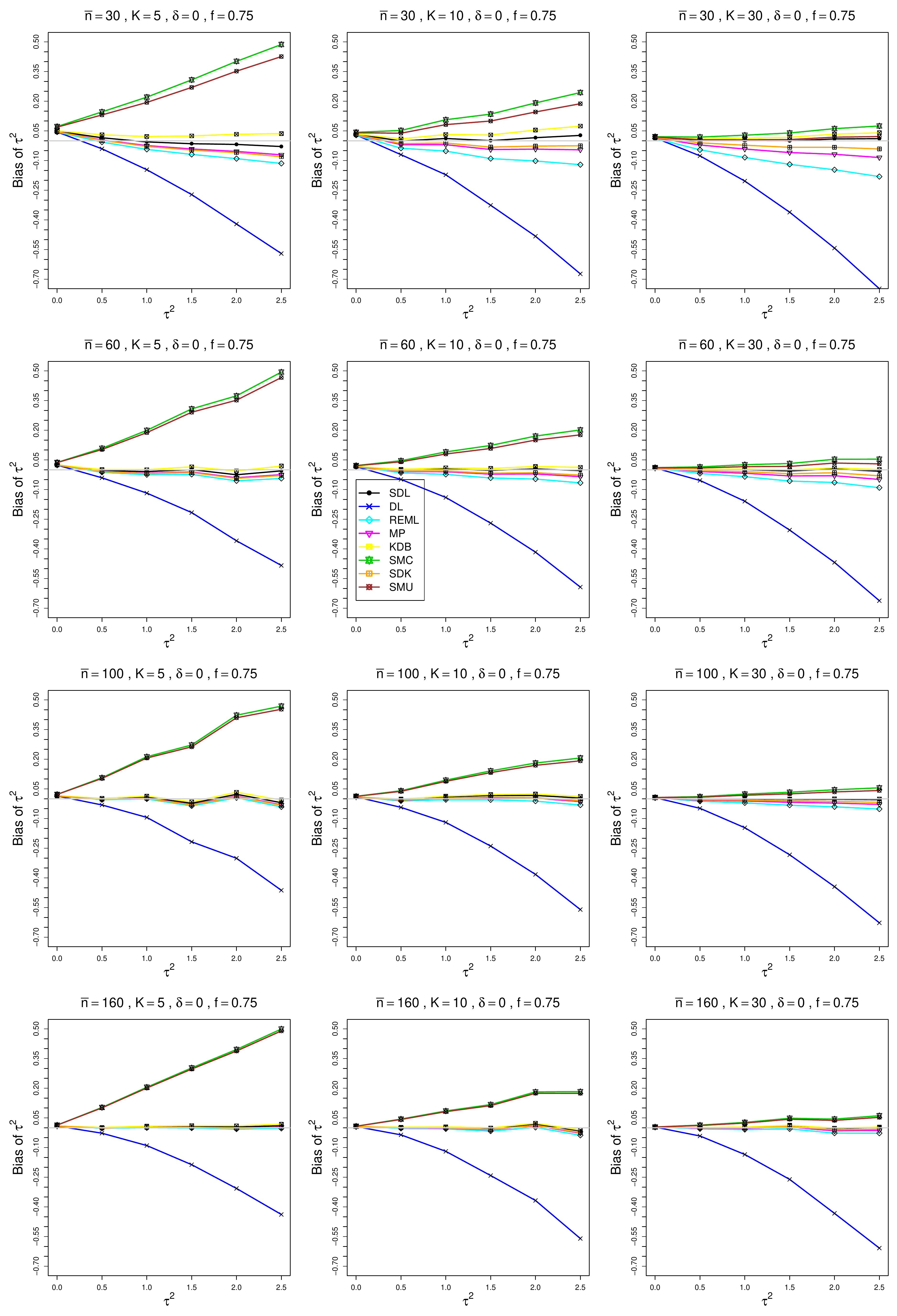}
	\caption{Bias of estimators of the between-studies variance $\tau^2$ for $\delta = 0$, $f = 0.75$, and unequal sample sizes
		\label{PlotBiasOfTau2delta0andq075_SMD_unequal}}
\end{figure}

\begin{figure}[t]
	\centering
	\includegraphics[scale=0.33]{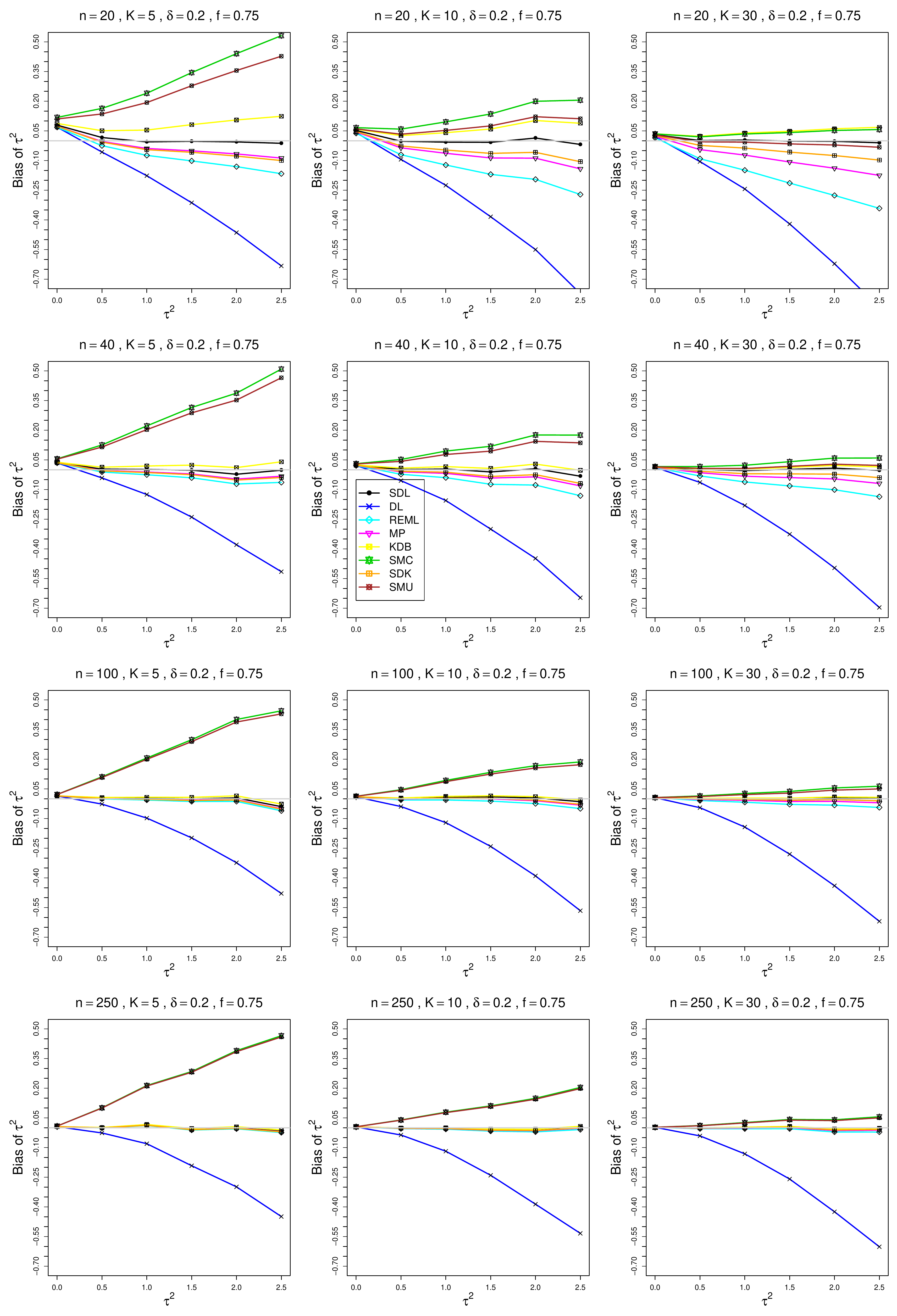}
	\caption{Bias of estimators of the between-studies variance $\tau^2$ for $\delta = 0.2$, $f = .75$, and equal sample sizes
		\label{PlotBiasOfTau2delta02andq075_SMD}}
\end{figure}

\begin{figure}[t]
	\centering
	\includegraphics[scale=0.33]{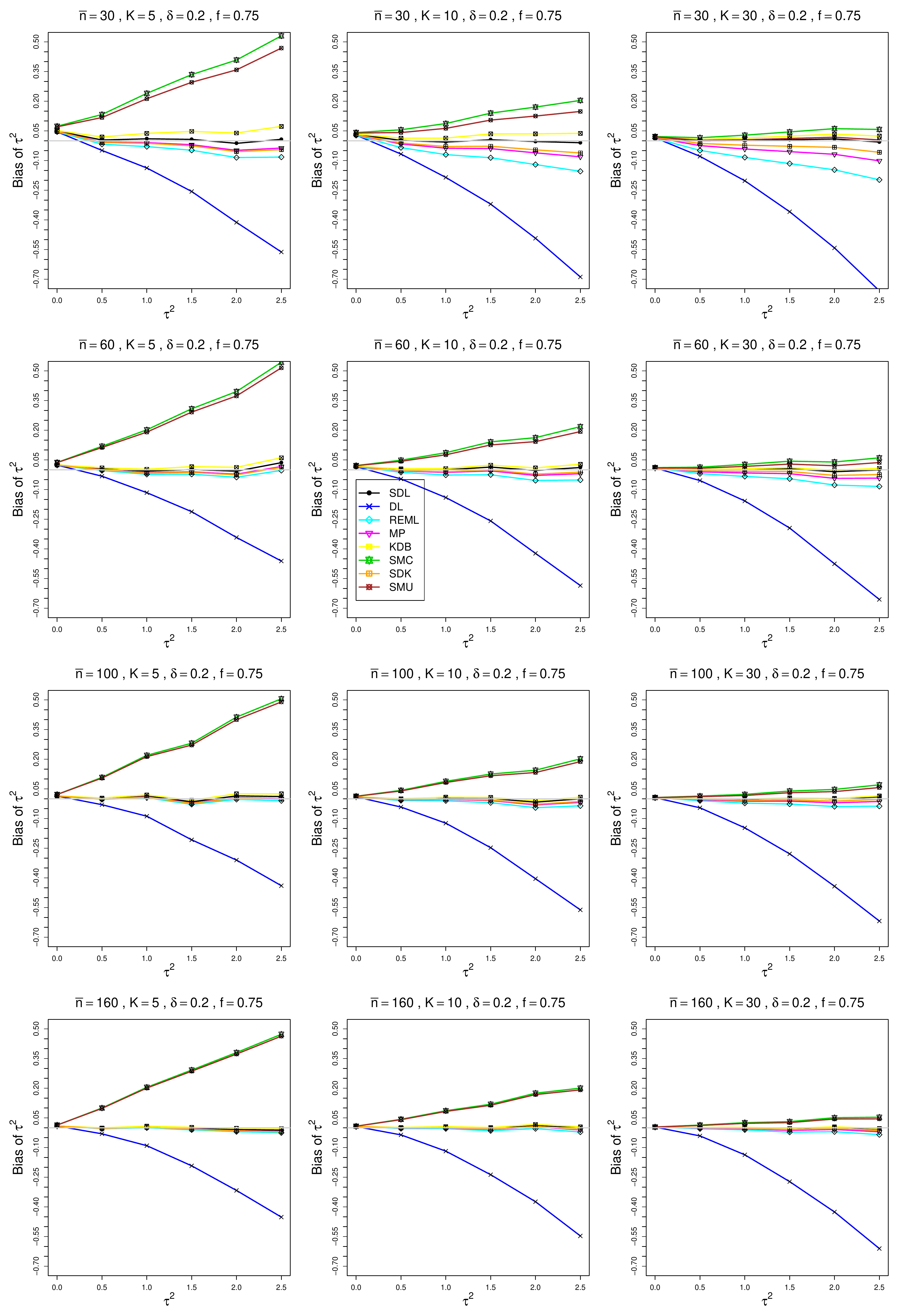}
	\caption{Bias of estimators of the between-studies variance $\tau^2$ for $\delta = 0.2$, $f = .75$, and unequal sample eizes
		\label{PlotBiasOfTau2delta02andq075_SMD_unequal}}
\end{figure}

\begin{figure}[t]
	\centering
	\includegraphics[scale=0.33]{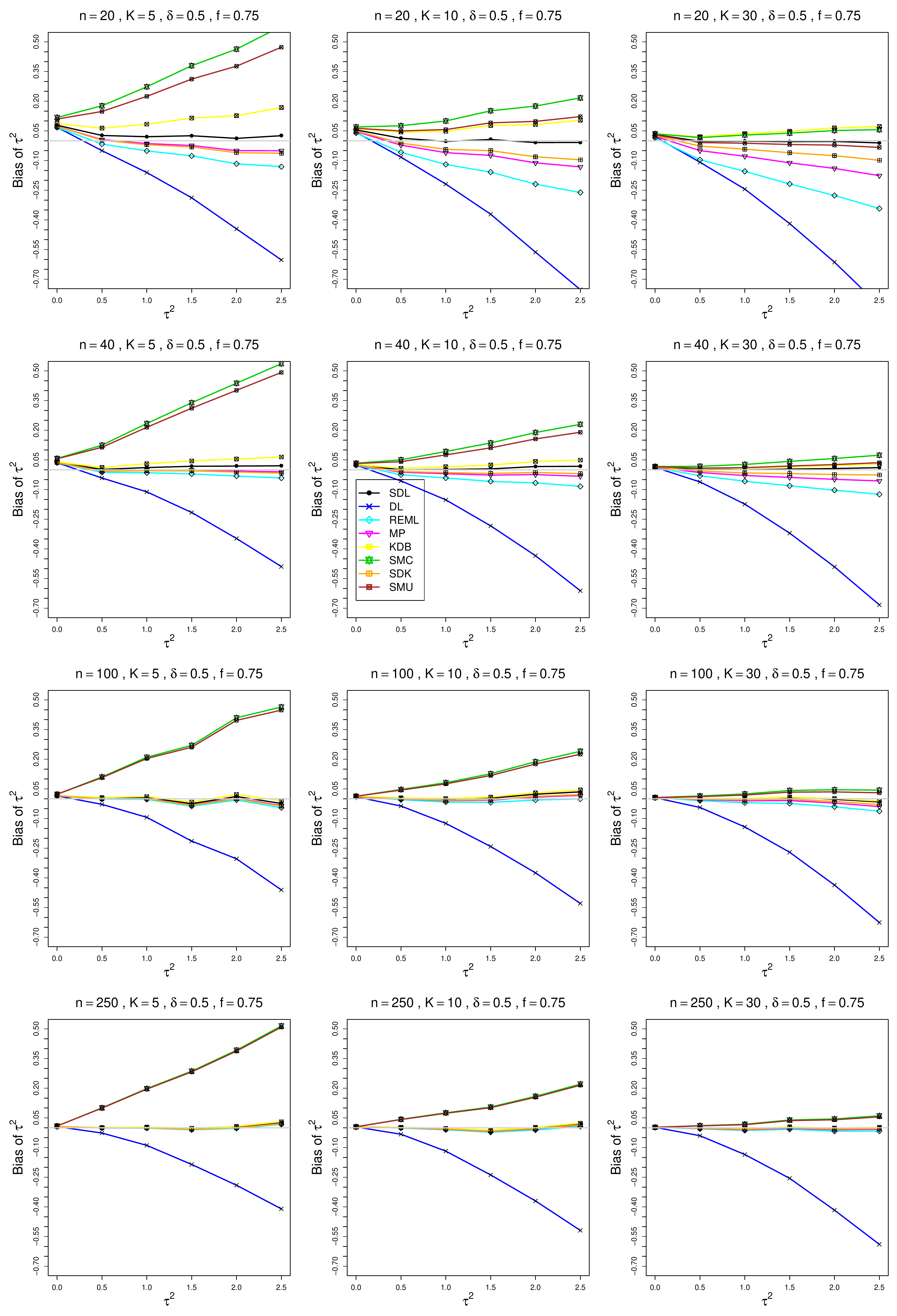}
	\caption{Bias of estimators of the between-studies variance $\tau^2$ for $\delta = 0.5$, $f = .75$, and equal sample sizes
		\label{PlotBiasOfTau2delta05andq075_SMD}}
\end{figure}

\begin{figure}[t]
	\centering
	\includegraphics[scale=0.33]{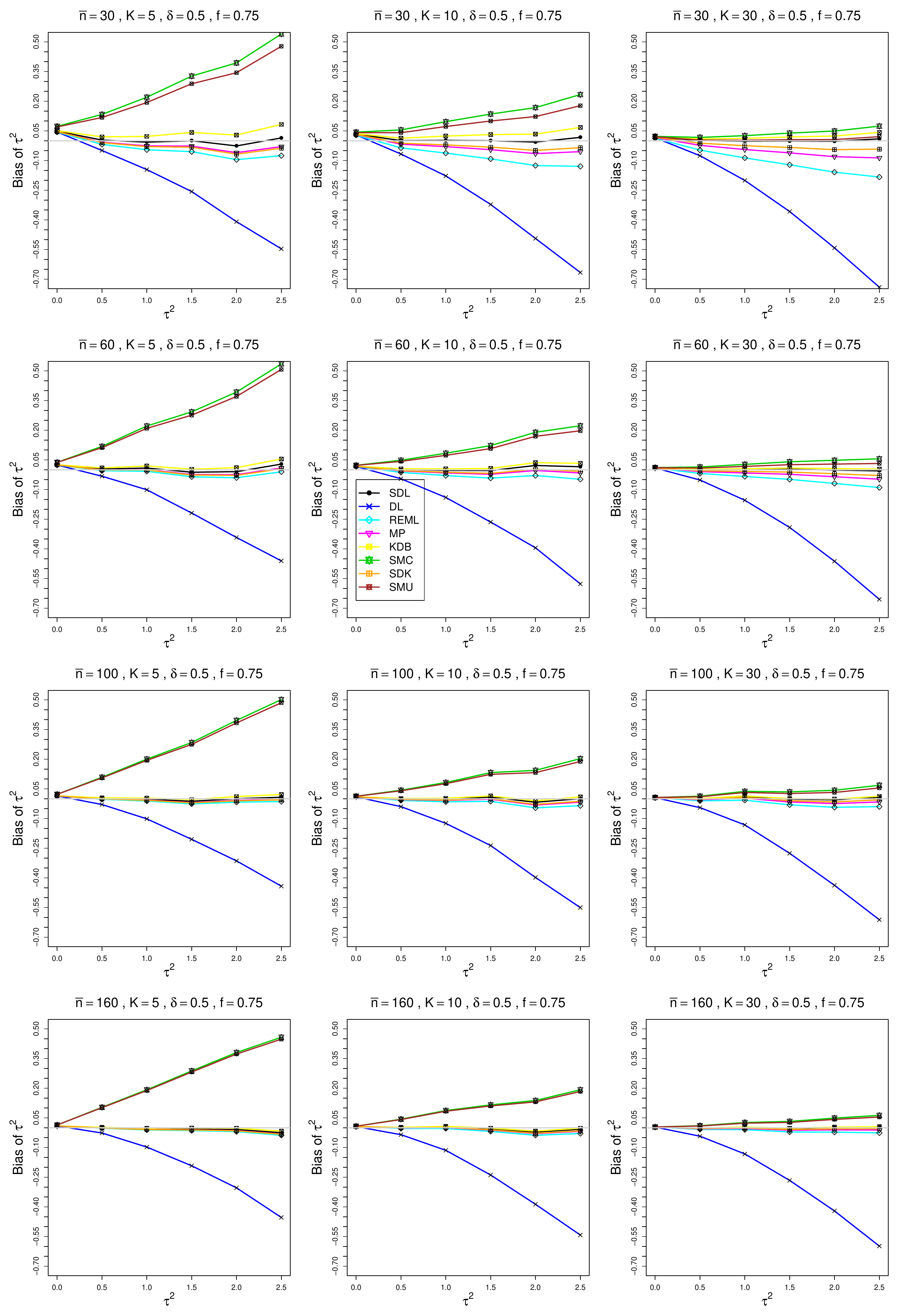}
	\caption{Bias of estimators of the between-studies variance $\tau^2$ for $\delta = 0.5$, $f = .75$, and unequal sample sizes
		\label{PlotBiasOfTau2delta05andq075_SMD_unequal} }
\end{figure}

\begin{figure}[t]
	\centering
	\includegraphics[scale=0.33]{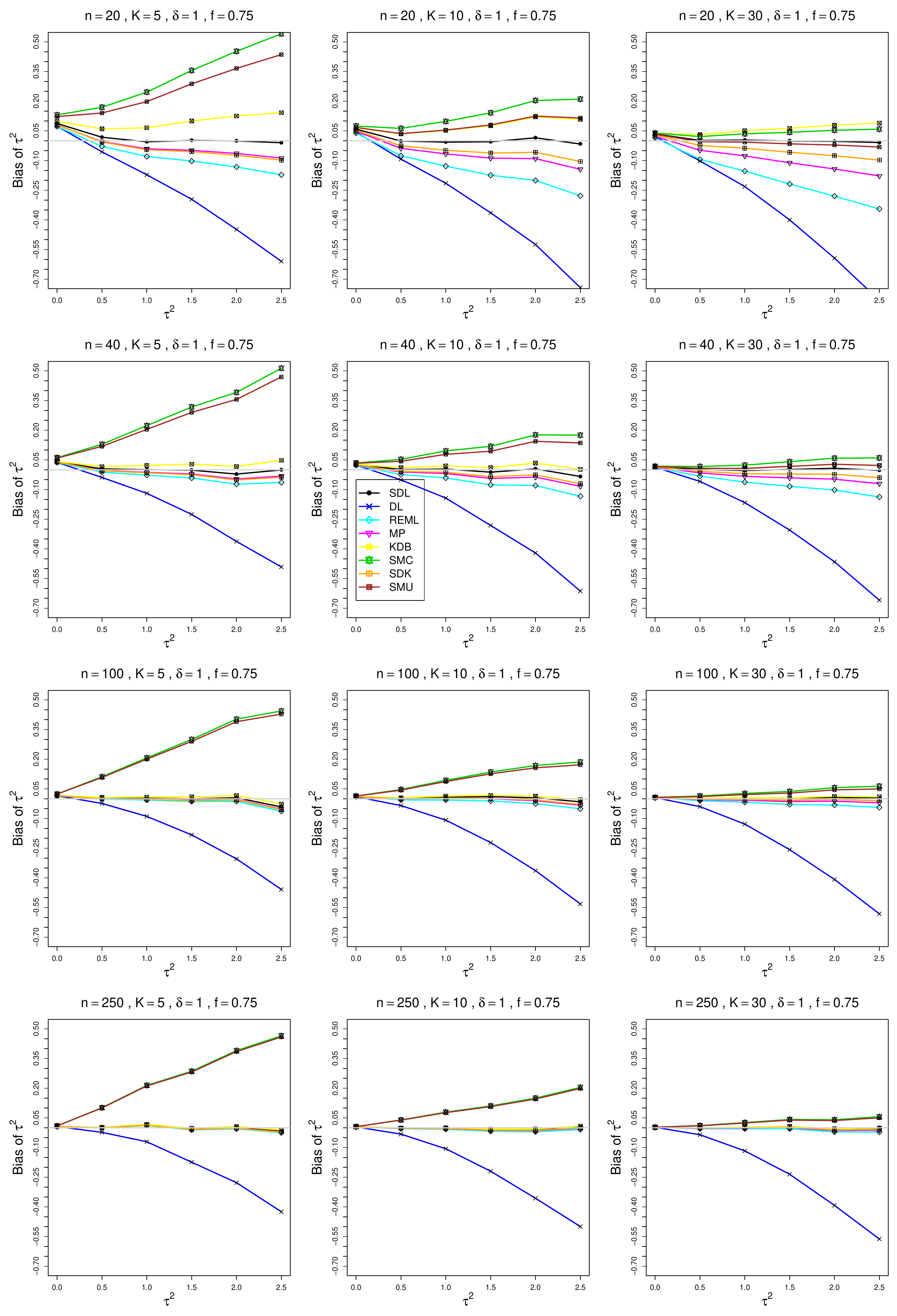}
	\caption{Bias of estimators of the between-studies variance $\tau^2$ for $\delta = 1$, $f = .75$, and equal sample sizes
		\label{PlotBiasOfTau2delta1andq075_SMD} }
\end{figure}

\begin{figure}[t]
	\centering
	\includegraphics[scale=0.33]{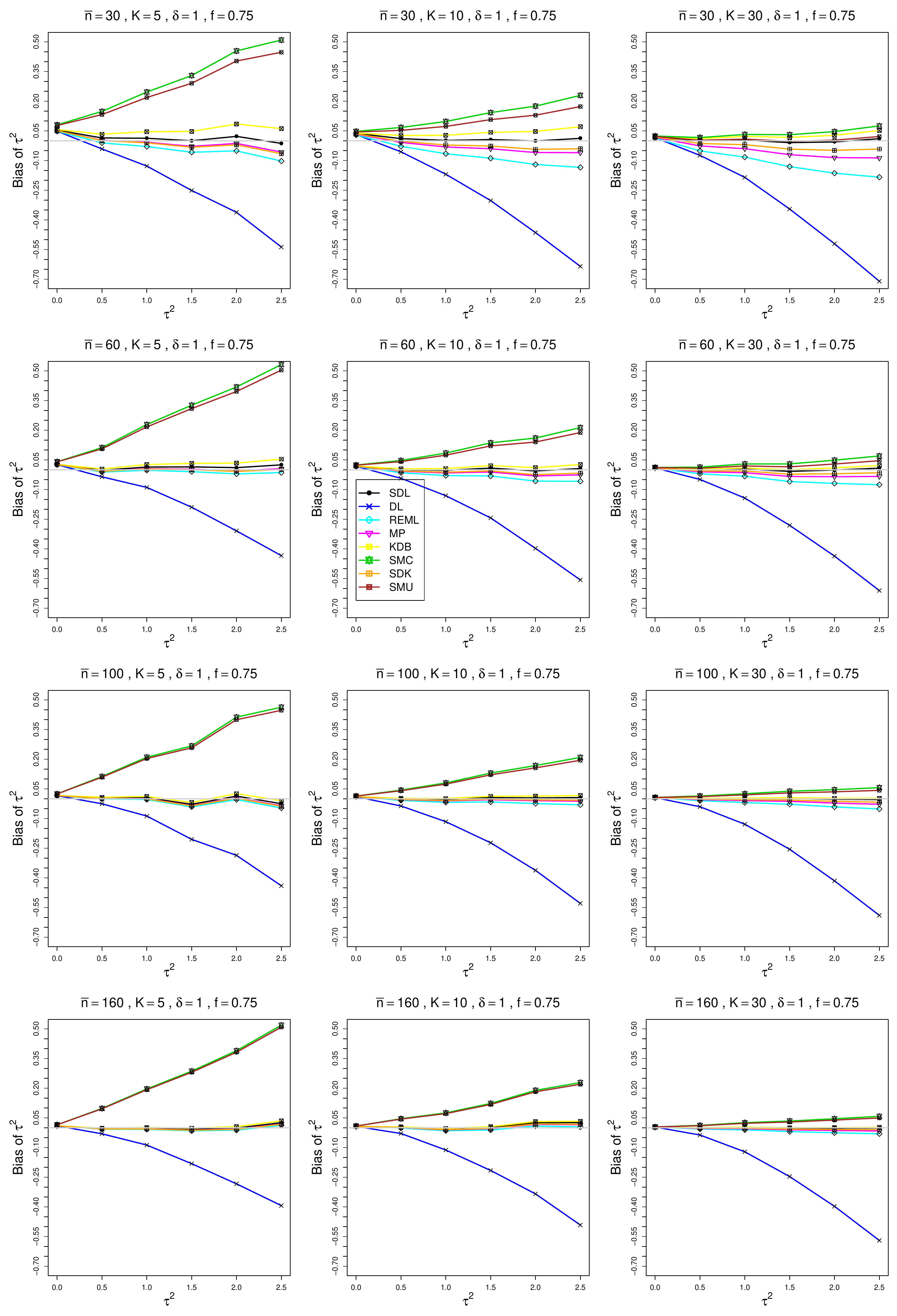}
	\caption{Bias of estimators of the between-studies variance $\tau^2$ for $\delta = 1$, $f = .75$, and unequal sample sizes
		\label{PlotBiasOfTau2delta1andq075_SMD_unequal} }
\end{figure}

\begin{figure}[t]
	\centering
	\includegraphics[scale=0.33]{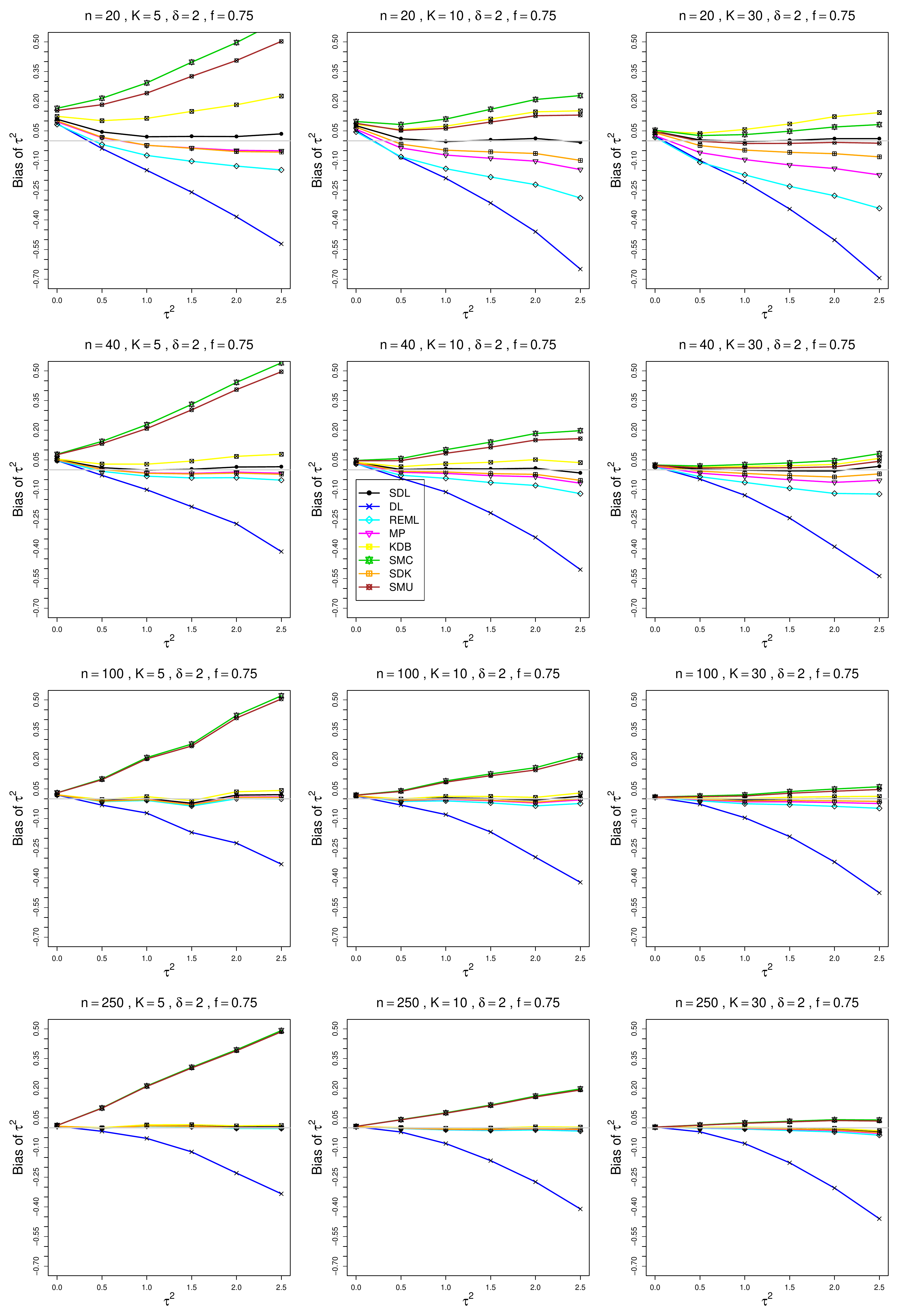}
	\caption{Bias of estimators of the between-studies variance $\tau^2$ for $\delta = 2$, $f = .75$, and equal sample sizes
		\label{PlotBiasOfTau2delta2andq075_SMD} }
\end{figure}

\begin{figure}[t]
	\centering
	\includegraphics[scale=0.33]{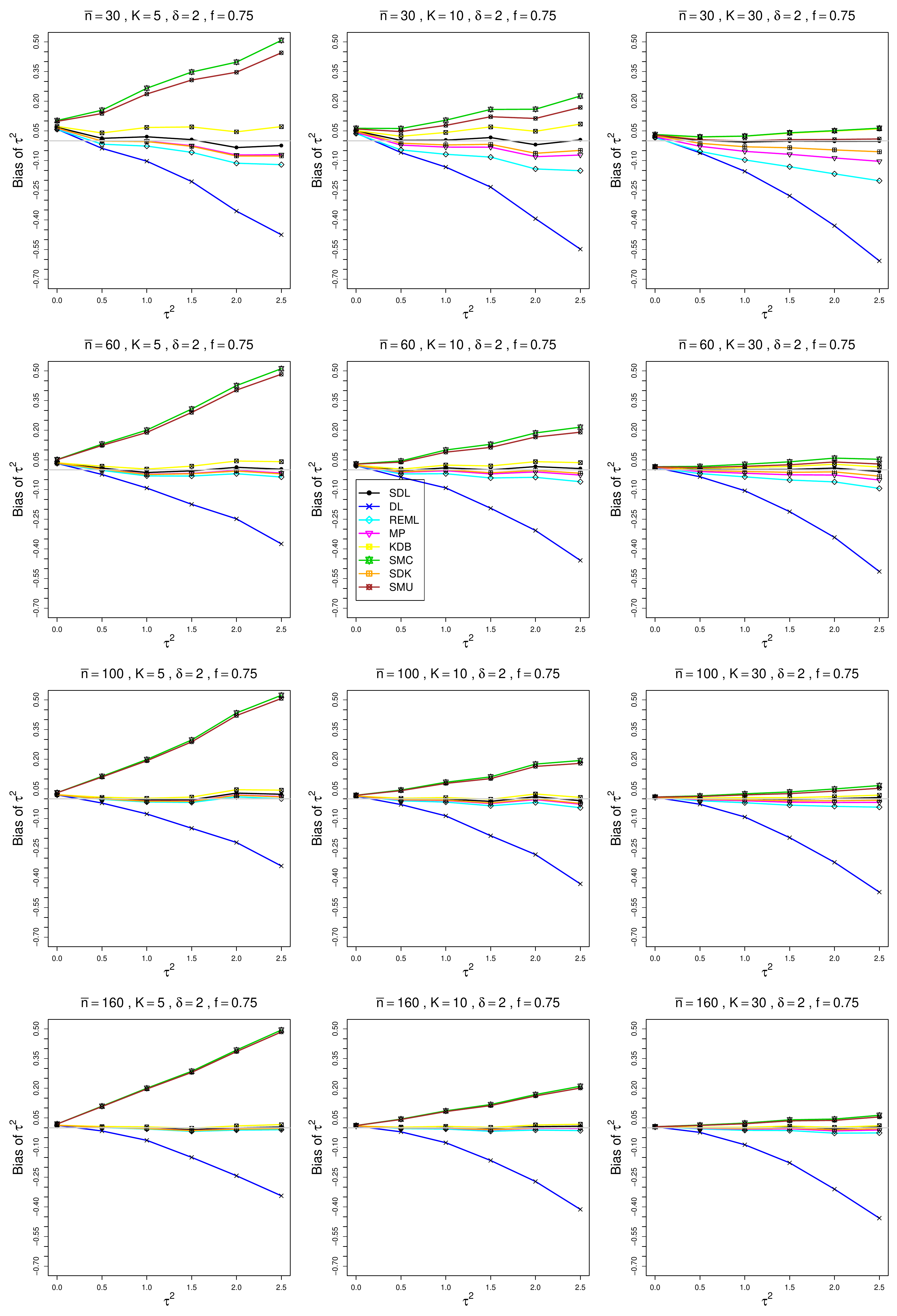}
	\caption{Bias of estimators of the between-studies variance $\tau^2$ for $\delta = 2$, $f = .75$, and unequal sample sizes
		\label{PlotBiasOfTau2delta2andq075_SMD_unequal} }
\end{figure}

\clearpage
\section{Coverage of interval estimators of the between-studies variance ($\tau^2$)}

The interval estimators of $\tau^2$ are
\begin{itemize}
	\item QP (Q profile interval)
	\item PL (Profile likelihood interval)
	\item KDB (inverse-variance method based on corrected first moment of null distribution of $Q$)
	\item FPC (profile based on Farebrother approximation with conditional moments)
	\item FPU (profile based on Farebrother approximation with unconditional moments)
\end{itemize}

\clearpage

\renewcommand{\thefigure}{A2.\arabic{figure}}
\setcounter{figure}{0}

\begin{figure}[t]
	\centering
	\includegraphics[scale=0.33]{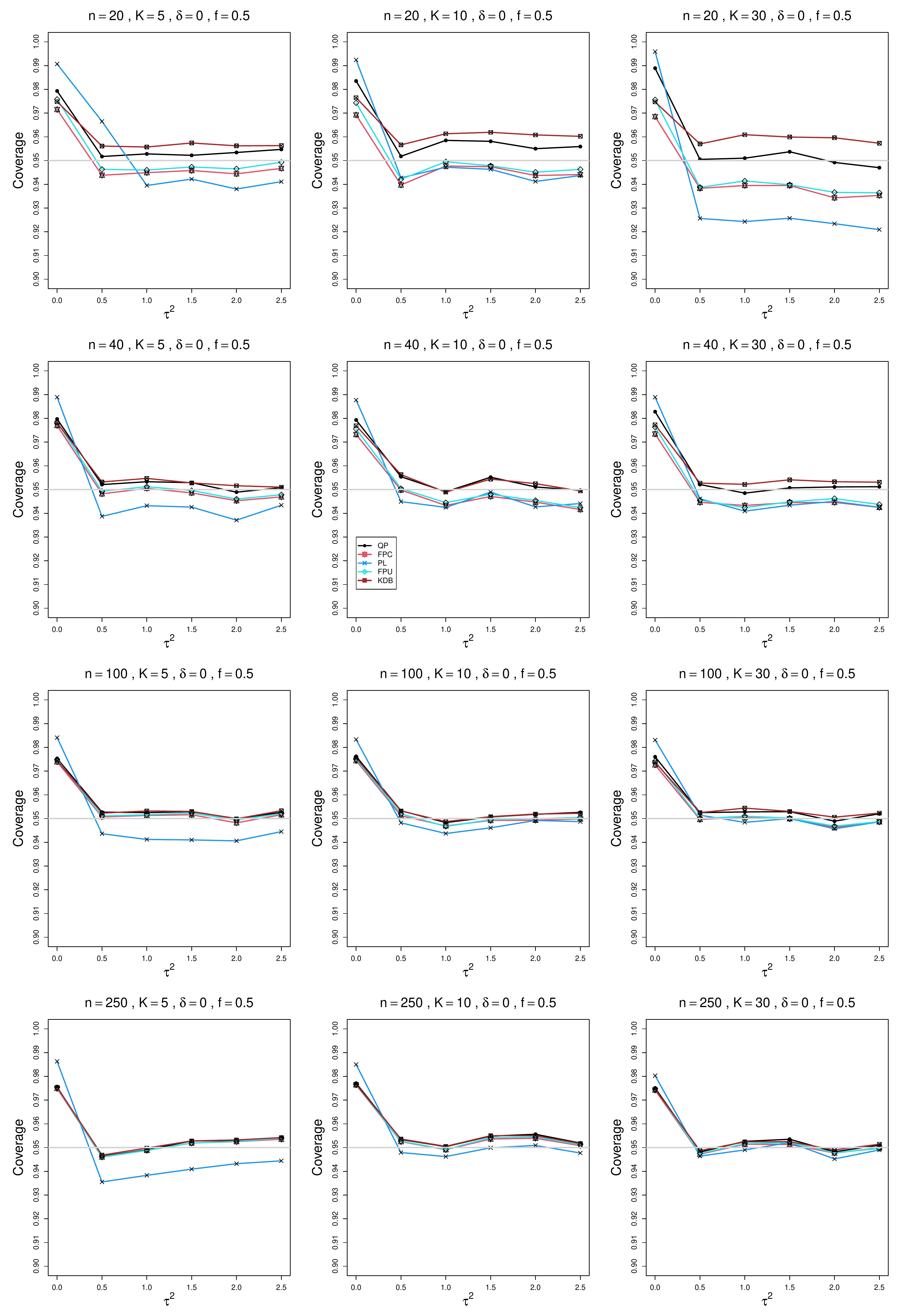}
	\caption{Coverage of $\tau^2$ for $\delta = 0$, $f = .5$, and equal sample sizes
		\label{PlotCovOfTau2delta0andq05_SMD}}
\end{figure}

\begin{figure}[t]
	\centering
	\includegraphics[scale=0.33]{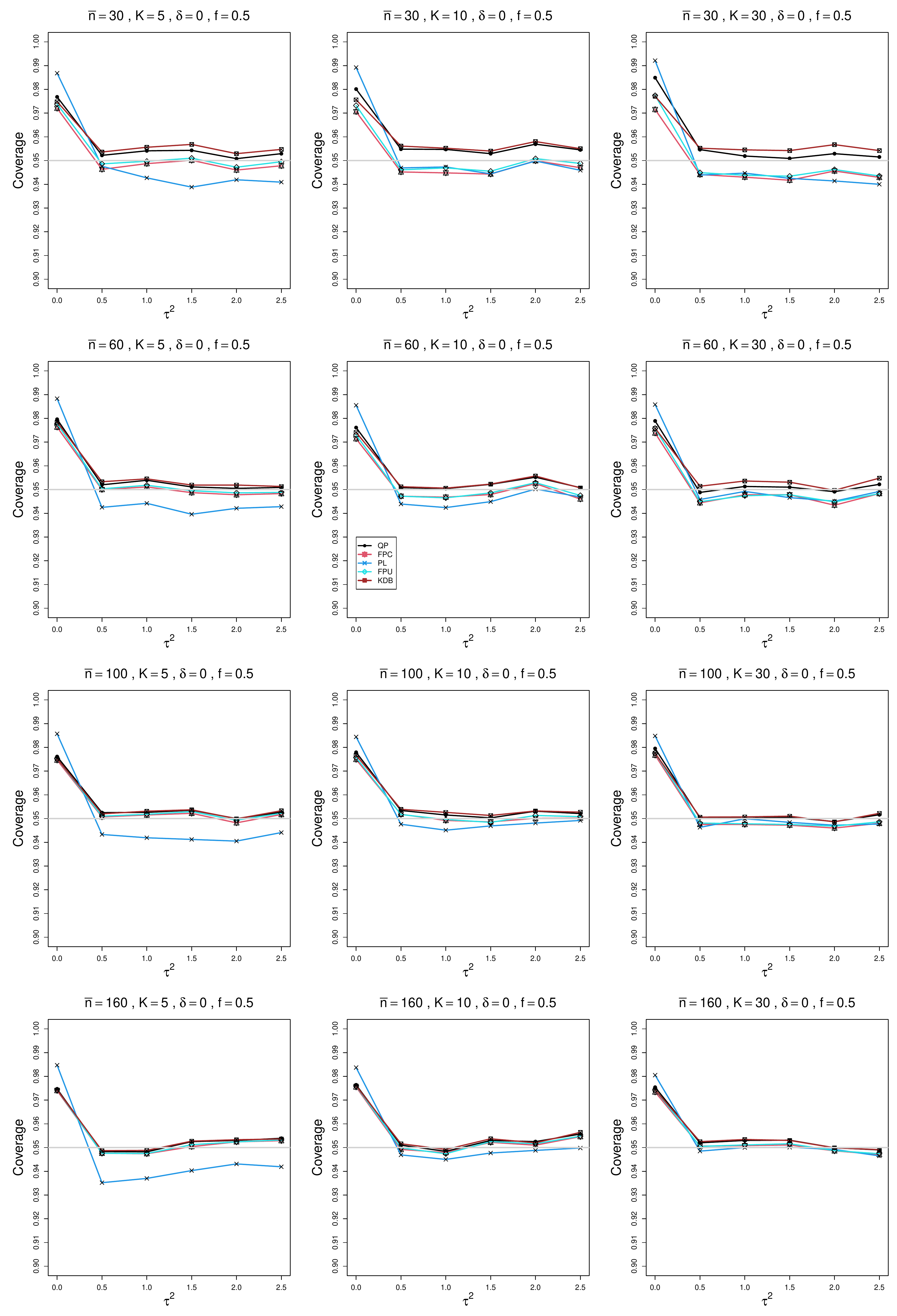}
	\caption{Coverage of $\tau^2$ for $\delta = 0$, $f = .5$, and unequal sample sizes
		\label{PlotCovOfTau2delta0andq05_SMD_unequal}}
\end{figure}

\begin{figure}[t]
	\centering
	\includegraphics[scale=0.33]{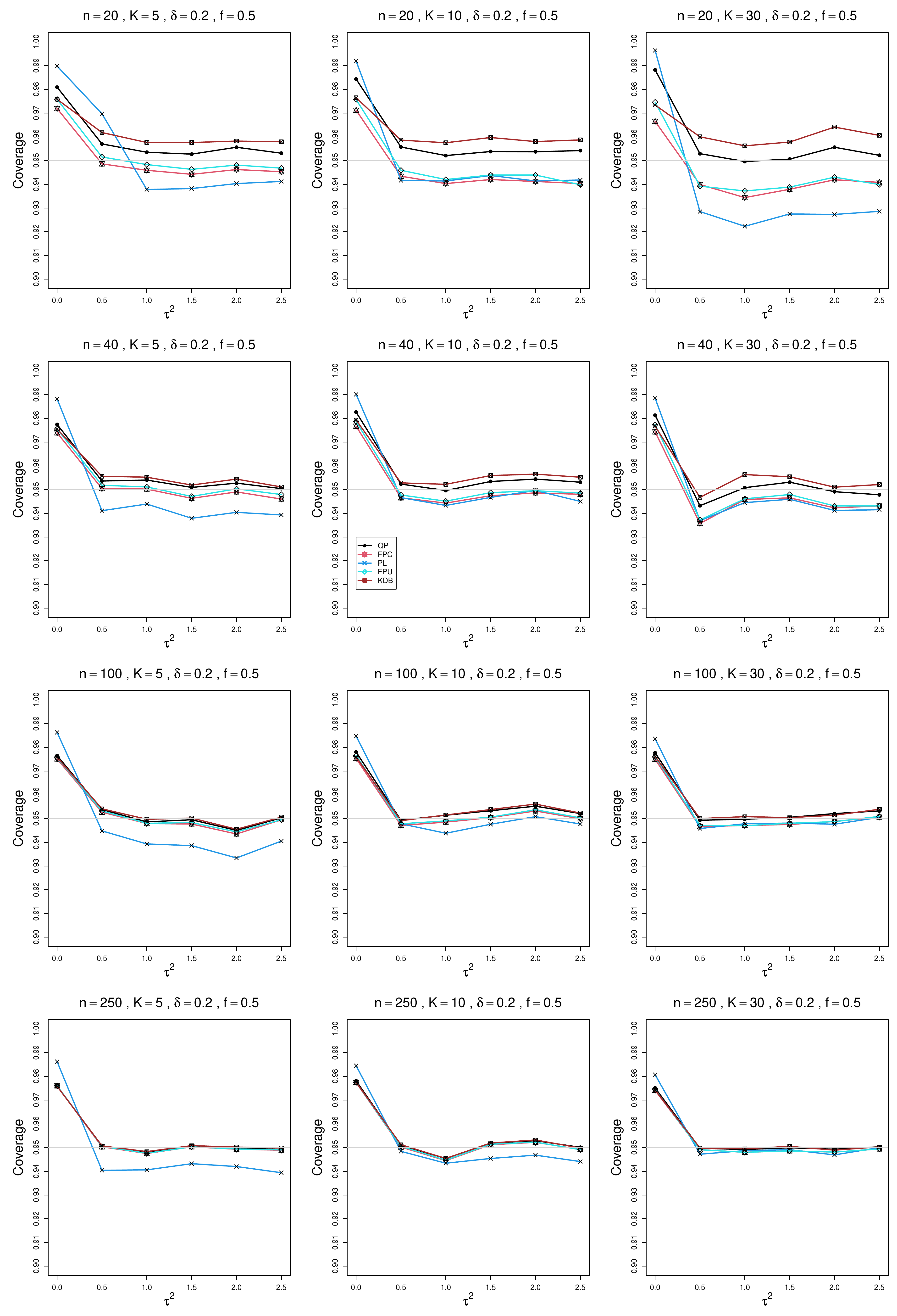}
	\caption{Coverage of $\tau^2$ for $\delta = 0.2$, $f = .5$, and equal sample sizes
		\label{PlotCovOfTau2delta02andq05_SMD}}
\end{figure}

\begin{figure}[t]
	\centering
	\includegraphics[scale=0.33]{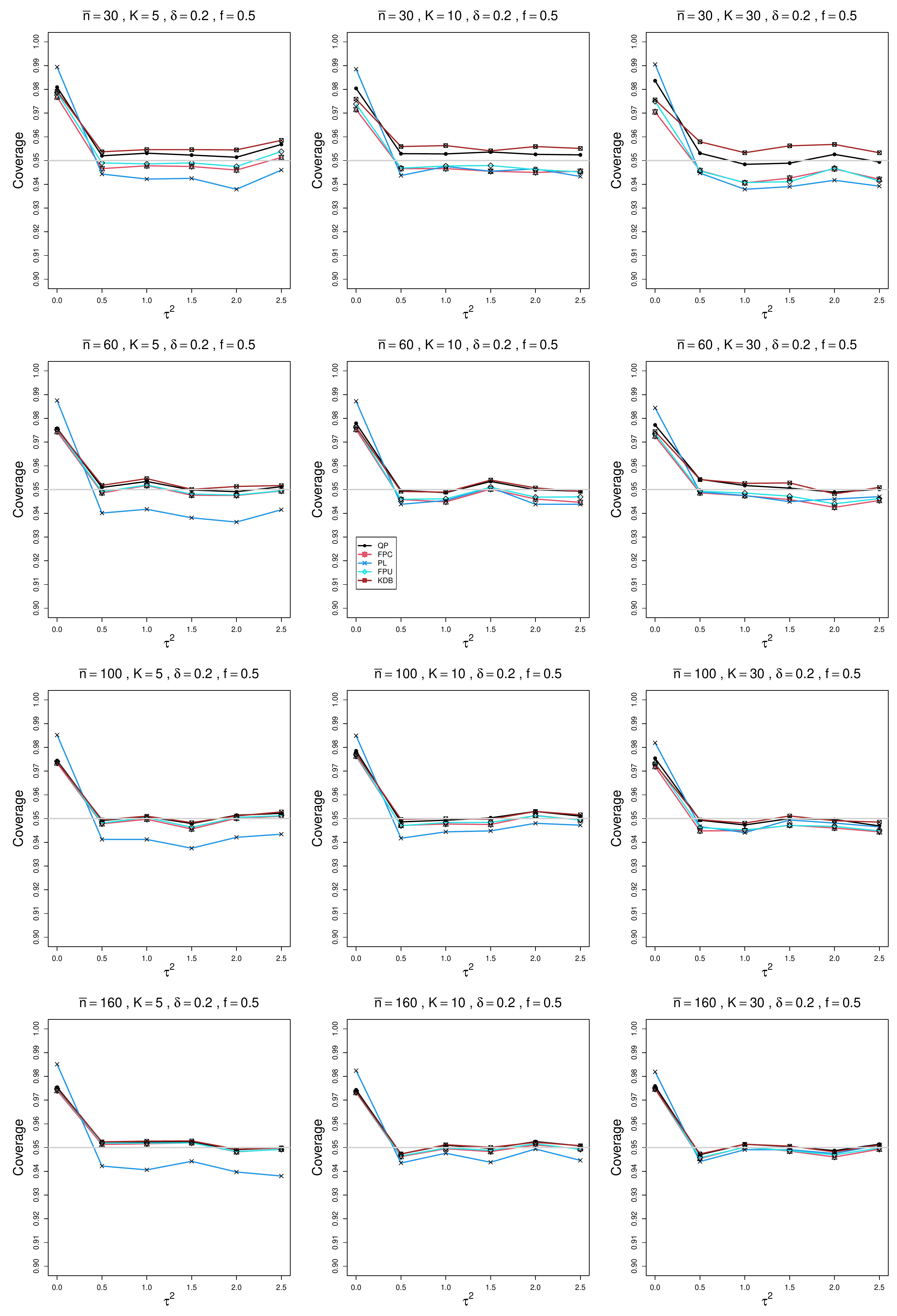}
	\caption{Coverage of $\tau^2$ for $\delta = 0.2$, $f = .5$, and unequal sample sizes
		\label{PlotCovOfTau2delta02andq05_SMD_unequal}}
\end{figure}

\begin{figure}[t]
	\centering
	\includegraphics[scale=0.33]{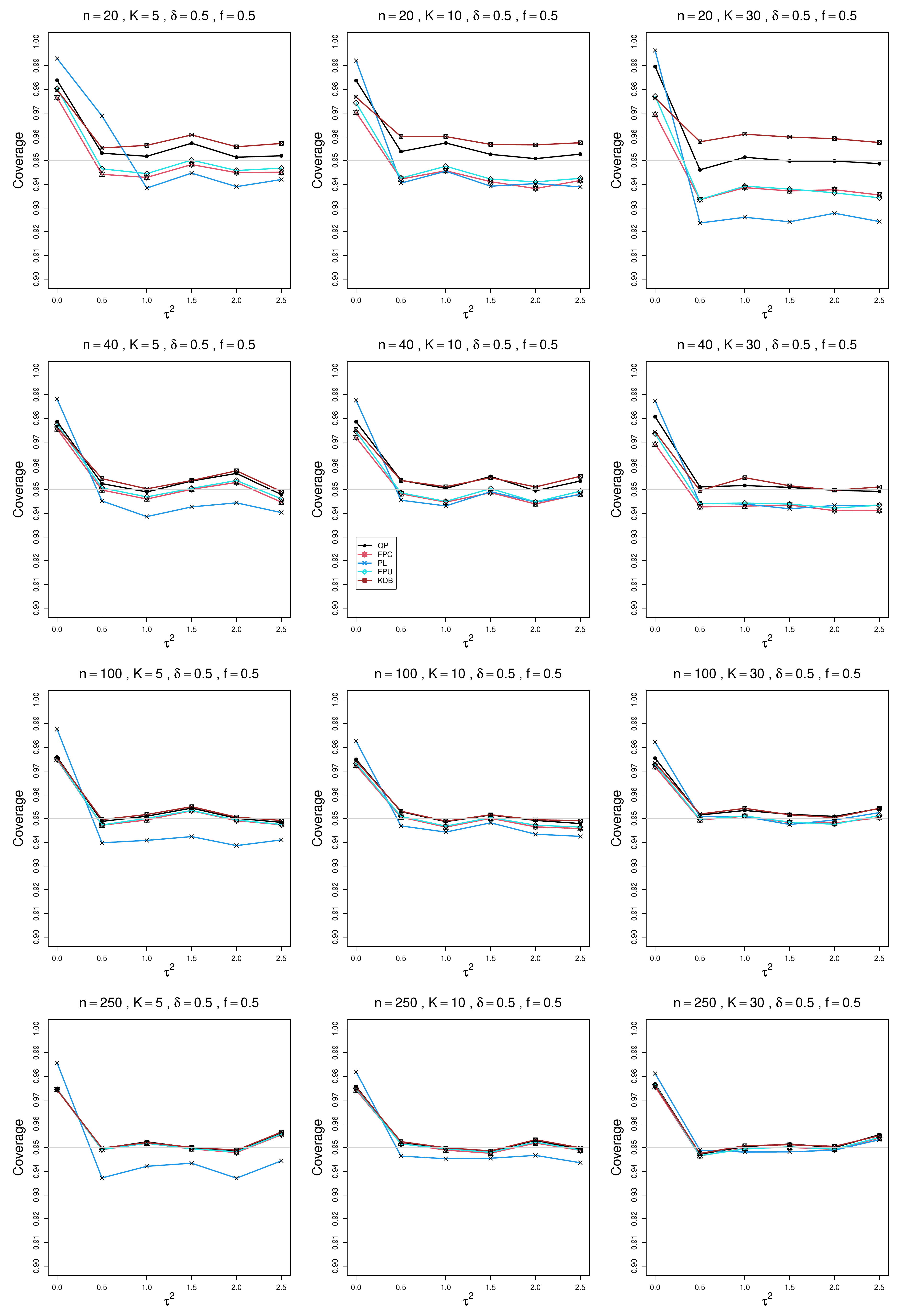}
	\caption{Coverage of $\tau^2$ for $\delta = 0.5$, $f = .5$, and equal sample sizes
		\label{PlotCovOfTau2delta05andq05_SMD}}
\end{figure}

\begin{figure}[t]
	\centering
	\includegraphics[scale=0.33]{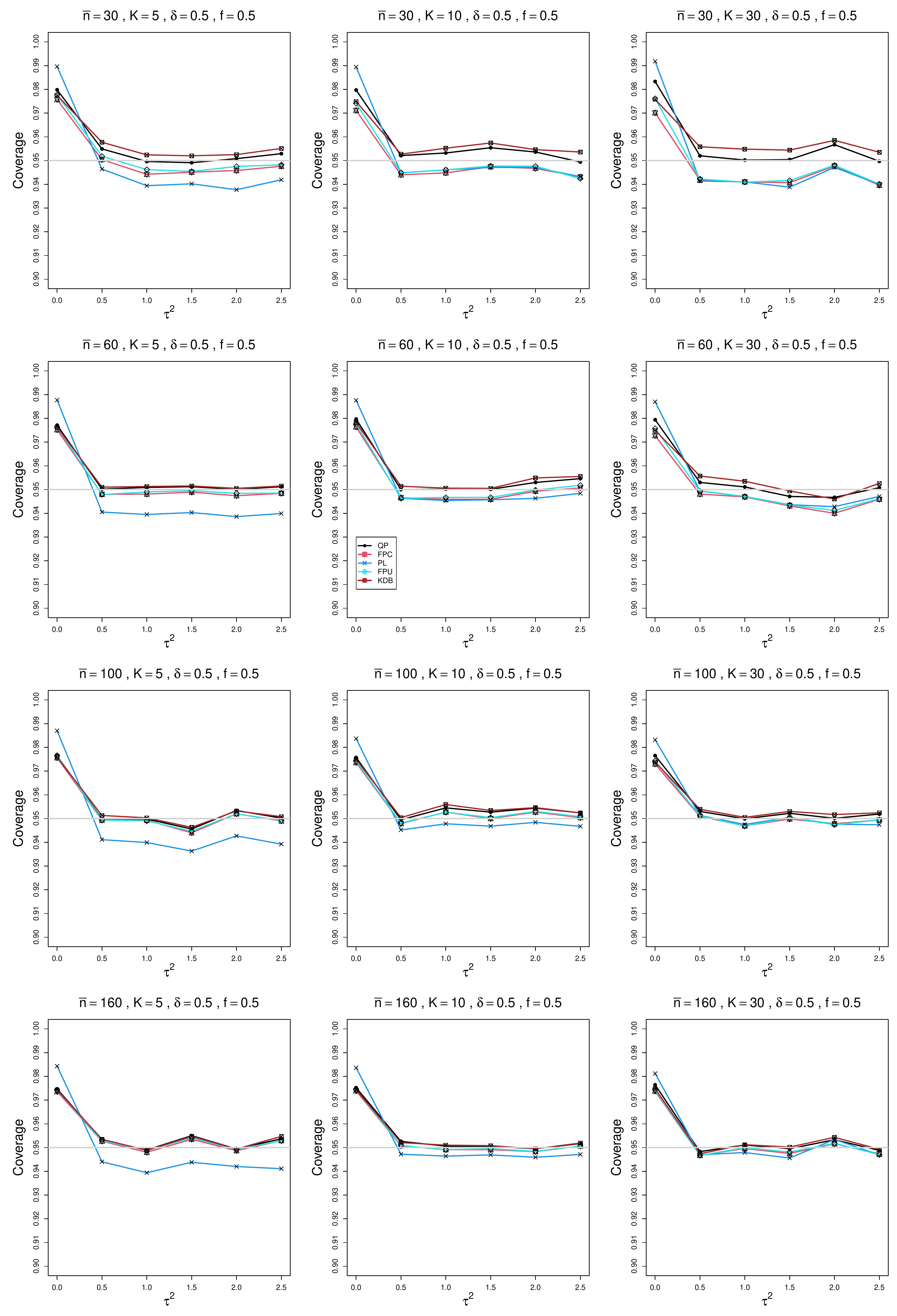}
	\caption{Coverage of $\tau^2$ for $\delta = 0.5$, $f = .5$, and unequal sample sizes
		\label{PlotCovOfTau2delta05andq05_SMD_unequal}}
\end{figure}

\begin{figure}[t]
	\centering
	\includegraphics[scale=0.33]{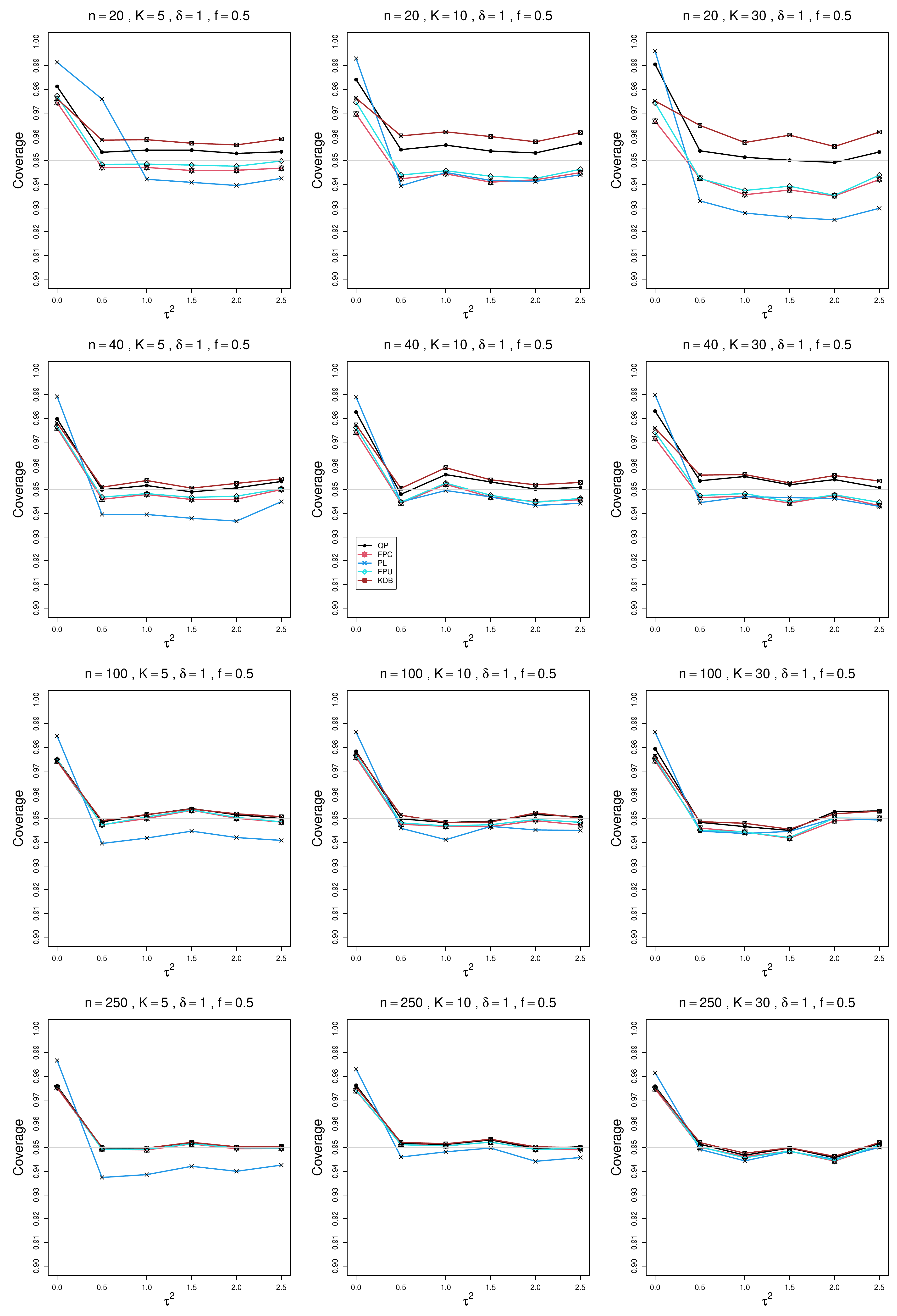}
	\caption{Coverage of $\tau^2$ for $\delta = 1$, $f = .5$, and equal sample sizes
		\label{PlotCovOfTau2delta1andq05_SMD}}
\end{figure}

\begin{figure}[t]
	\centering
	\includegraphics[scale=0.33]{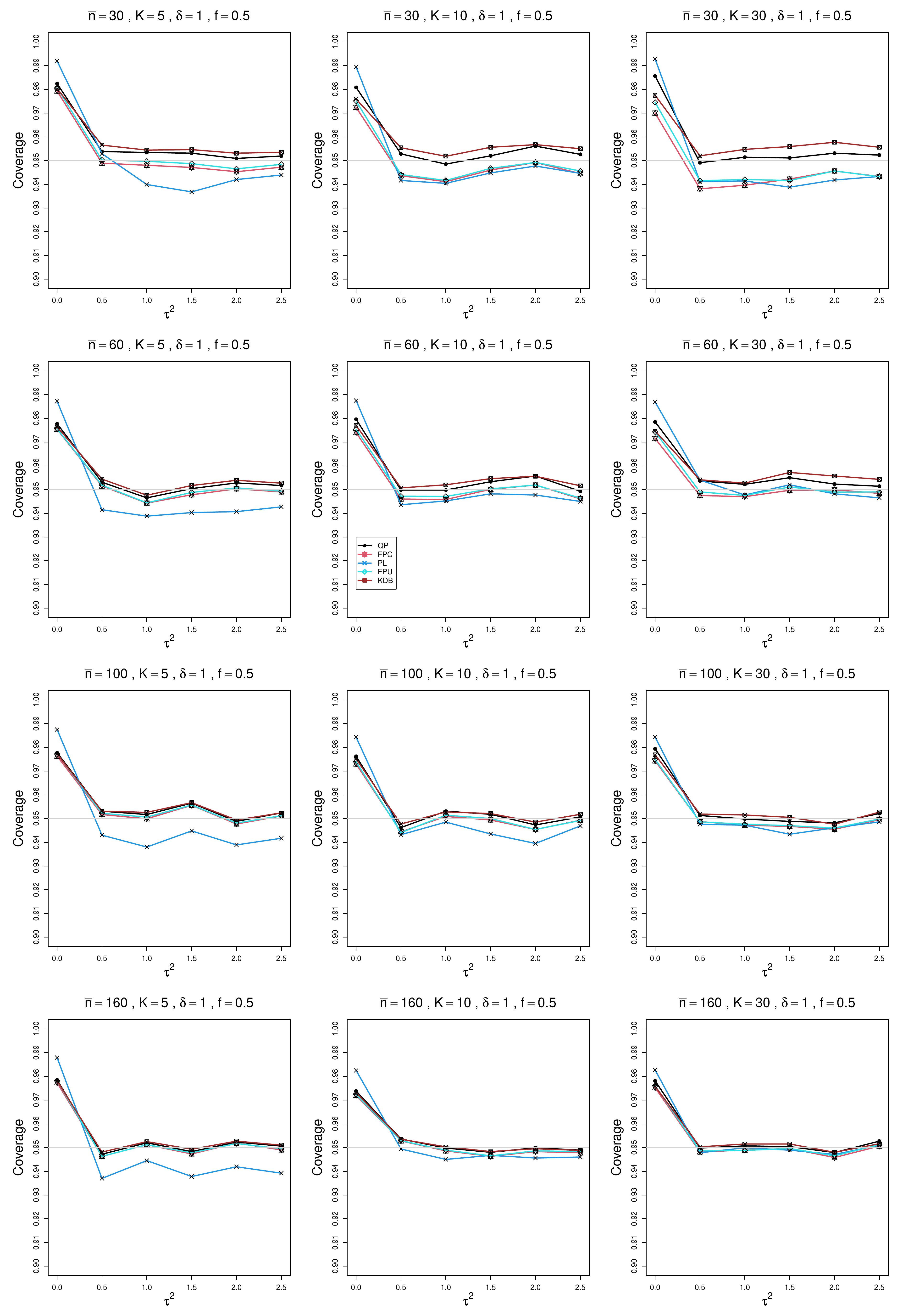}
	\caption{Coverage of $\tau^2$ for $\delta = 1$, $f = .5$, and unequal sample sizes
		\label{PlotCovOfTau2delta1andq05_SMD_unequal}}
\end{figure}

\begin{figure}[t]
	\centering
	\includegraphics[scale=0.33]{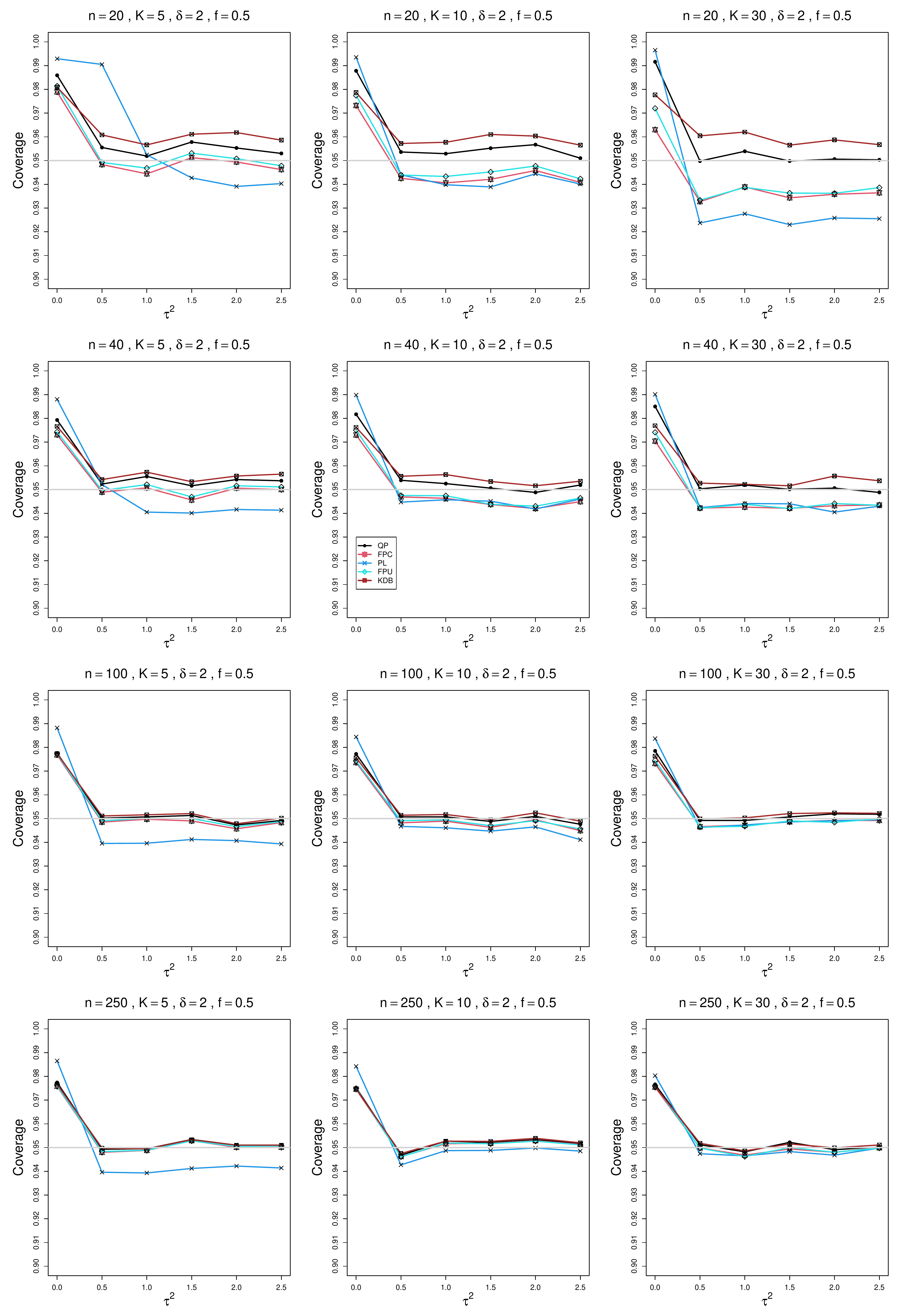}
	\caption{Coverage of $\tau^2$ for $\delta = 2$, $f = .5$, and equal sample sizes
		\label{PlotCovOfTau2delta2andq05_SMD}}
\end{figure}

\begin{figure}[t]
	\centering
	\includegraphics[scale=0.33]{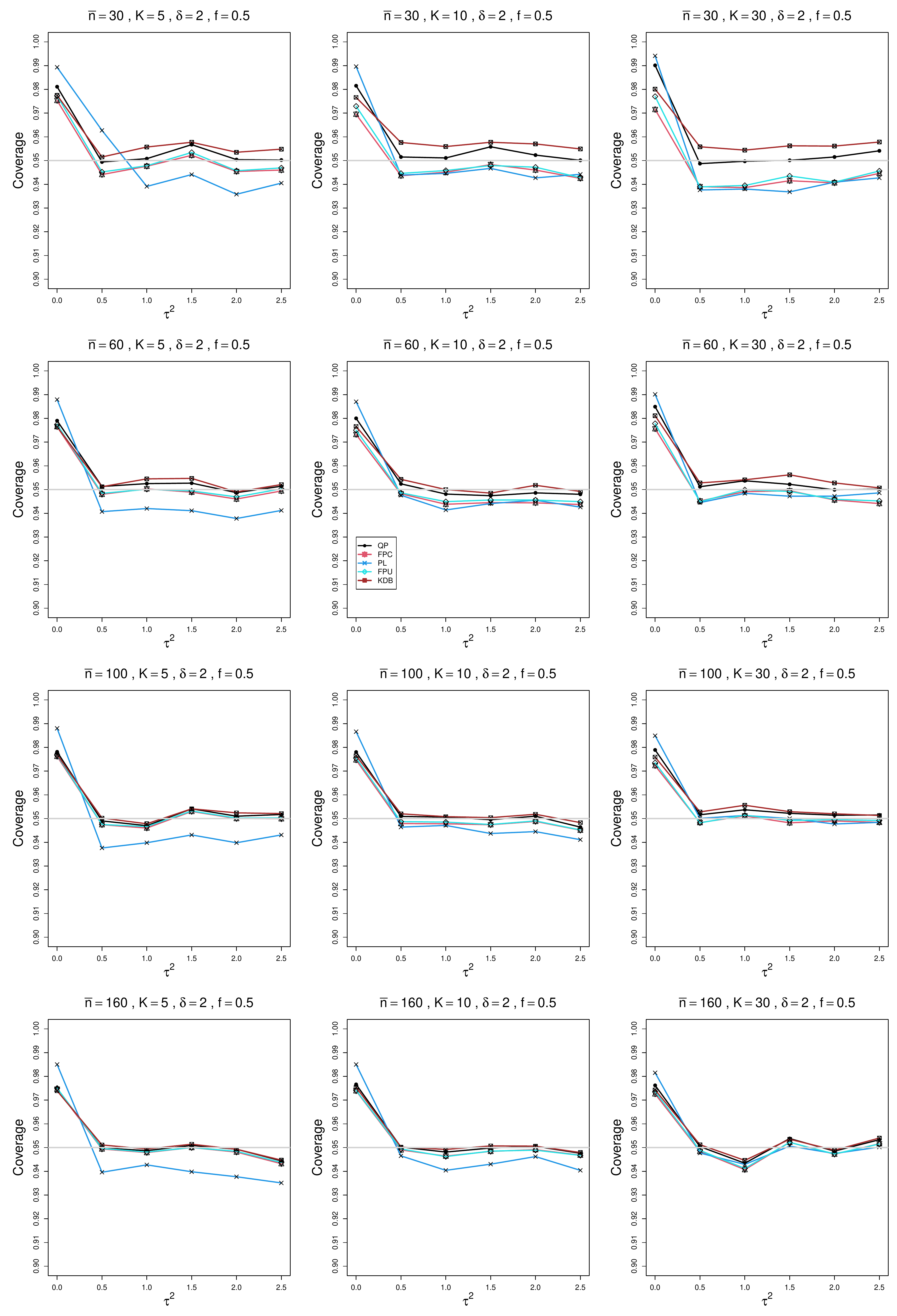}
	\caption{Coverage of $\tau^2$ for $\delta = 2$, $f = .5$, and unequal sample sizes
		\label{PlotCovOfTau2delta2andq05_SMD_unequal}}
\end{figure}

%%%%%%%%%%%%%%%%%%%%%%%%%%%%%%%%%%%%%%%%%%%%%%%%%%%%%%%%%%%%%%%%%%%%%%%%%%%%%%%%%%%%%%%%%%%%%%%%%%%%%%%%%%%%%%%%%%%%%%%%%%%%%%
%q=0.75

\begin{figure}[t]
	\centering
	\includegraphics[scale=0.33]{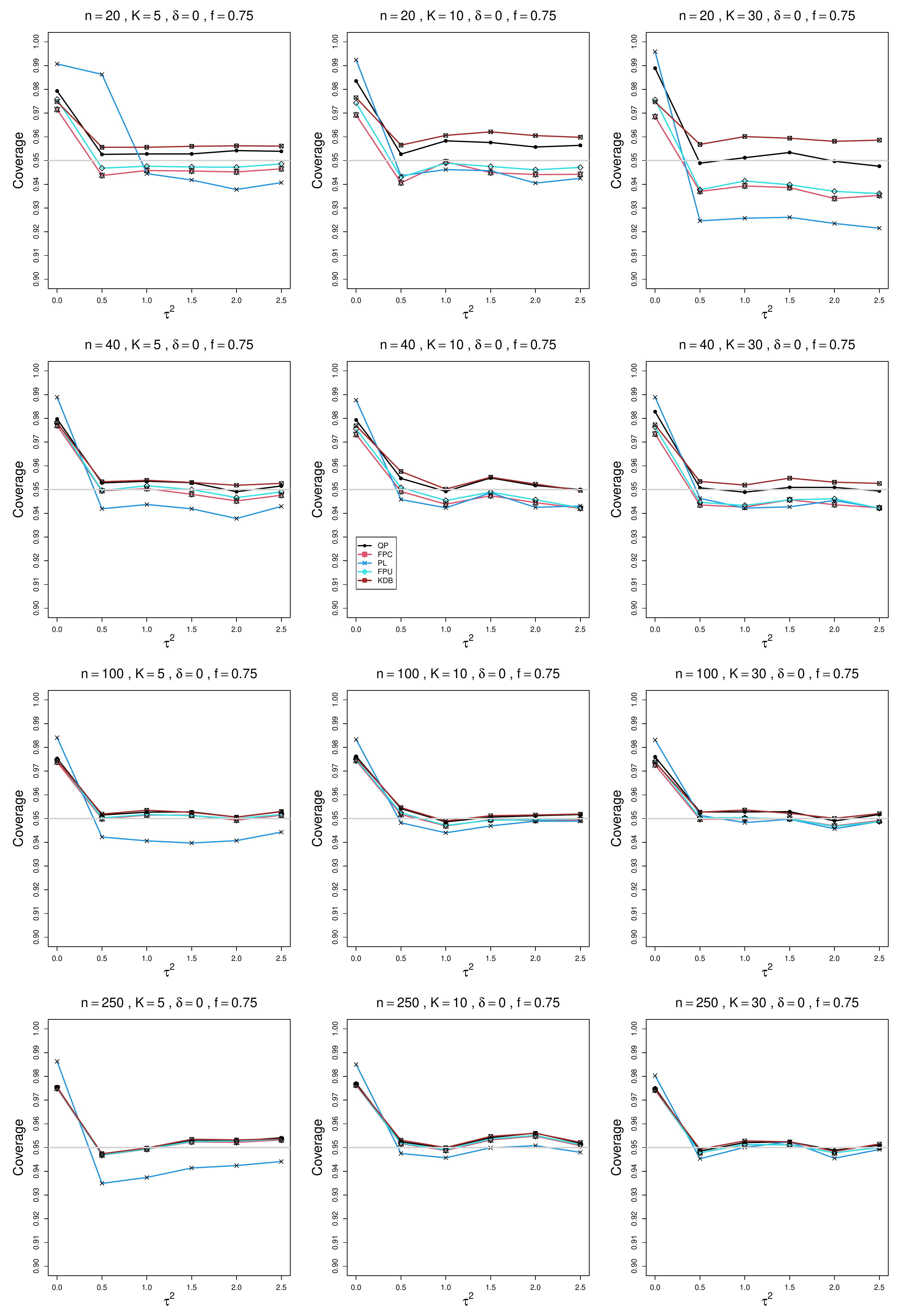}
	\caption{Coverage of $\tau^2$ for $\delta = 0$, $f = .75$, and equal sample sizes
		\label{PlotCovOfTau2delta0andq075_SMD}}
\end{figure}

\begin{figure}[t]
	\centering
	\includegraphics[scale=0.33]{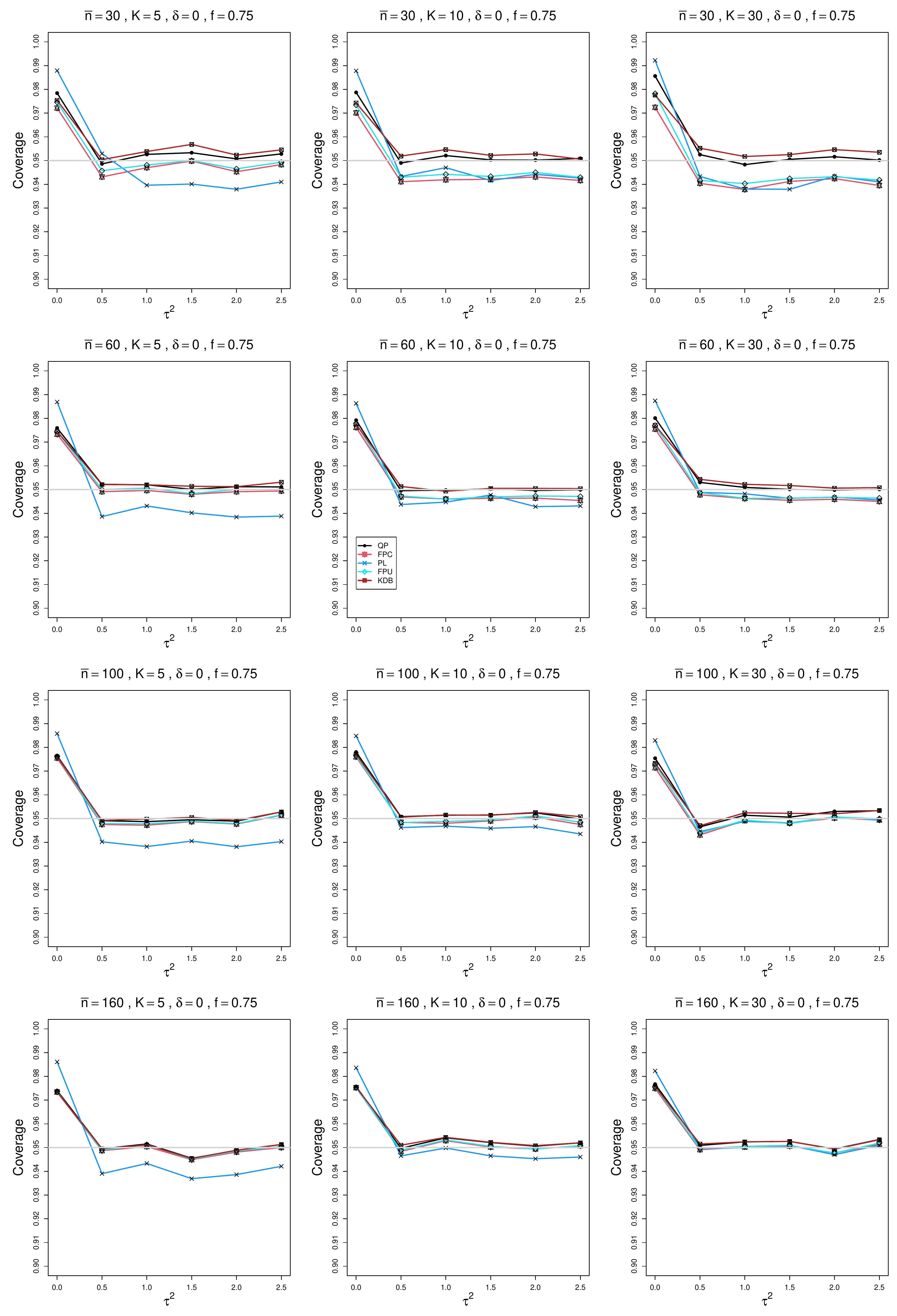}
	\caption{Coverage of $\tau^2$ for $\delta = 0$, $f = .75$, and unequal sample sizes
		\label{PlotCovOfTau2delta0andq075_SMD_unequal}}
\end{figure}

\begin{figure}[t]
	\centering
	\includegraphics[scale=0.33]{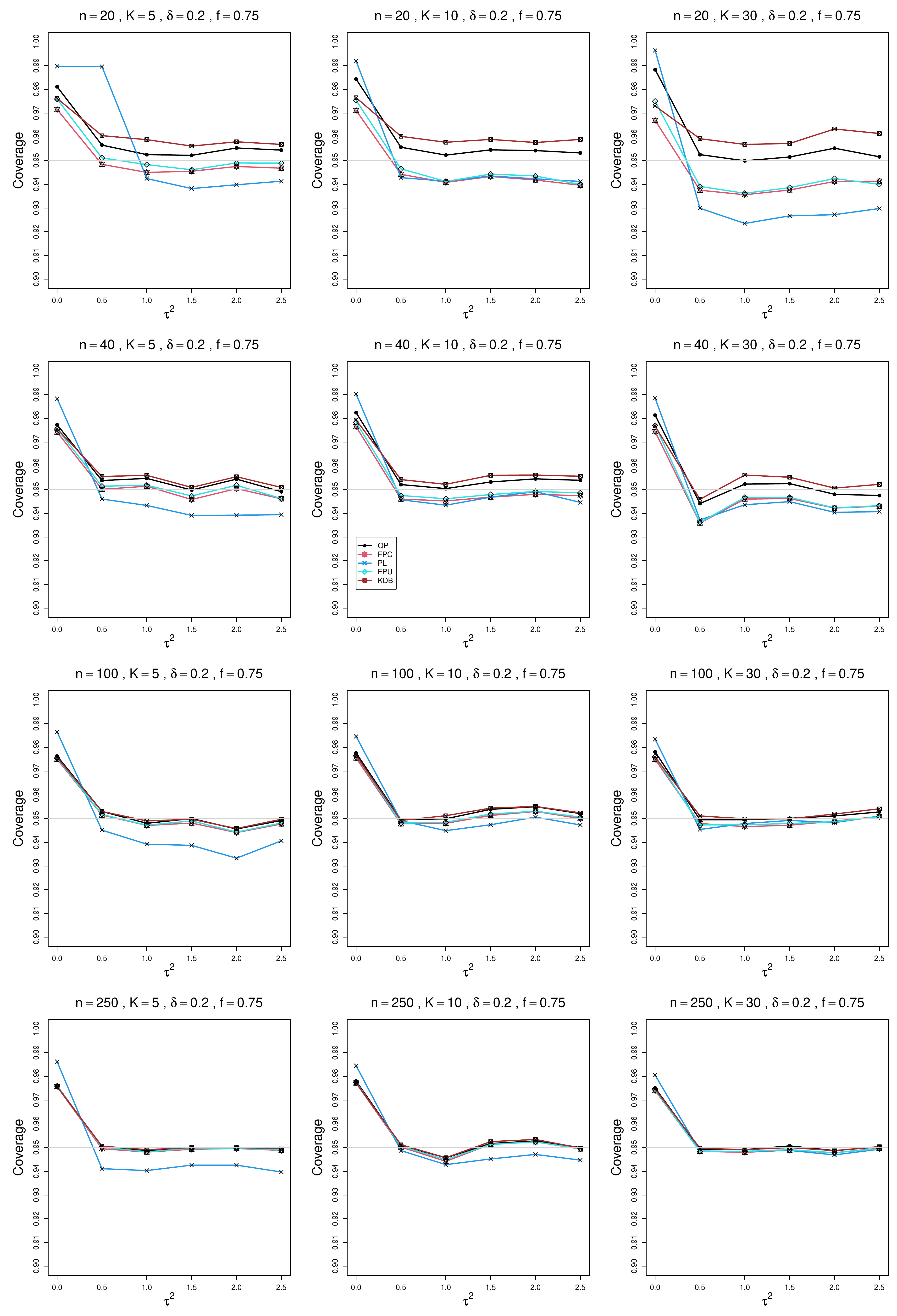}
	\caption{Coverage of $\tau^2$ for $\delta = 0.2$, $f = .75$, and equal sample sizes
		\label{PlotCovOfTau2delta02andq075_SMD}}
\end{figure}

\begin{figure}[t]
	\centering
	\includegraphics[scale=0.33]{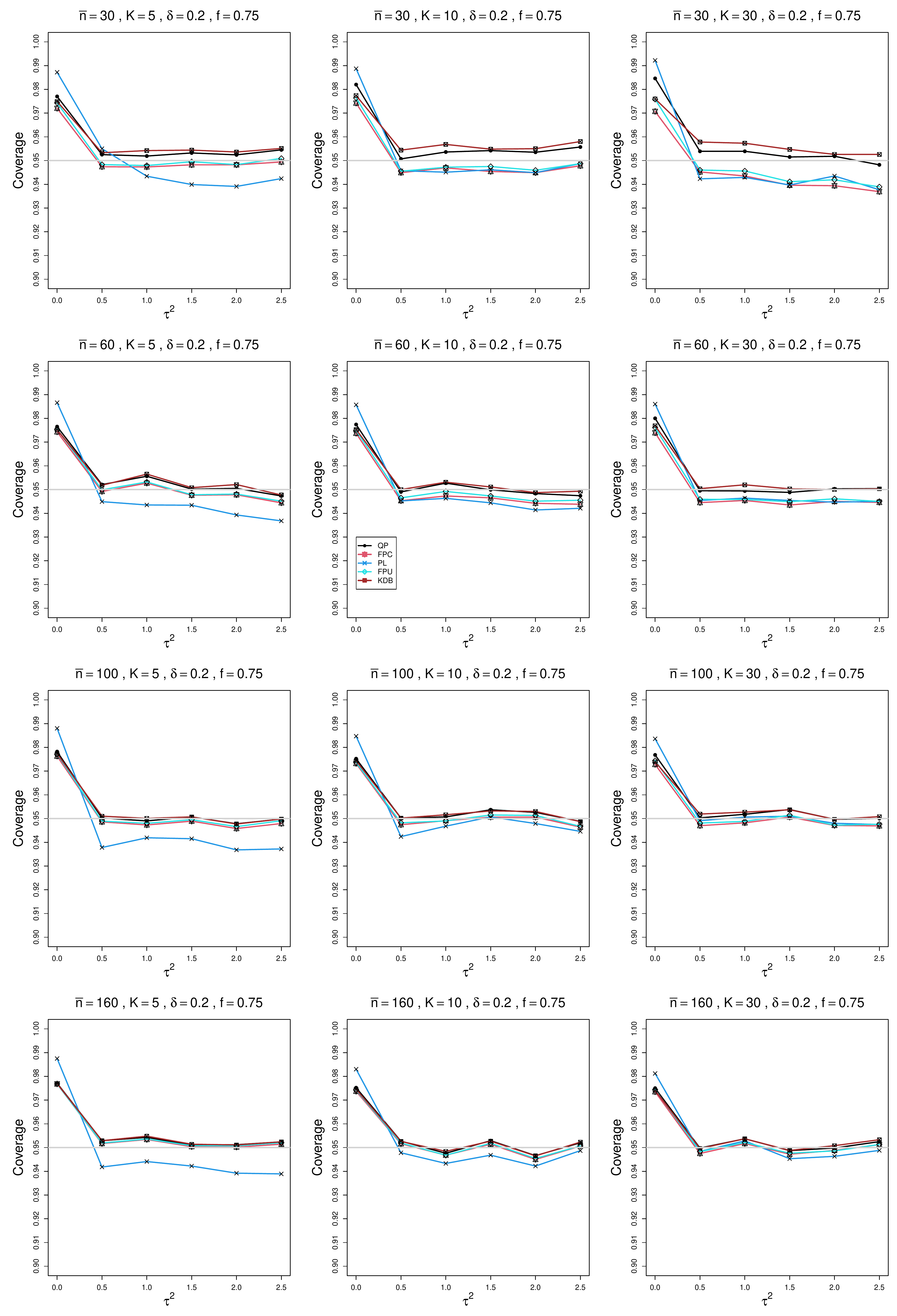}
	\caption{Coverage of $\tau^2$ for $\delta = 0.2$, $f = .75$, and unequal sample sizes
		\label{PlotCovOfTau2delta02andq075_SMD_unequal}}
\end{figure}

\begin{figure}[t]
	\centering
	\includegraphics[scale=0.33]{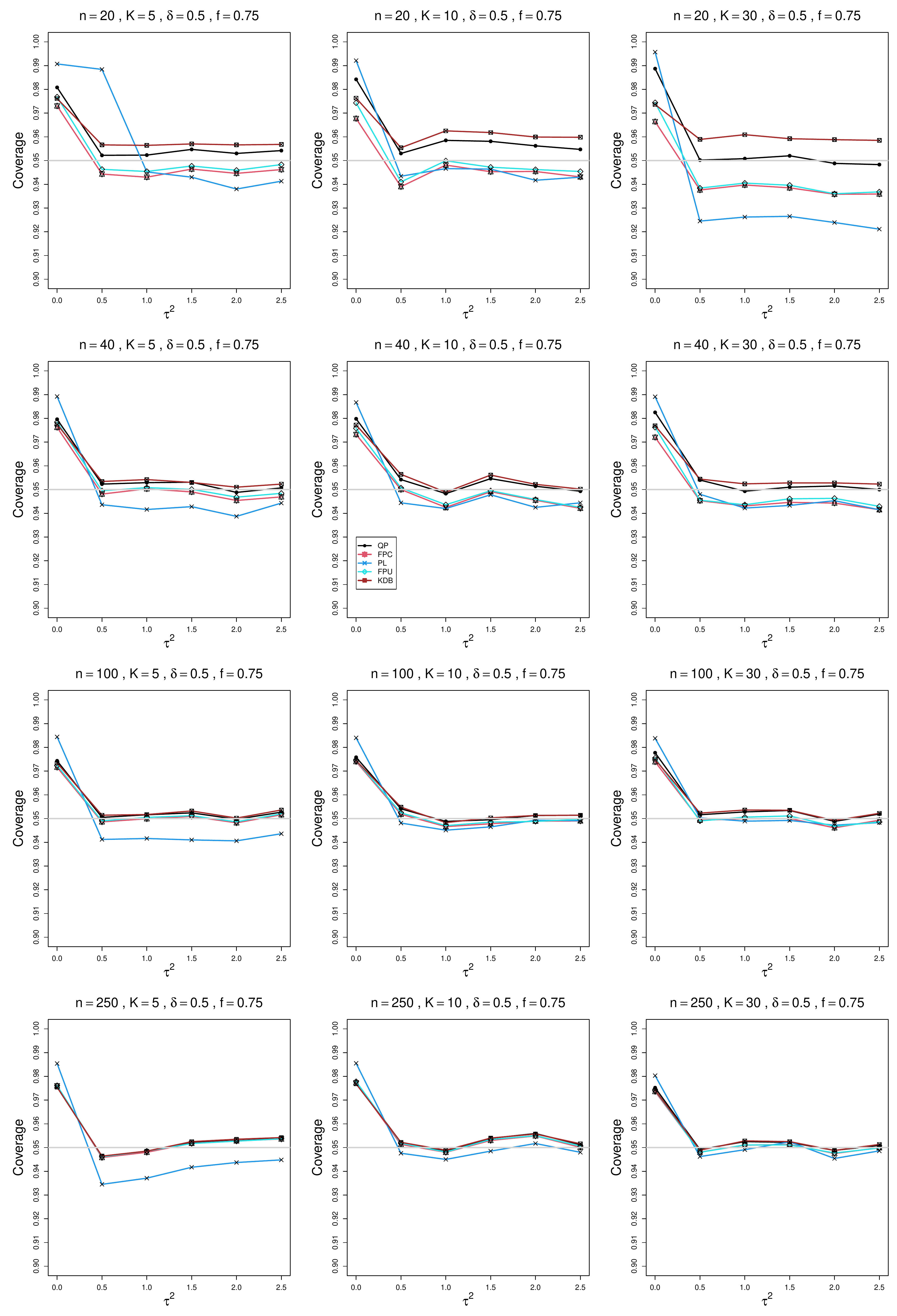}
	\caption{Coverage of $\tau^2$ for $\delta = 0.5$, $f = .75$, and equal sample sizes
		\label{PlotCovOfTau2delta05andq075_SMD}}
\end{figure}

\begin{figure}[t]
	\centering
	\includegraphics[scale=0.33]{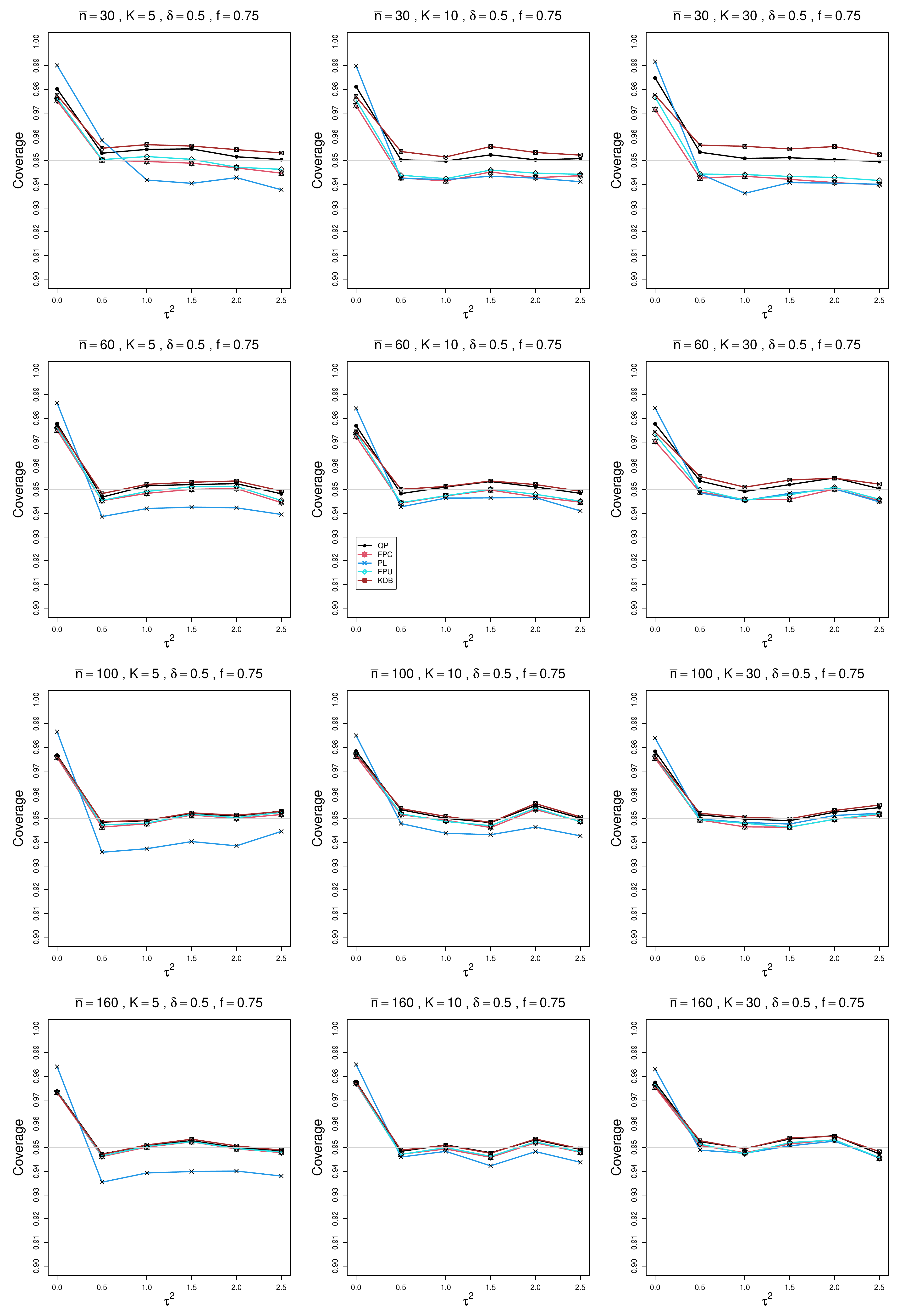}
	\caption{Coverage of $\tau^2$ for $\delta = 0.5$, $f = .75$, and unequal sample sizes
		\label{PlotCovOfTau2delta05andq075_SMD_unequal}}
\end{figure}

\begin{figure}[t]
	\centering
	\includegraphics[scale=0.33]{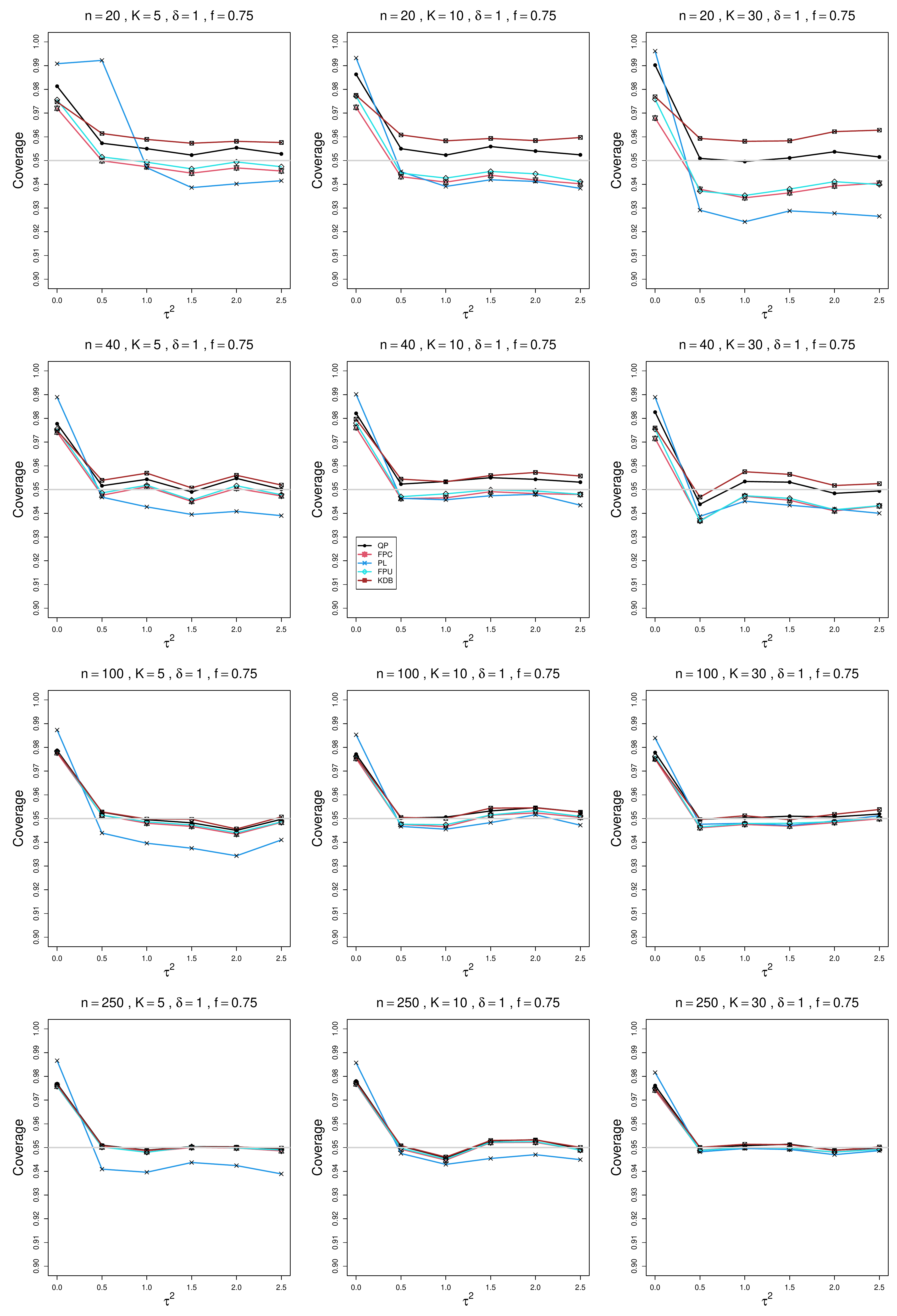}
	\caption{Coverage of $\tau^2$ for $\delta = 1$, $f = .75$, and equal sample sizes
		\label{PlotCovOfTau2delta1andq075_SMD}}
\end{figure}

\begin{figure}[t]
	\centering
	\includegraphics[scale=0.33]{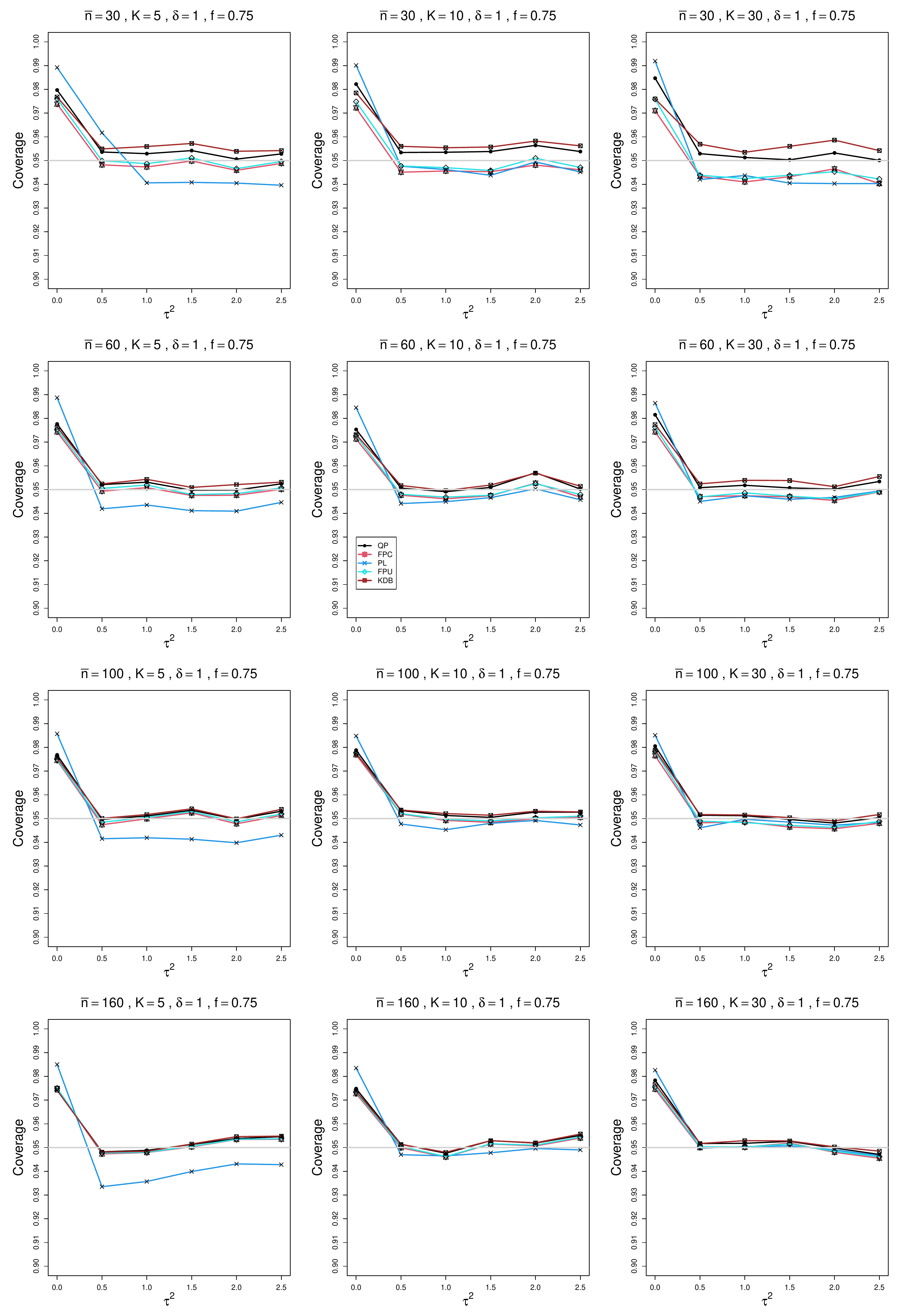}
	\caption{Coverage of $\tau^2$ for $\delta = 1$, $f = .75$, and unequal sample sizes
		\label{PlotCovOfTau2delta1andq075_SMD_unequal}}
\end{figure}

\begin{figure}[t]
	\centering
	\includegraphics[scale=0.33]{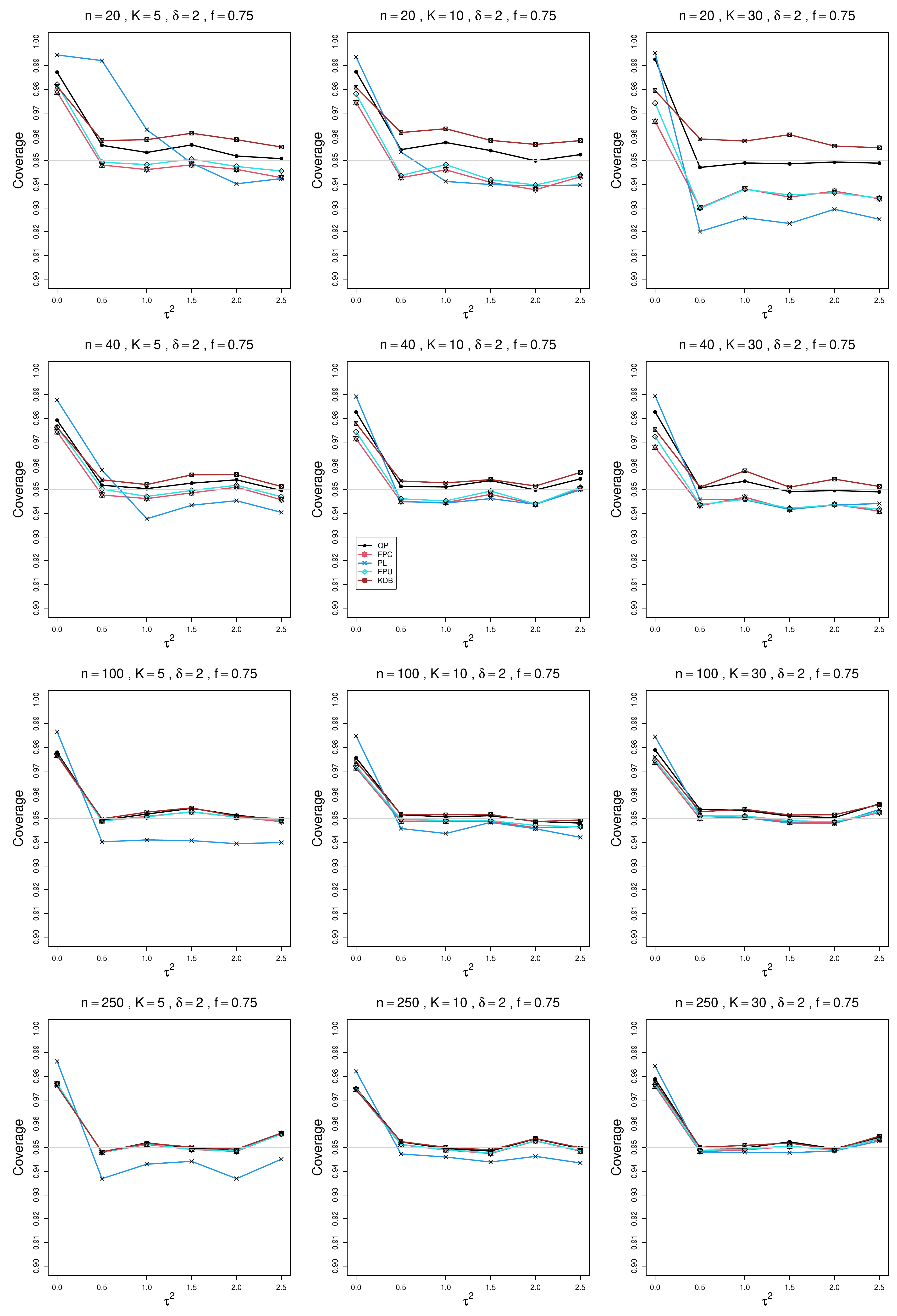}
	\caption{Coverage of $\tau^2$ for $\delta = 2$, $f = .75$, and equal sample sizes
		\label{PlotCovOfTau2delta2andq075_SMD}}
\end{figure}

\begin{figure}[t]
	\centering
	\includegraphics[scale=0.33]{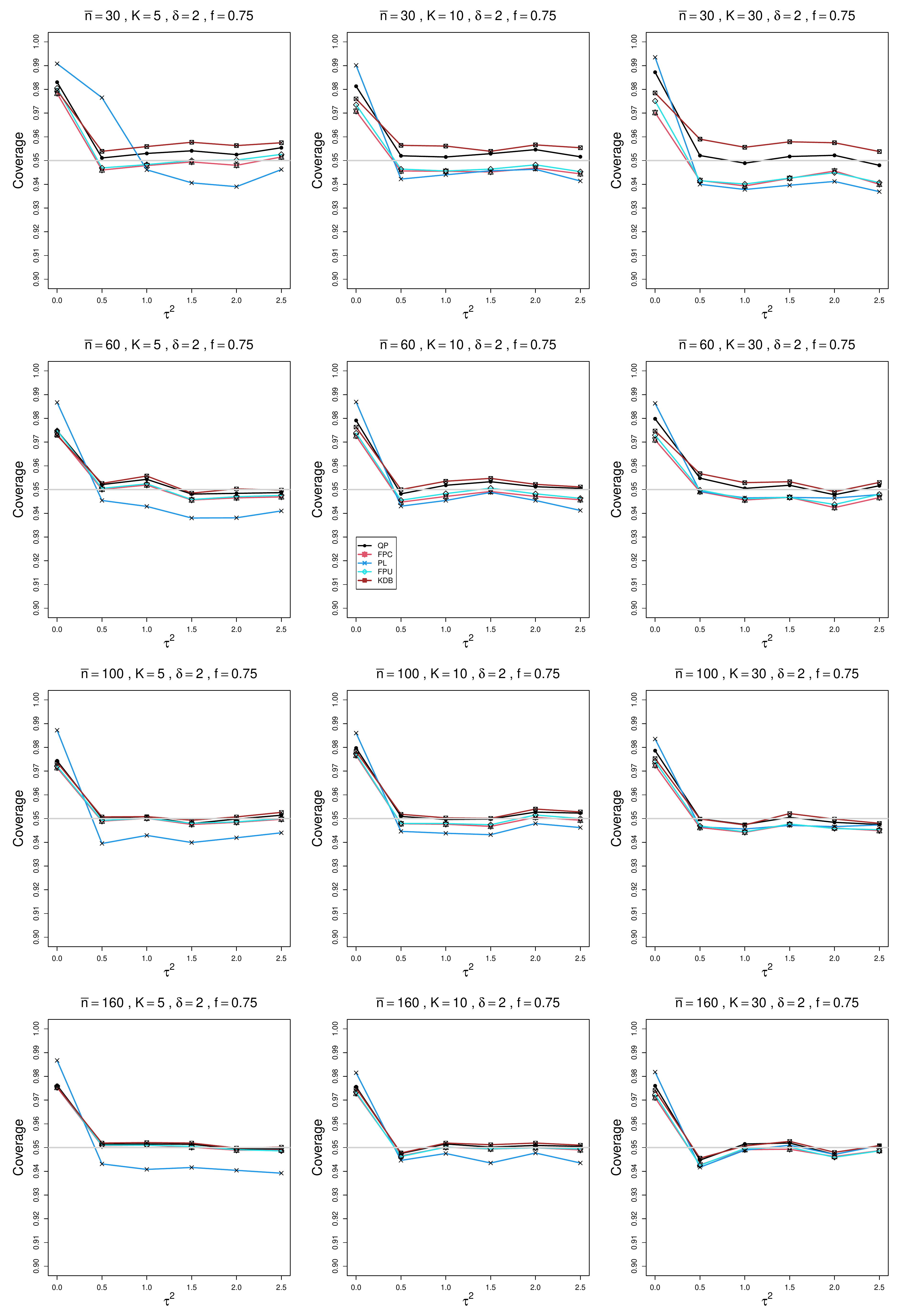}
	\caption{Coverage of $\tau^2$ for $\delta = 2$, $f = .75$, and unequal sample sizes
		\label{PlotCovOfTau2delta2andq075_SMD_unequal}}
\end{figure}

\clearpage
\setcounter{section}{0}
\setcounter{figure}{0}
\renewcommand{\thesection}{B.\arabic{section}}
\section*{B. Error in approximations for null distribution of $Q$}
Each figure corresponds to a value of $\delta$, a value of $f$, and either equal or unequal sample sizes. \\
For each combination of a value of $n$ or $\bar{n}$ and a value of $K$, a panel plots the error, $\hat{P}(\hat{F}(Q) > 1 - p) - p$, for the upper-tail probabilities $p$ = .001, .0025, .005, .01, .025, .05, .1, .25, .5, .75, .9, .95, .975, .99, .995, .9975, .999. \\
The approximations for the distribution of $Q$ are
\begin{itemize}
	\item F SW (Farebrother approximation, effective-sample-size weights)
	\item M2 SW (Two-moment approximation, effective-sample-size weights)
	\item $\chi_{K - 1}^2$ (Chi-square, IV weights)
	\item KDB (Chi-square approximation based on corrected first moment, IV weights)
\end{itemize}

\clearpage
\renewcommand{\thefigure}{B.\arabic{figure}}
%q=0.5
\begin{figure}[t]
	\centering
	\includegraphics[scale=0.33]{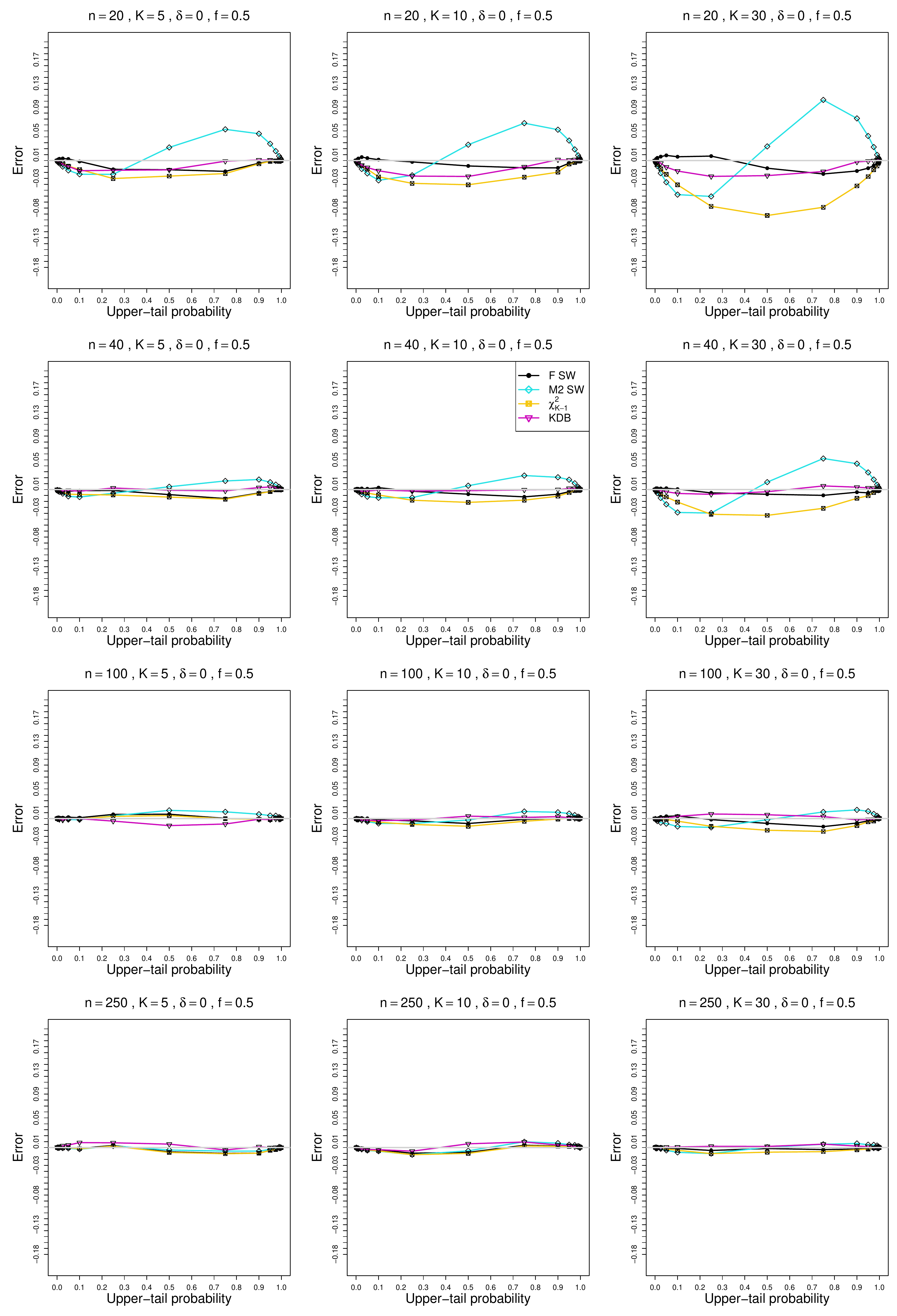}
	\caption{Approximation error for $\delta = 0$, $f = .5$, and equal sample sizes
		\label{PPplotNominalAgainstEstimatedAt095OfEQ1AgainstNsigma2T0andq05_SMD}}
\end{figure}

\begin{figure}[t]
	\centering
	\includegraphics[scale=0.33]{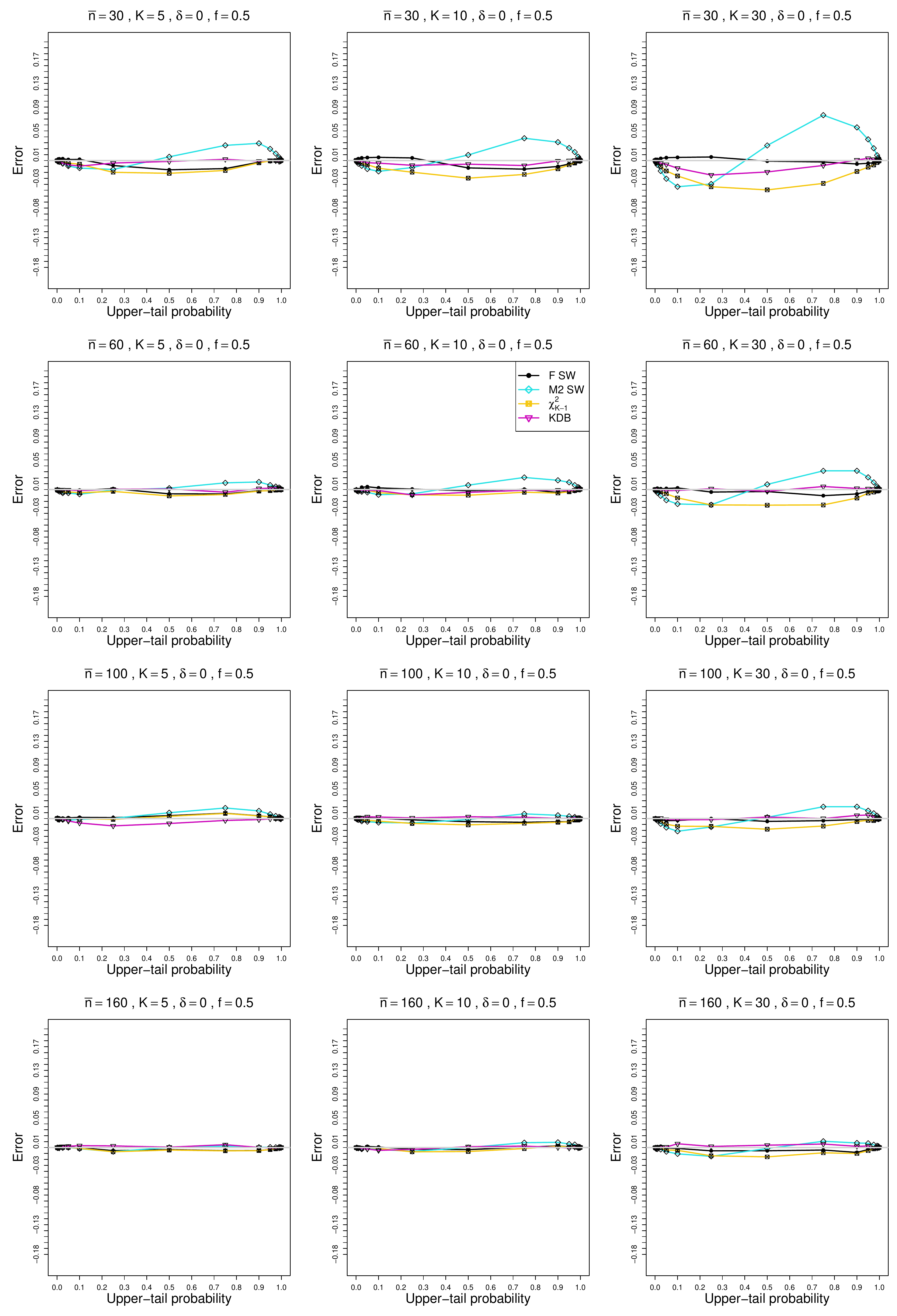}
	\caption{Approximation error for $\delta = 0$, $f = .5$, and unequal sample sizes
		\label{PPplotNominalAgainstEstimatedAt095OfEQ1AgainstNsigma2T0andq05_SMD_unequal}}
\end{figure}

\begin{figure}[t]
	\centering
	\includegraphics[scale=0.33]{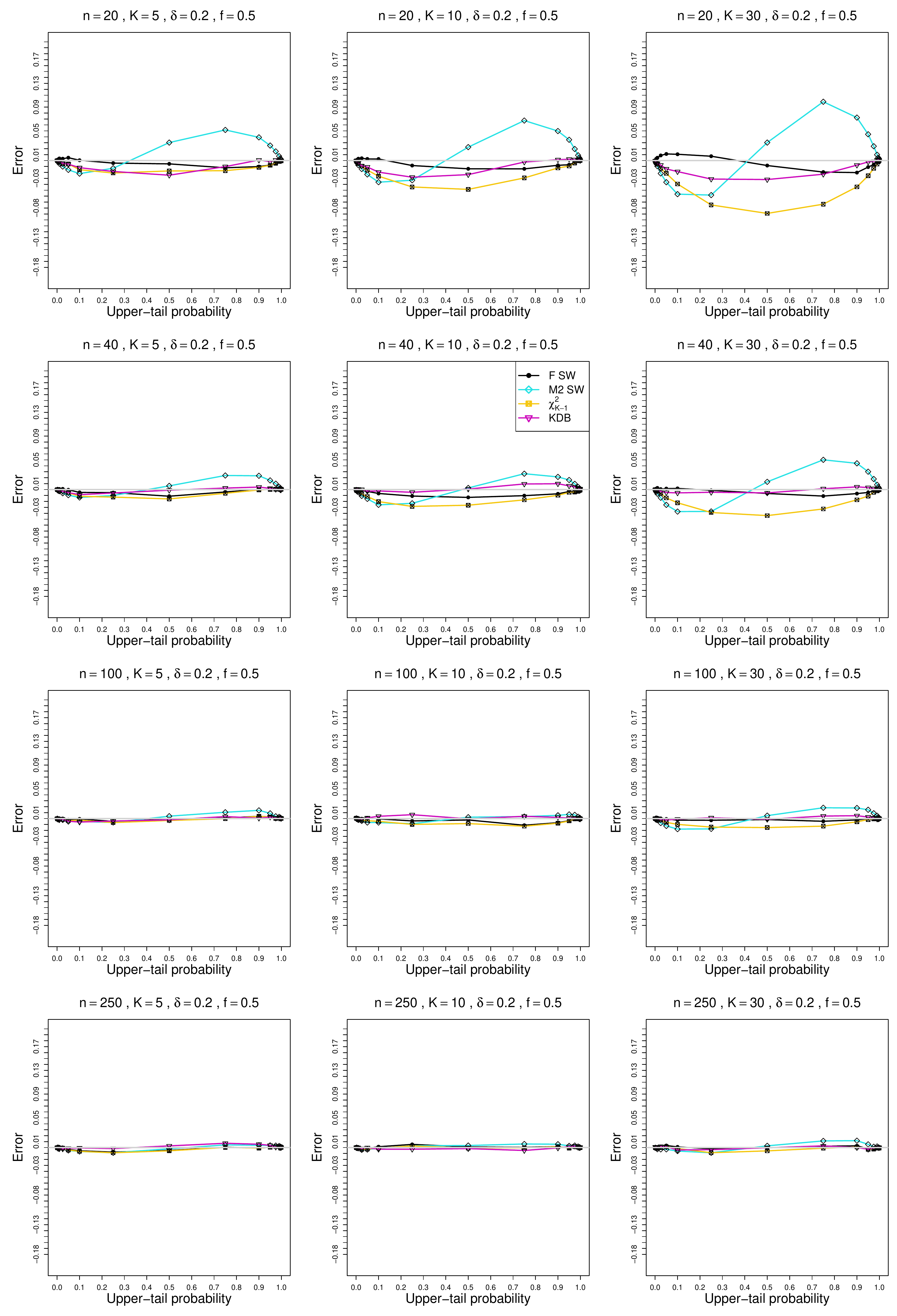}
	\caption{Approximation error for $\delta = 0.2$, $f = .5$, and equal sample sizes
		\label{PPplotNominalAgainstEstimatedAt095OfEQ1AgainstNdelta02andq05_SMD}}
\end{figure}

\begin{figure}[t]
	\centering
	\includegraphics[scale=0.33]{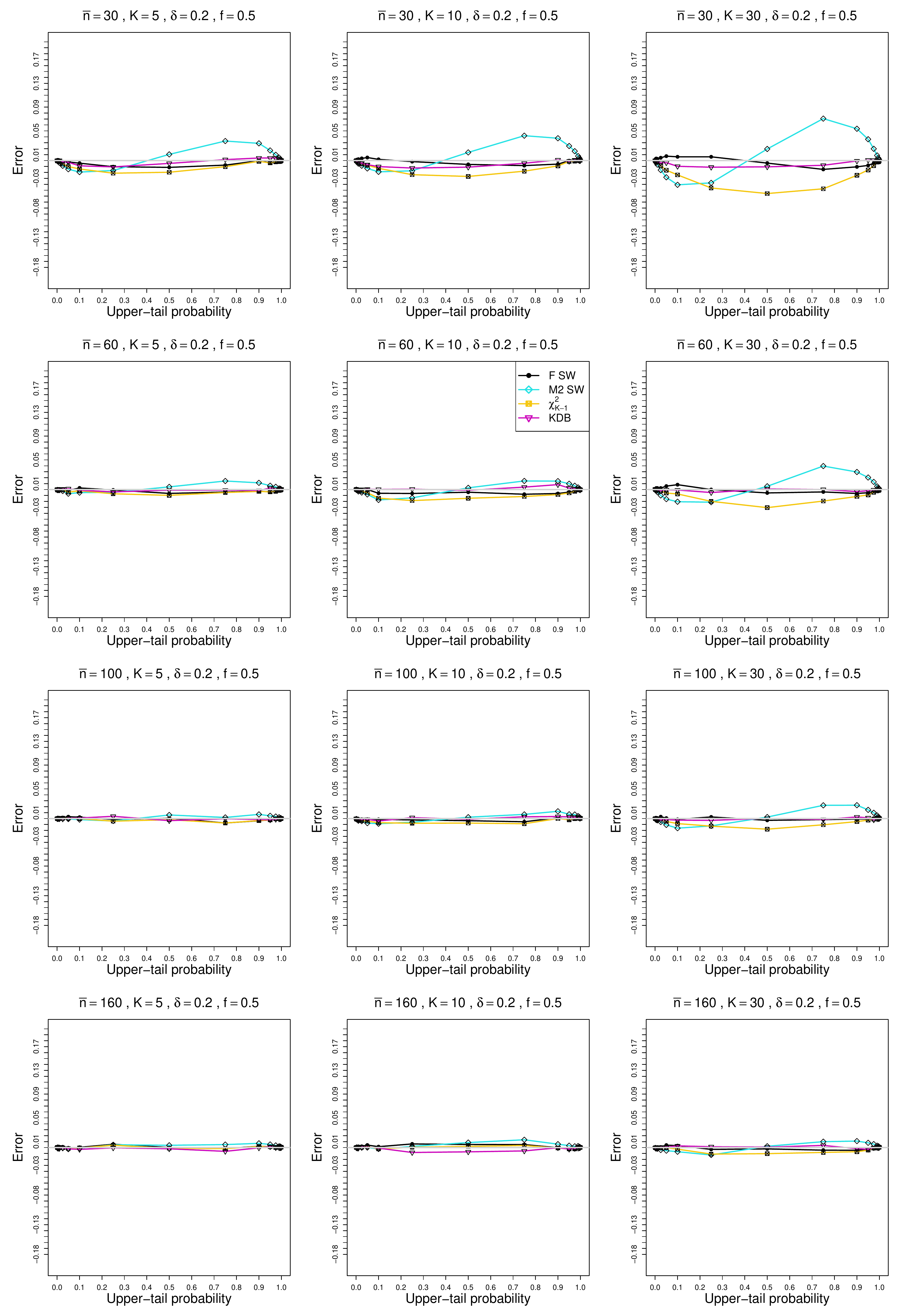}
	\caption{Approximation error $\delta = 0.2$, $f = .5$, and unequal sample sizes
		\label{PPplotNominalAgainstEstimatedAt095OfEQ1AgainstNdelta02andq05_SMD_unequal}}
\end{figure}

\begin{figure}[t]
	\centering
	\includegraphics[scale=0.33]{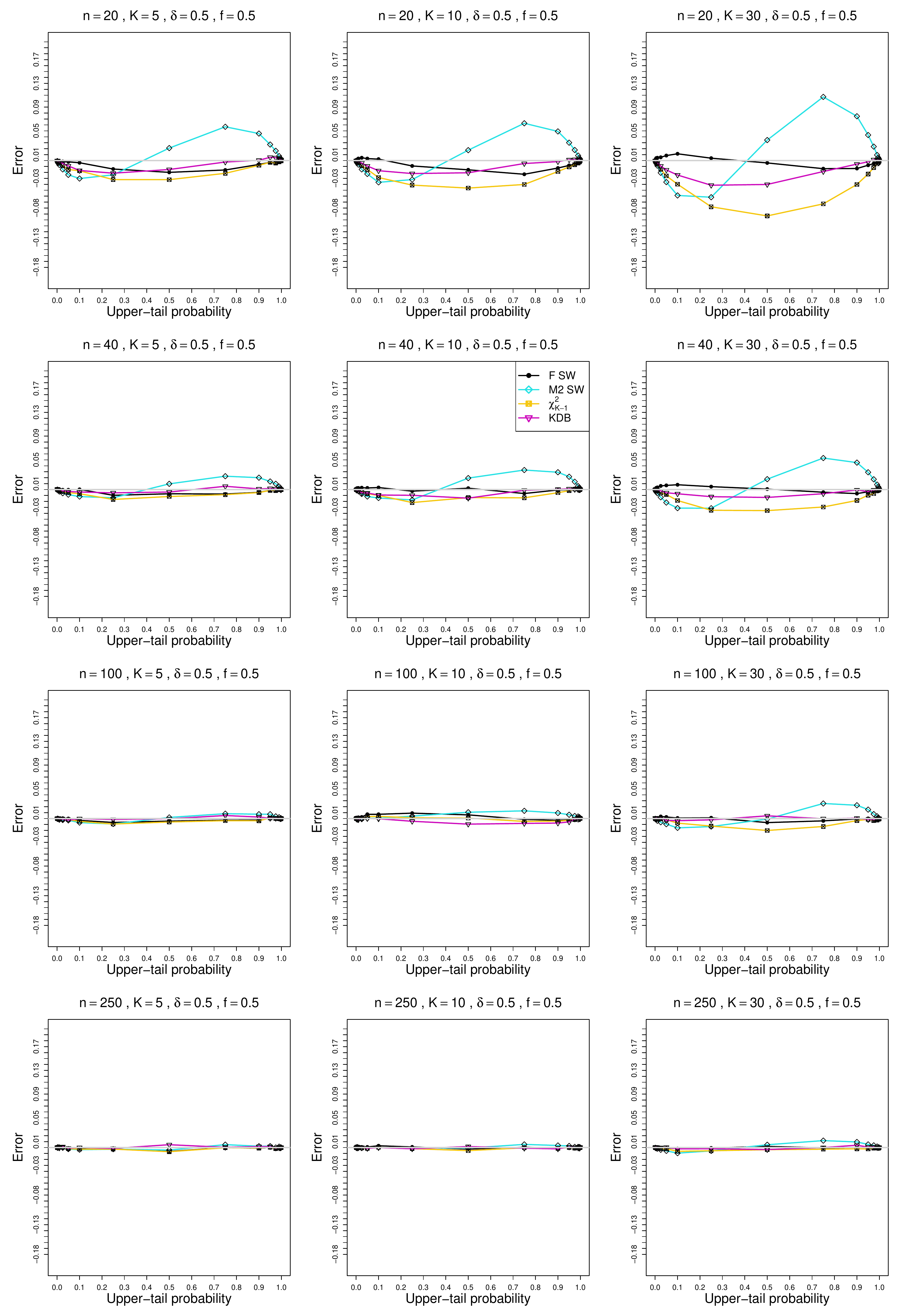}
	\caption{Approximation error $\delta = 0.5$, $f = .5$, and equal sample sizes
		\label{PPplotNominalAgainstEstimatedAt095OfEQ1AgainstNdelta05andq05_SMD}}
\end{figure}

\begin{figure}[t]
	\centering
	\includegraphics[scale=0.33]{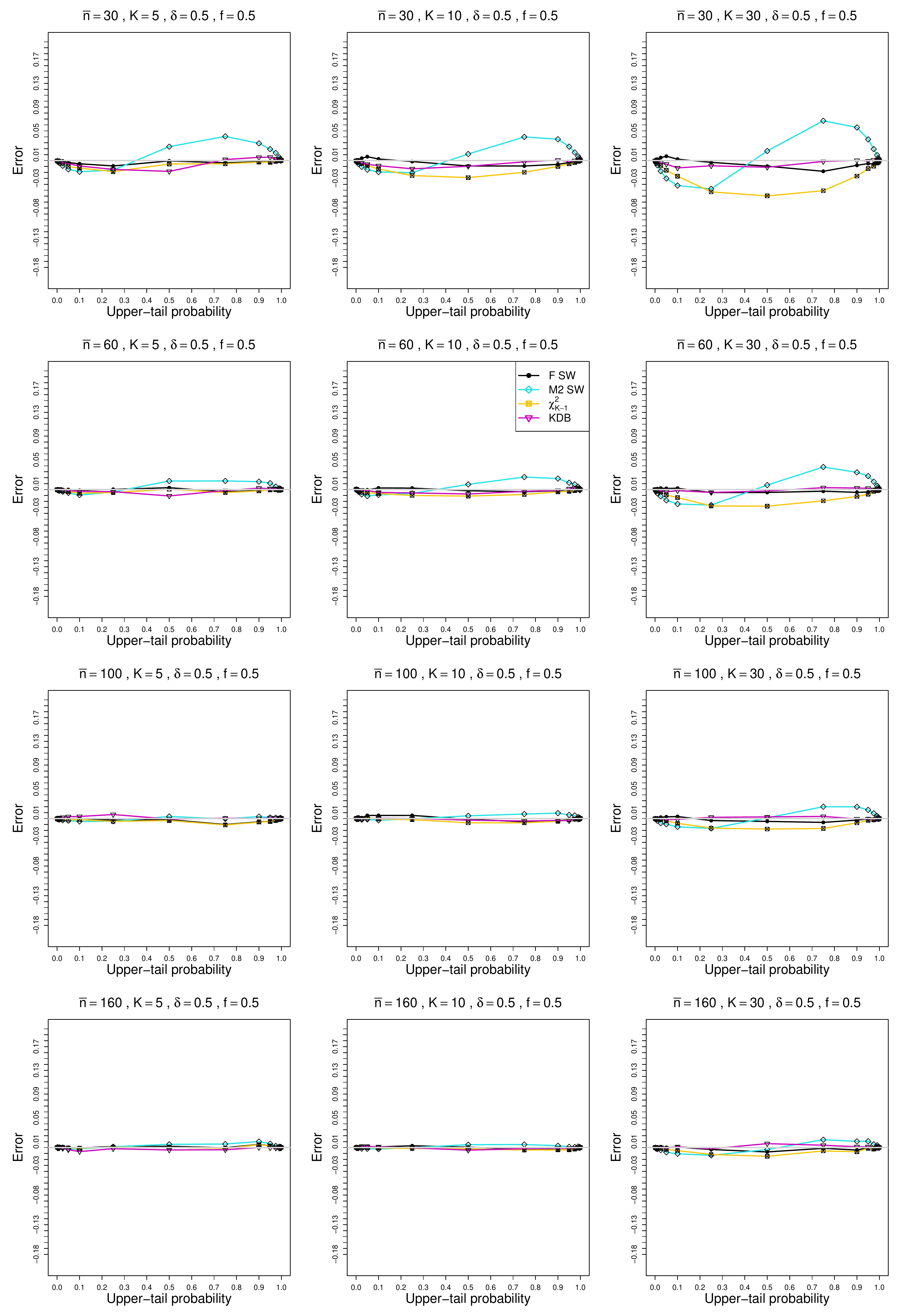}
	\caption{Approximation error for $\delta = 0.5$, $f = .5$, and unequal sample sizes
		\label{PPplotNominalAgainstEstimatedAt095OfEQ1AgainstNdelta05andq05_SMD_unequal}}
\end{figure}

\begin{figure}[t]
	\centering
	\includegraphics[scale=0.33]{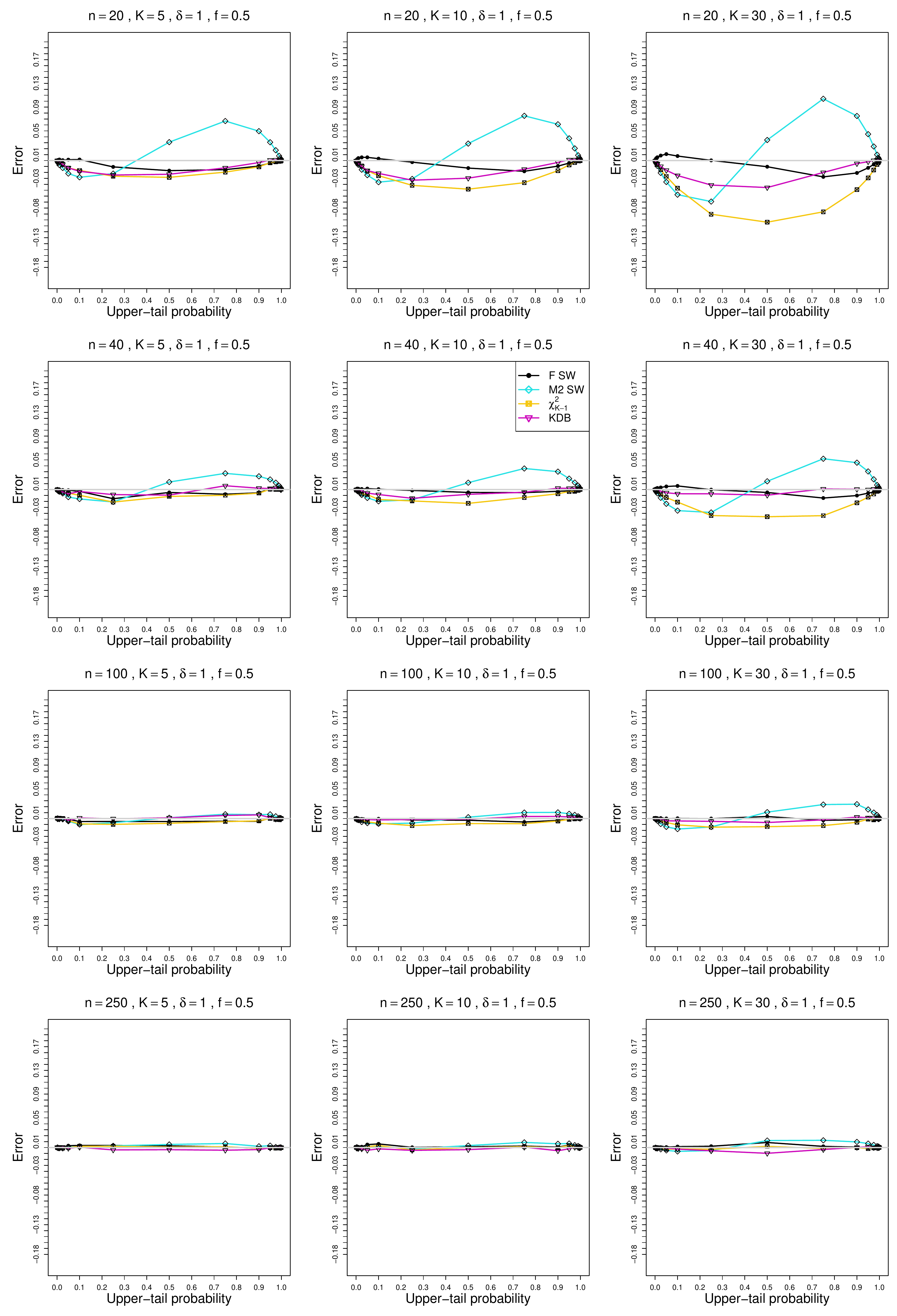}
	\caption{Approximation error for $\delta = 1$, $f = .5$, and equal sample sizes
		\label{PPplotNominalAgainstEstimatedAt095OfEQ1AgainstNdelta1andq05_SMD}}
\end{figure}

\begin{figure}[t]
	\centering
	\includegraphics[scale=0.33]{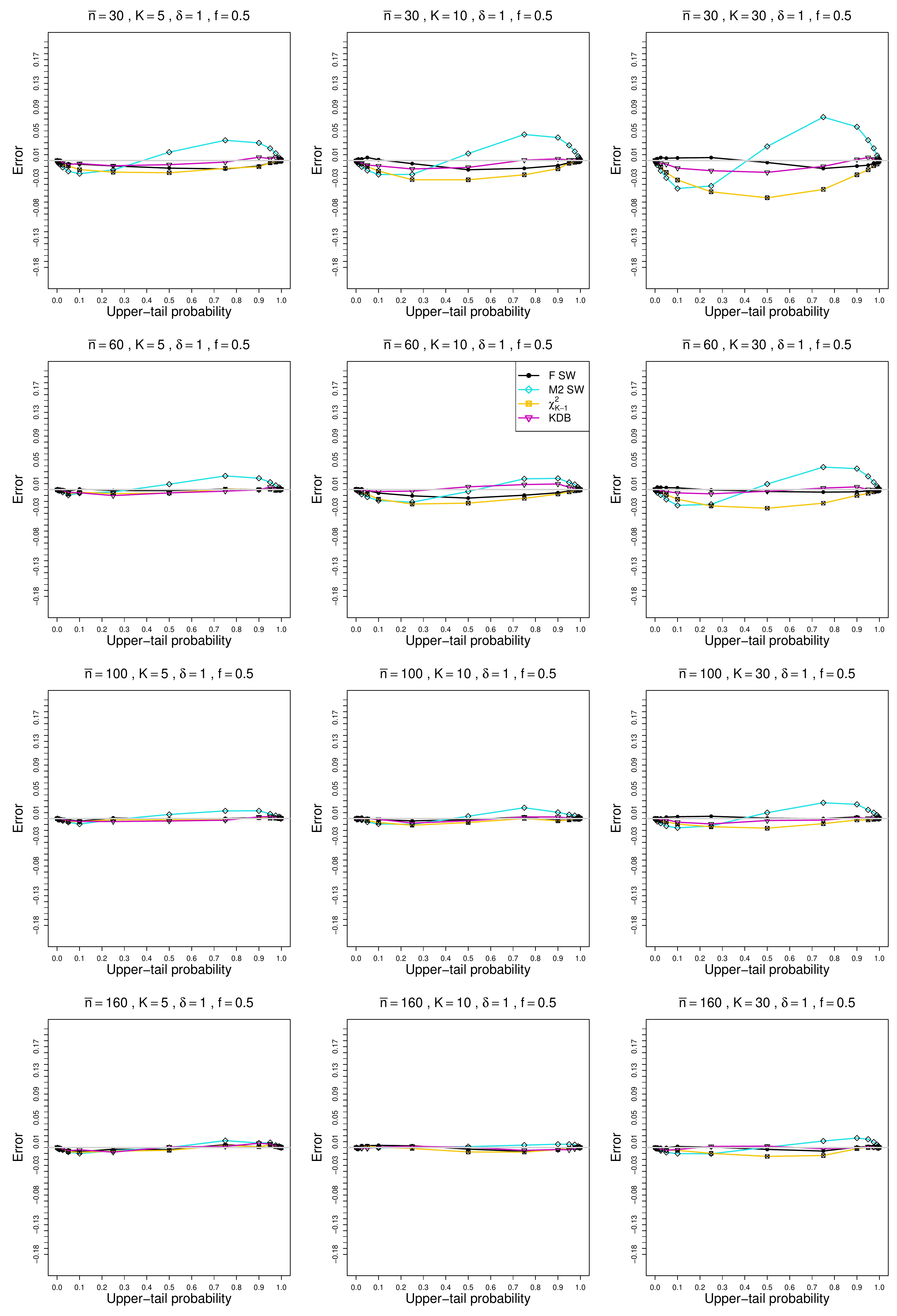}
	\caption{Approximation error for $\delta = 1$, $f = .5$, and unequal sample sizes
		\label{PPplotNominalAgainstEstimatedAt095OfEQ1AgainstNdelta1andq05_SMD_unequal}}
\end{figure}

\begin{figure}[t]
	\centering
	\includegraphics[scale=0.33]{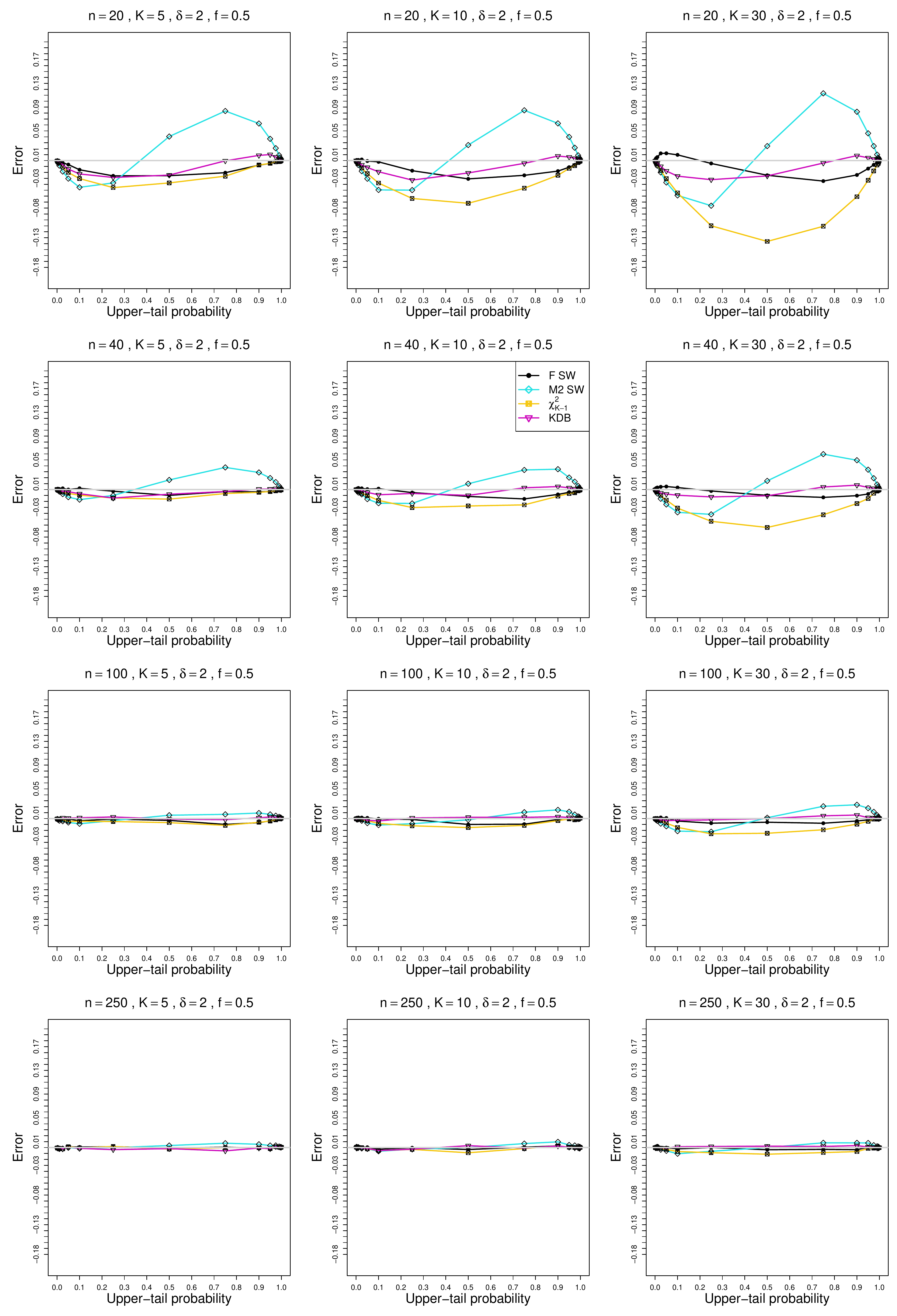}
	\caption{Approximation error for $\delta = 2$, $f = .5$, and equal sample sizes
		\label{PPplotNominalAgainstEstimatedAt095OfEQ1AgainstNdelta2andq05_SMD}}
\end{figure}

\begin{figure}[t]
	\centering
	\includegraphics[scale=0.33]{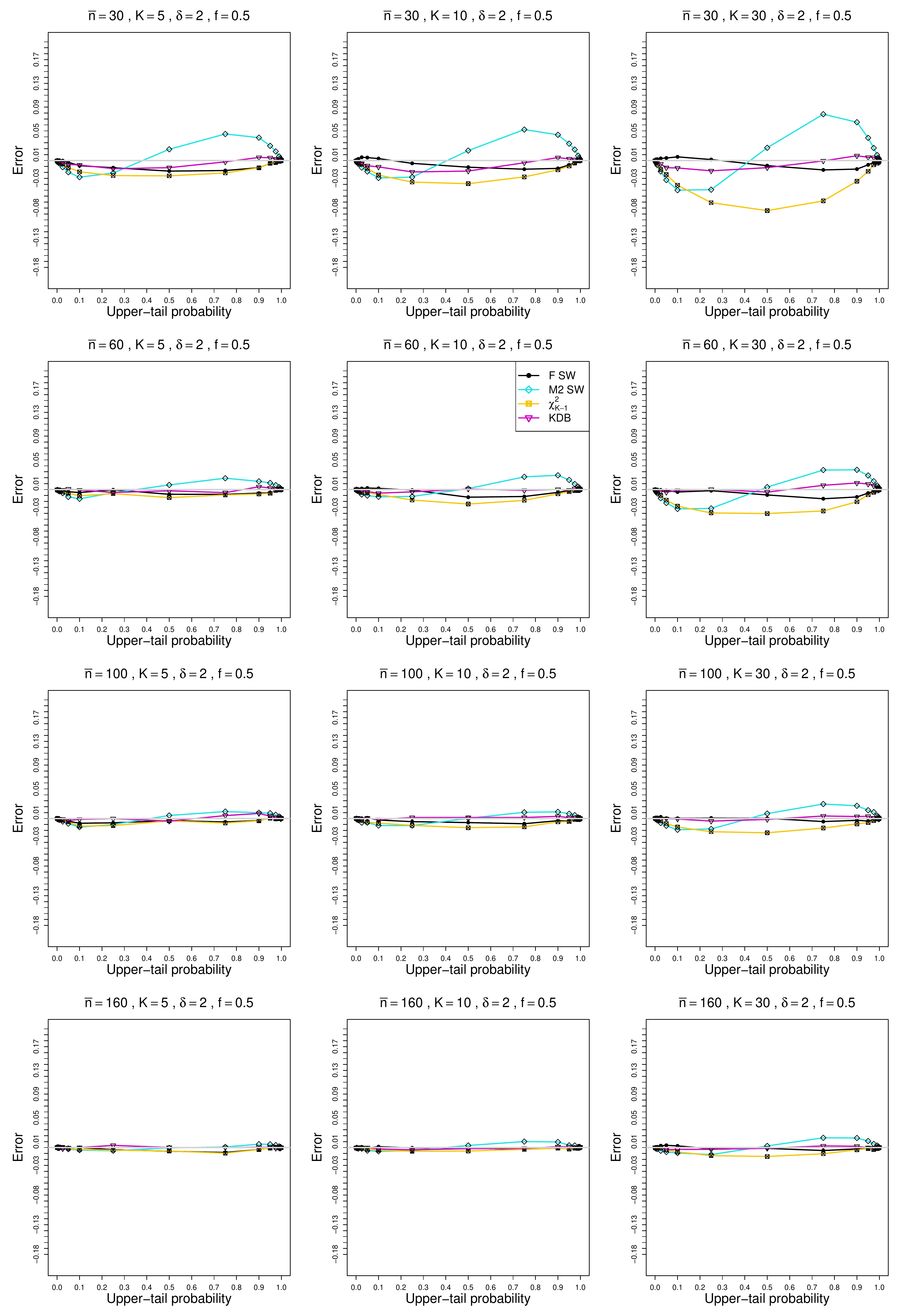}
	\caption{Approximation error for $\delta = 2$, $f = .5$, and unequal sample sizes
		\label{PPplotNominalAgainstEstimatedAt095OfEQ1AgainstNdelta2andq05_SMD_unequal}}
\end{figure}

%%%%%%%%%%%%%%%%%%%%%%%%%%%%%%%%%%%%%%%%%%%%%%%%%%%%%%%%%%%%%%%%%%%%%%%%%%%%%%%%%%%%%%%%%%%%%%%%%%%%%%%%%%%%%%%%%%%%%%%%%%%%%%
%q=0.75

\begin{figure}[t]
	\centering
	\includegraphics[scale=0.33]{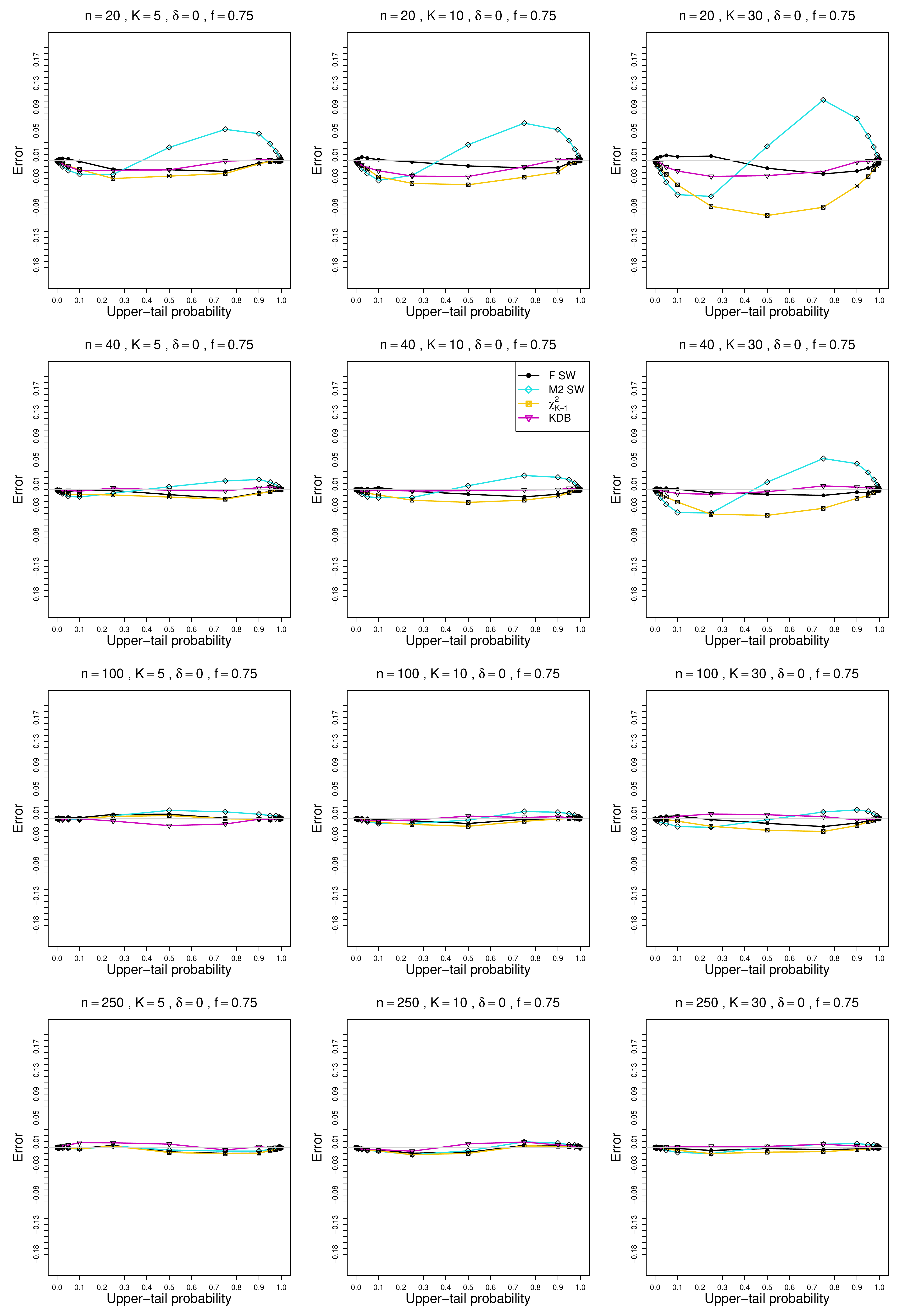}
	\caption{Approximation error for $\delta = 0$, $f = .75$, and equal sample sizes
		\label{PPplotNominalAgainstEstimatedAt095OfEQ1AgainstNsigma2T0andq075_SMD}}
\end{figure}

\begin{figure}[t]
	\centering
	\includegraphics[scale=0.33]{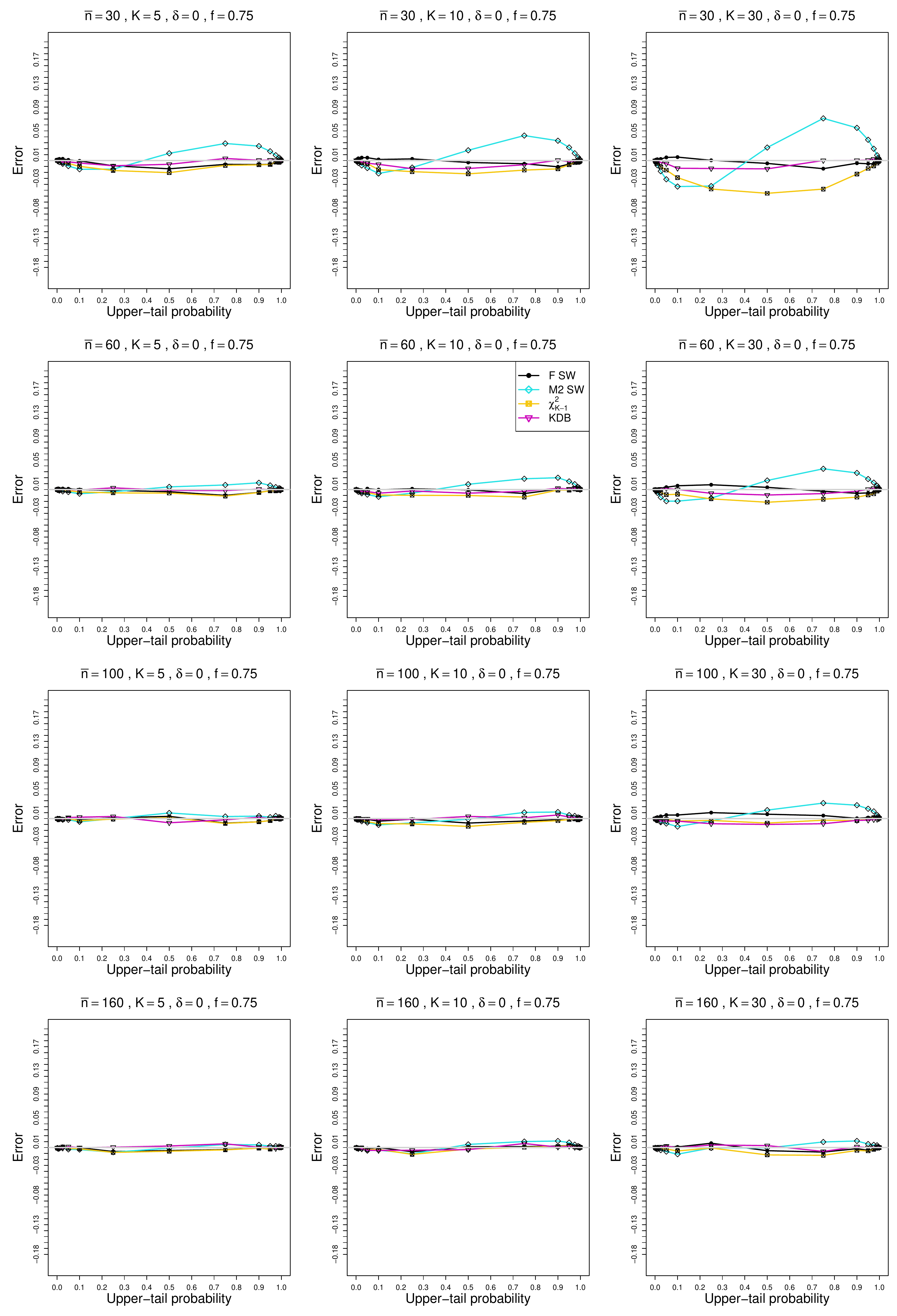}
	\caption{Approximation error for $\delta = 0$, $f = .75$, and unequal sample sizes
		\label{PPplotNominalAgainstEstimatedAt095OfEQ1AgainstNsigma2T0andq075_SMD_unequal}}
\end{figure}

\begin{figure}[t]
	\centering
	\includegraphics[scale=0.33]{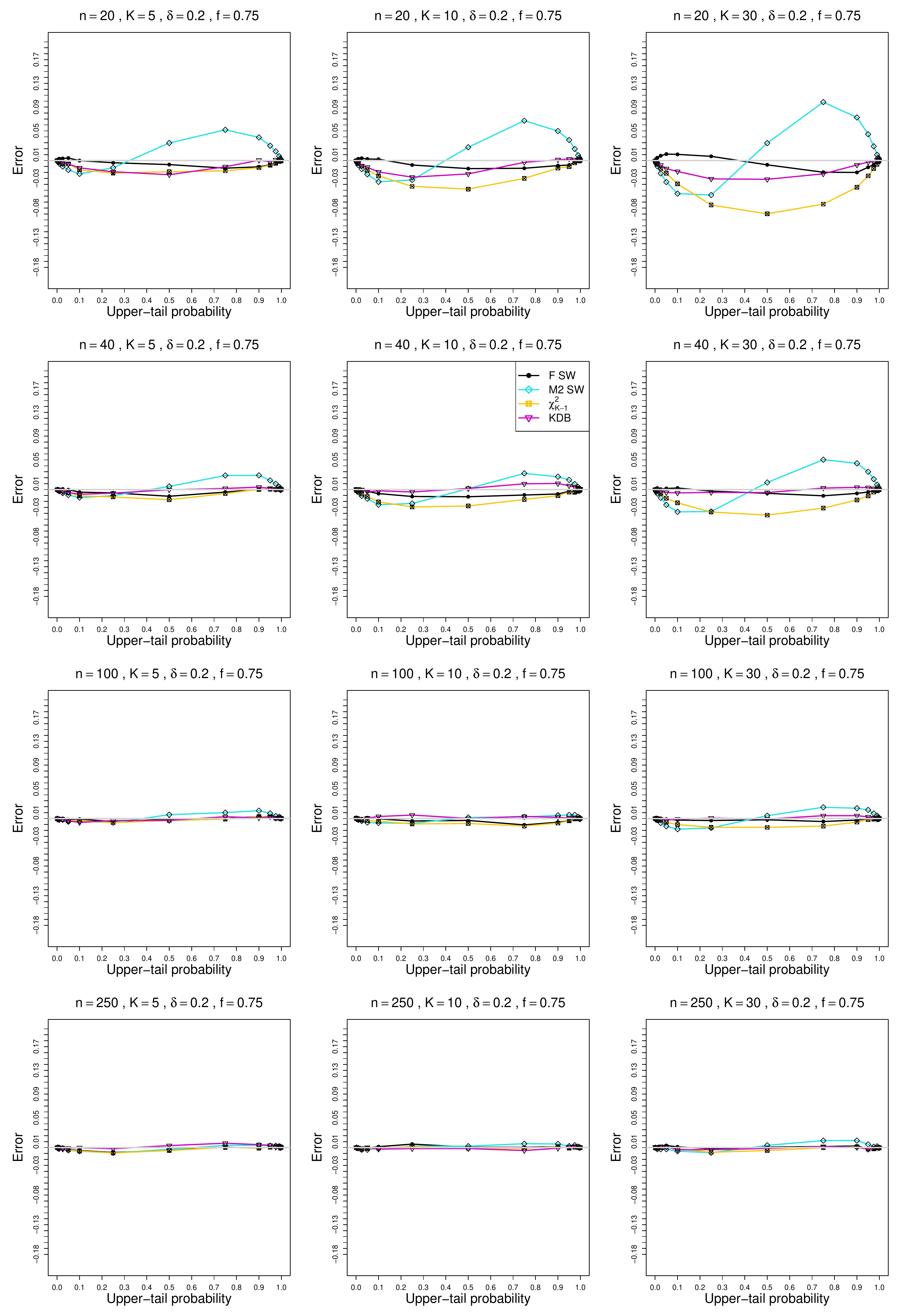}
	\caption{Approximation error for $\delta = 0.2$, $f = .75$, and equal sample sizes
		\label{PPplotNominalAgainstEstimatedAt095OfEQ1AgainstNdelta02andq075_SMD}}
\end{figure}

\begin{figure}[t]
	\centering
	\includegraphics[scale=0.33]{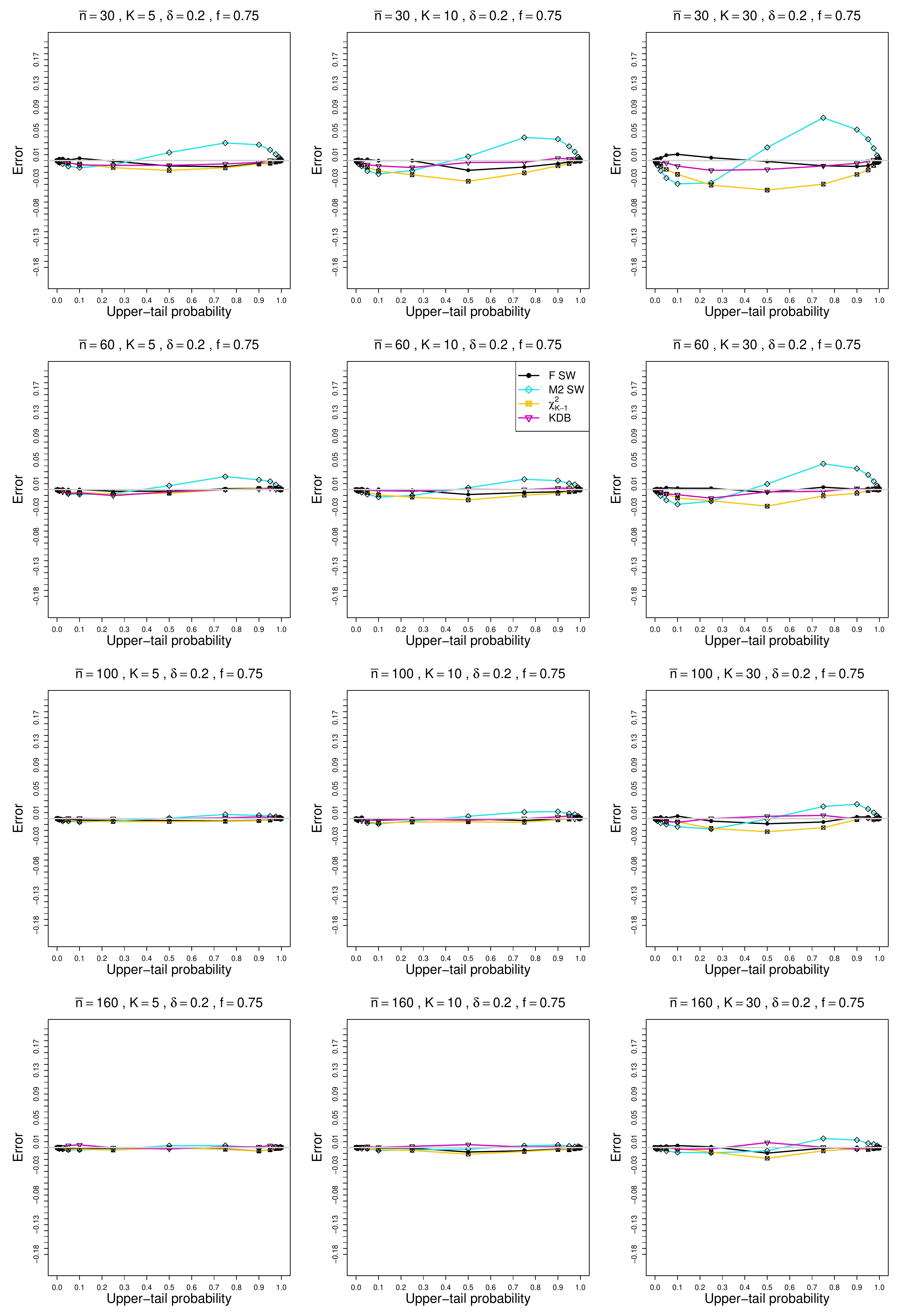}
	\caption{Approximation error for $\delta = 0.2$, $f = .75$, and unequal sample sizes
		\label{PPplotNominalAgainstEstimatedAt095OfEQ1AgainstNdelta02andq075_SMD_unequal}}
\end{figure}

\begin{figure}[t]
	\centering
	\includegraphics[scale=0.33]{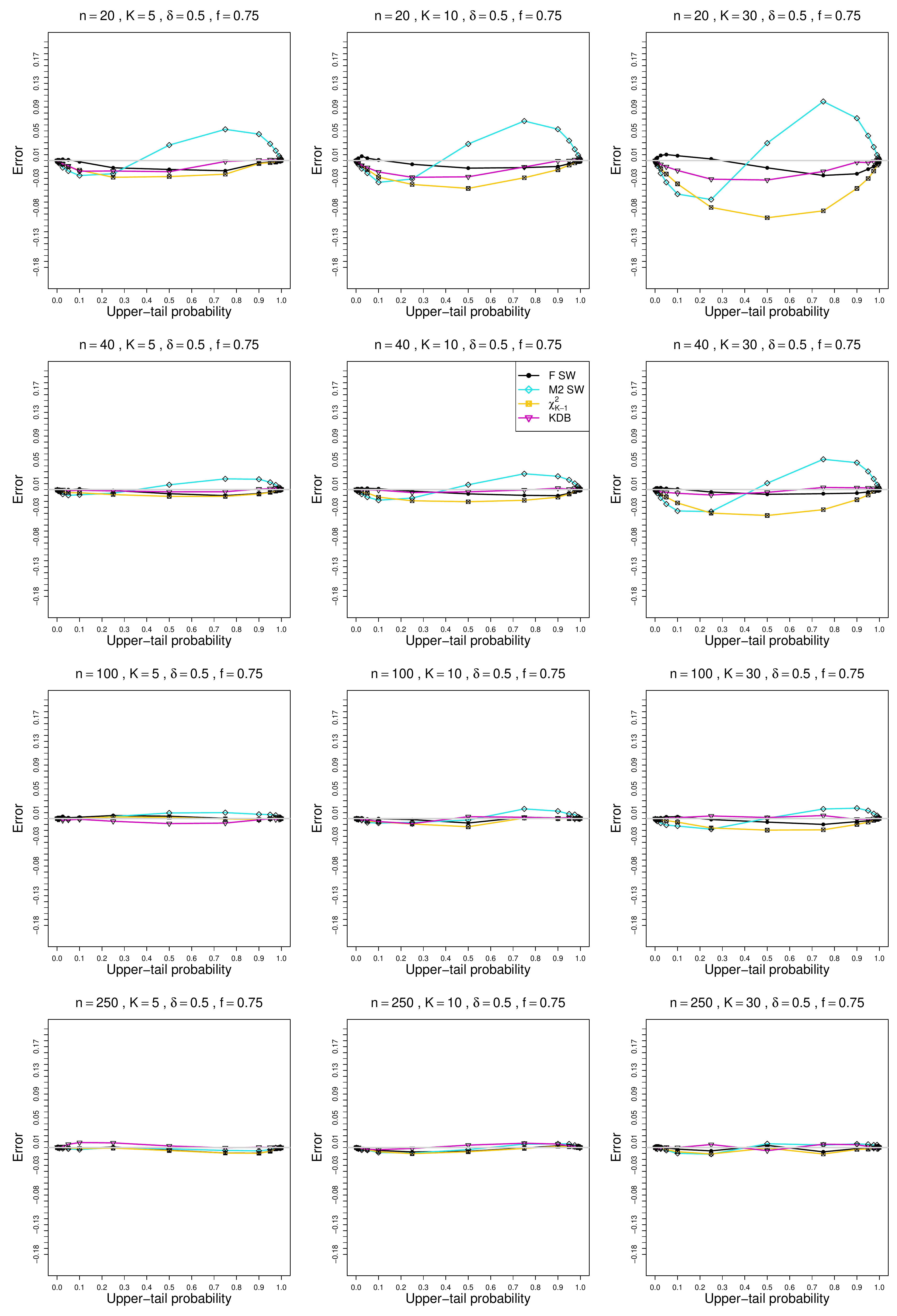}
	\caption{Approximation error for $\delta = 0.5$, $f = .75$, and equal sample sizes
		\label{PPplotNominalAgainstEstimatedAt095OfEQ1AgainstNdelta05andq075_SMD}}
\end{figure}

\begin{figure}[t]
	\centering
	\includegraphics[scale=0.33]{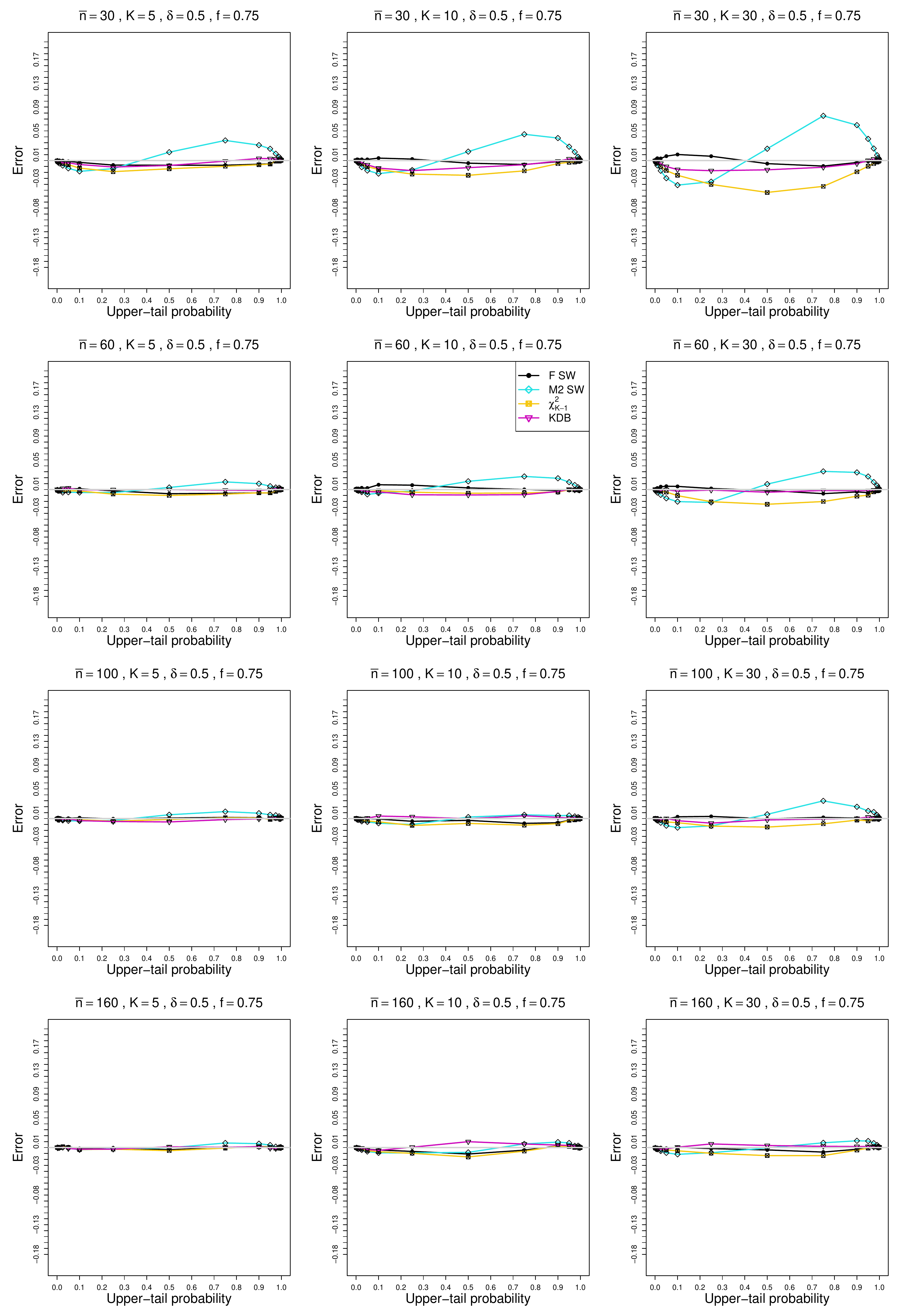}
	\caption{Approximation error for $\delta = 0.5$, $f = .75$, and unequal sample sizes
		\label{PPplotNominalAgainstEstimatedAt095OfEQ1AgainstNdelta05andq075_SMD_unequal}}
\end{figure}

\begin{figure}[t]
	\centering
	\includegraphics[scale=0.33]{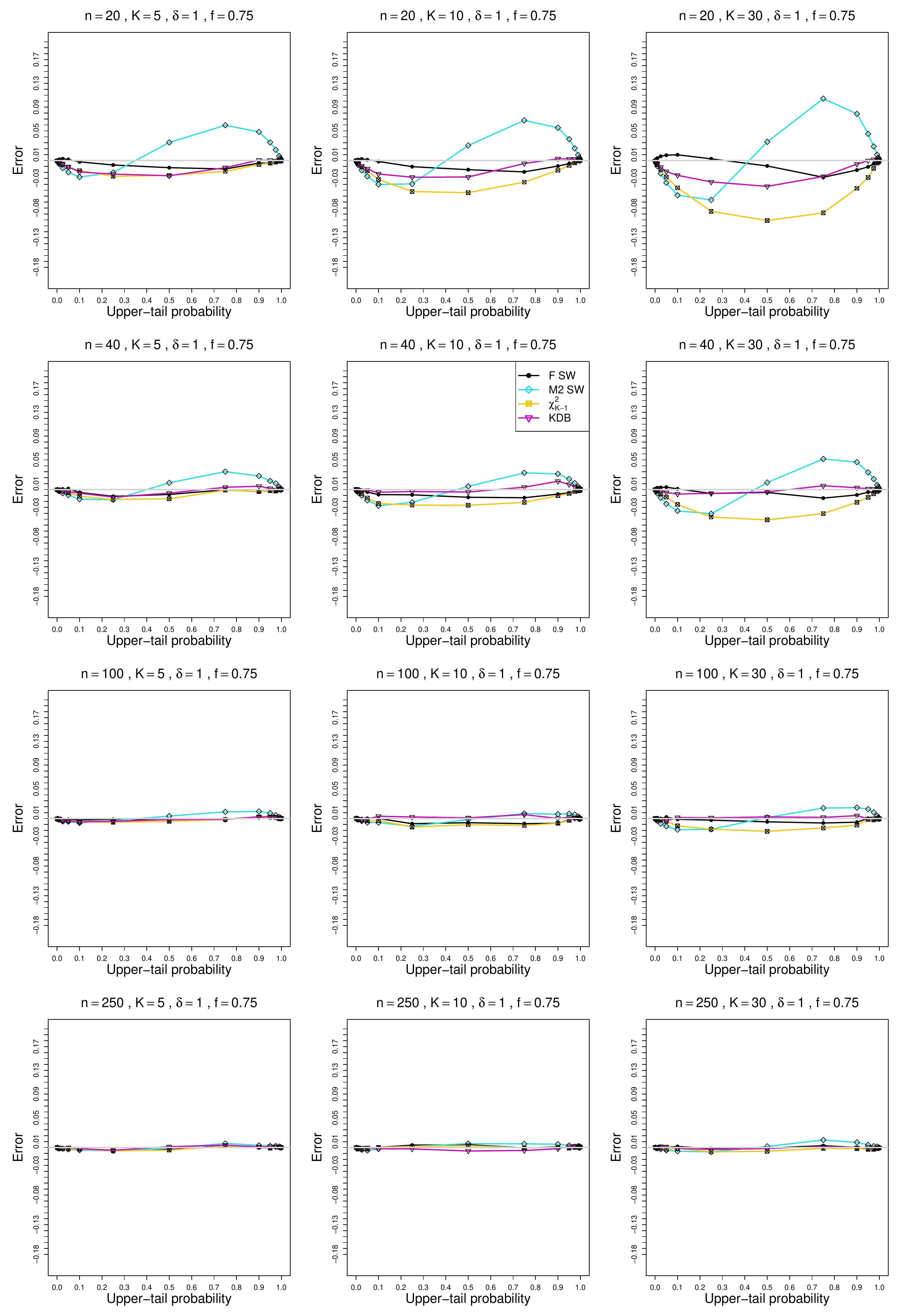}
	\caption{Approximation error for $\delta = 1$, $f = .75$, and equal sample sizes
		\label{PPplotNominalAgainstEstimatedAt095OfEQ1AgainstNdelta1andq075_SMD}}
\end{figure}

\begin{figure}[t]
	\centering
	\includegraphics[scale=0.33]{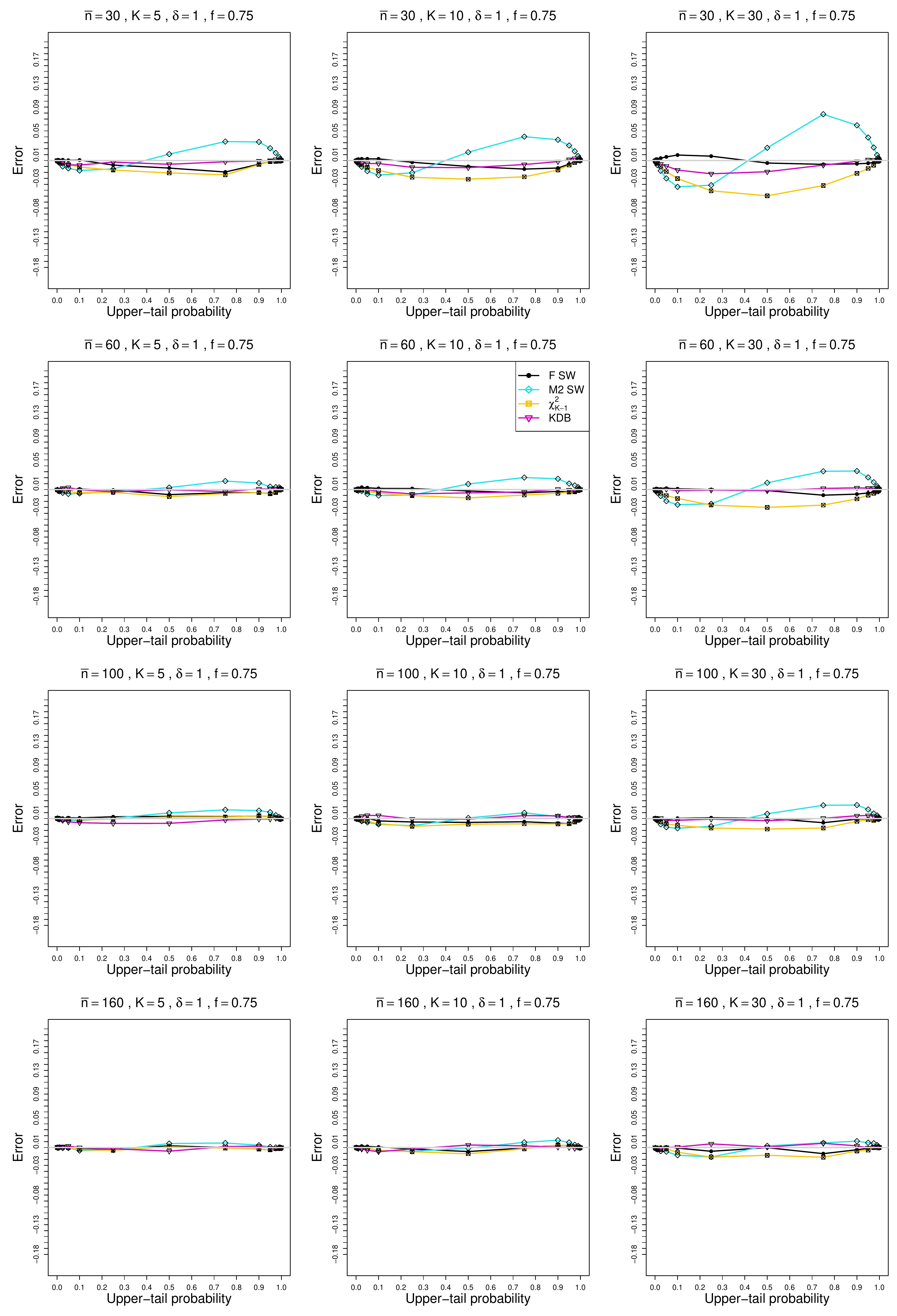}
	\caption{Approximation error for $\delta = 1$, $f = .75$, and unequal sample sizes
		\label{PPplotNominalAgainstEstimatedAt095OfEQ1AgainstNdelta1andq075_SMD_unequal}}
\end{figure}

\begin{figure}[t]
	\centering
	\includegraphics[scale=0.33]{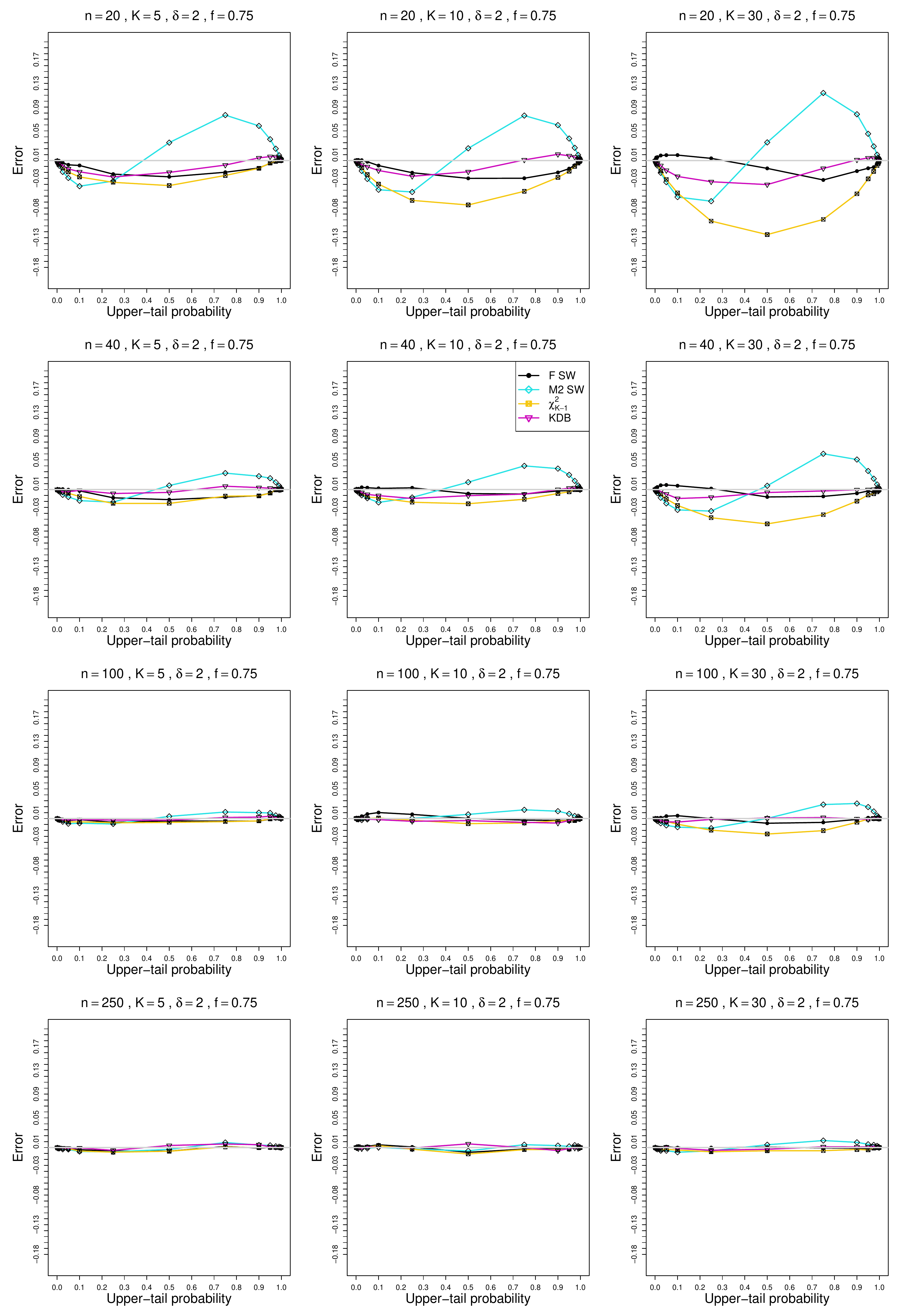}
	\caption{Approximation error for $\delta = 2$, $f = .75$, and equal sample sizes
		\label{PPplotNominalAgainstEstimatedAt095OfEQ1AgainstNdelta2andq075_SMD}}
\end{figure}

\begin{figure}[t]
	\centering
	\includegraphics[scale=0.33]{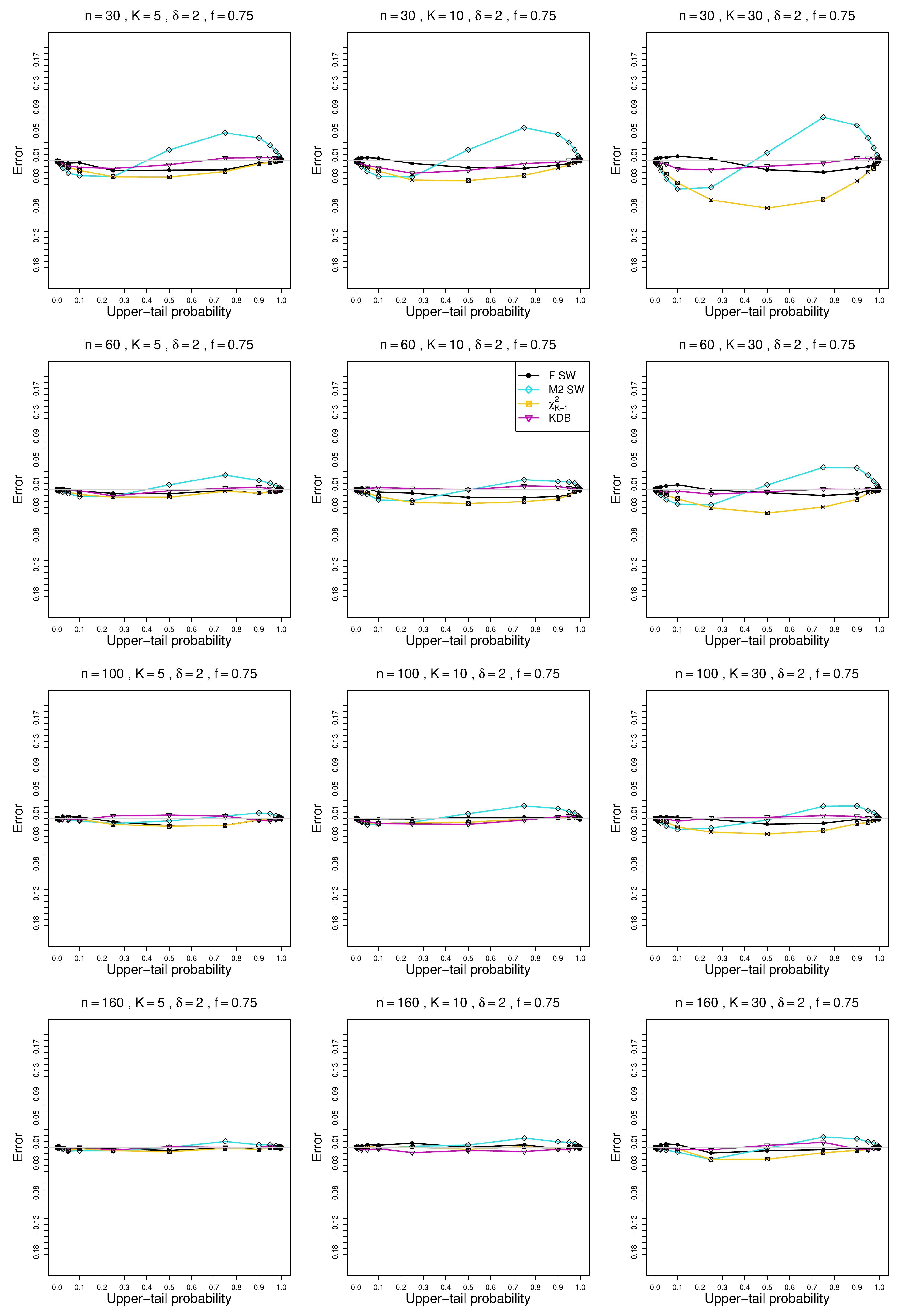}
	\caption{Approximation error for $\delta = 2$, $g = .75$, and unequal sample sizes
		\label{PPplotNominalAgainstEstimatedAt095OfEQ1AgainstNdelta2andq075_SMD_unequal}}
\end{figure}

\clearpage

\setcounter{section}{0}
\renewcommand{\thesection}{C.\arabic{section}}

\section*{C. Empirical level of the test for heterogeneity ($\tau^2$ = 0 versus $\tau^2 > 0$) based on approximations for the distribution of $Q$, plotted vs sample size}
In sets of figures for $f = .5$ and $f = .75$, each figure corresponds to a value of $\alpha$ (= .001, .005, .01, .05) and either equal sample sizes or unequal sample sizes. (For all figures, $\tau^2 = 0$.) \\
For each combination of a value of $\delta$ (omitting $\delta = 0.2$) and a value of $K$, a panel plots the empirical level versus $n$ or $\bar{n}$.\\
The approximations for the distribution of $Q$ are
\begin{itemize}
	\item F SW (Farebrother approximation, effective-sample-size weights)
	\item M2 SW (Two-moment approximation, effective-sample-size weights)
	\item $\chi^2_{K - 1}$ (Chi-square, IV weights)
	\item KDB (Chi-square approximation based on corrected first moment, IV weights)
\end{itemize}

\setcounter{figure}{0}
\clearpage
\renewcommand{\thefigure}{C1.\arabic{figure}}
%%sigma2T=1
%q=0.5and p=0.001
\subsection*{C1. $f = .5$}
\clearpage
\begin{figure}[t]
	\centering
	\includegraphics[scale=0.33]{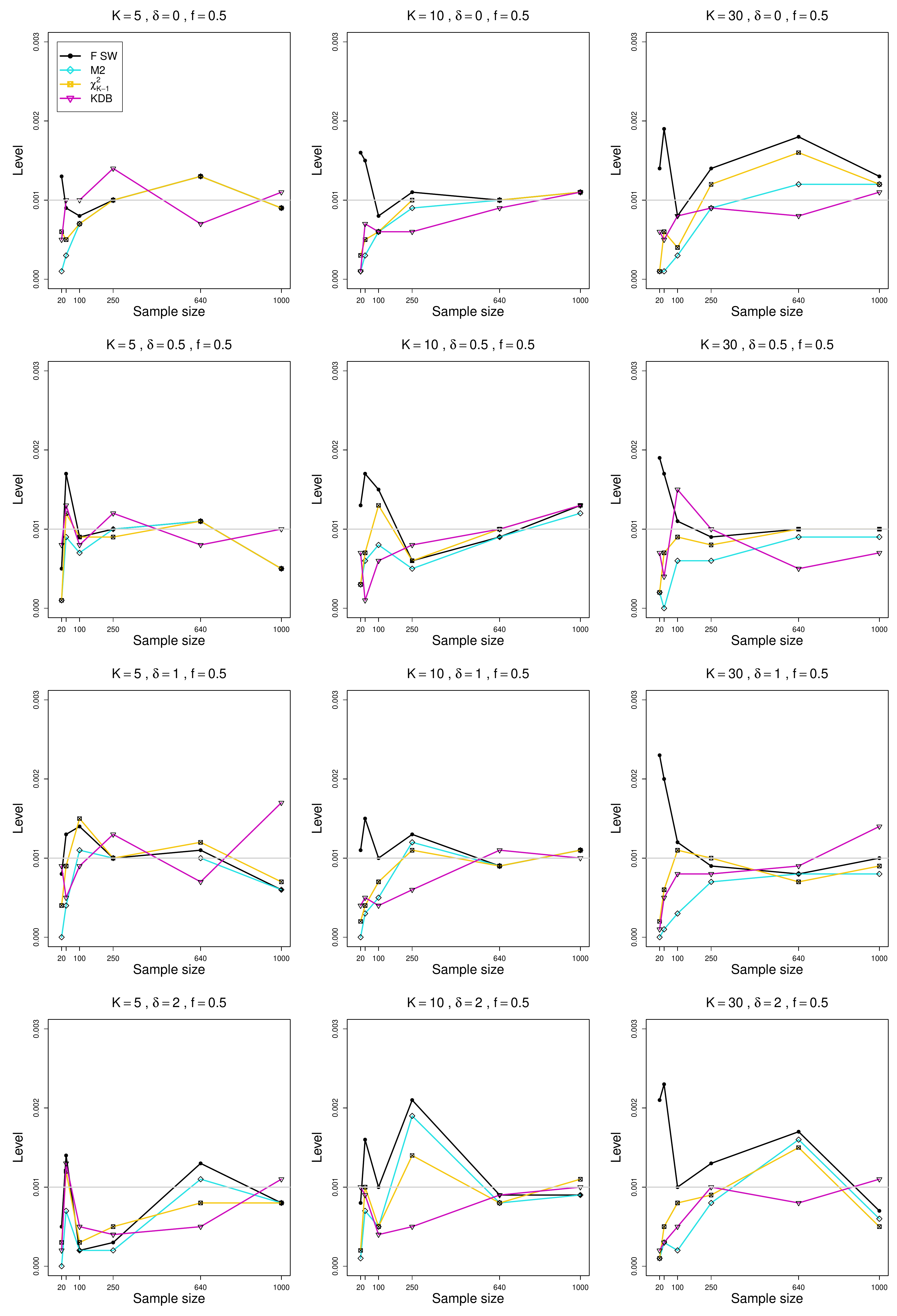}
	\caption{Empirical level at $\alpha = .001$ of test for heterogeneity vs sample size
		\label{PlotOfPvaluesAgainstN_0001level_q05}}
\end{figure}

\begin{figure}[t]
	\centering
	\includegraphics[scale=0.33]{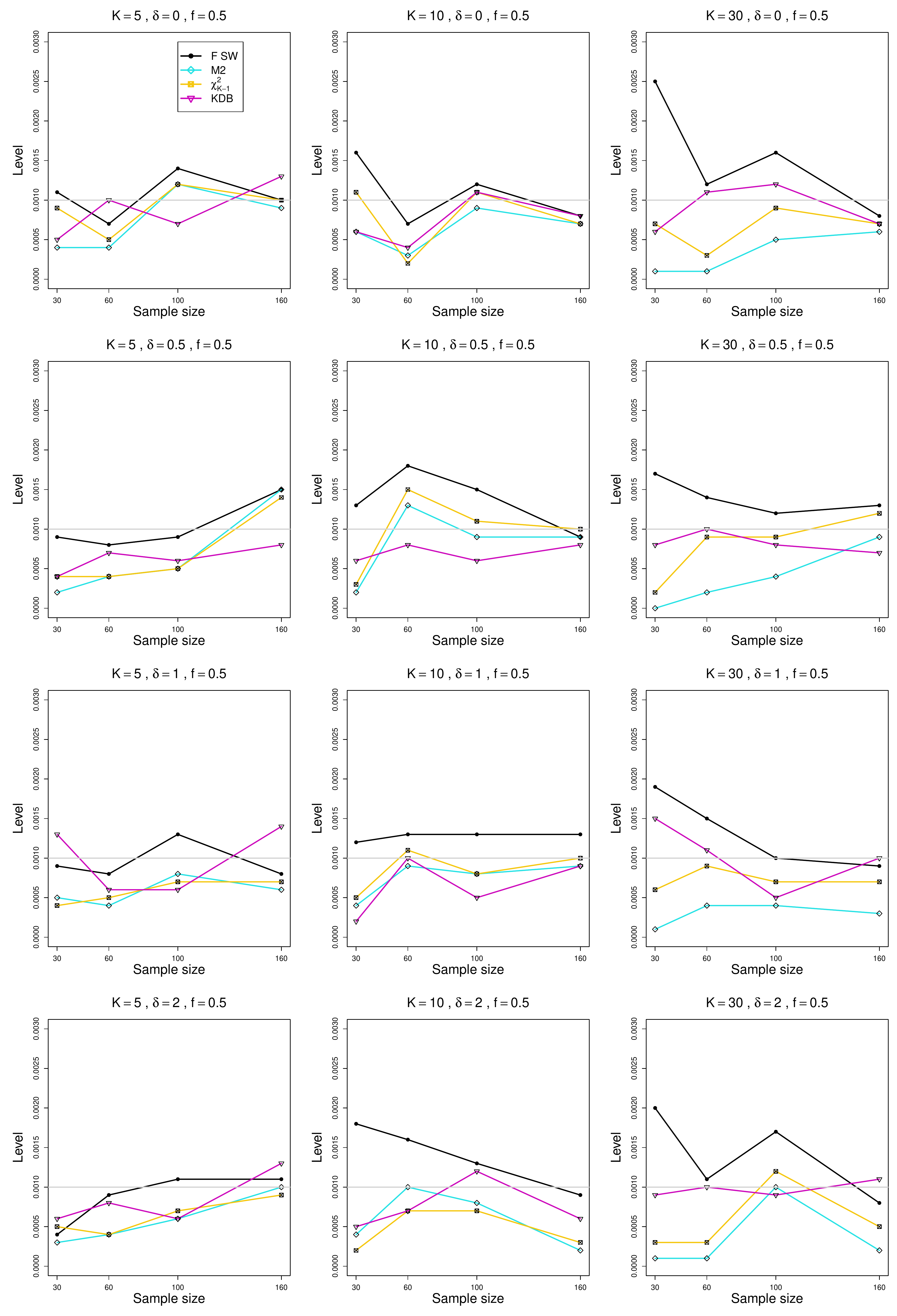}
	\caption{Empirical level at $\alpha = .001$ of test for heterogeneity vs average sample size
		\label{PlotOfPvaluesAgainstN_0001level_unequal_q05}}
\end{figure}

\begin{figure}[t]
	\centering
	\includegraphics[scale=0.33]{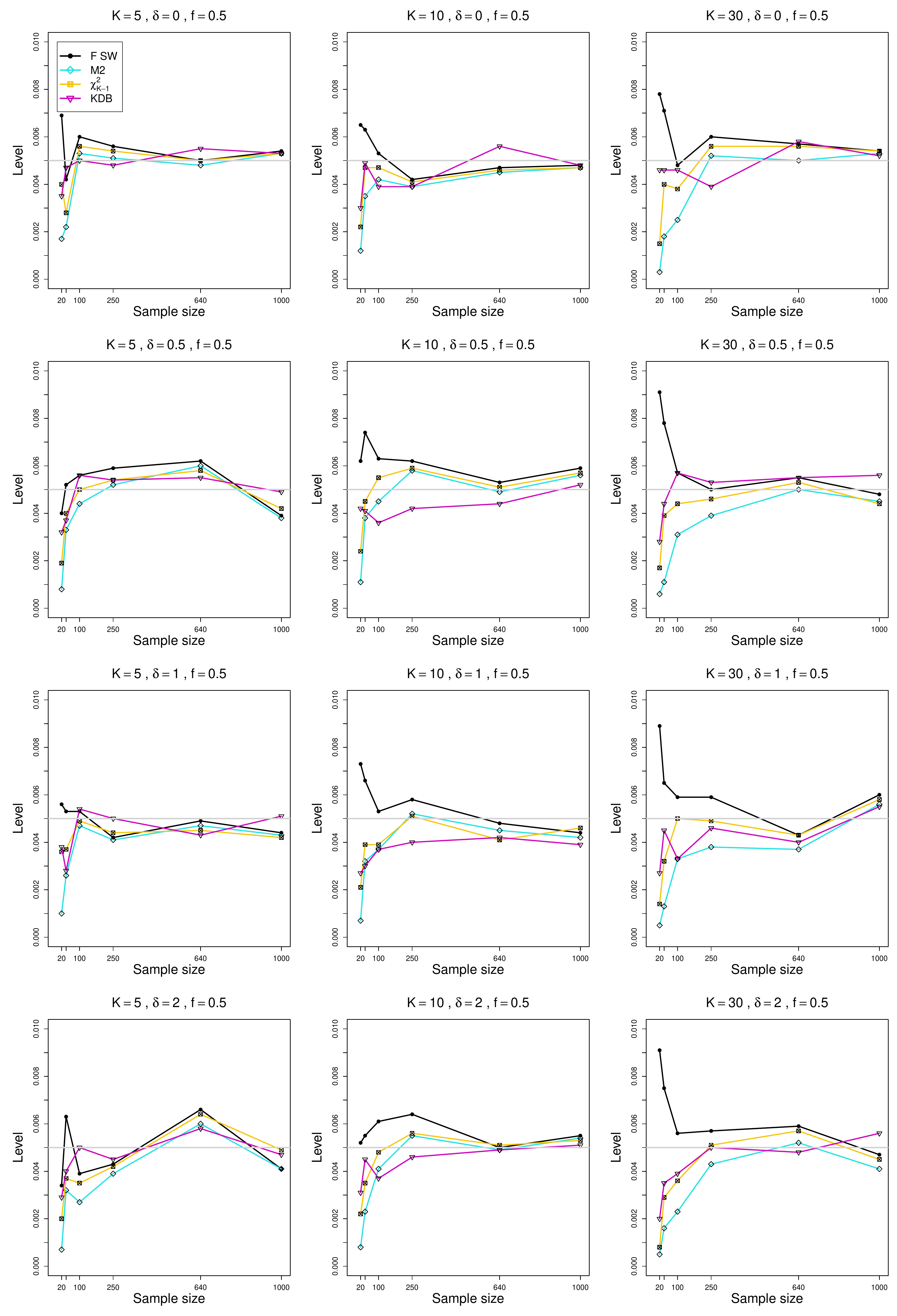}
	\caption{Empirical level at $\alpha = .005$ of test for heterogeneity vs sample size
		\label{PlotOfPvaluesAgainstN_0005level_q05}}
\end{figure}

\begin{figure}[t]
	\centering
	\includegraphics[scale=0.33]{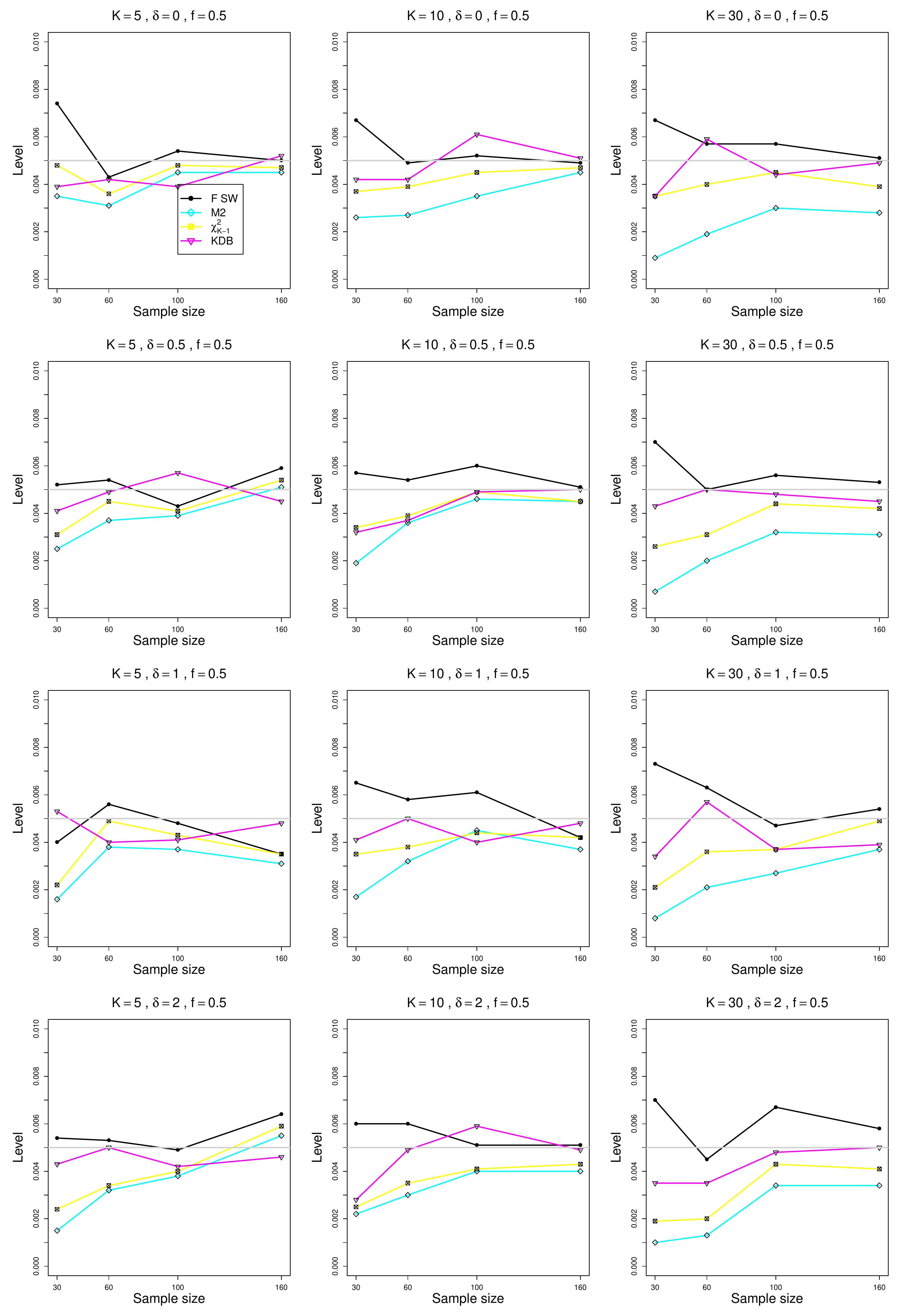}
	\caption{Empirical level at $\alpha = .005$ of test for heterogeneity vs average sample size
		\label{PlotOfPvaluesAgainstN_0005level_unequal_q05}}
\end{figure}

\begin{figure}[t]
	\centering
	\includegraphics[scale=0.33]{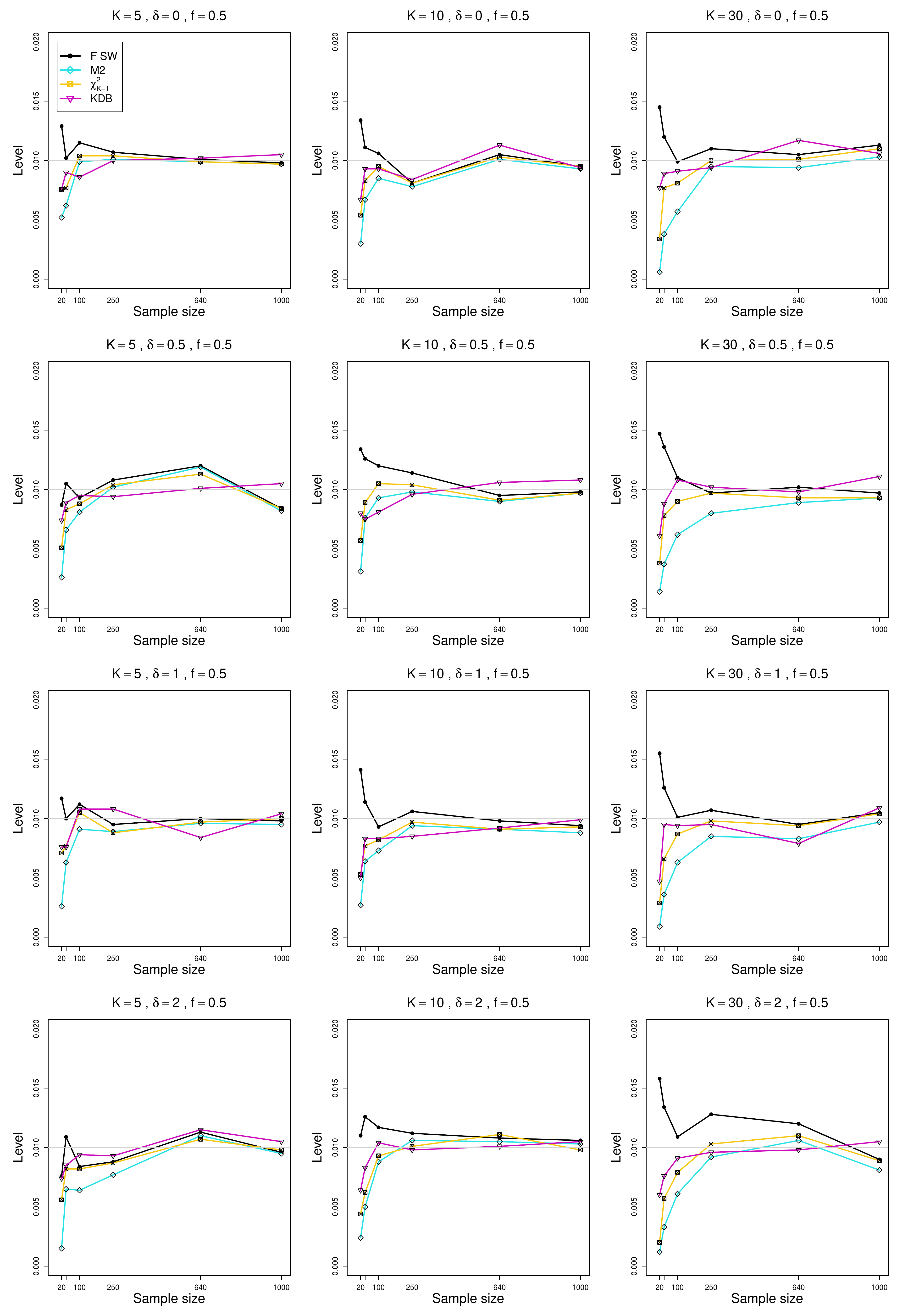}
	\caption{Empirical level at $\alpha = .01$ of test for heterogeneity vs sample size
		\label{PlotOfPvaluesAgainstN_001level_q05}}
\end{figure}

\begin{figure}[t]
	\centering
	\includegraphics[scale=0.33]{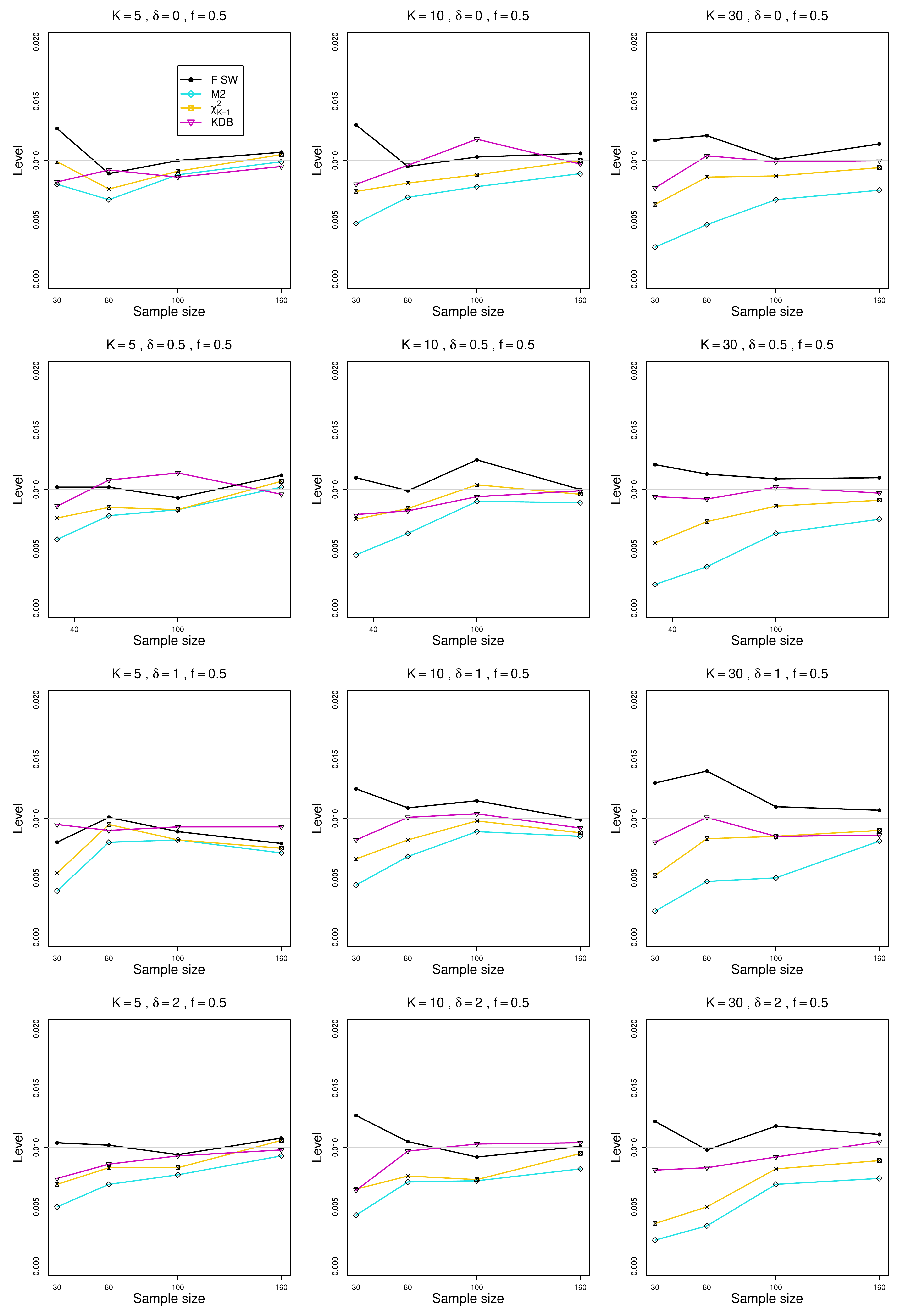}
	\caption{Empirical level at $\alpha = .01$ of test for heterogeneity vs average sample size
		\label{PlotOfPvaluesAgainstN_001level_unequal_q05}}
\end{figure}

\begin{figure}[t]
	\centering
	\includegraphics[scale=0.33]{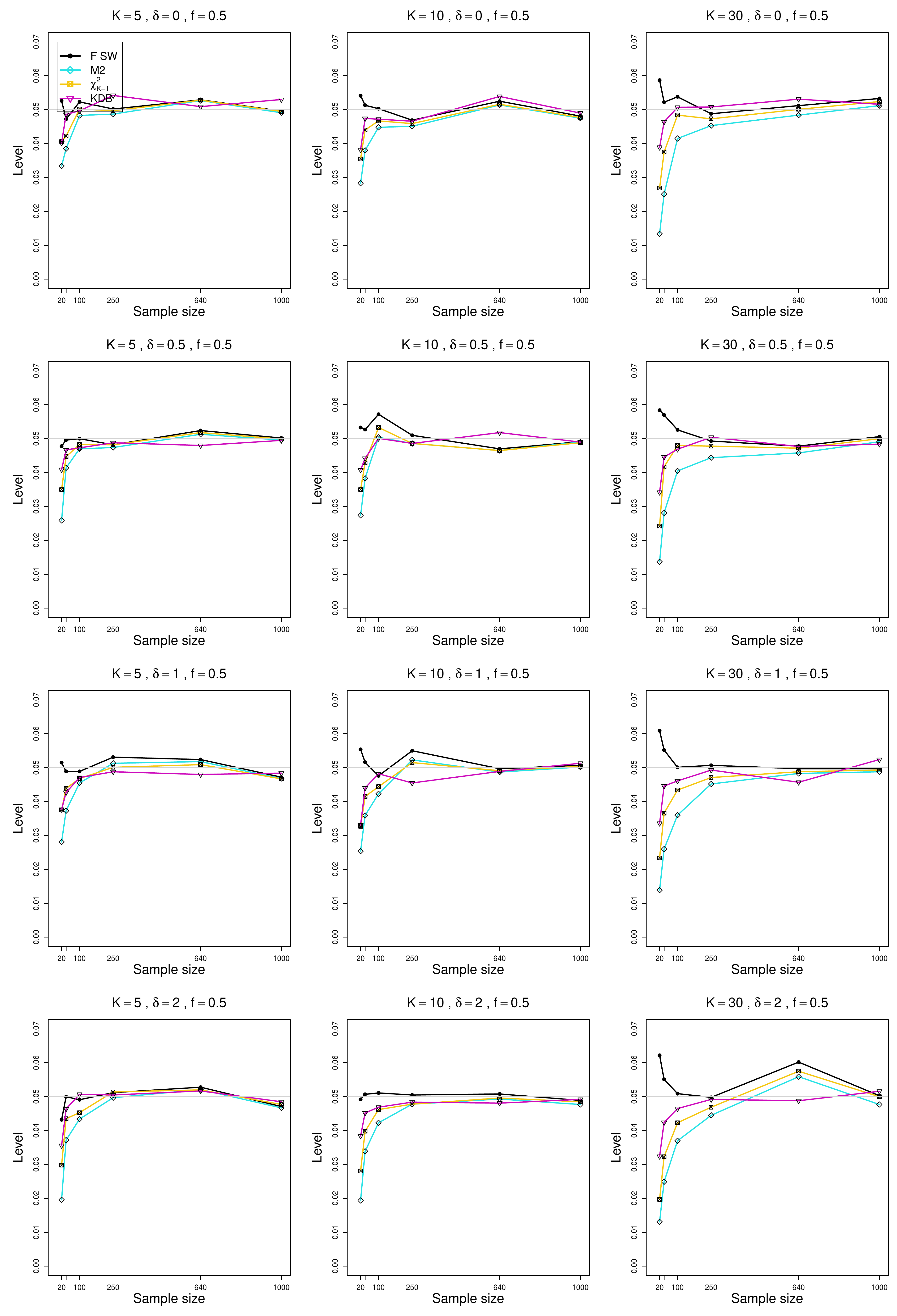}
	\caption{Empirical level at $\alpha = .05$ of test for heterogeneity vs sample size
		\label{PlotOfPvaluesAgainstN_005level_q05}}
\end{figure}

\begin{figure}[t]
	\centering
	\includegraphics[scale=0.33]{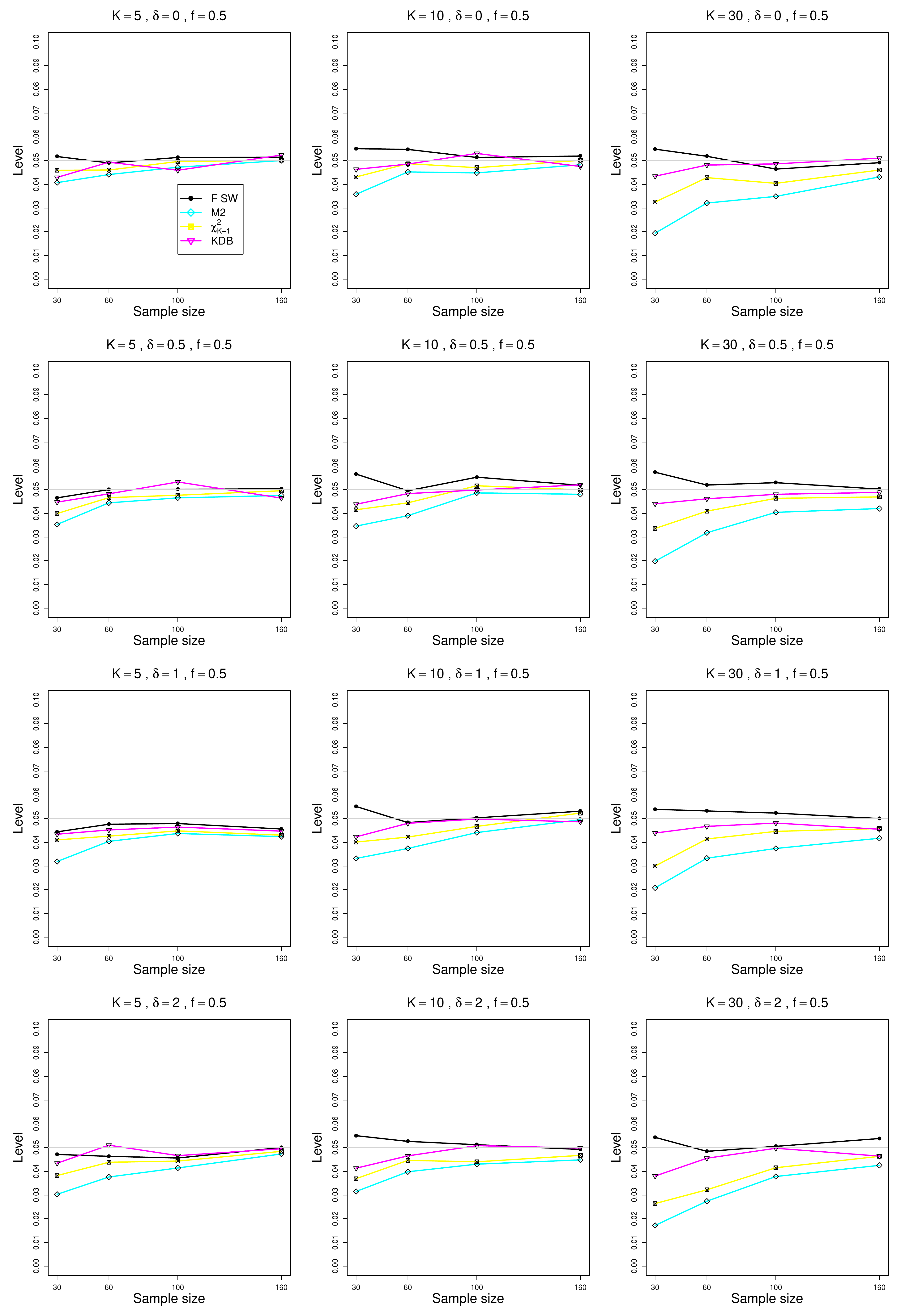}
	\caption{Empirical level at $\alpha = .05$ of test for heterogeneity vs average sample size
		\label{PlotOfPvaluesAgainstN_005level_unequal_q05}}
\end{figure}

\clearpage
\subsection*{C2. $f = .75$}
\setcounter{figure}{0}
\renewcommand{\thefigure}{C2.\arabic{figure}}
\clearpage

\begin{figure}[t]
	\centering
	\includegraphics[scale=0.33]{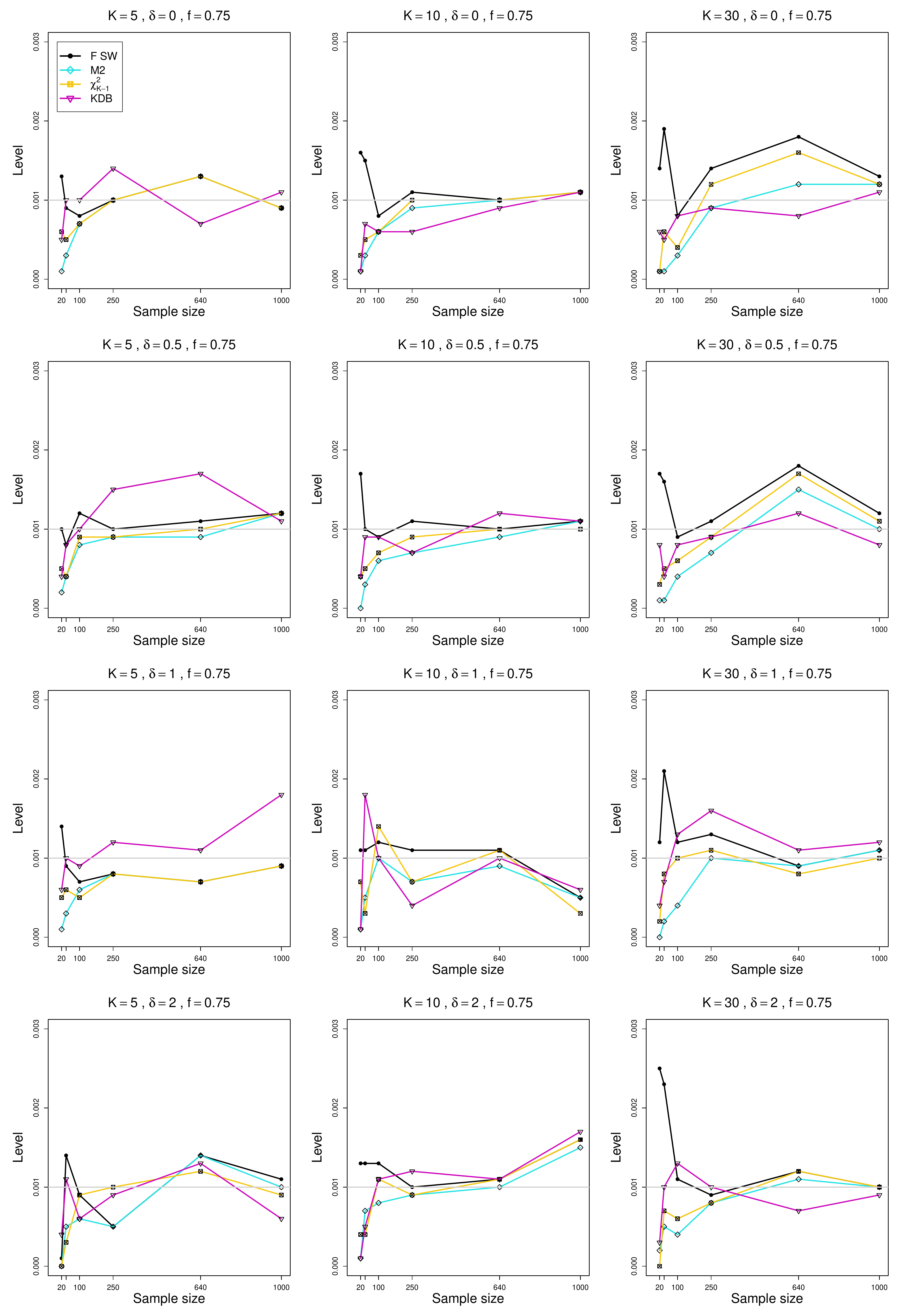}
	\caption{Empirical level at $\alpha = .001$ of test for heterogeneity vs sample size
		\label{PlotOfPvaluesAgainstN_0001level_q075}}
\end{figure}

\begin{figure}[t]
	\centering
	\includegraphics[scale=0.33]{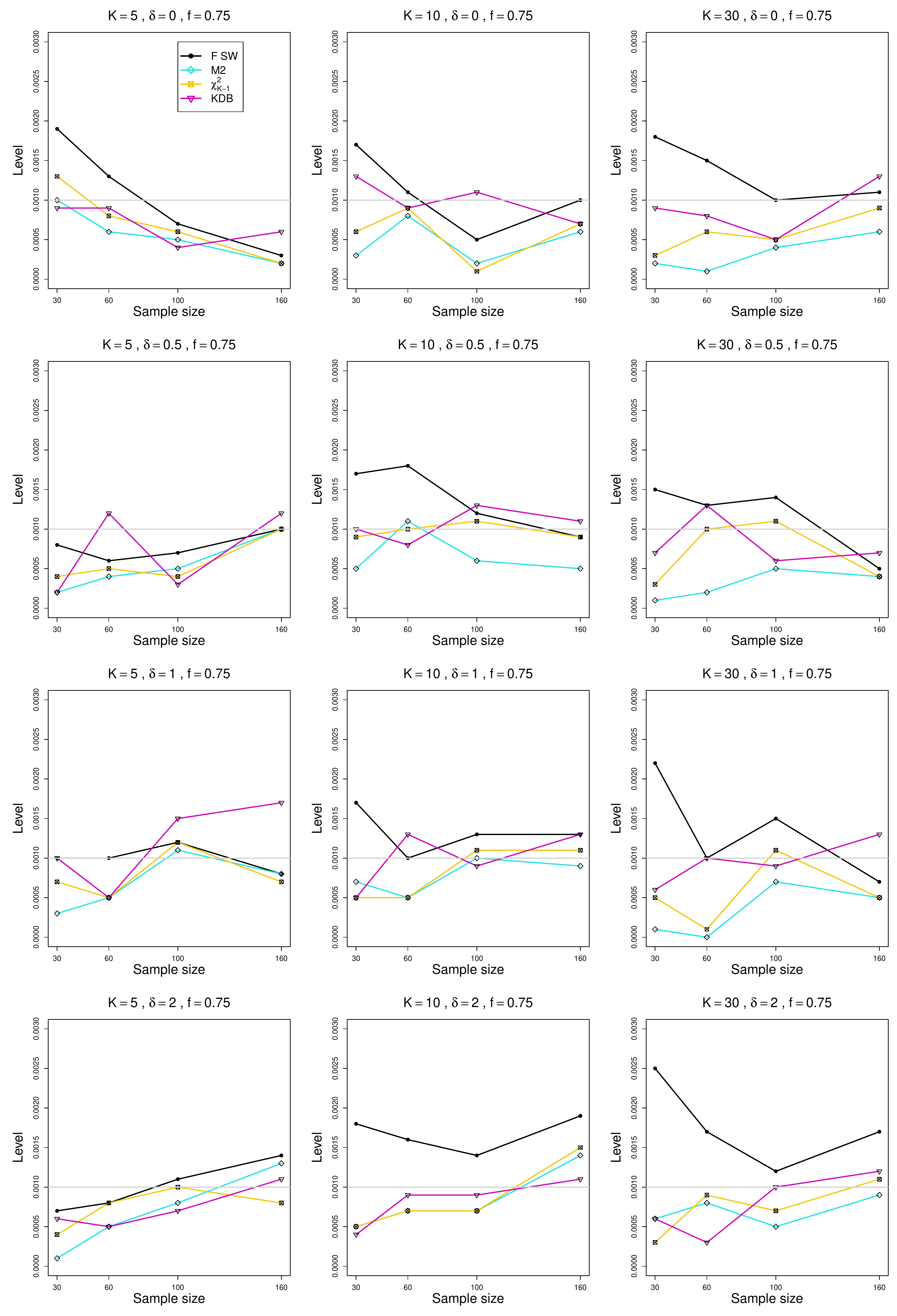}
	\caption{Empirical level at $\alpha = .001$ of test for heterogeneity vs average sample size
		\label{PlotOfPvaluesAgainstN_0001levelunequal_q075}}
\end{figure}

\begin{figure}[t]
	\centering
	\includegraphics[scale=0.33]{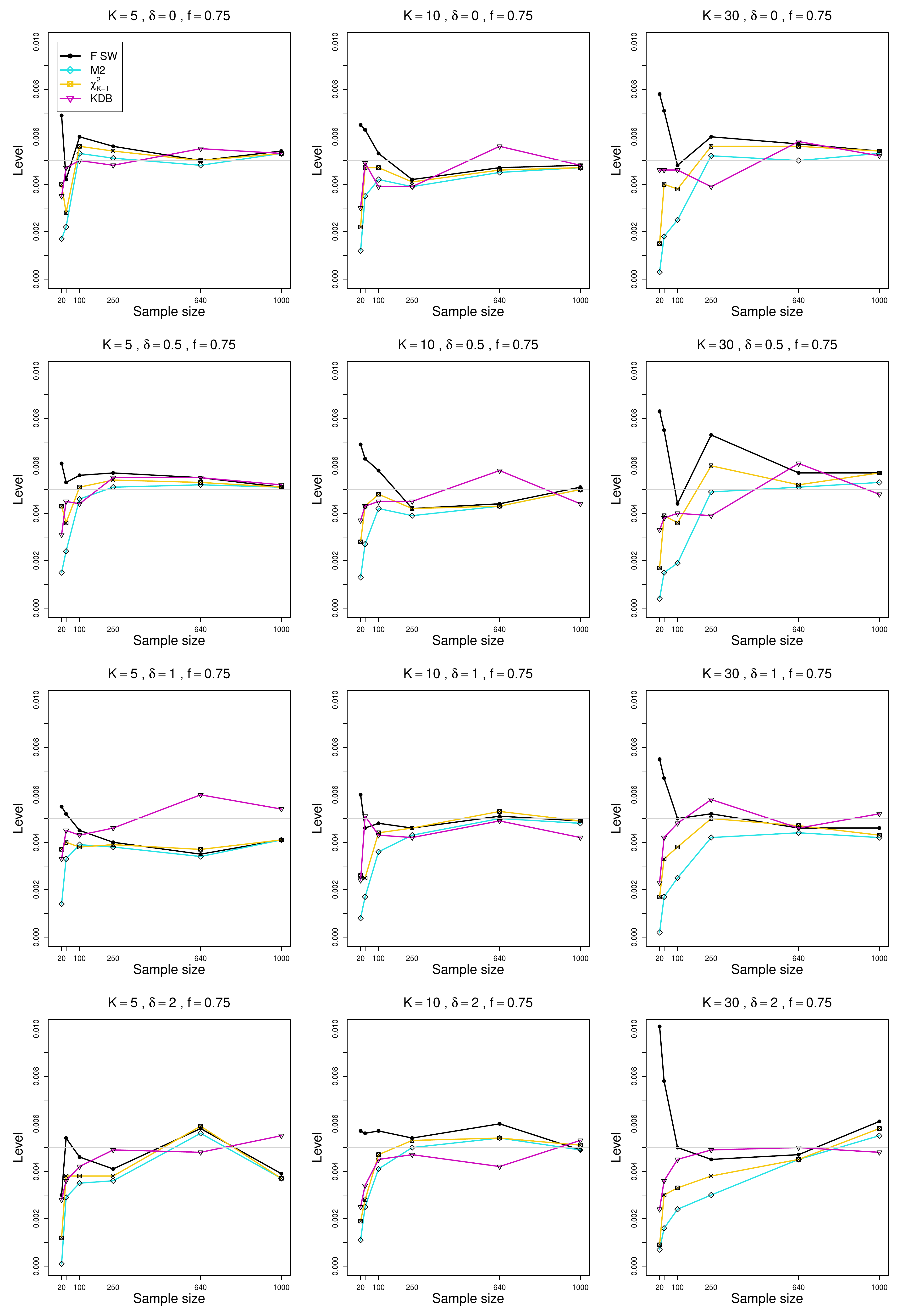}
	\caption{Empirical level at $\alpha = .005$ of test for heterogeneity vs sample size
		\label{PlotOfPvaluesAgainstN_0005level_q075}}
\end{figure}

\begin{figure}[t]
	\centering
	\includegraphics[scale=0.33]{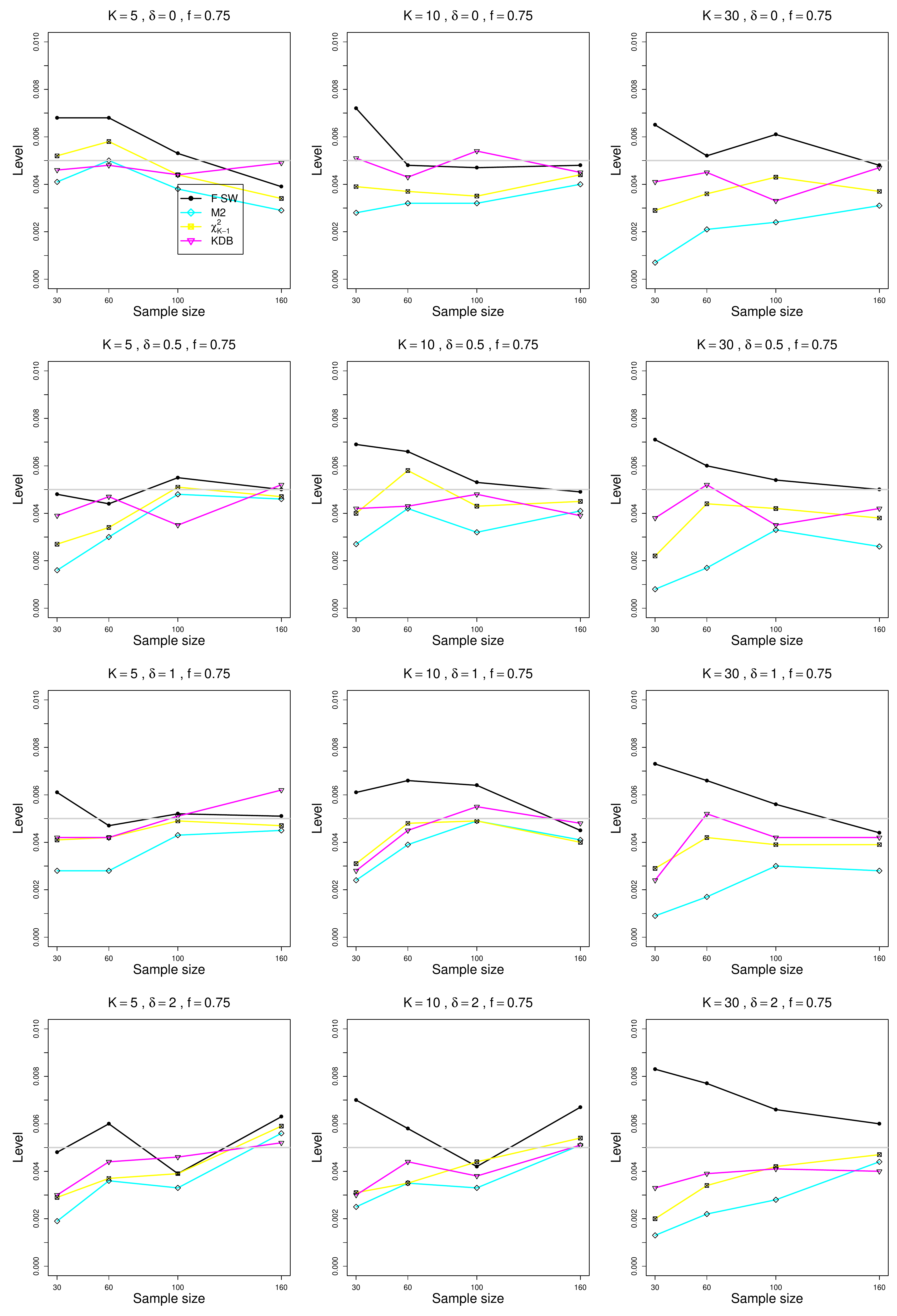}
	\caption{Empirical level at $\alpha = .005$ of test for heterogeneity vs average sample size
		\label{PlotOfPvaluesAgainstN_0005level_unequal_q075}}
\end{figure}

\begin{figure}[t]
	\centering
	\includegraphics[scale=0.33]{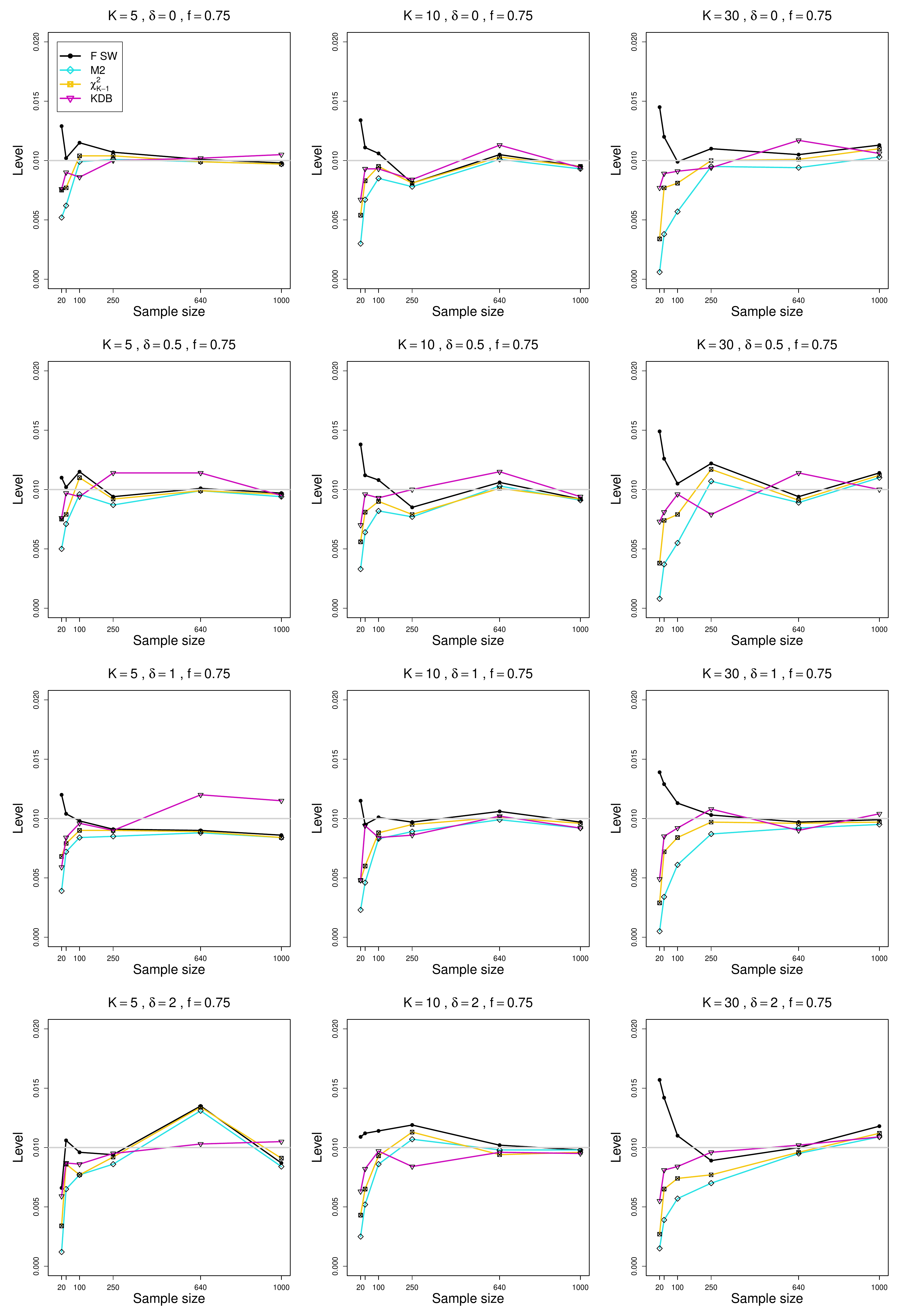}
	\caption{Empirical level at $\alpha = .01$ of test for heterogeneity vs sample size
		\label{PlotOfPvaluesAgainstN_001level_q075}}
\end{figure}

\begin{figure}[t]
	\centering
	\includegraphics[scale=0.33]{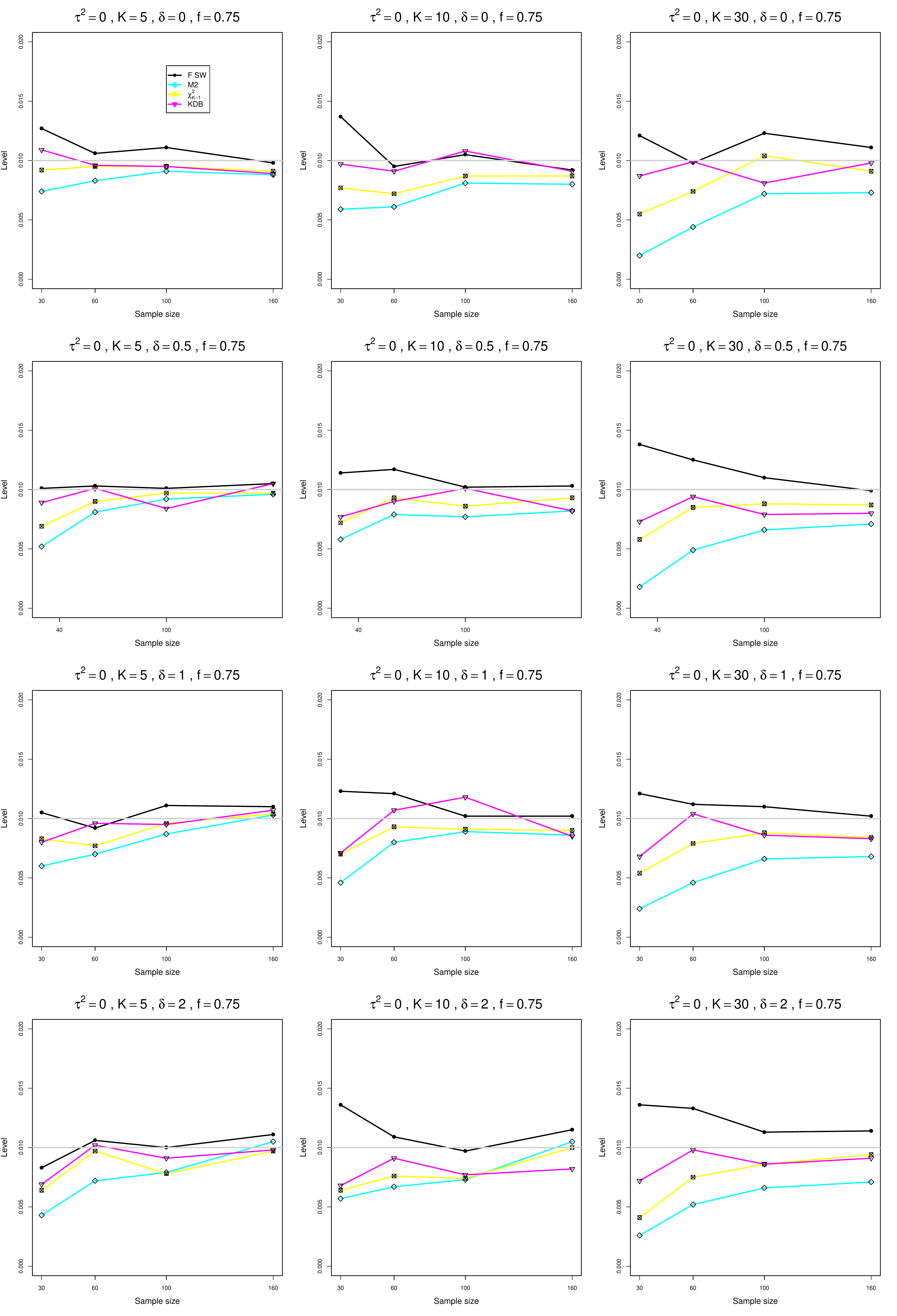}
	\caption{Empirical level at $\alpha = .01$ of test for heterogeneity vs average sample size
		\label{PlotOfPvaluesAgainstN_001level_unequal_q075}}
\end{figure}

\begin{figure}[t]
	\centering
	\includegraphics[scale=0.33]{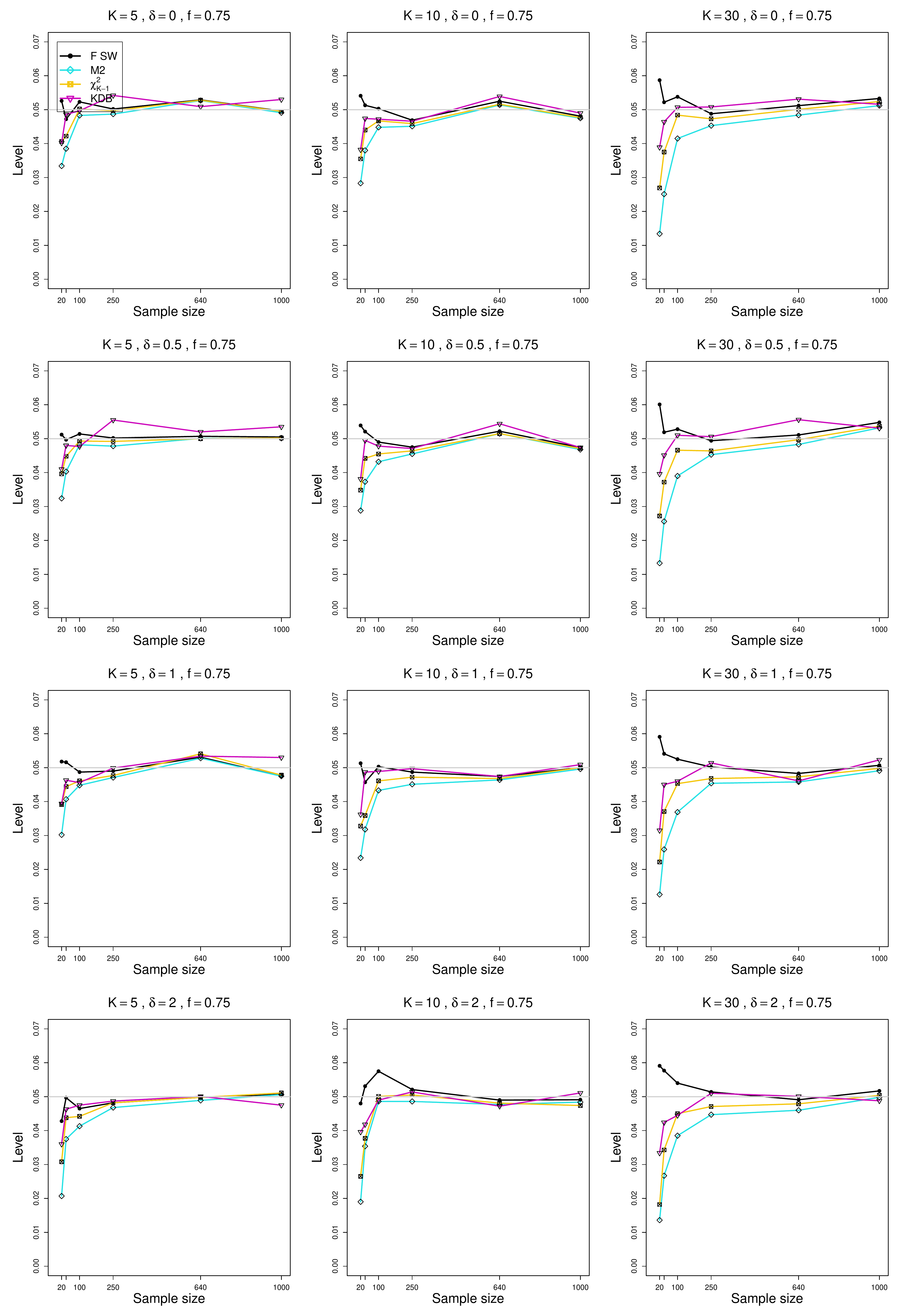}
	\caption{Empirical level at $\alpha = .05$ of test for heterogeneity vs sample size
		\label{PlotOfPvaluesAgainstN_005level_q075}}
\end{figure}

\begin{figure}[t]
	\centering
	\includegraphics[scale=0.33]{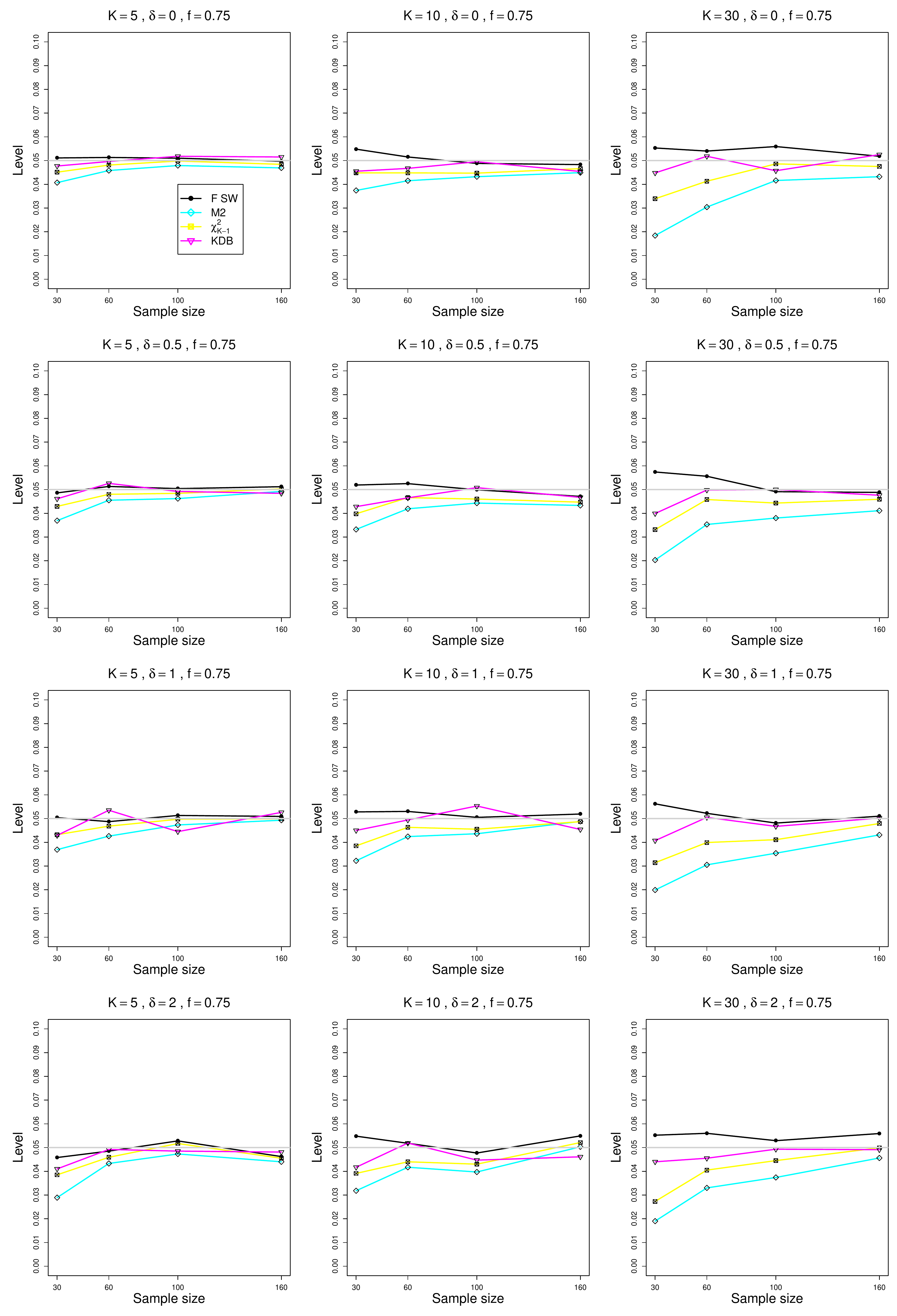}
	\caption{Empirical level at $\alpha = .05$ of test for heterogeneity vs average sample size
		\label{PlotOfPvaluesAgainstN_005level_unequal_q075}}
\end{figure}

\clearpage
\setcounter{figure}{0}
\setcounter{section}{0}
\renewcommand{\thesection}{D.\arabic{section}}

\section*{D. Empirical level of the test for heterogeneity ($\tau^2 \leq \tau_{0}^2$ versus $\tau^2 > \tau_{0}^2$) based on approximations for the distribution of $Q$, plotted vs $\tau_{0}^2$}

In sets of figures for $\alpha = .01$ and $\alpha =  .05$, each figure corresponds to a value of $\delta$, a value of $f$, and either equal sample sizes or unequal sample sizes. \\
For each combination of a value of $n$ or $\bar{n}$ and a value of $K$, a panel plots the empirical level versus $\tau_{0}^2$. \\
The approximations for the distribution of $Q$ are
\begin{itemize}
	\item F SW (Farebrother approximation, effective-sample-size weights)
	\item M2 SW (Two-moment approximation, effective-sample-size weights)
	\item BJ (Biggerstaff and Jackson approximation, IV weights)
\end{itemize}

\clearpage
\renewcommand{\thefigure}{D1.\arabic{figure}}
%alpha 0.01 under H1

\subsection*{D1. Empirical level at $\alpha = .01$}

%q=0.5
\begin{figure}[t]
	\centering
	\includegraphics[scale=0.33]{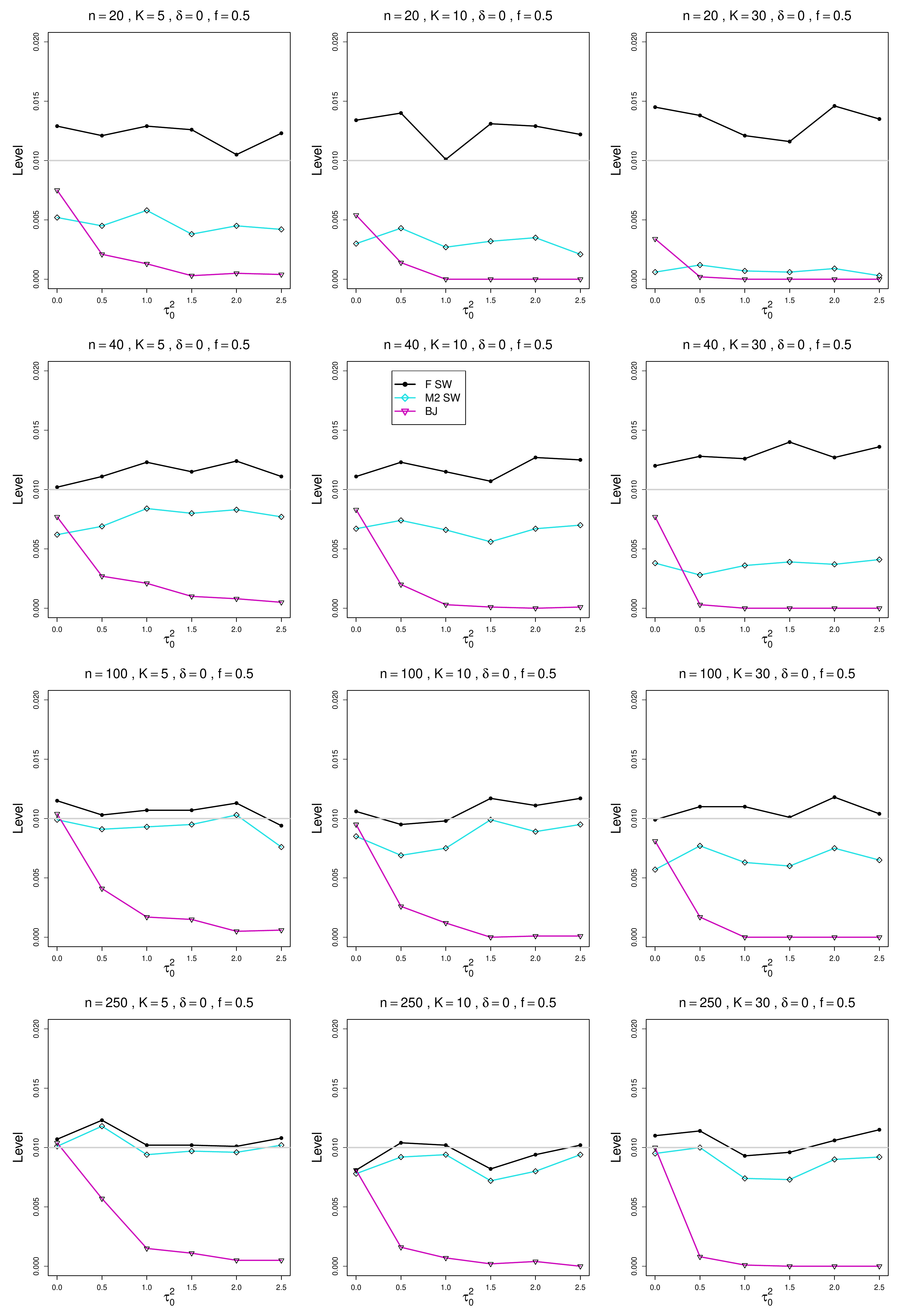}
	\caption{Empirical level for $\delta = 0$, $f = .5$, and equal sample sizes
		\label{PlotOfPhatAt001delta0andq05SMD_underH1}}
\end{figure}

\begin{figure}[t]
	\centering
	\includegraphics[scale=0.33]{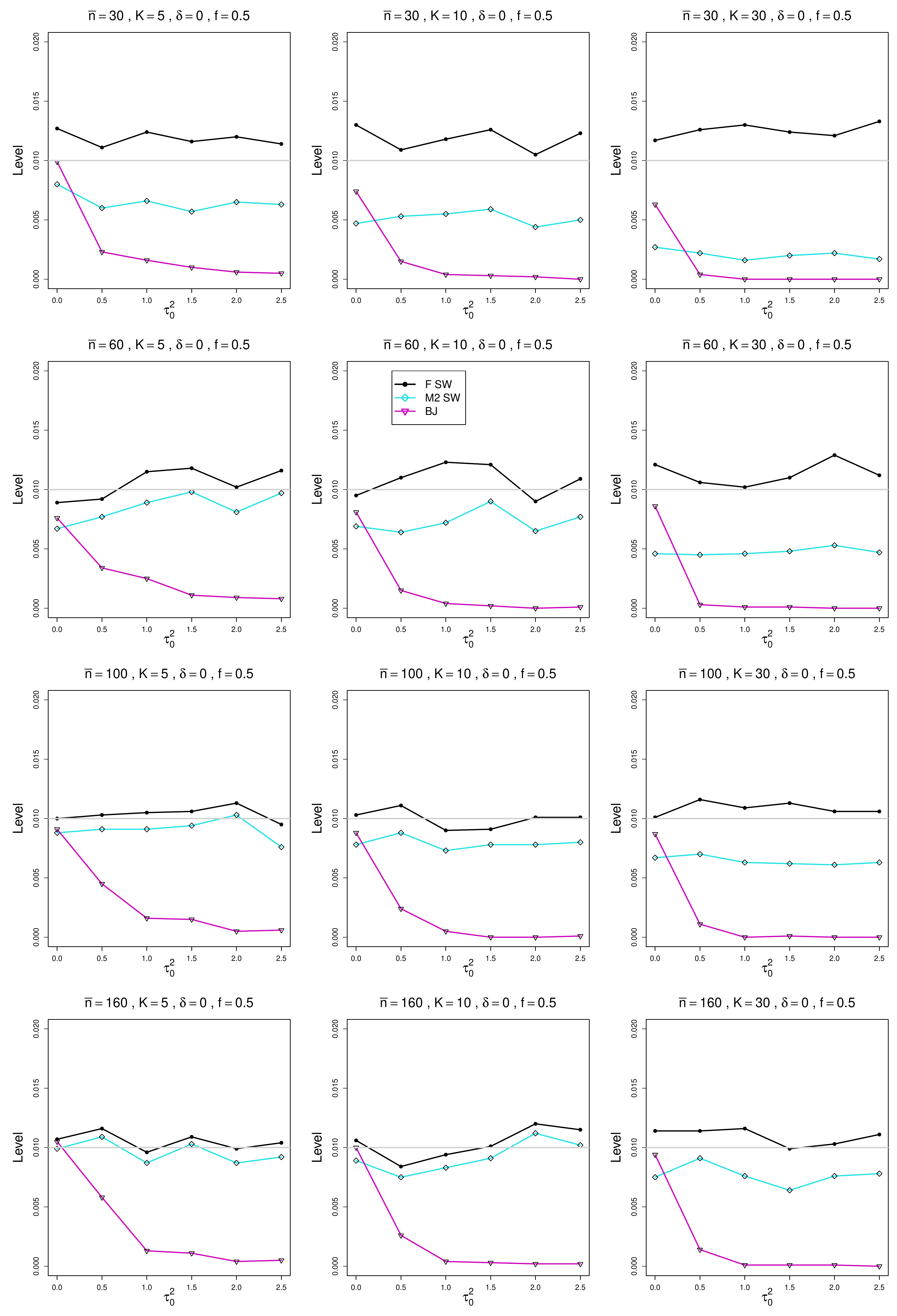}
	\caption{Empirical level for $\delta = 0$, $f = .5$, and unequal sample sizes
		\label{PlotOfPhatAt001delta0andq05SMD_underH1_unequal}}
\end{figure}

\begin{figure}[t]
	\centering
	\includegraphics[scale=0.33]{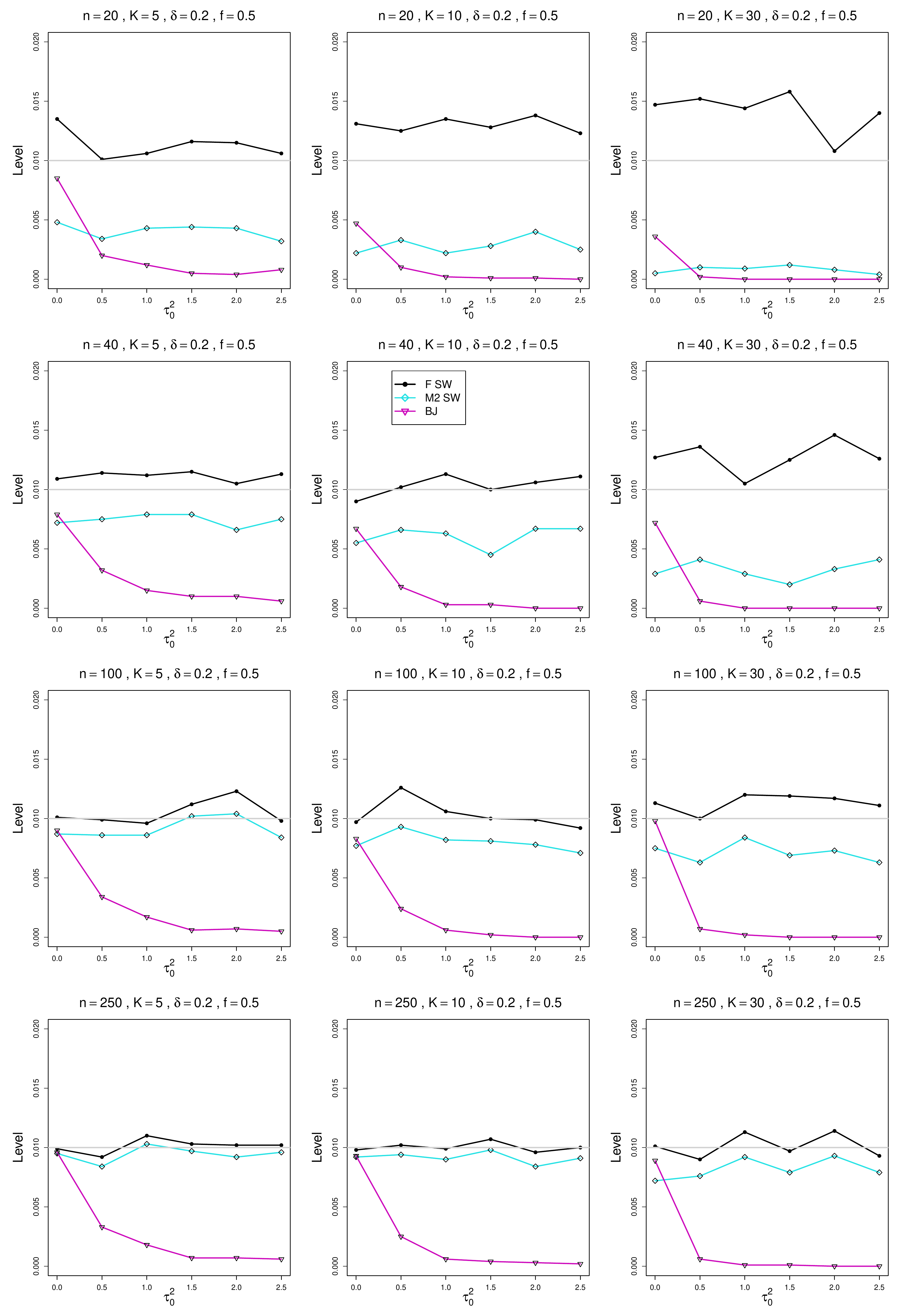}
	\caption{Empirical level for $\delta = 0.2$, $f = .5$, and equal sample sizes
		\label{PlotOfPhatAt001delta02andq05SMD_underH1}}
\end{figure}

\begin{figure}[t]
	\centering
	\includegraphics[scale=0.33]{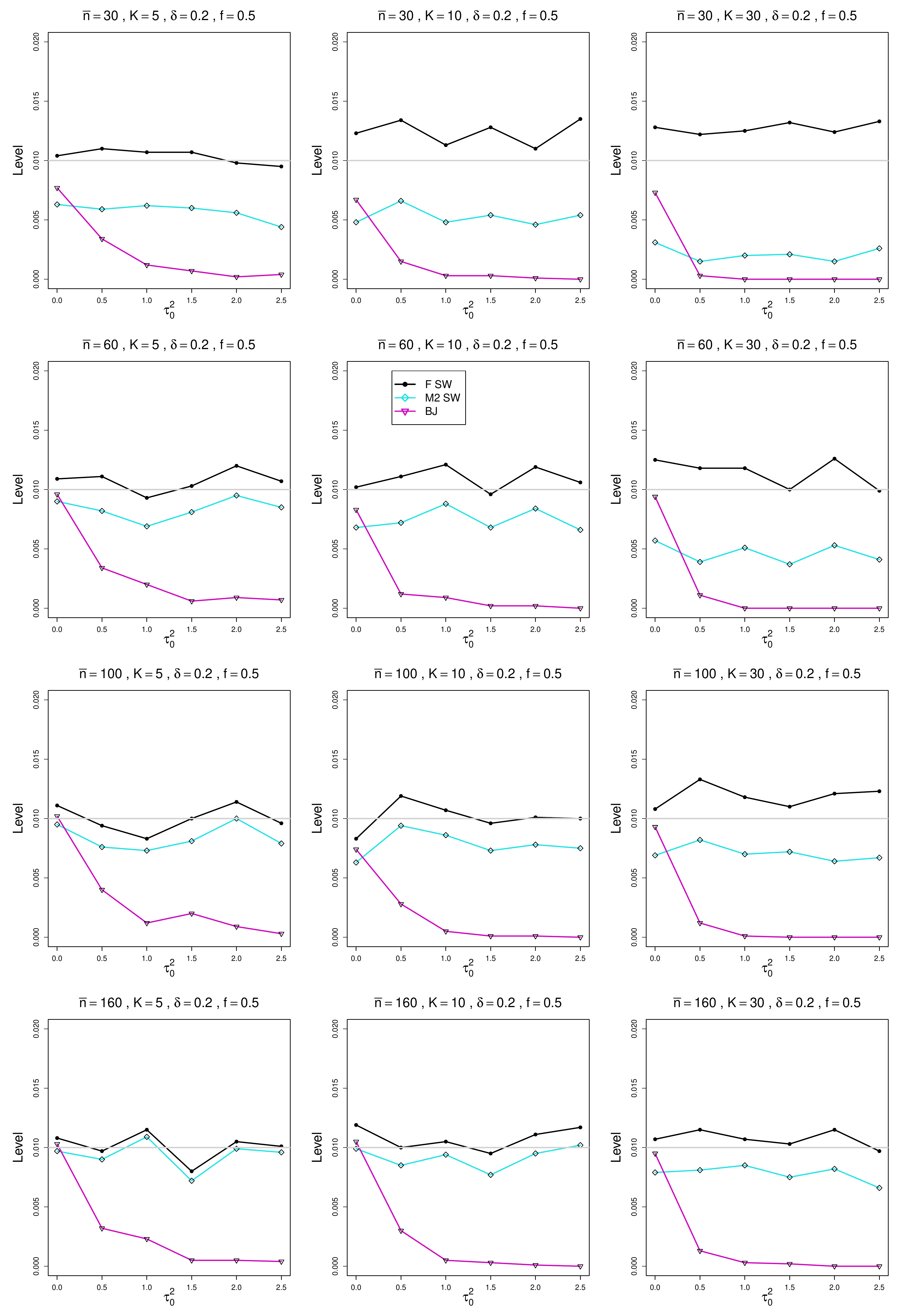}
	\caption{Empirical level for $\delta = 0.2$, $f = .5$, and unequal sample sizes
		\label{PlotOfPhatAt001delta02andq05SMD_underH1_unequal}}
\end{figure}

\begin{figure}[t]
	\centering
	\includegraphics[scale=0.33]{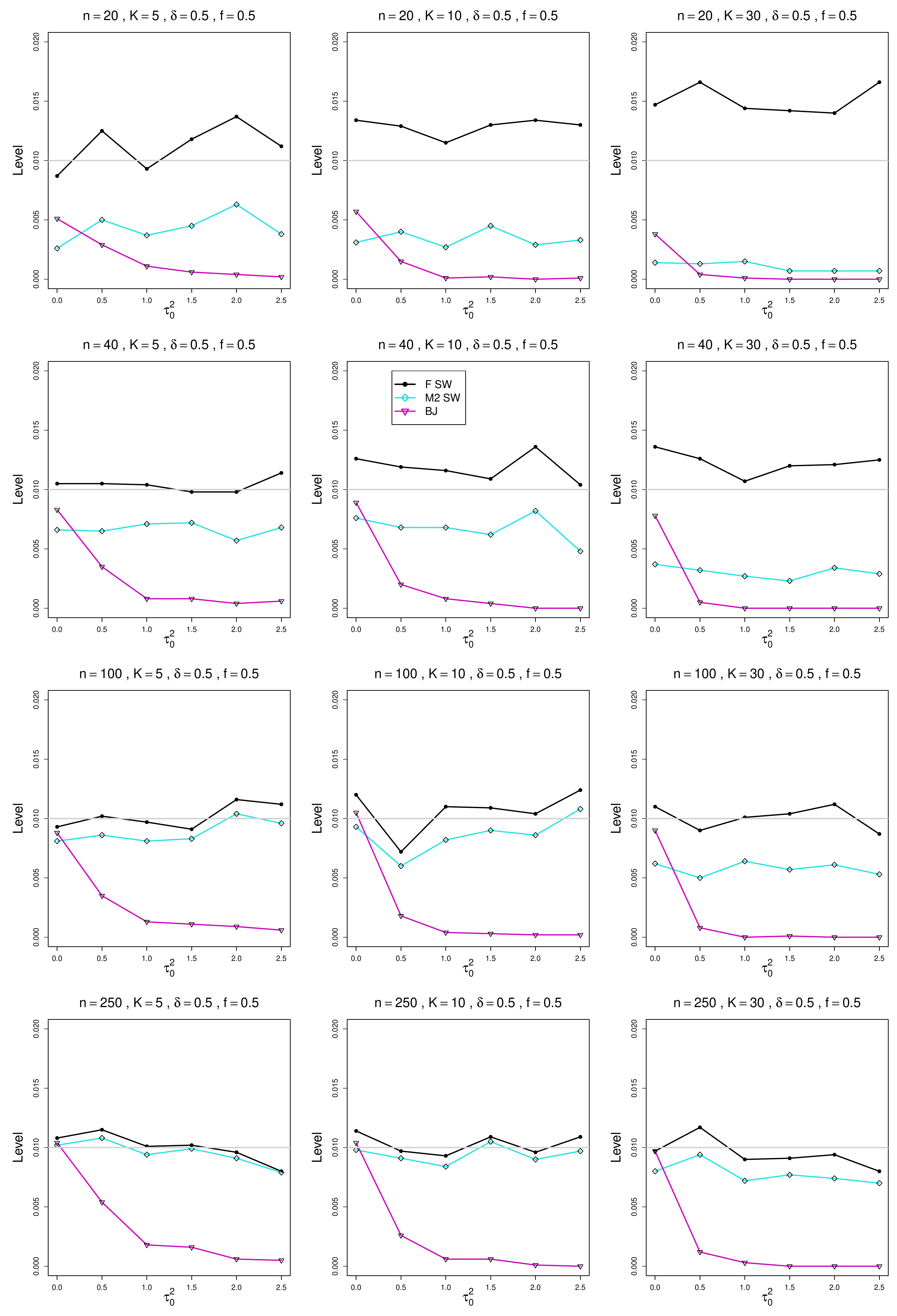}
	\caption{Empirical level for $\delta = 0.5$, $f = .5$, and equal sample sizes
		\label{PlotOfPhatAt001delta05andq05SMD_underH1}}
\end{figure}

\begin{figure}[t]
	\centering
	\includegraphics[scale=0.33]{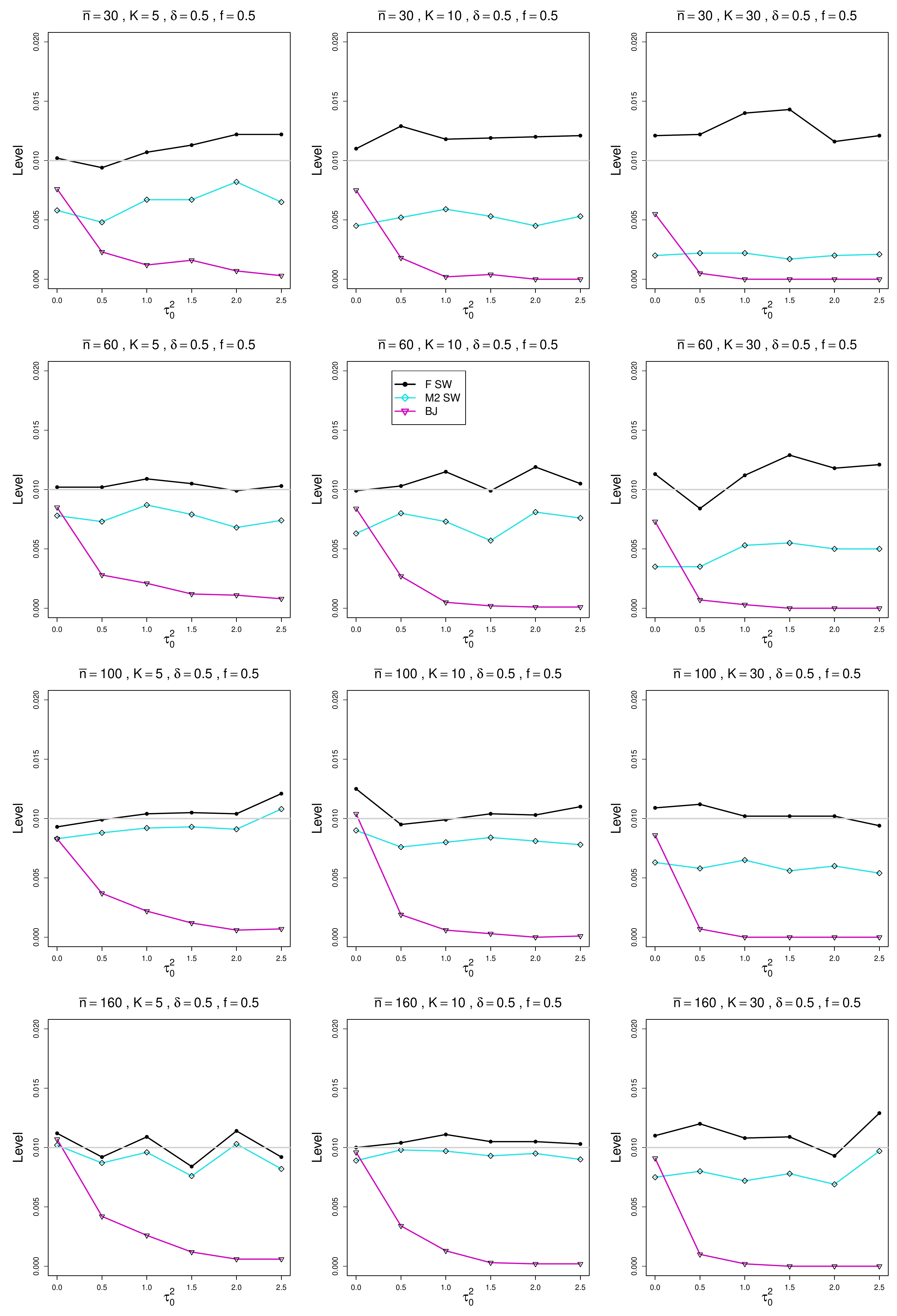}
	\caption{Empirical level for $\delta = 0.5$, $f = .5$, and unequal sample sizes
		\label{PlotOfPhatAt001delta05andq05SMD_underH1_unequal}}
\end{figure}

\begin{figure}[t]
	\centering
	\includegraphics[scale=0.33]{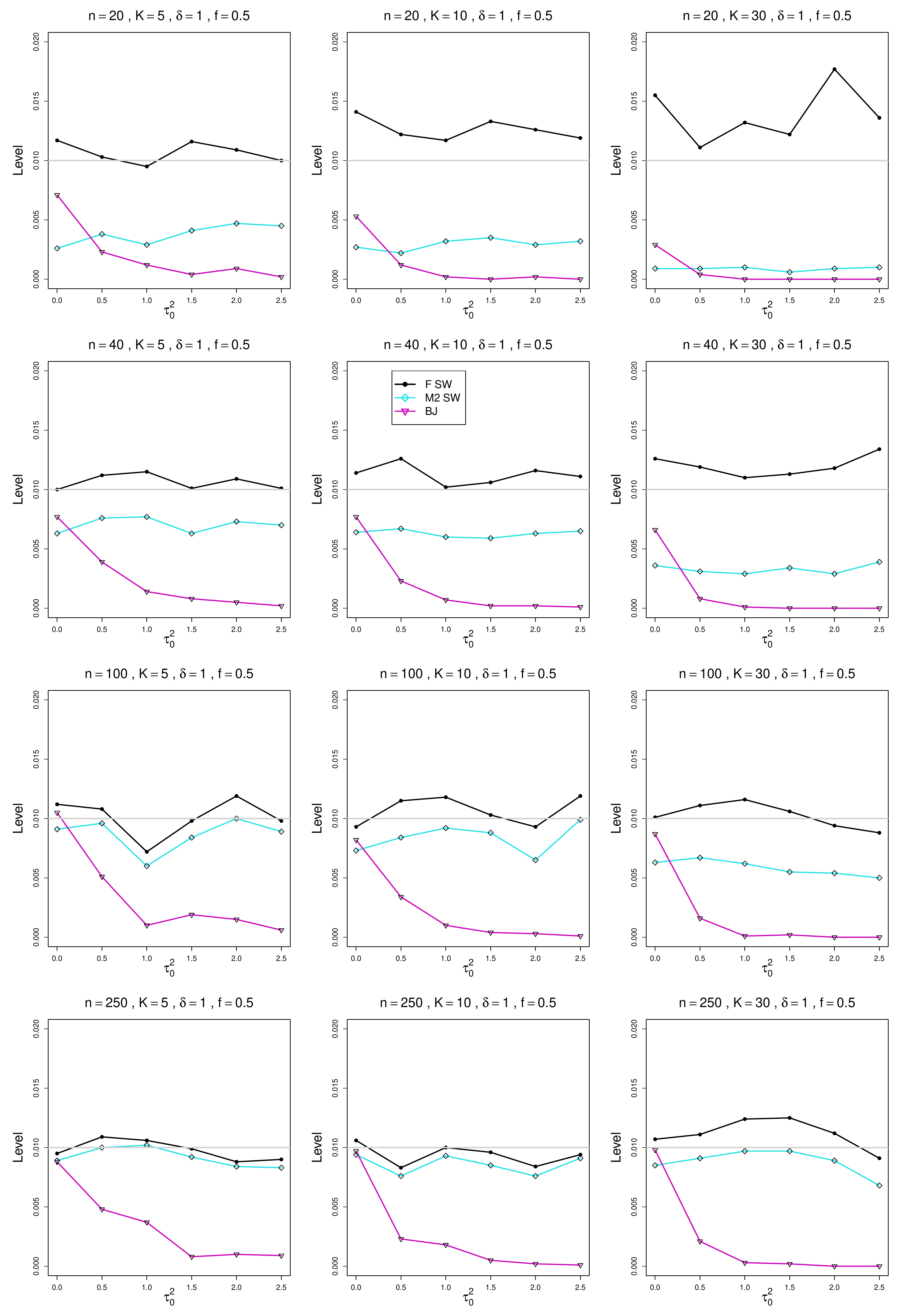}
	\caption{Empirical level for $\delta = 1$, $f = .5$, and equal sample sizes
		\label{PlotOfPhatAt001delta1andq05SMD_underH1}}
\end{figure}

\begin{figure}[t]
	\centering
	\includegraphics[scale=0.33]{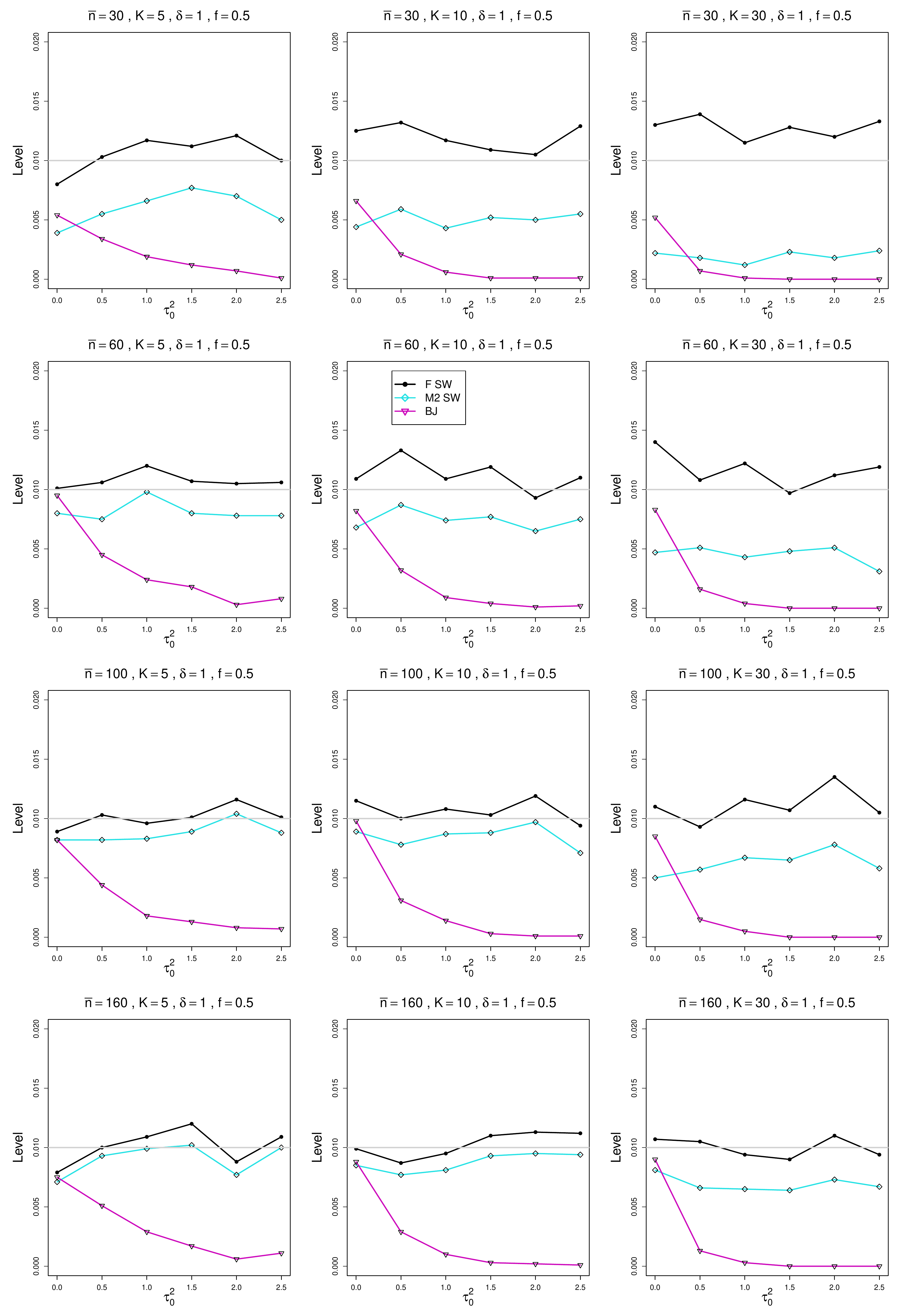}
	\caption{Empirical level for $\delta = 1$, $f = .5$, and unequal sample sizes
		\label{PlotOfPhatAt001delta1andq05SMD_underH1_unequal}}
\end{figure}

\begin{figure}[t]
	\centering
	\includegraphics[scale=0.33]{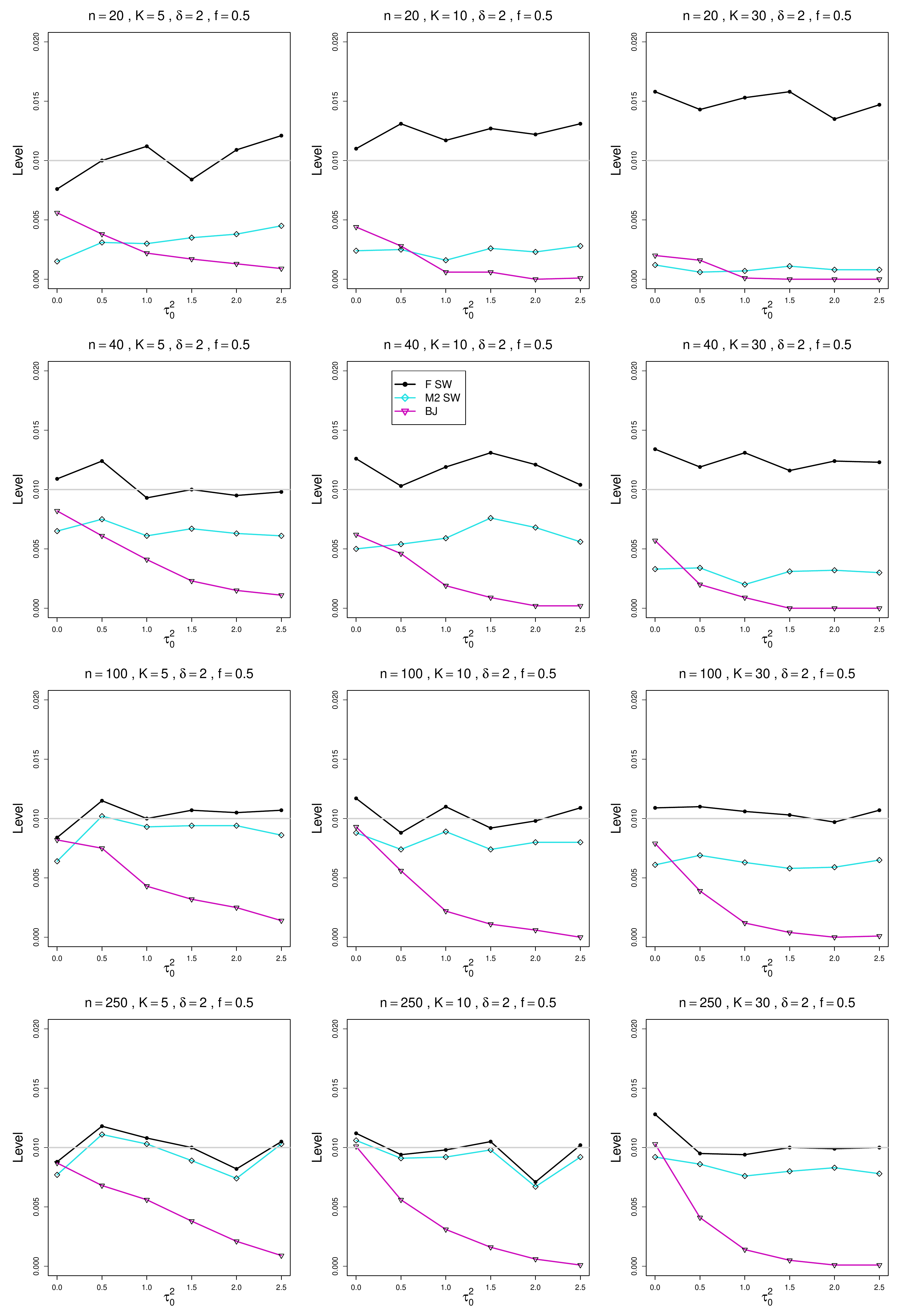}
	\caption{Empirical level for $\delta = 2$, $f = .5$, and equal sample sizes
		\label{PlotOfPhatAt001delta2andq05SMD_underH1}}
\end{figure}

\begin{figure}[t]
	\centering
	\includegraphics[scale=0.33]{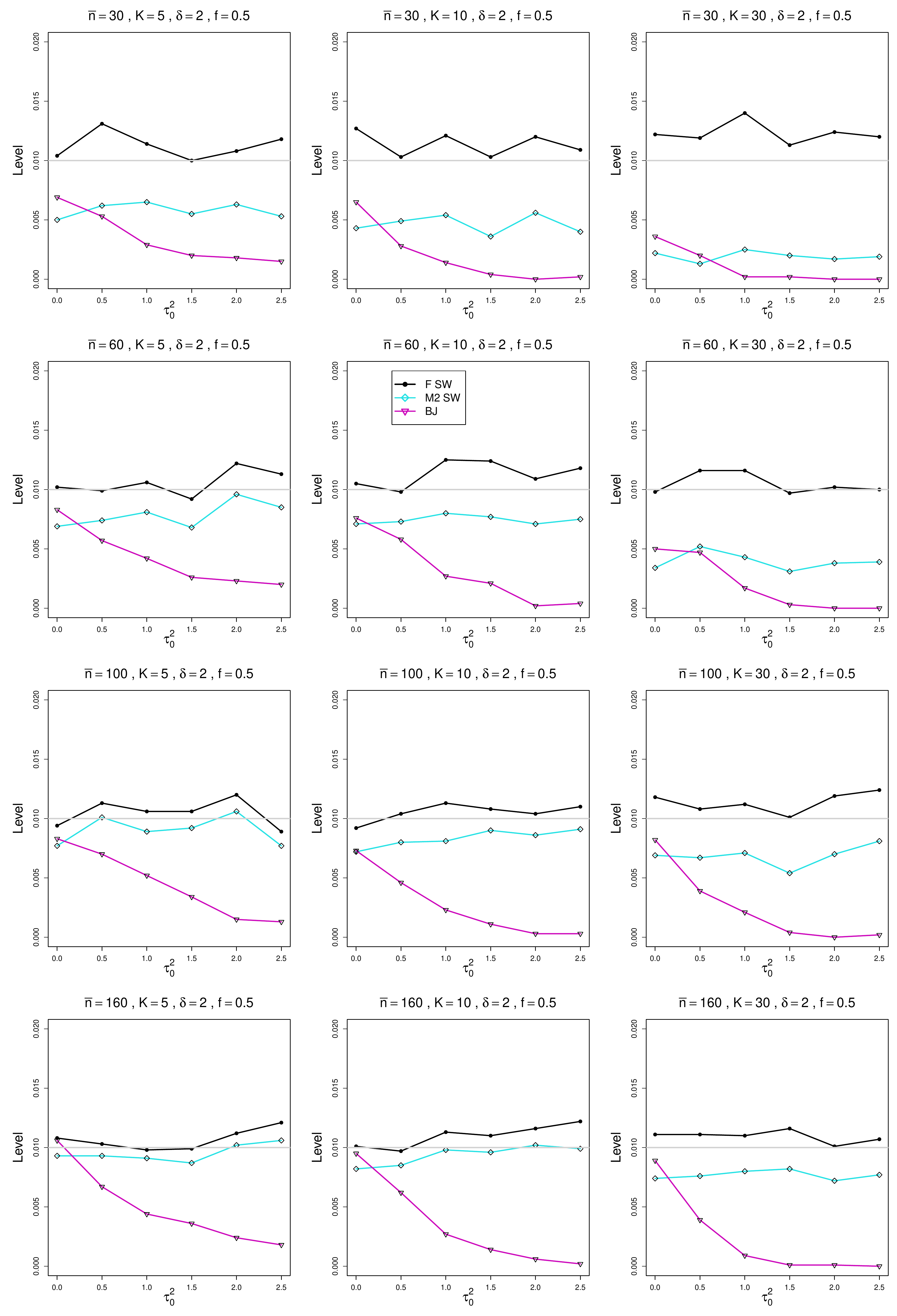}
	\caption{Empirical level for $\delta = 2$, $f = .5$, and unequal sample sizes
		\label{PlotOfPhatAt001delta2andq05SMD_underH1_unequal}}
\end{figure}

\begin{figure}[t]
	\centering
	\includegraphics[scale=0.33]{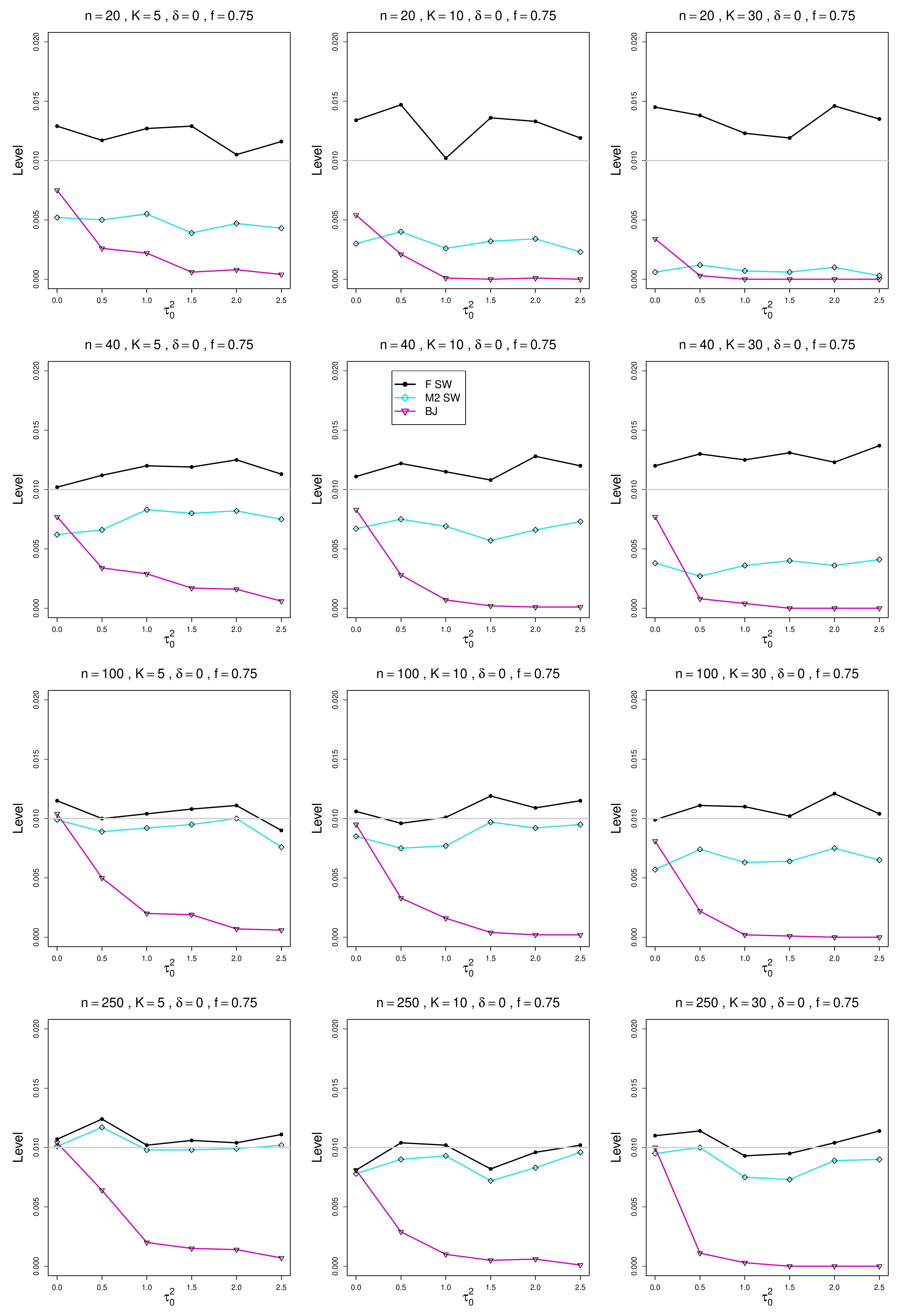}
	\caption{Empirical level for $\delta = 0$, $f = .75$, and equal sample sizes
		\label{PlotOfPhatAt001delta0andq075SMD_underH1}}
\end{figure}

\begin{figure}[t]
	\centering
	\includegraphics[scale=0.33]{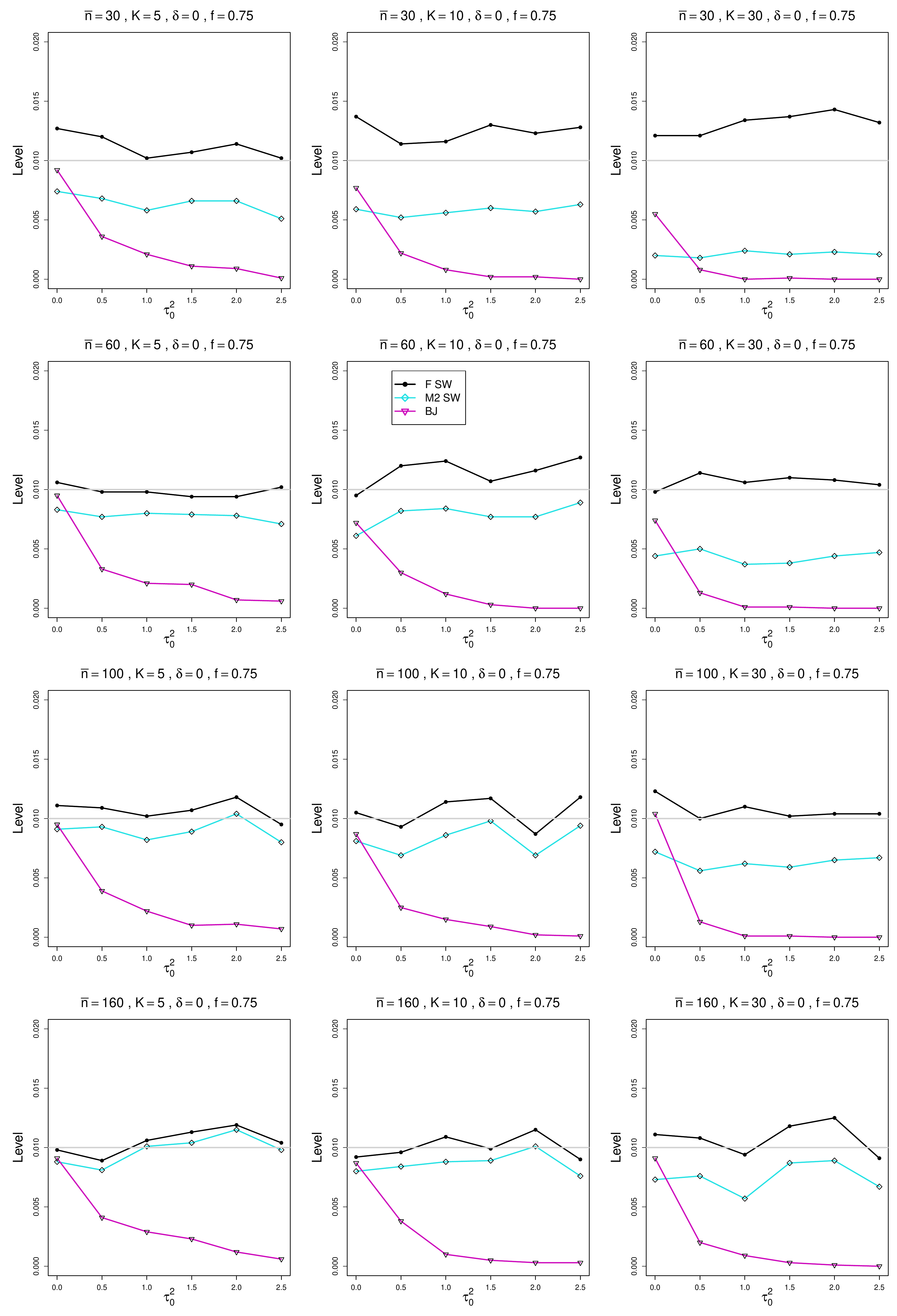}
	\caption{Empirical level for $\delta = 0$, $f = .75$, and unequal sample sizes
		\label{PlotOfPhatAt001delta0andq075SMD_underH1_unequal}}
\end{figure}

\begin{figure}[t]
	\centering
	\includegraphics[scale=0.33]{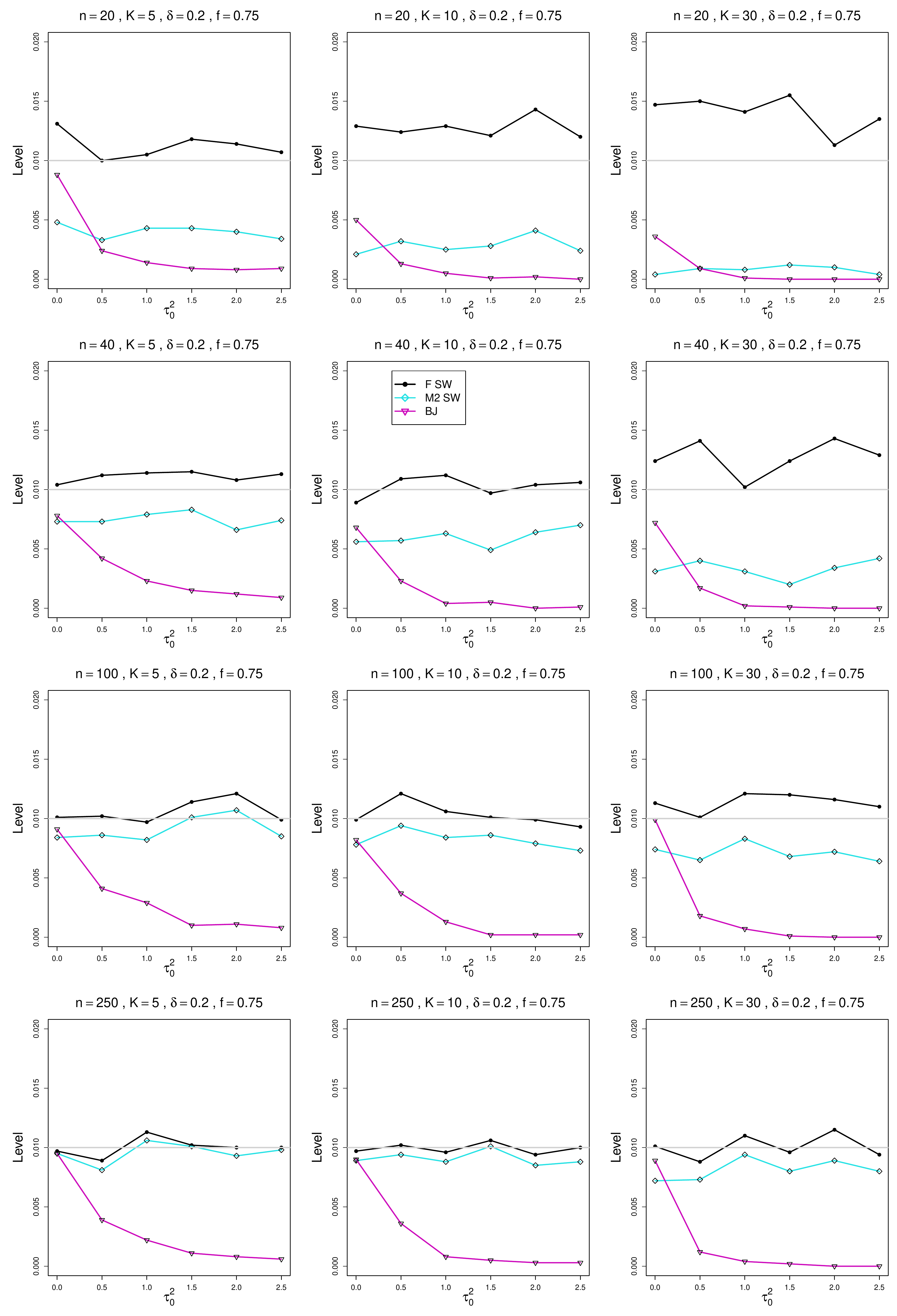}
	\caption{Empirical level for $\delta = 0.2$, $f = .75$, and equal sample sizes
		\label{PlotOfPhatAt001delta02andq075SMD_underH1}}
\end{figure}

\begin{figure}[t]
	\centering
	\includegraphics[scale=0.33]{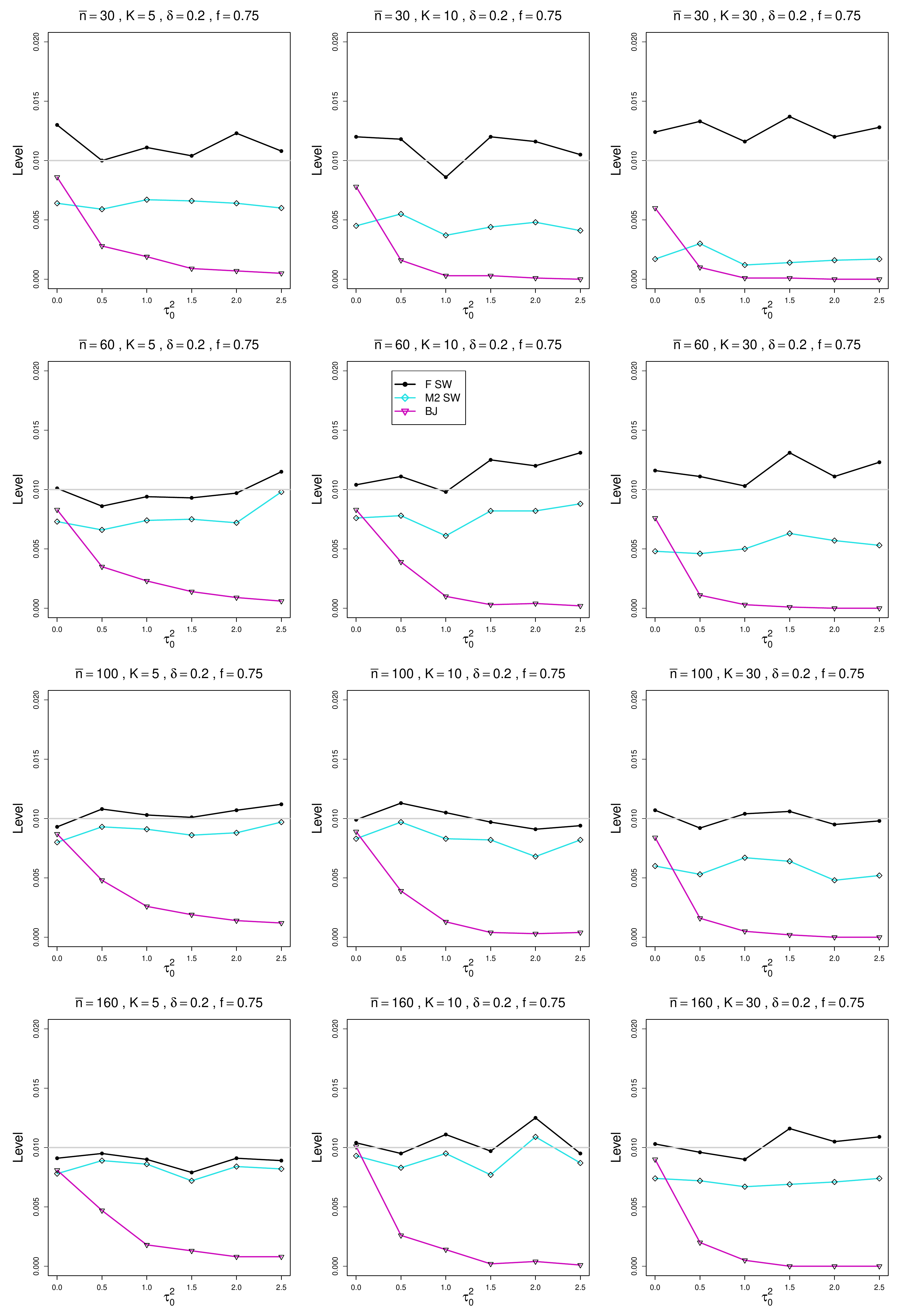}
	\caption{Empirical level for $\delta = 0.2$, $f = .75$, and unequal sample sizes
		\label{PlotOfPhatAt001delta02andq075SMD_underH1_unequal}}
\end{figure}

\begin{figure}[t]
	\centering
	\includegraphics[scale=0.33]{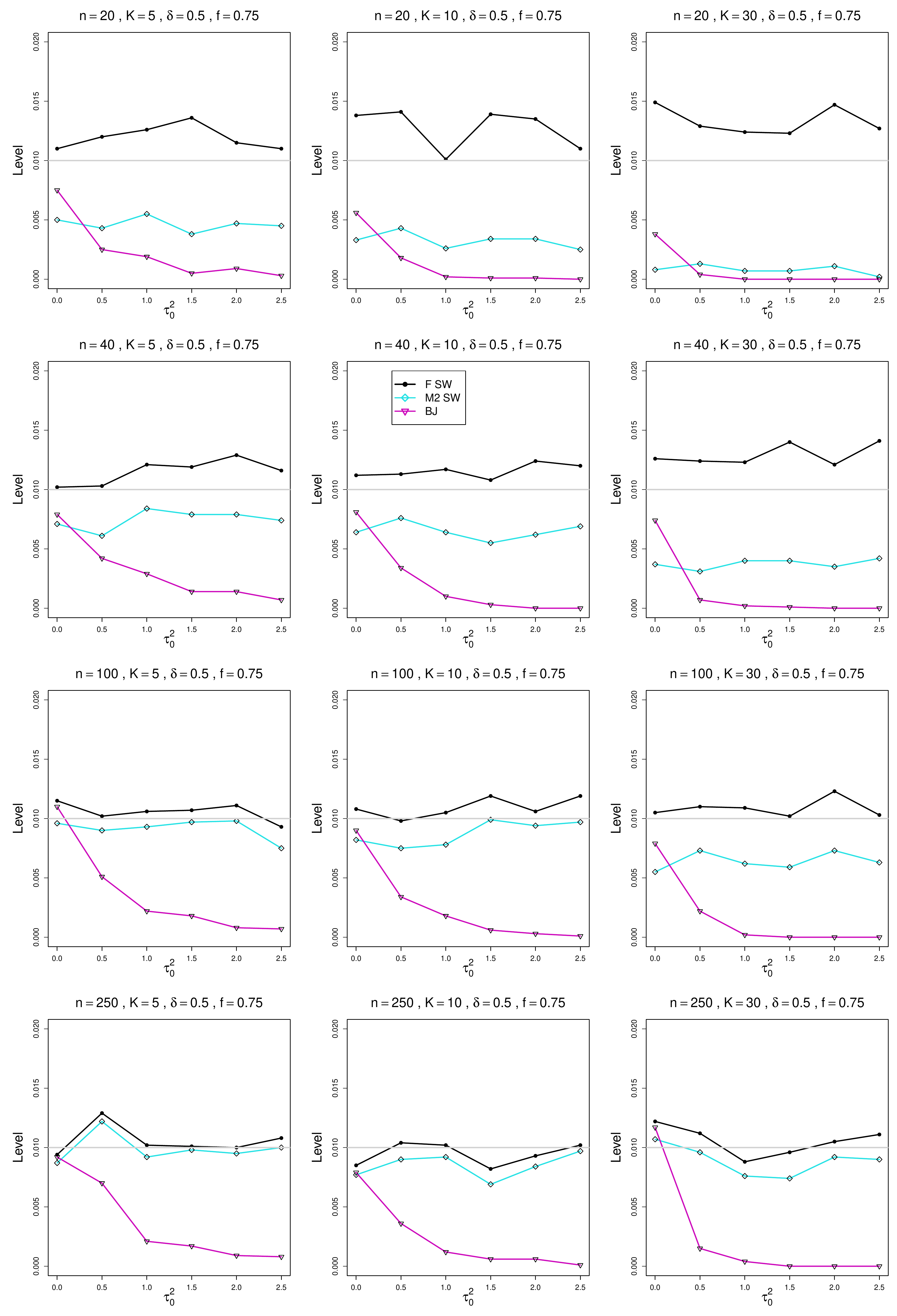}
	\caption{Empirical level for $\delta = 0.5$, $f = .75$, and equal sample sizes
		\label{PlotOfPhatAt001delta05andq075SMD_underH1}}
\end{figure}

\begin{figure}[t]
	\centering
	\includegraphics[scale=0.33]{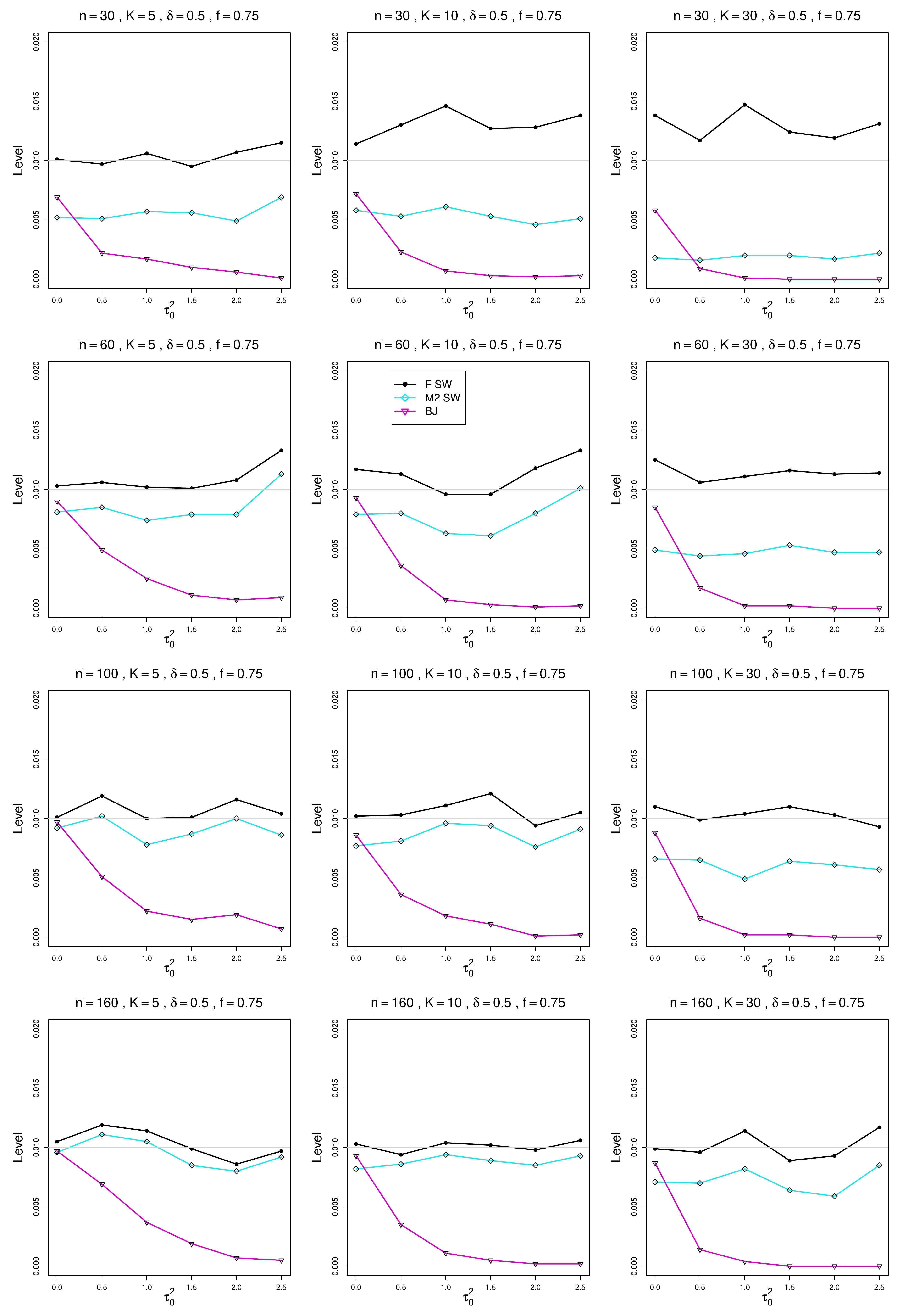}
	\caption{Empirical level for $\delta = 0.5$, $f = .75$, and unequal sample sizes
		\label{PlotOfPhatAt001delta05andq075SMD_underH1_unequal}}
\end{figure}

\begin{figure}[t]
	\centering
	\includegraphics[scale=0.33]{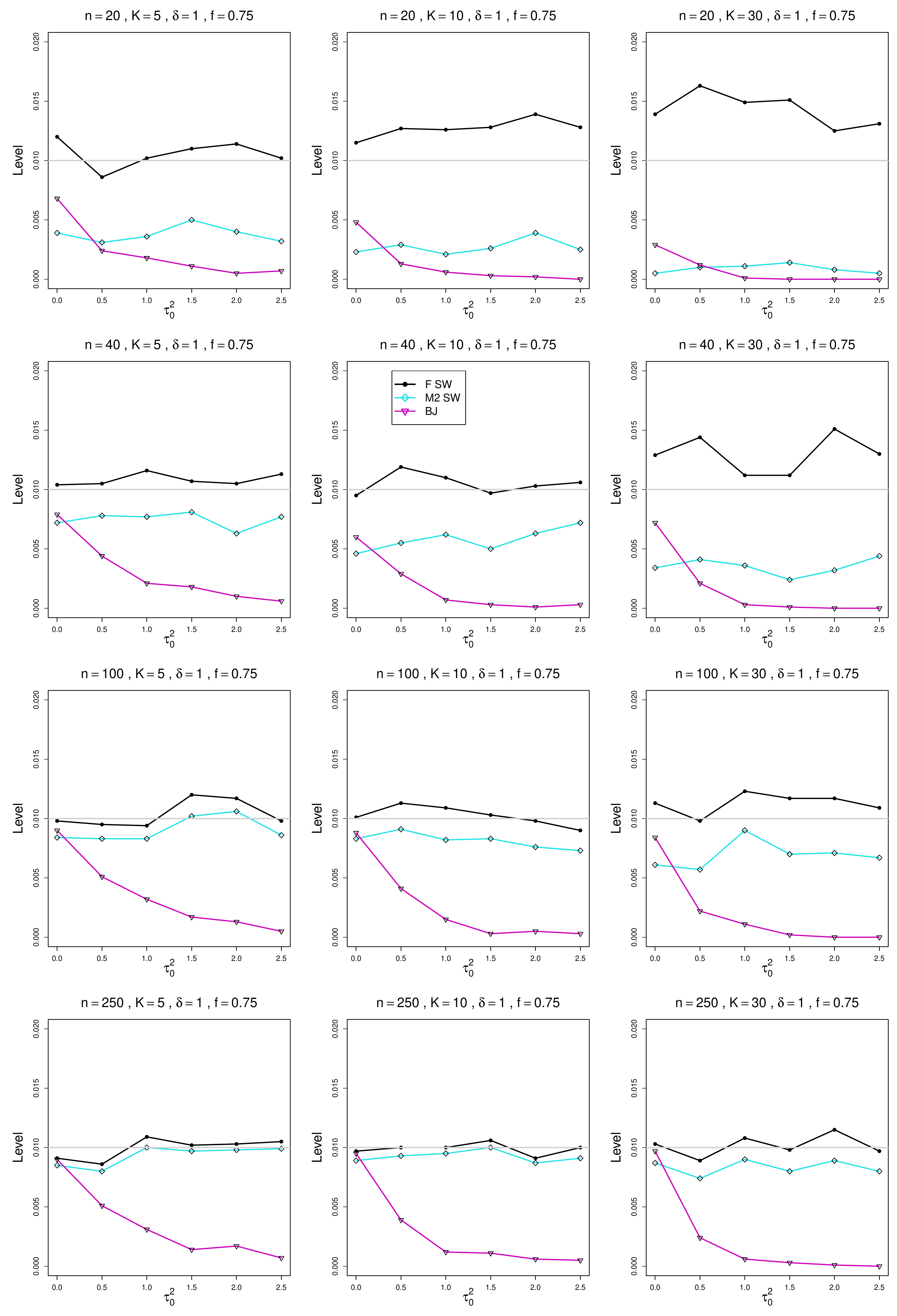}
	\caption{Empirical level for $\delta = 1$, $f = .75$, and equal sample sizes
		\label{PlotOfPhatAt001delta1andq075SMD_underH1}}
\end{figure}

\begin{figure}[t]
	\centering
	\includegraphics[scale=0.33]{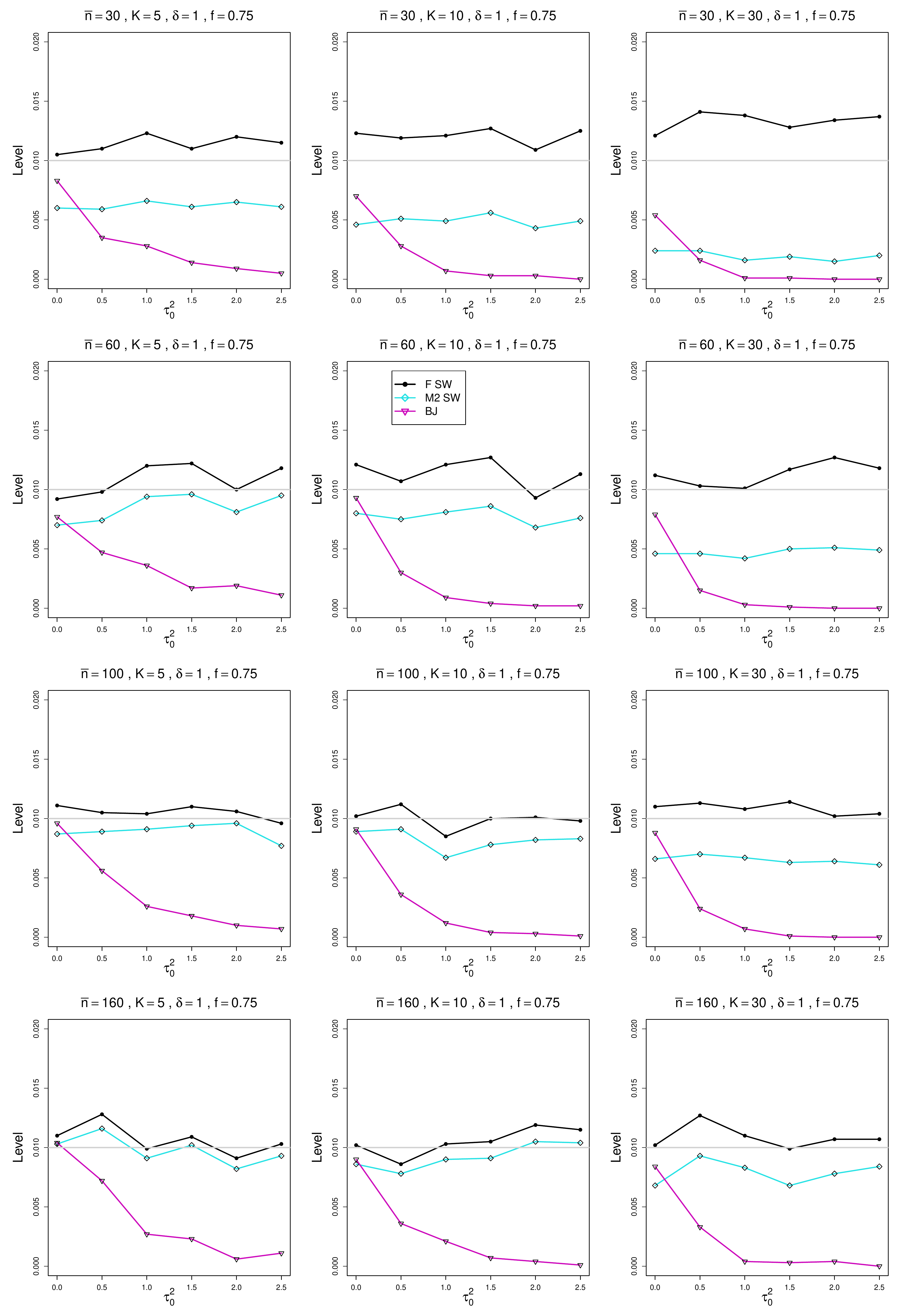}
	\caption{Empirical level for $\delta = 1$, $f = .75$, and unequal sample sizes
		\label{PlotOfPhatAt001delta1andq075SMD_underH1_unequal}}
\end{figure}

\begin{figure}[t]
	\centering
	\includegraphics[scale=0.33]{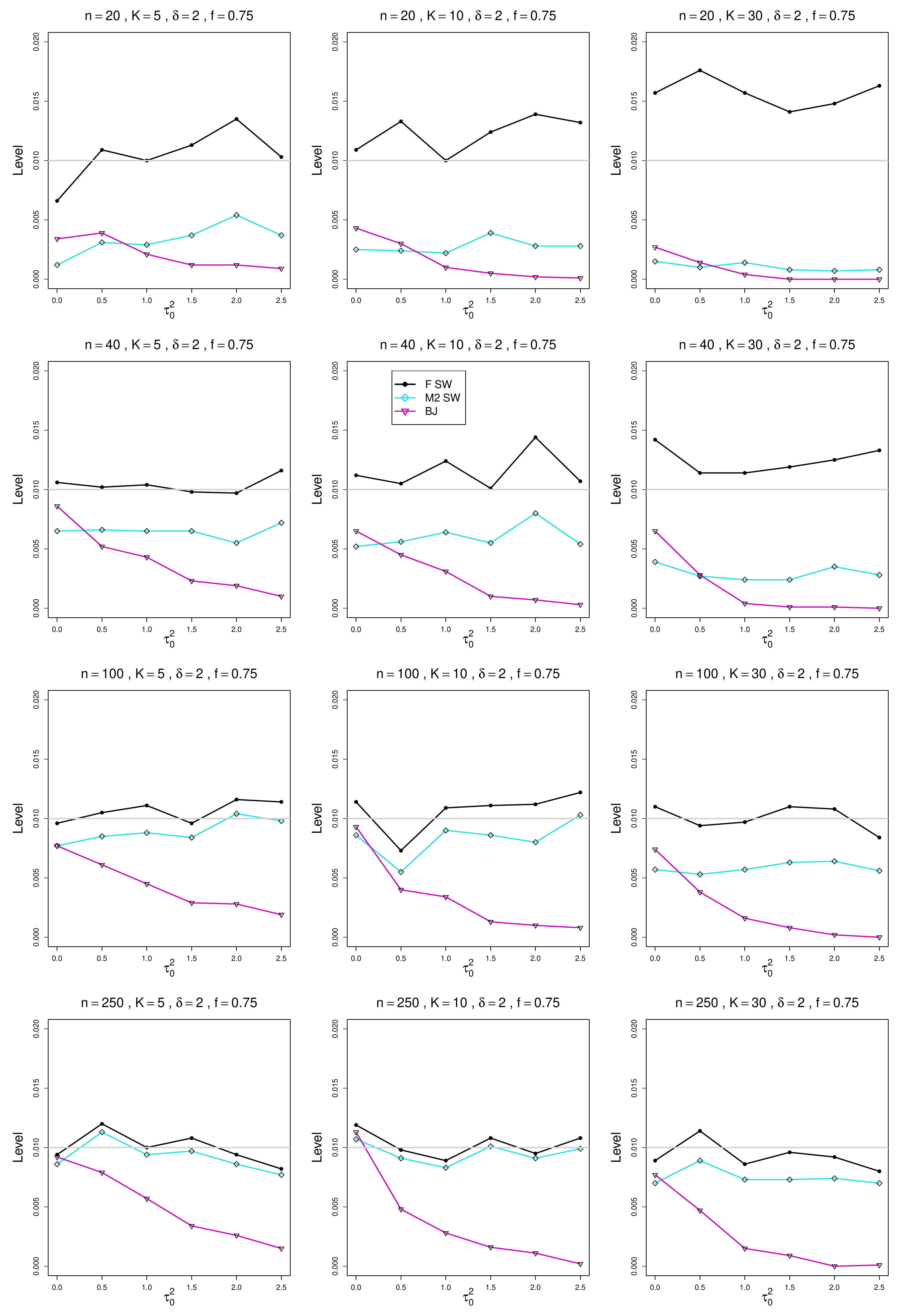}
	\caption{Empirical level for $\delta = 2$, $f = .75$, and equal sample sizes
		\label{PlotOfPhatAt001delta2andq075SMD_underH1}}
\end{figure}

\begin{figure}[t]
	\centering
	\includegraphics[scale=0.33]{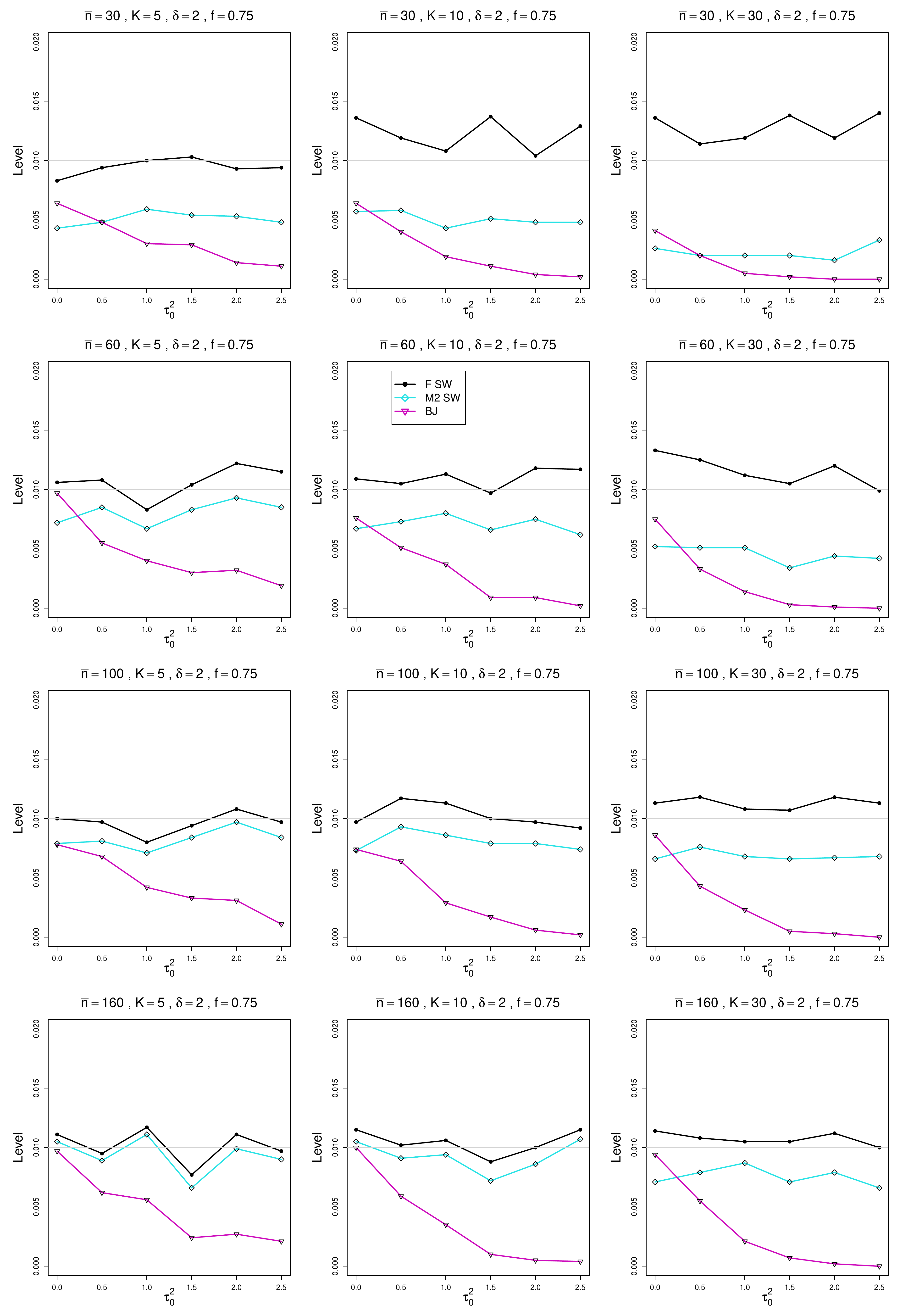}
	\caption{Empirical level for $\delta = 2$, $f = .75$, and unequal sample sizes
		\label{PlotOfPhatAt001delta2andq075SMD_underH1_unequal}}
\end{figure}

\clearpage

\subsection*{D2. Empirical level at $\alpha = .05$ (only $f = .5$)}
\setcounter{figure}{0}

\renewcommand{\thefigure}{D2.\arabic{figure}}

\begin{figure}[t]
	\centering
	\includegraphics[scale=0.33]{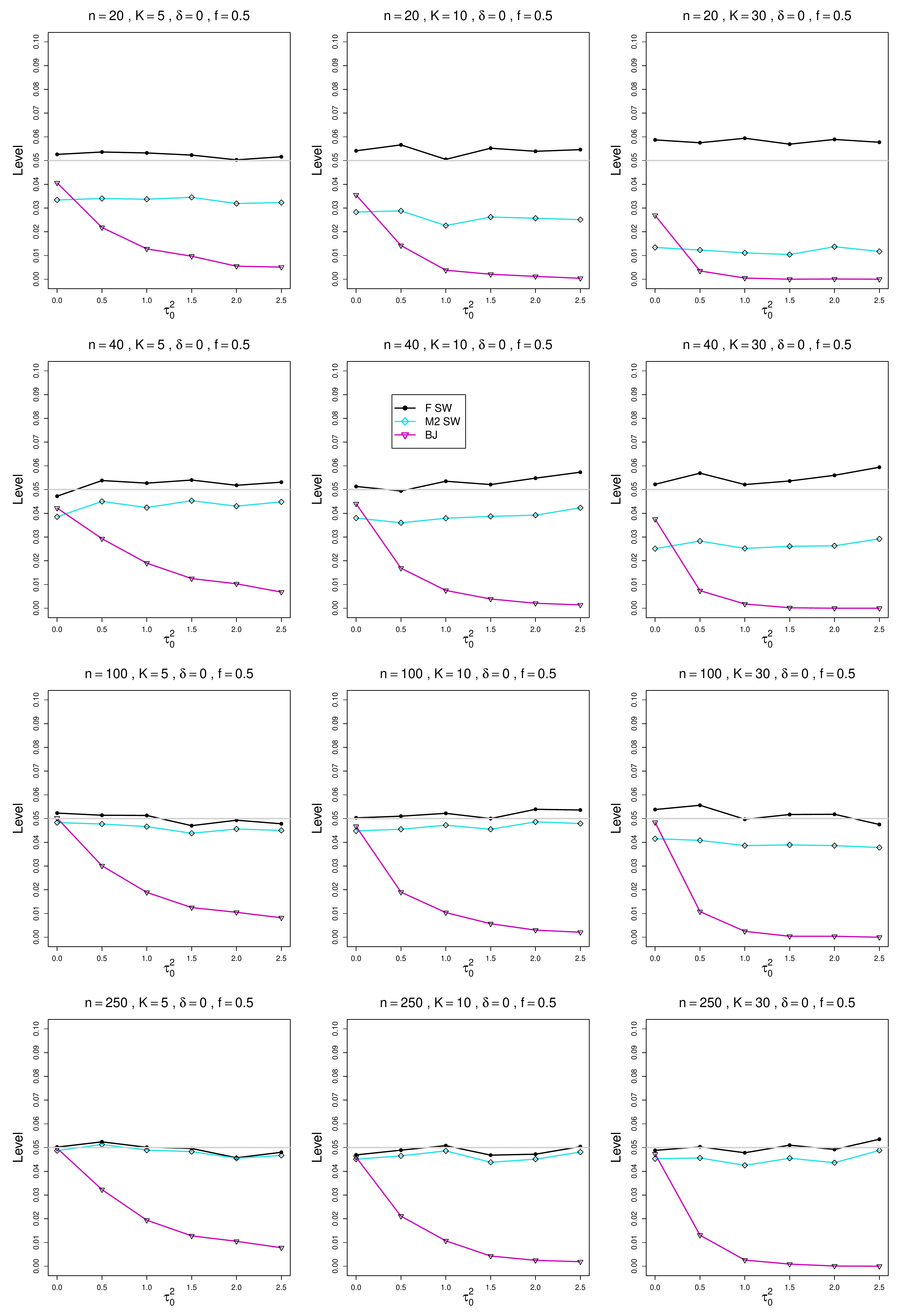}
	\caption{Empirical level for $\delta = 0$, $f = .5$, and equal sample sizes
		\label{PlotOfPhatAt005delta0andq05SMD_underH1}}
\end{figure}

\begin{figure}[t]
	\centering
	\includegraphics[scale=0.33]{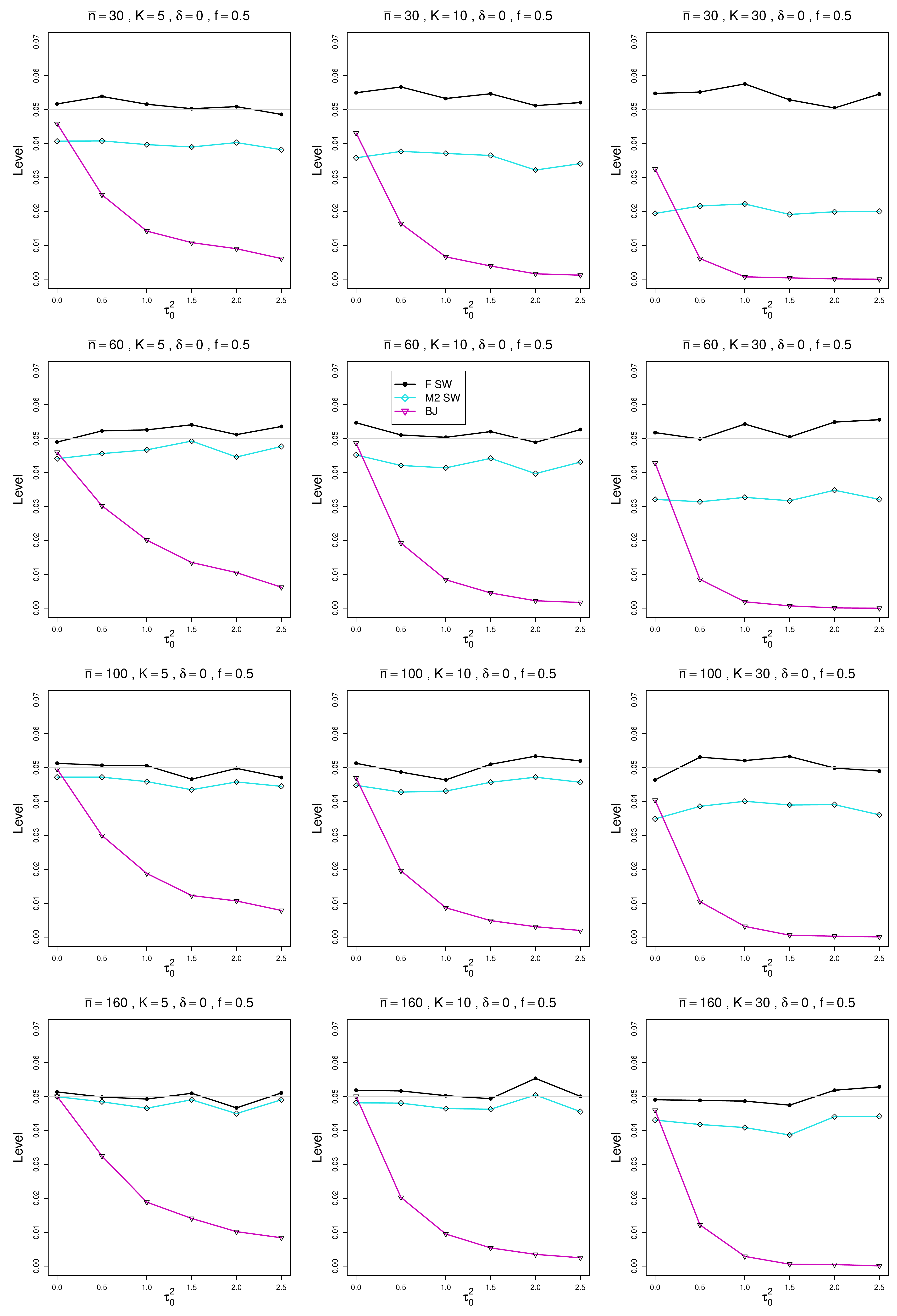}
	\caption{Empirical level for $\delta = 0$, $f = .5$, and unequal sample sizes
		\label{PlotOfPhatAt005delta0andq05SMD_underH1_unequal}}
\end{figure}

\begin{figure}[t]
	\centering
	\includegraphics[scale=0.33]{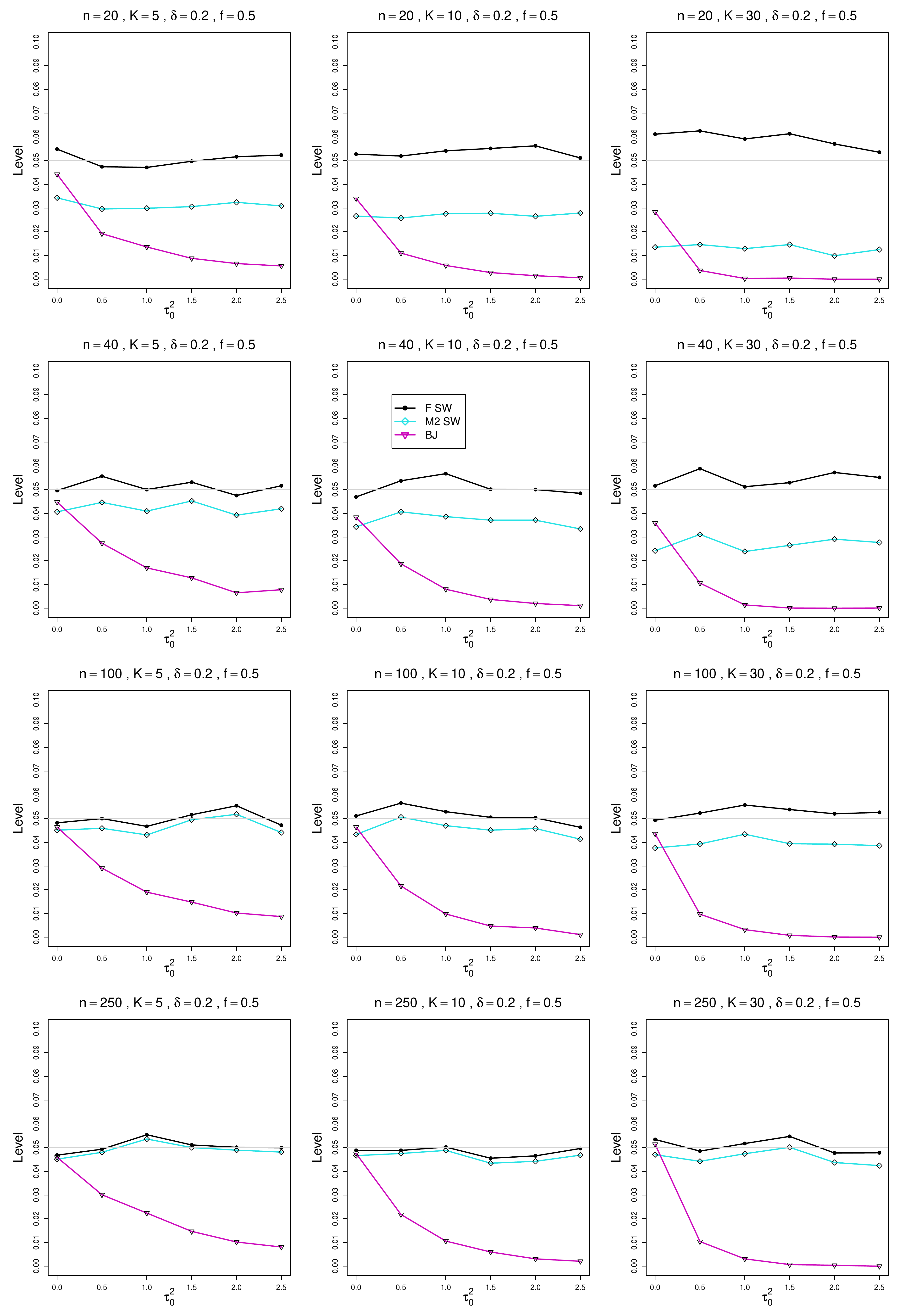}
	\caption{Empirical level for $\delta = 0.2$, $f = .5$, and equal sample sizes
		\label{PlotOfPhatAt005delta02andq05SMD_underH1}}
\end{figure}

\begin{figure}[t]
	\centering
	\includegraphics[scale=0.33]{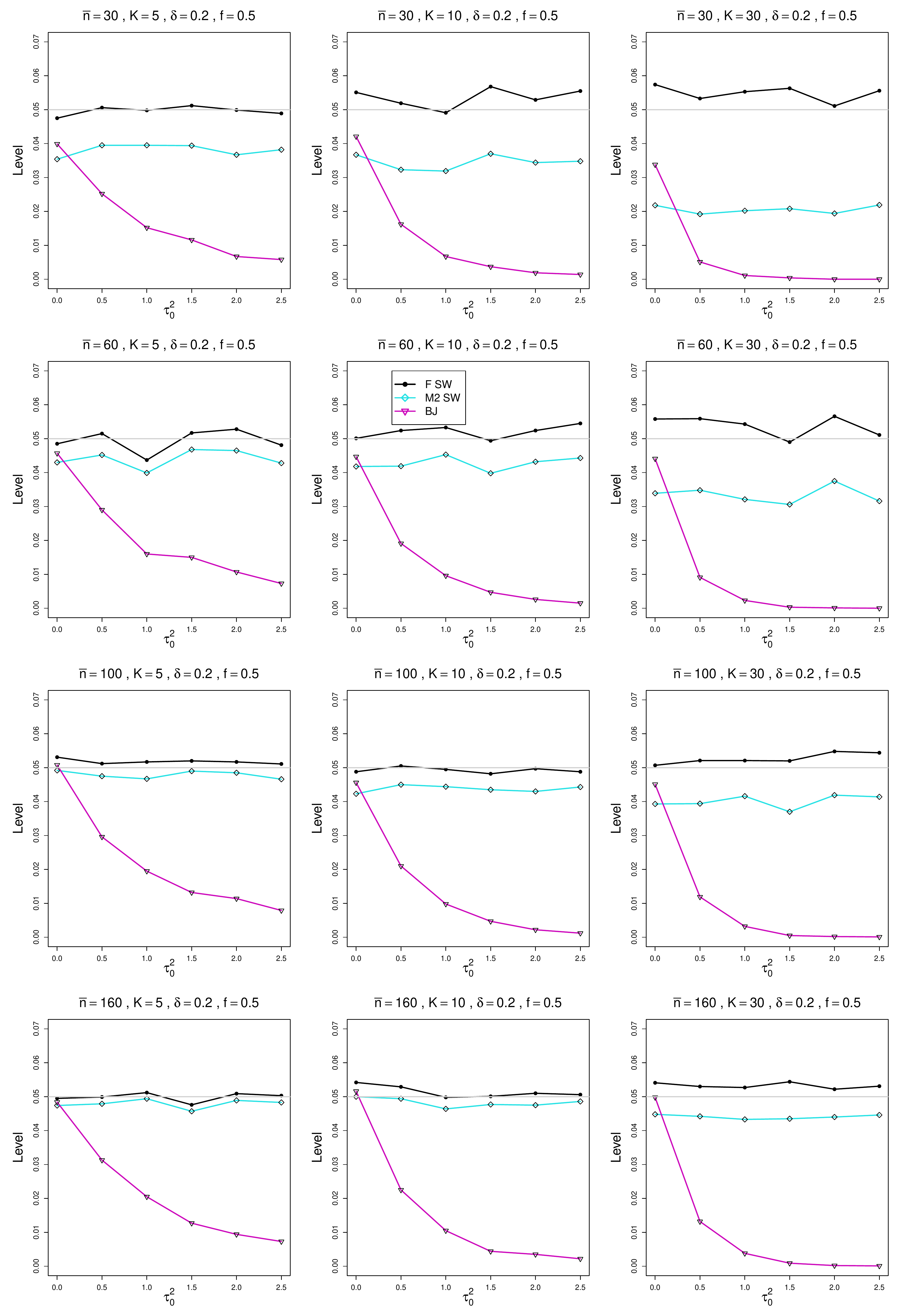}
	\caption{Empirical level for $\delta = 0.2$, $f = .5$, and unequal sample sizes
		\label{PlotOfPhatAt005delta02andq05SMD_underH1_unequal}}
\end{figure}

\begin{figure}[t]
	\centering
	\includegraphics[scale=0.33]{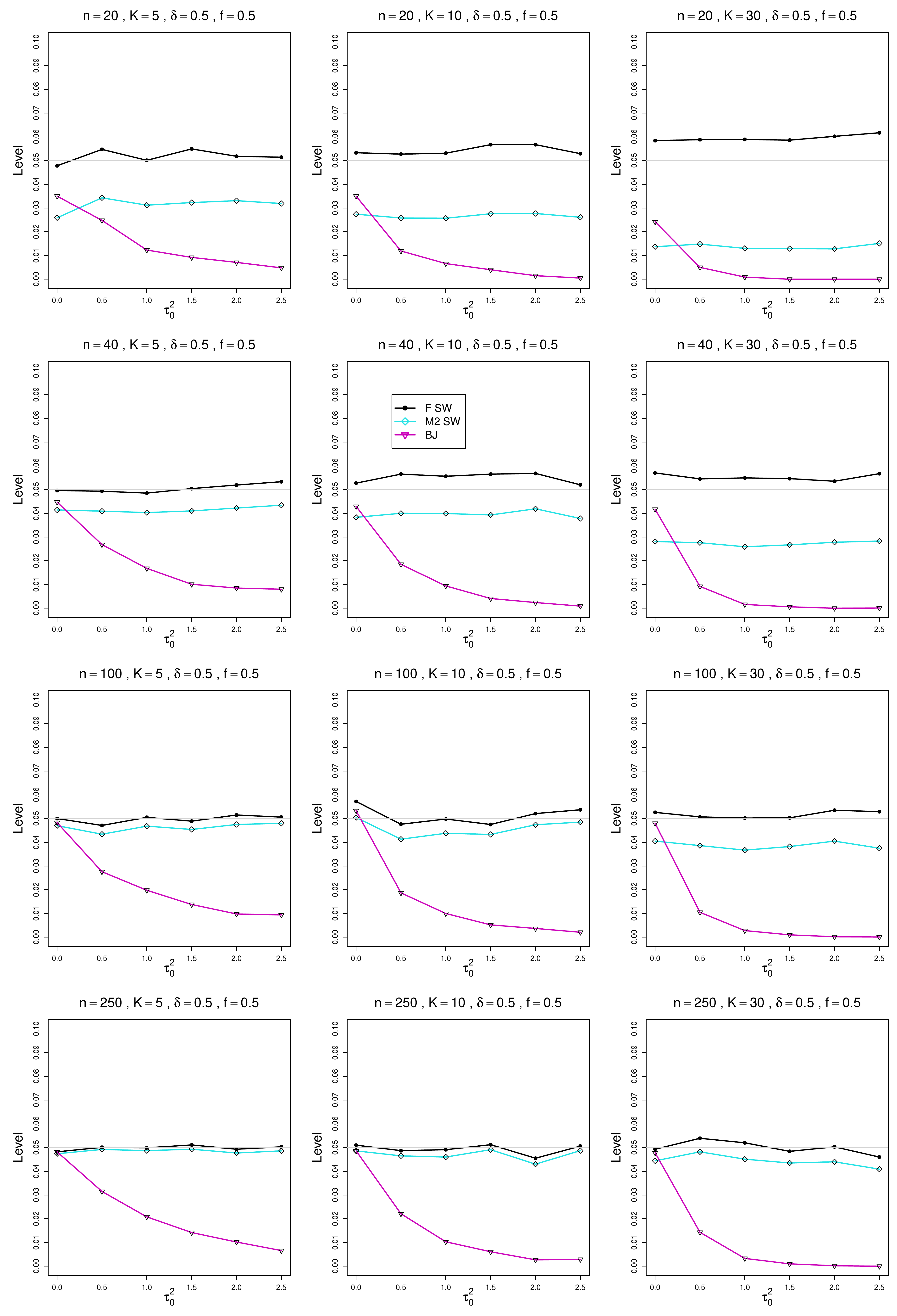}
	\caption{Empirical level for $\delta = 0.5$, $f = .5$, and equal sample sizes
		\label{PlotOfPhatAt005delta05andq05SMD_underH1}}
\end{figure}

\begin{figure}[t]
	\centering
	\includegraphics[scale=0.33]{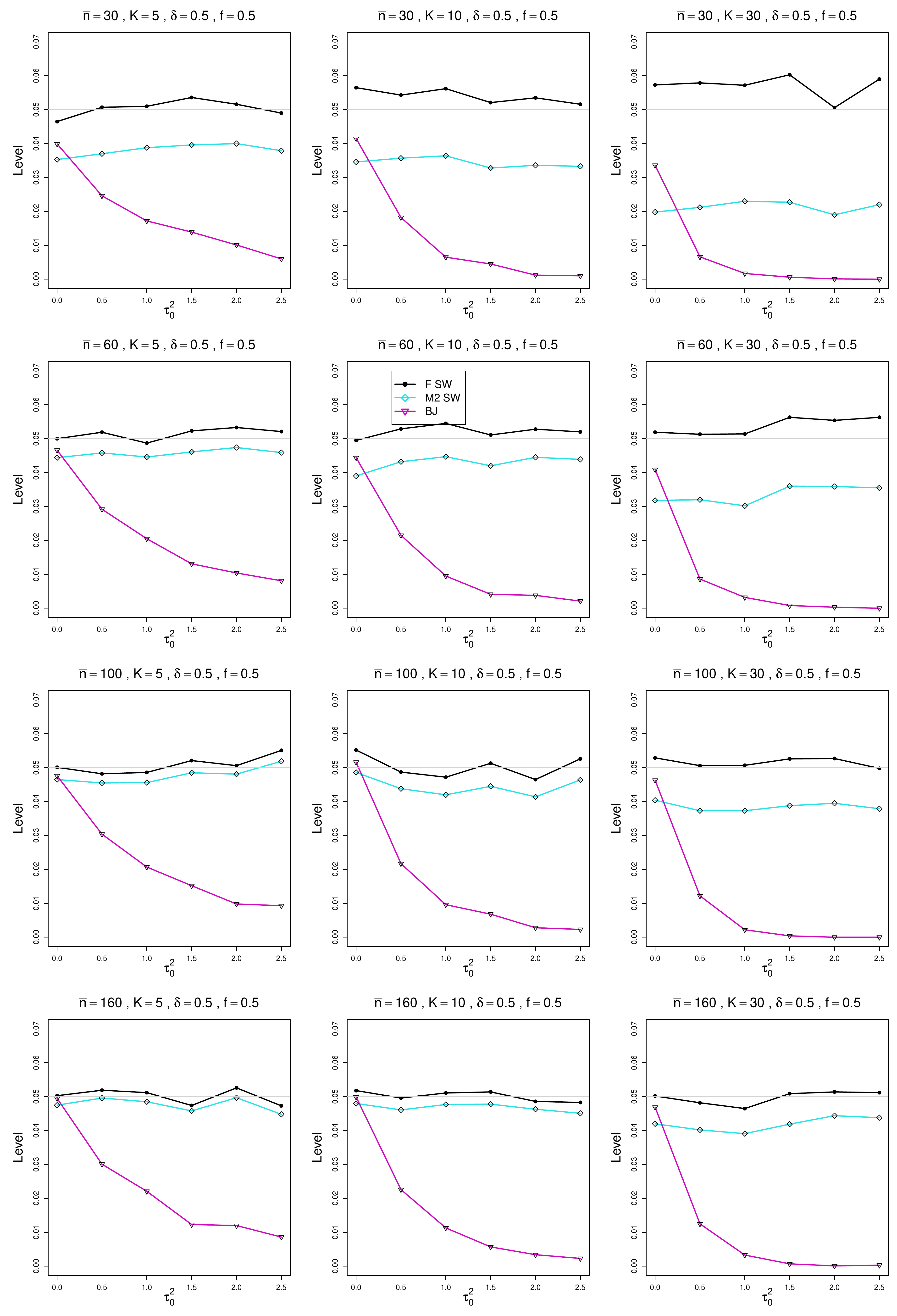}
	\caption{Empirical level for $\delta = 0.5$, $f = .5$, and unequal sample sizes
		\label{PlotOfPhatAt005delta05andq05SMD_underH1_unequal}}
\end{figure}

\begin{figure}[t]
	\centering
	\includegraphics[scale=0.33]{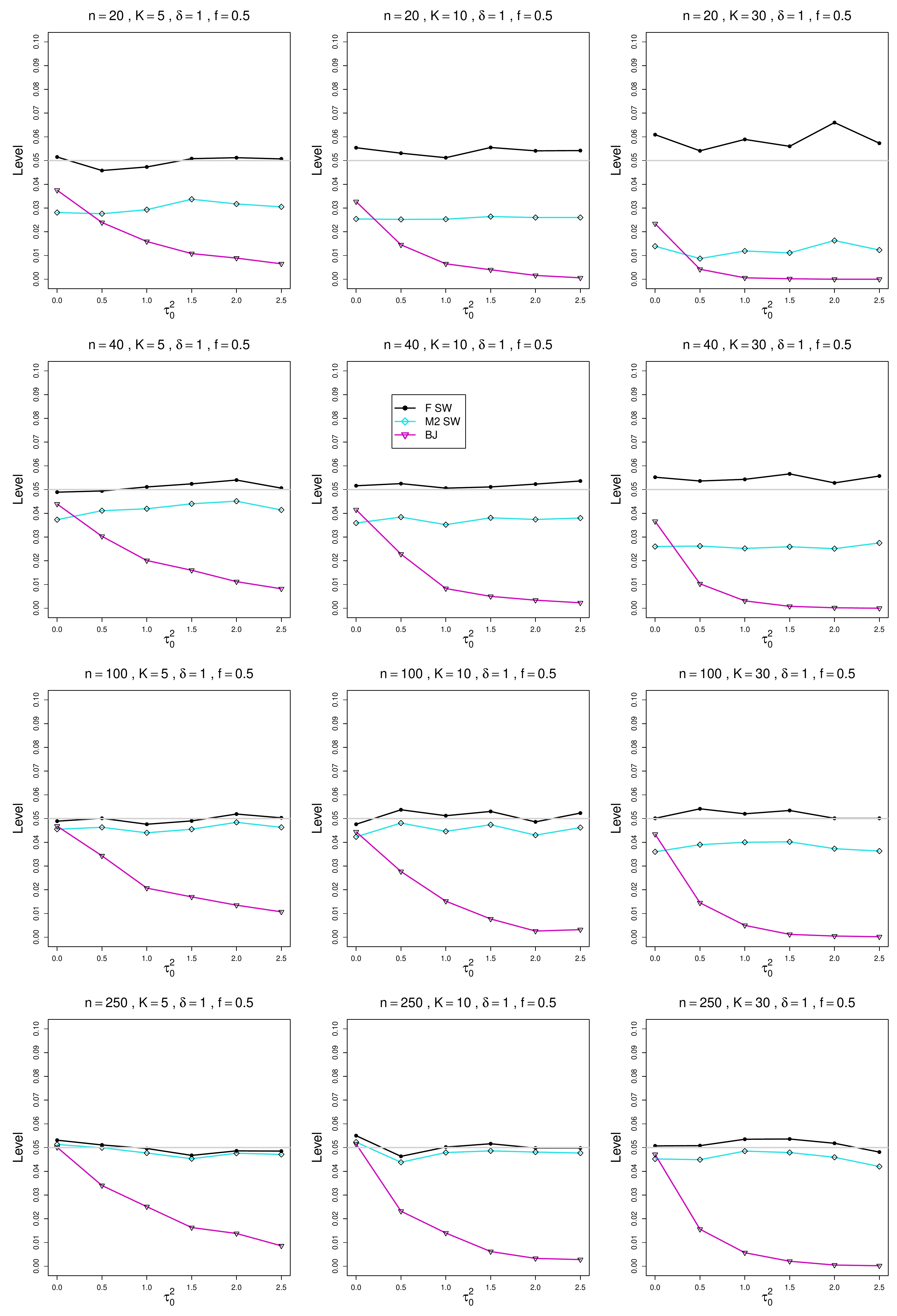}
	\caption{Empirical level for $\delta = 1$, $f = .5$, and equal sample sizes
		\label{PlotOfPhatAt005delta1andq05SMD_underH1}}
\end{figure}

\begin{figure}[t]
	\centering
	\includegraphics[scale=0.33]{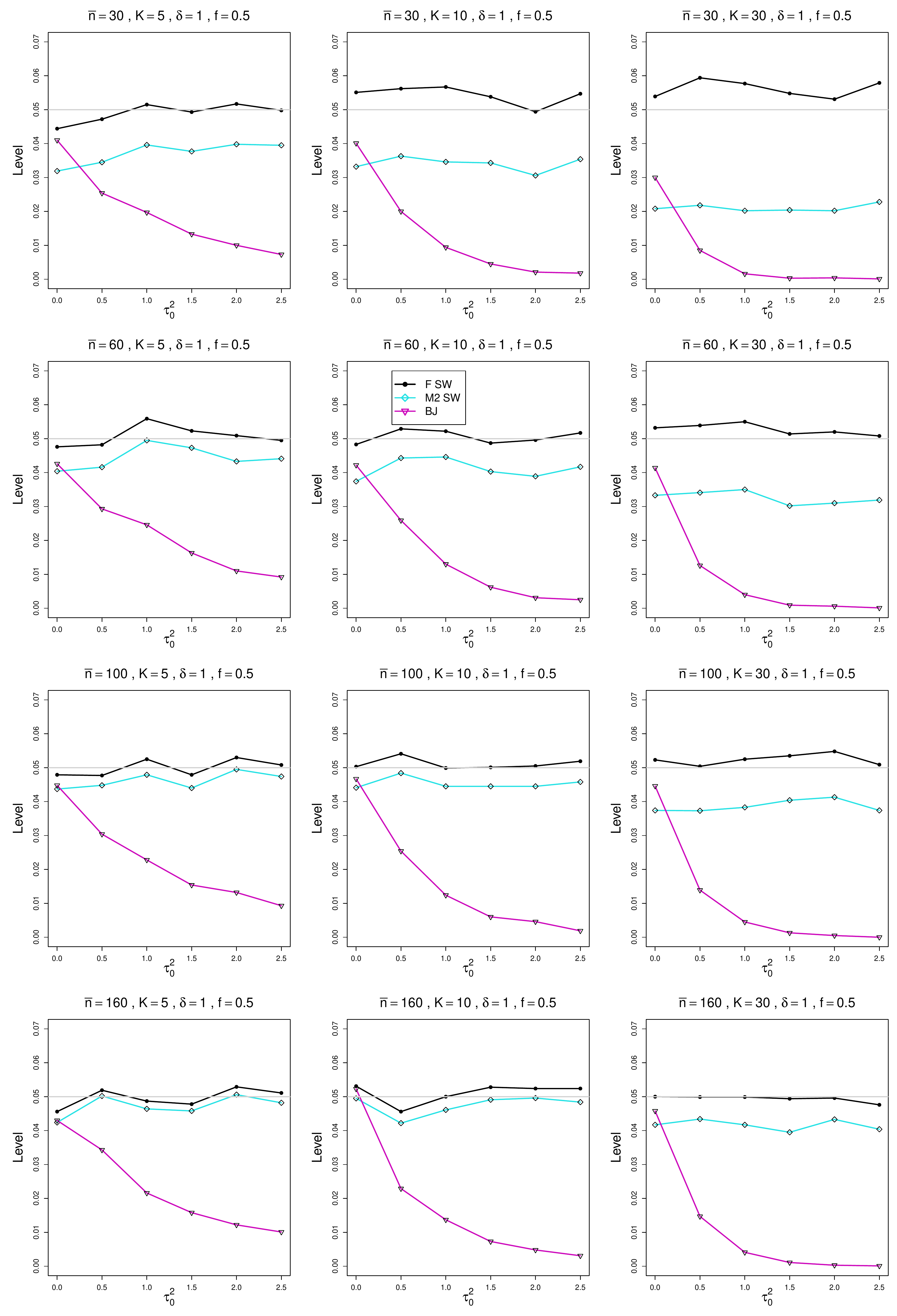}
	\caption{Empirical level for $\delta = 1$, $f = .5$, and unequal sample sizes
		\label{PlotOfPhatAt005delta1andq05SMD_underH1_unequal}}
\end{figure}

\begin{figure}[t]
	\centering
	\includegraphics[scale=0.33]{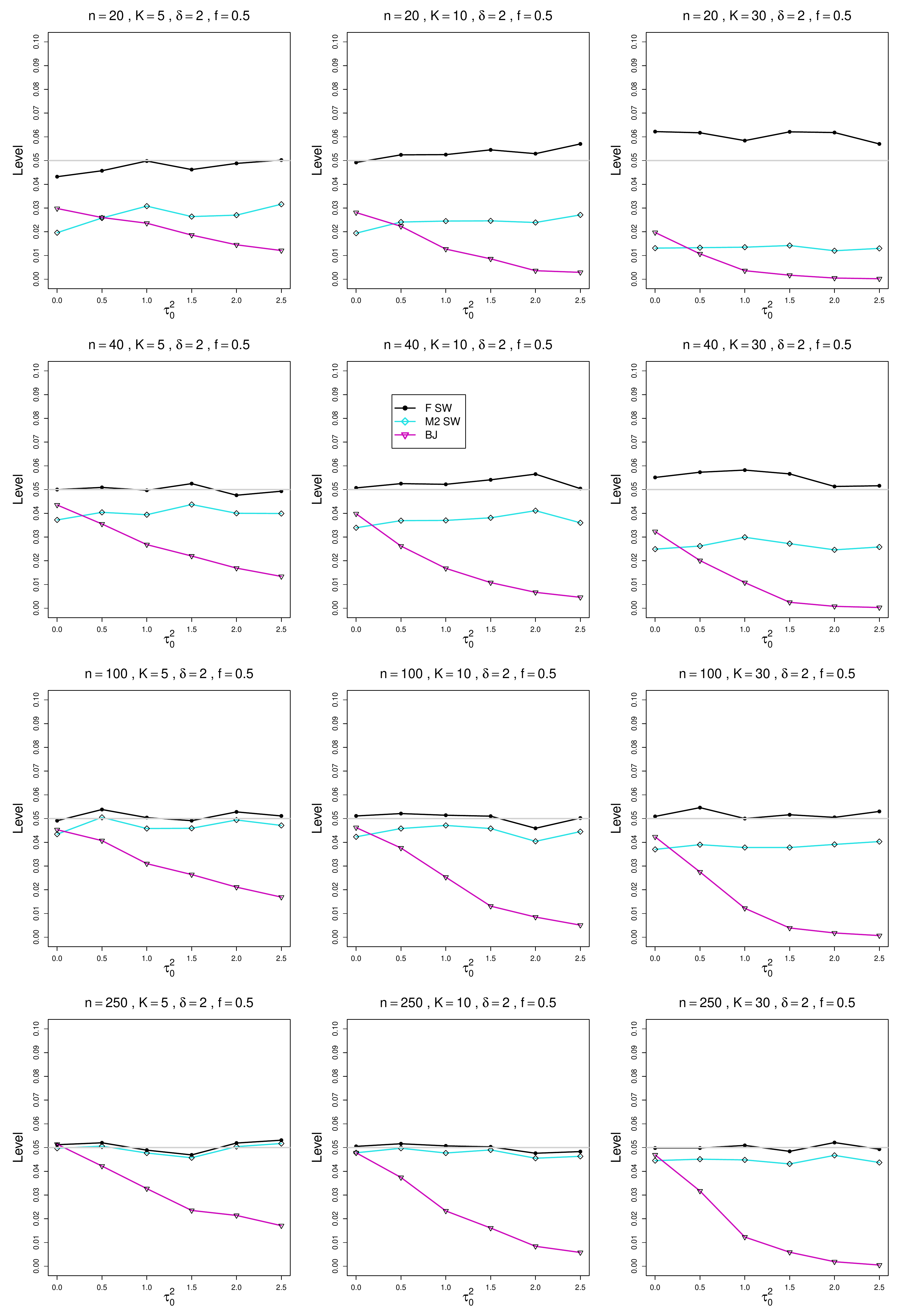}
	\caption{Empirical level for $\delta = 2$, $f = .5$, and equal sample sizes
		\label{PlotOfPhatAt005delta2andq05SMD_underH1}}
\end{figure}

\begin{figure}[t]
	\centering
	\includegraphics[scale=0.33]{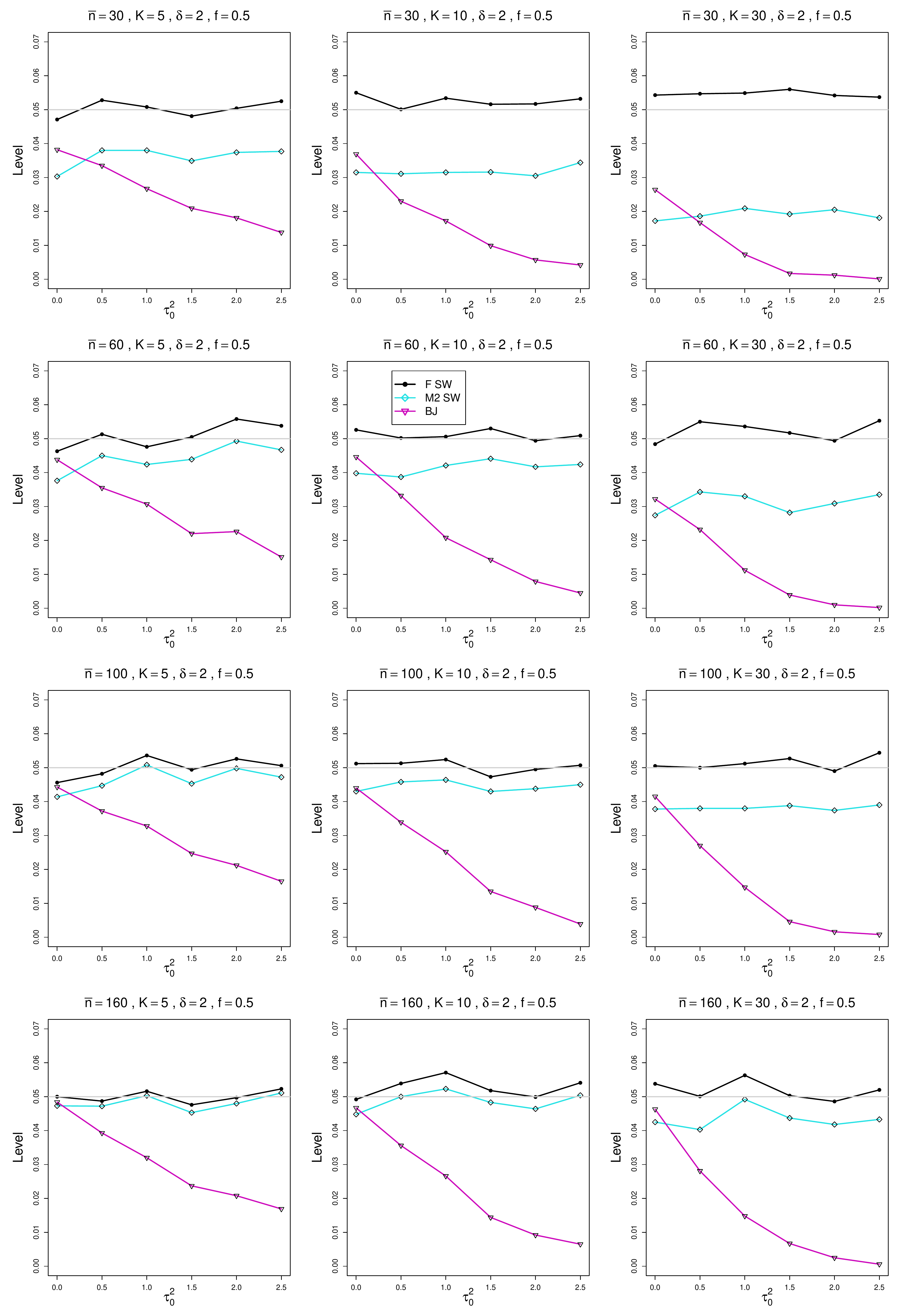}
	\caption{Empirical level for $\delta = 2$, $f = .5$, and unequal sample sizes
		\label{PlotOfPhatAt005delta2andq05SMD_underH1_unequal}}
\end{figure}

\clearpage
\setcounter{figure}{0}
\setcounter{section}{0}
\renewcommand{\thesection}{E.\arabic{section}}
\section*{E. Power of the test for heterogeneity ($\tau^2 = 0$ versus $\tau^2 > 0$) based on approximations for the distribution of $Q$}
In sets of figures for $\alpha = .01$ and $\alpha = .05$, each figure corresponds to a value of $\delta$, a value of $f$, and either equal sample sizes or unequal sample sizes. \\
For each combination of a value of $n$ or $\bar{n}$ and a value of $K$, a panel plots power versus $\tau^2$. \\
The approximations for the distribution of $Q$ are
\begin{itemize}
	\item F SW (Farebrother approximation, effective-sample-size weights)
	\item M2 SW (Two-moment approximation, effective-sample-size weights)
	\item $\chi_{K-1}^2$ (Chi-square, IV weights)
	\item KDB (Chi-square approximation based on corrected first moment, IV weights)
\end{itemize}

\clearpage

\subsection*{E1. Power when $\alpha = .01$}

\renewcommand{\thefigure}{E1.\arabic{figure}}
%alpha 0.01 under H0
%q=0.5
\begin{figure}[t]
	\centering
	\includegraphics[scale=0.33]{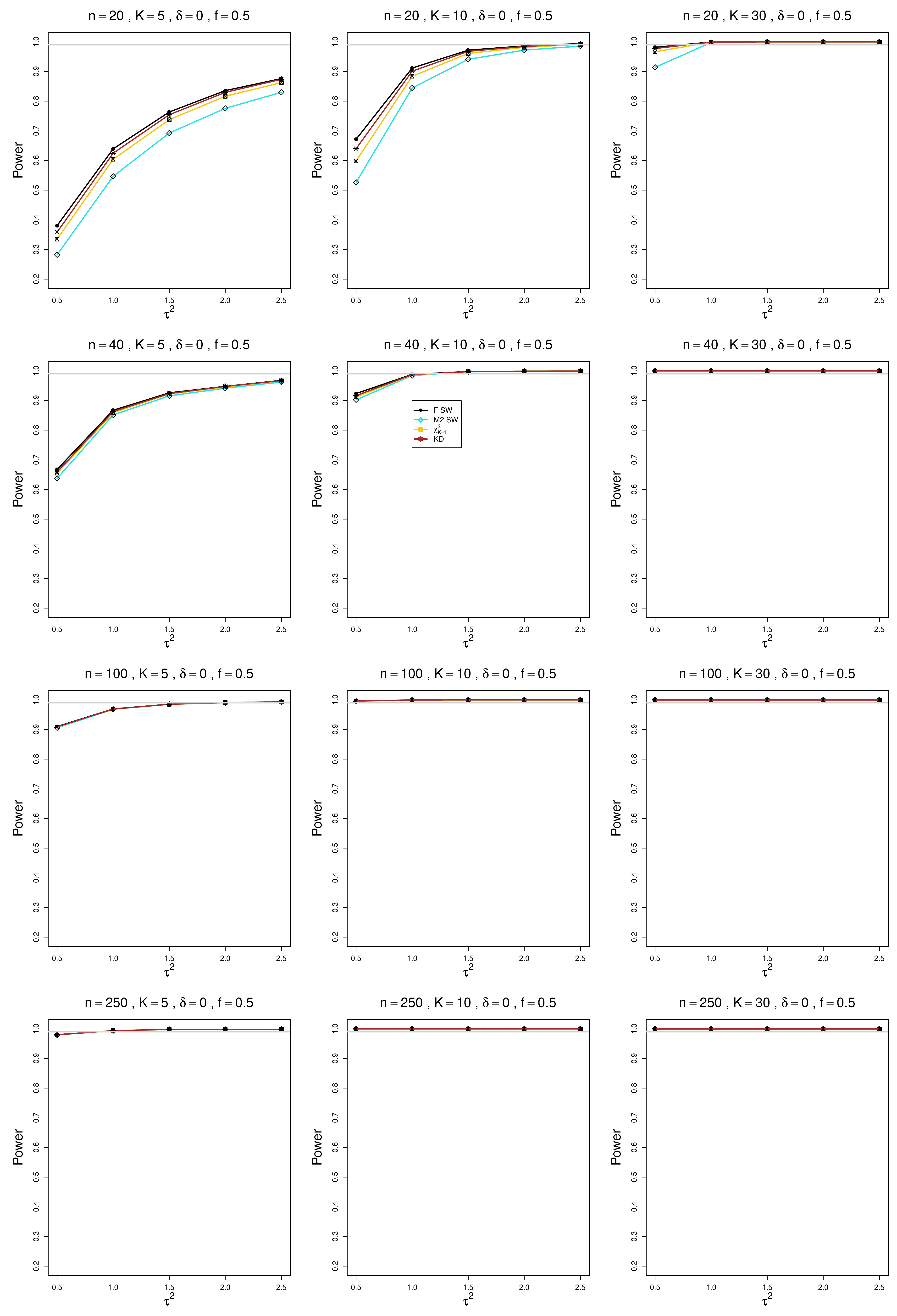}
	\caption{Power for $\delta = 0$, $f = .5$, and equal sample sizes
		\label{PlotOfPhatAt001delta0andq05SMD_underH0}}
\end{figure}

\begin{figure}[t]
	\centering
	\includegraphics[scale=0.33]{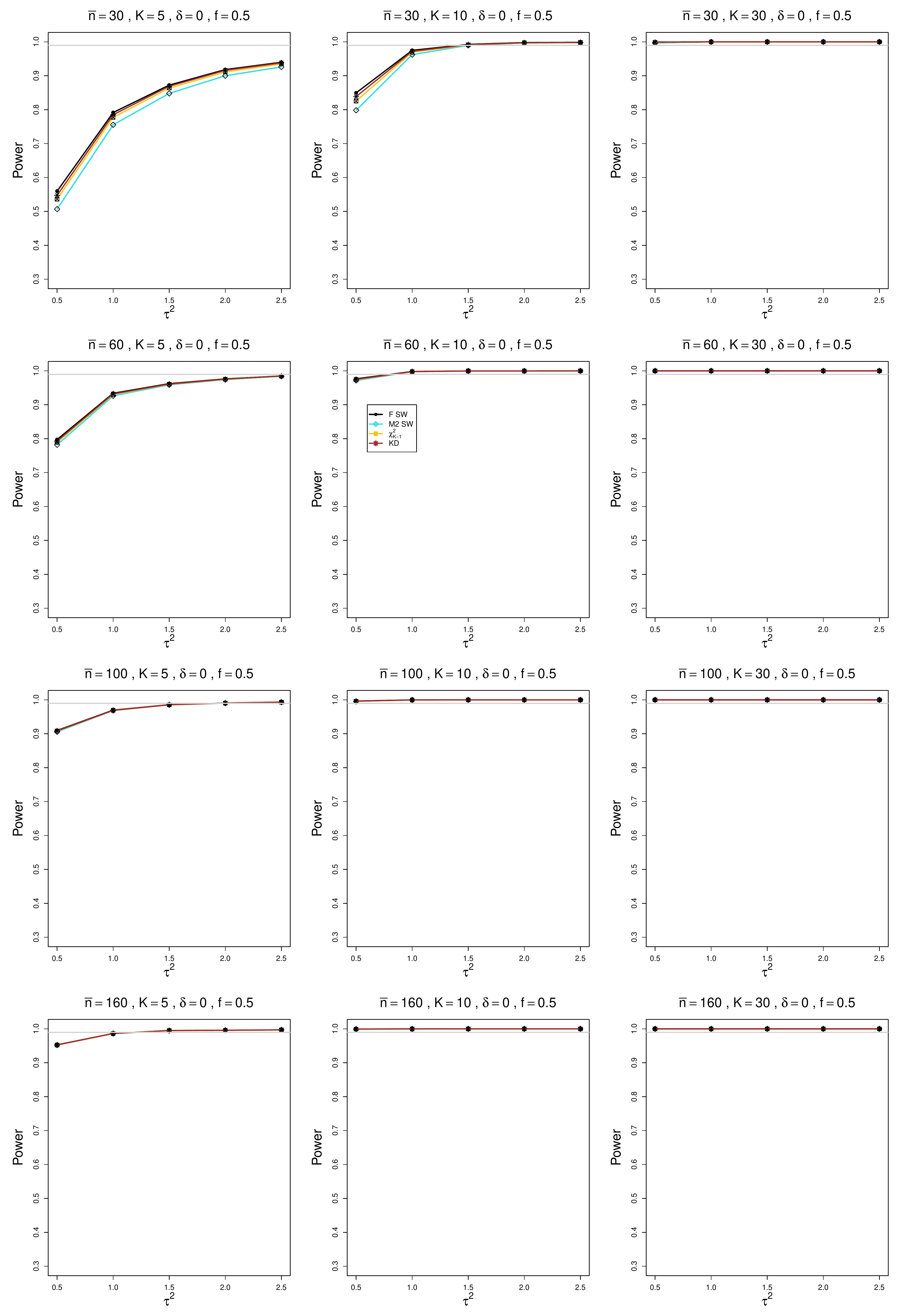}
	\caption{Power for $\delta = 0$, $f = .5$, and unequal sample sizes
		\label{PlotOfPhatAt001delta0andq05SMD_underH0_unequal}}
\end{figure}

\begin{figure}[t]
	\centering
	\includegraphics[scale=0.33]{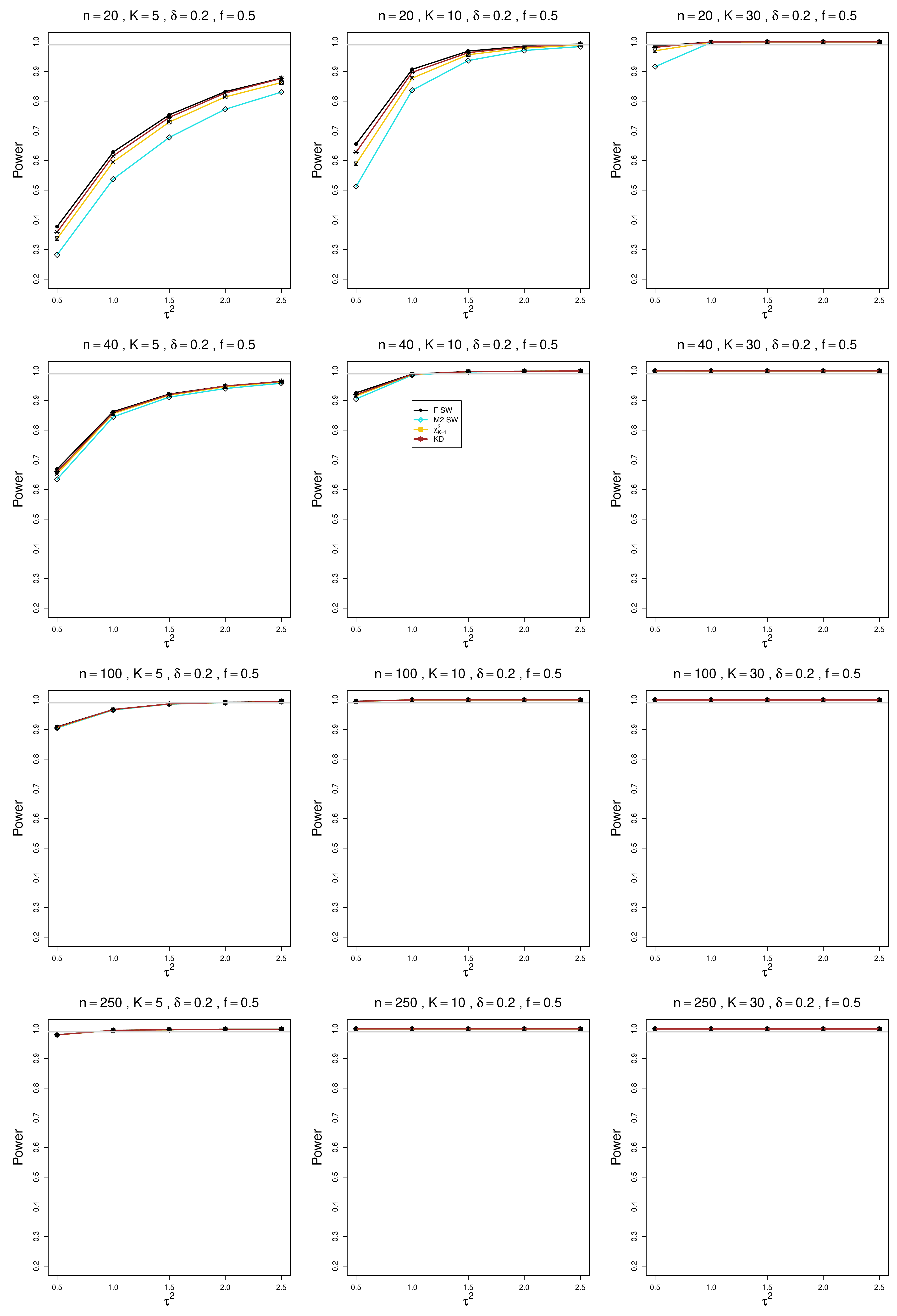}
	\caption{Power for $\delta = 0.2$, $f = .5$, and equal sample sizes
		\label{PlotOfPhatAt001delta02andq05SMD_underH0}}
\end{figure}

\begin{figure}[t]
	\centering
	\includegraphics[scale=0.33]{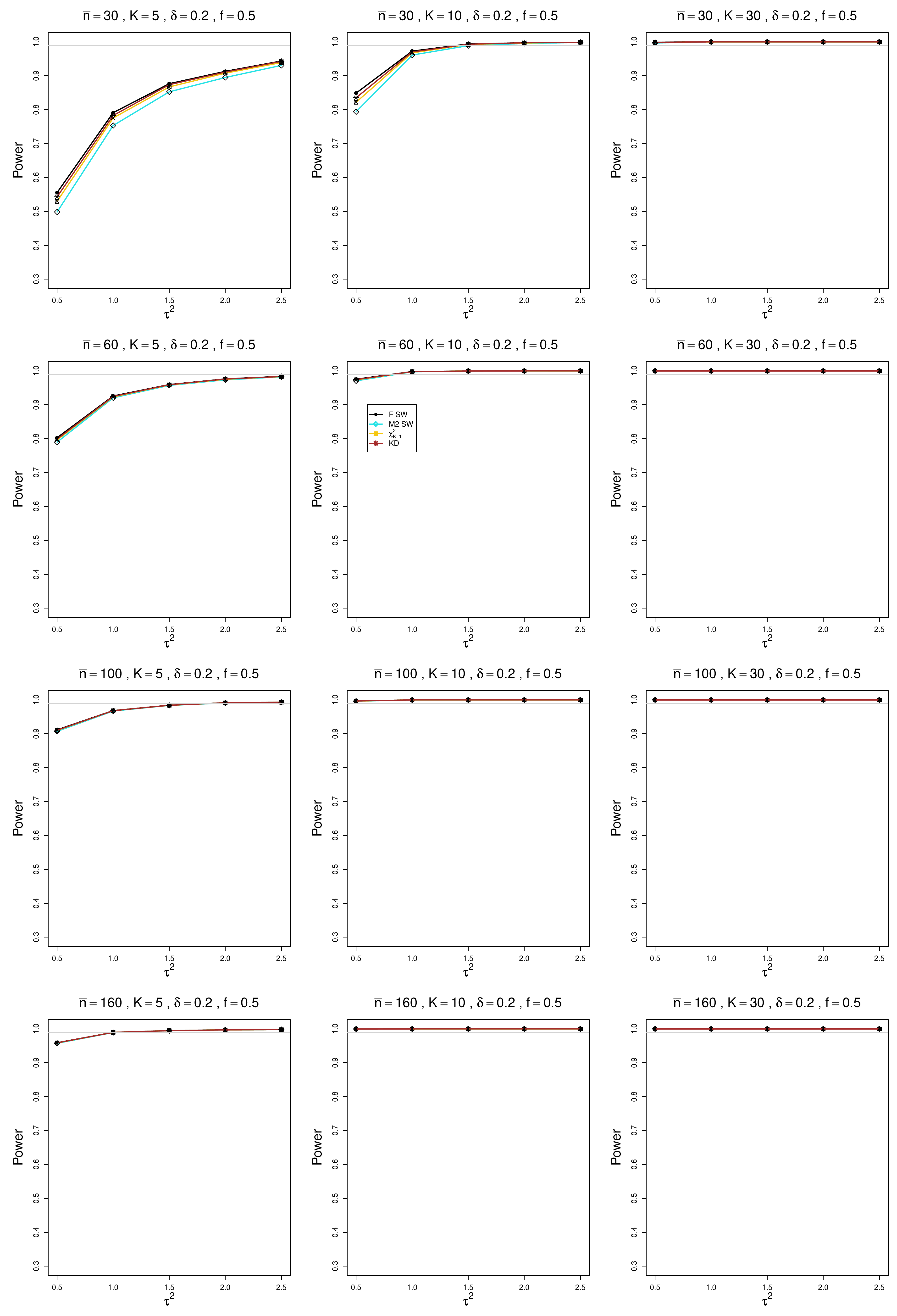}
	\caption{Power for $\delta = 0.2$, $f = .5$, and unequal sample sizes
		\label{PlotOfPhatAt001delta02andq05SMD_underH0_unequal}}
\end{figure}

\begin{figure}[t]
	\centering
	\includegraphics[scale=0.33]{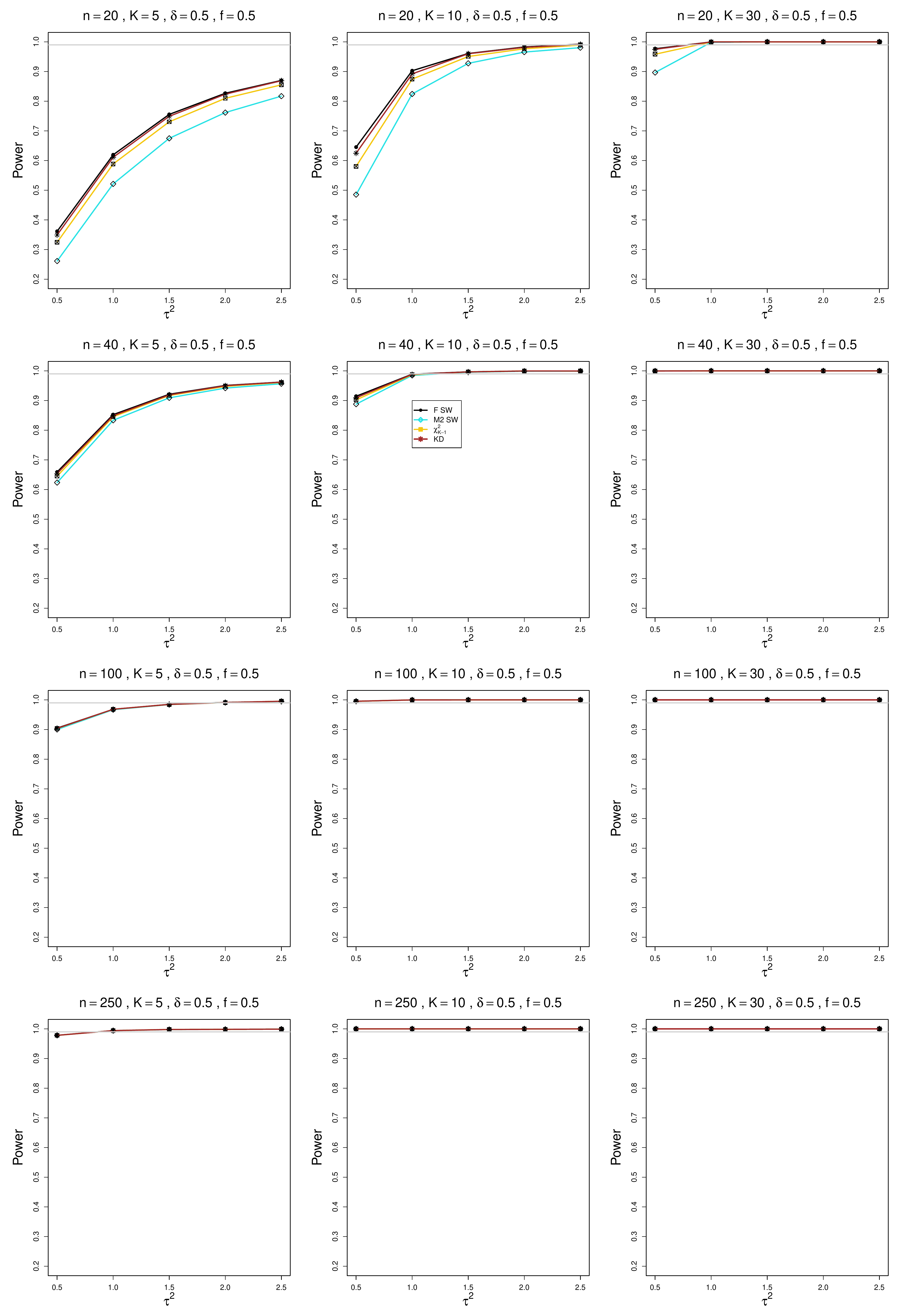}
	\caption{Power for $\delta = 0.5$, $f = .5$, and equal sample sizes
		\label{PlotOfPhatAt001delta05andq05SMD_underH0}}
\end{figure}

\begin{figure}[t]
	\centering
	\includegraphics[scale=0.33]{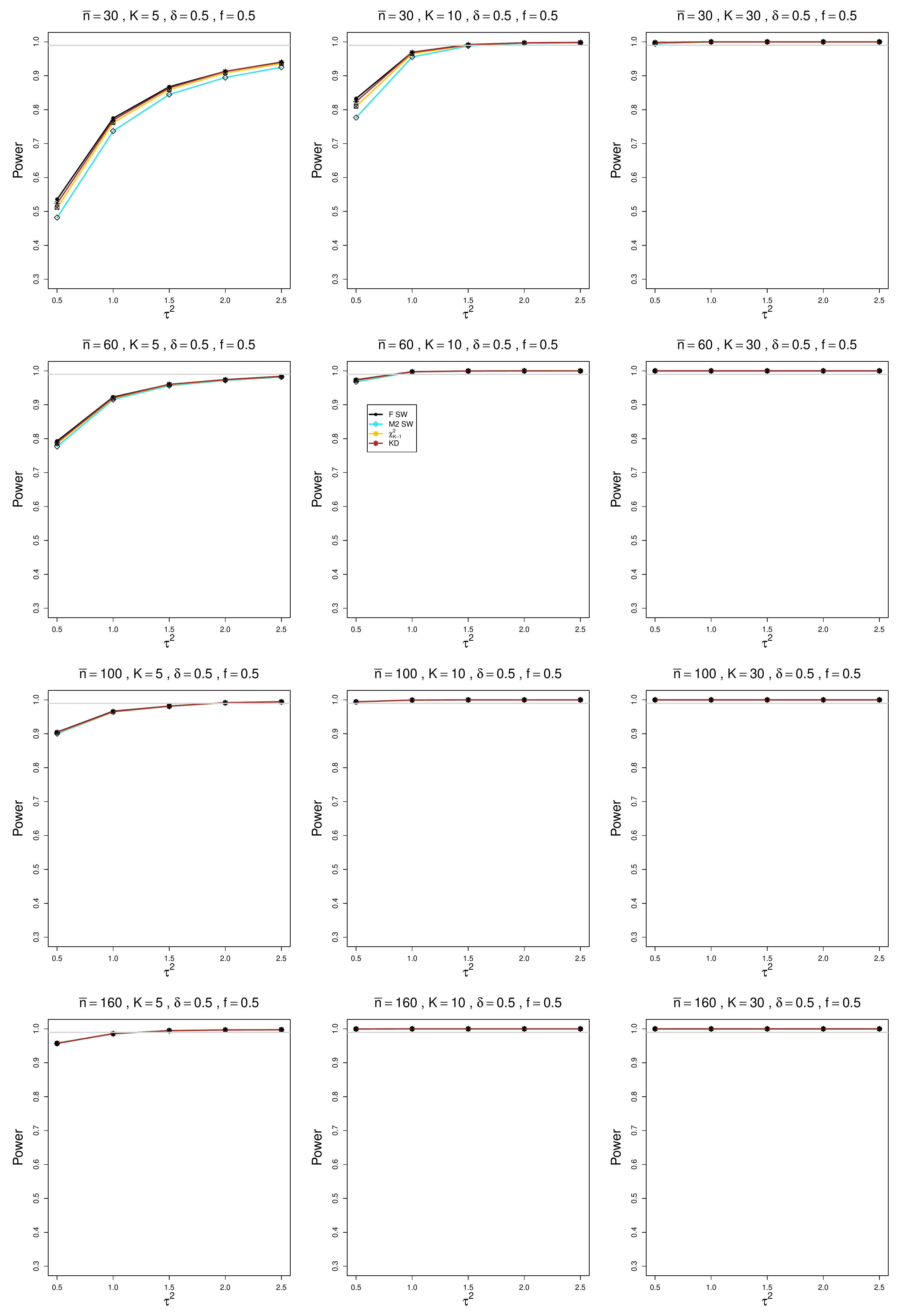}
	\caption{Power for $\delta = 0.5$, $f = .5$, and unequal sample sizes
		\label{PlotOfPhatAt001delta05andq05SMD_underH0_unequal}}
\end{figure}

\begin{figure}[t]
	\centering
	\includegraphics[scale=0.33]{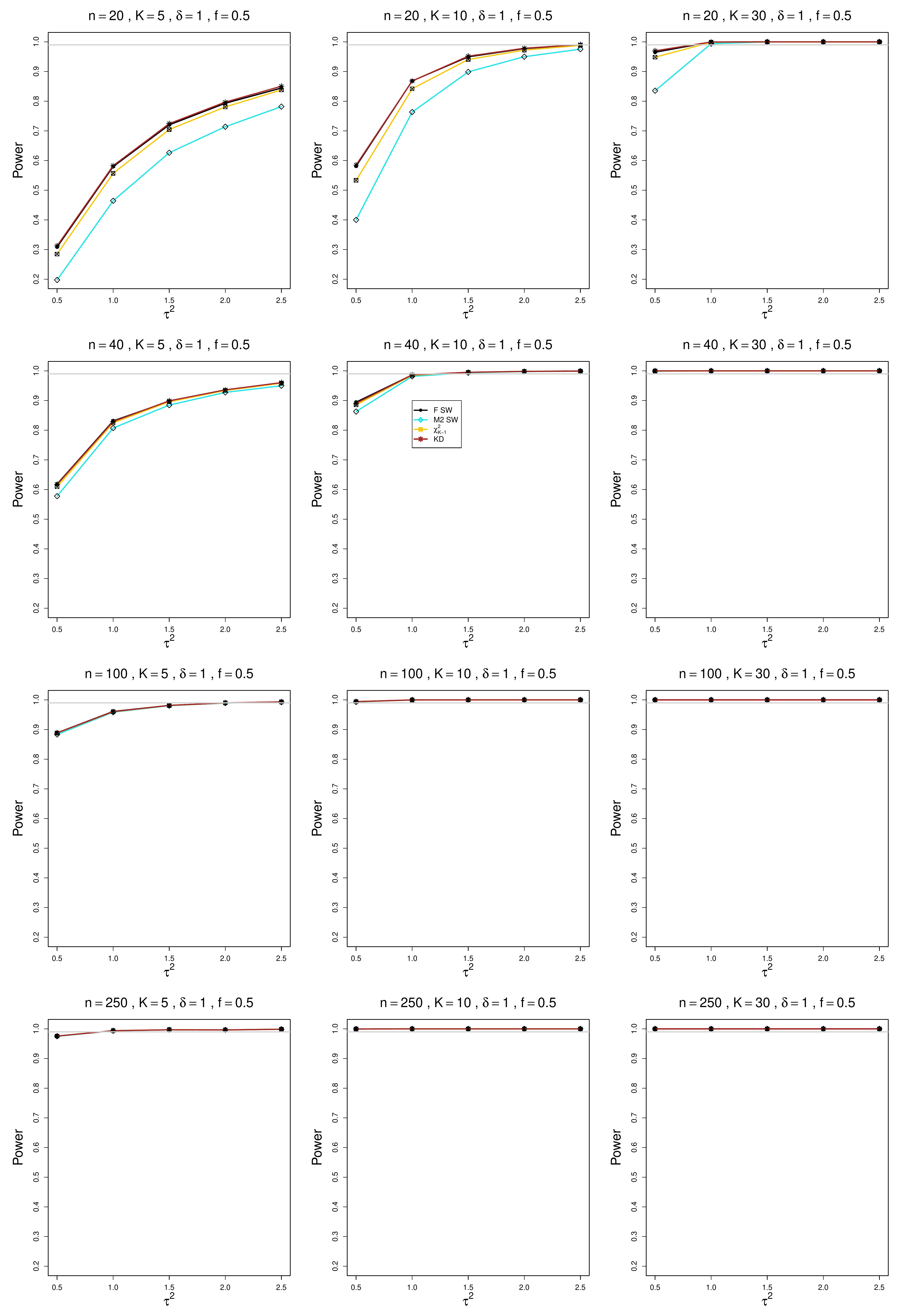}
	\caption{Power for $\delta = 1$, $f = .5$, and equal sample sizes
		\label{PlotOfPhatAt001delta1andq05SMD_underH0}}
\end{figure}

\begin{figure}[t]
	\centering
	\includegraphics[scale=0.33]{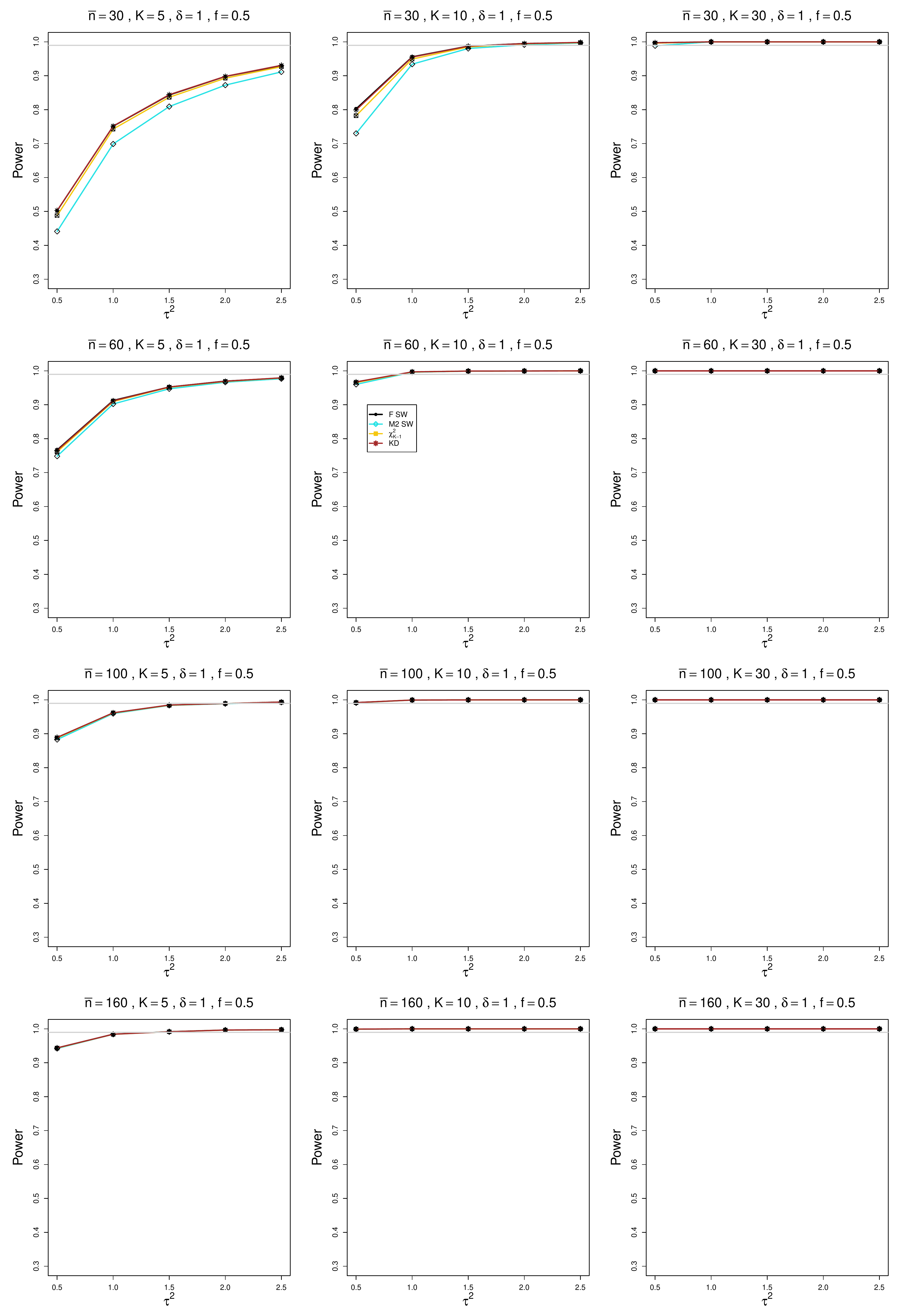}
	\caption{Power for $\delta = 1$, $f = .5$, and unequal sample sizes
		\label{PlotOfPhatAt001delta1andq05SMD_underH0_unequal}}
\end{figure}

\begin{figure}[t]
	\centering
	\includegraphics[scale=0.33]{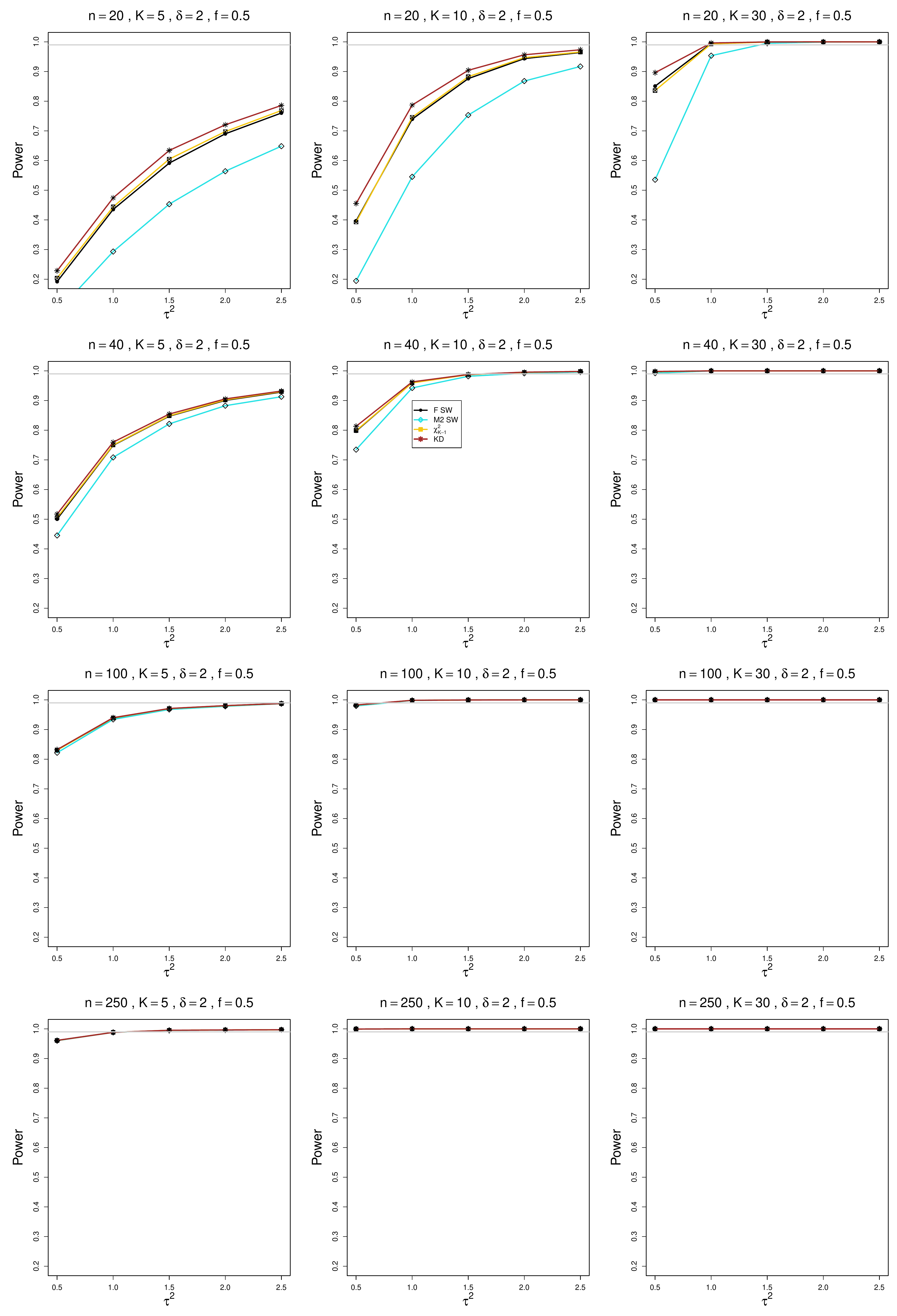}
	\caption{Power for $\delta = 2$, $f = .5$, and equal sample sizes
		\label{PlotOfPhatAt001delta2andq05SMD_underH0}}
\end{figure}

\begin{figure}[t]
	\centering
	\includegraphics[scale=0.33]{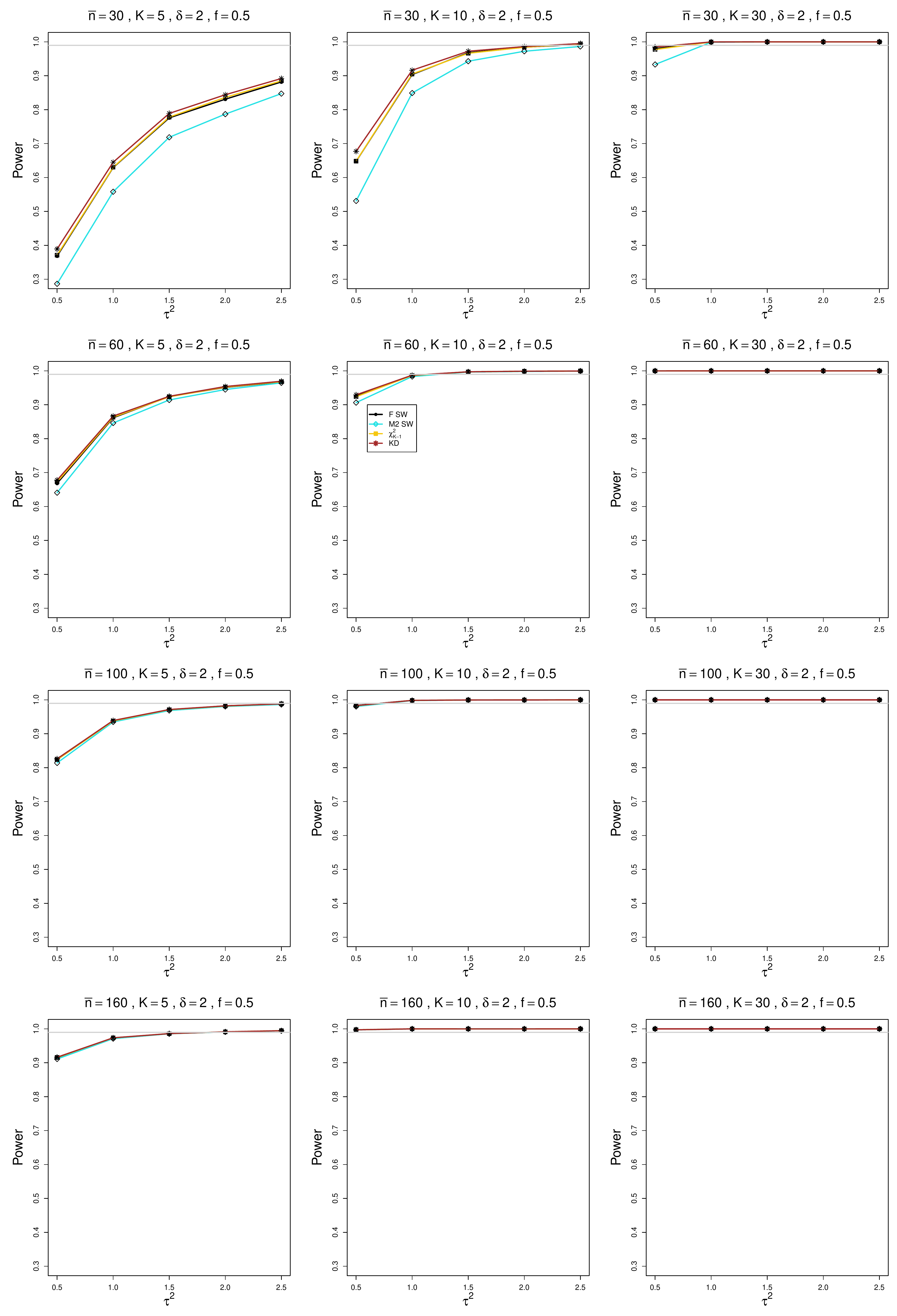}
	\caption{Power for $\delta = 2$, $f = .5$, and unequal sample sizes
		\label{PlotOfPhatAt001delta2andq05SMD_underH0_unequal}}
\end{figure}

%%%%%%%%%%%%%%%%%%%%%%%%%%%%%%%%%%%%%%%%%%%%%%%%%%%%%%%%%%%%%%%%%%%%%%%%%%%%%%%%%%%%%%%%%%%%%%%%%%%%%%%%%%%%%%%%%%%%%%%%%%%%%%
%q=0.75

\begin{figure}[t]
	\centering
	\includegraphics[scale=0.33]{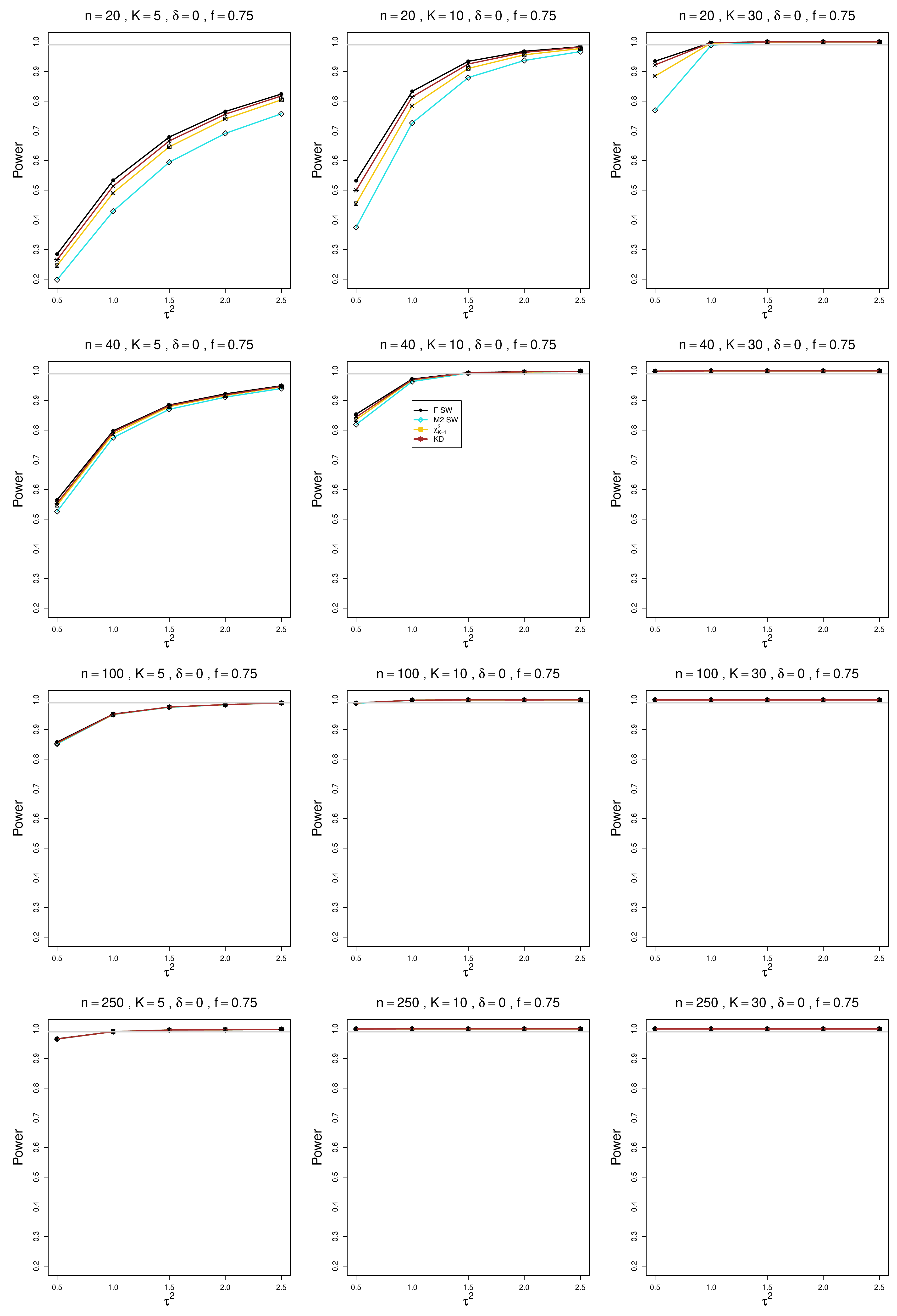}
	\caption{Power for $\delta = 0$, $f = .75$, and equal sample sizes
		\label{PlotOfPhatAt001delta0andq075SMD_underH0}}
\end{figure}

\begin{figure}[t]
	\centering
	\includegraphics[scale=0.33]{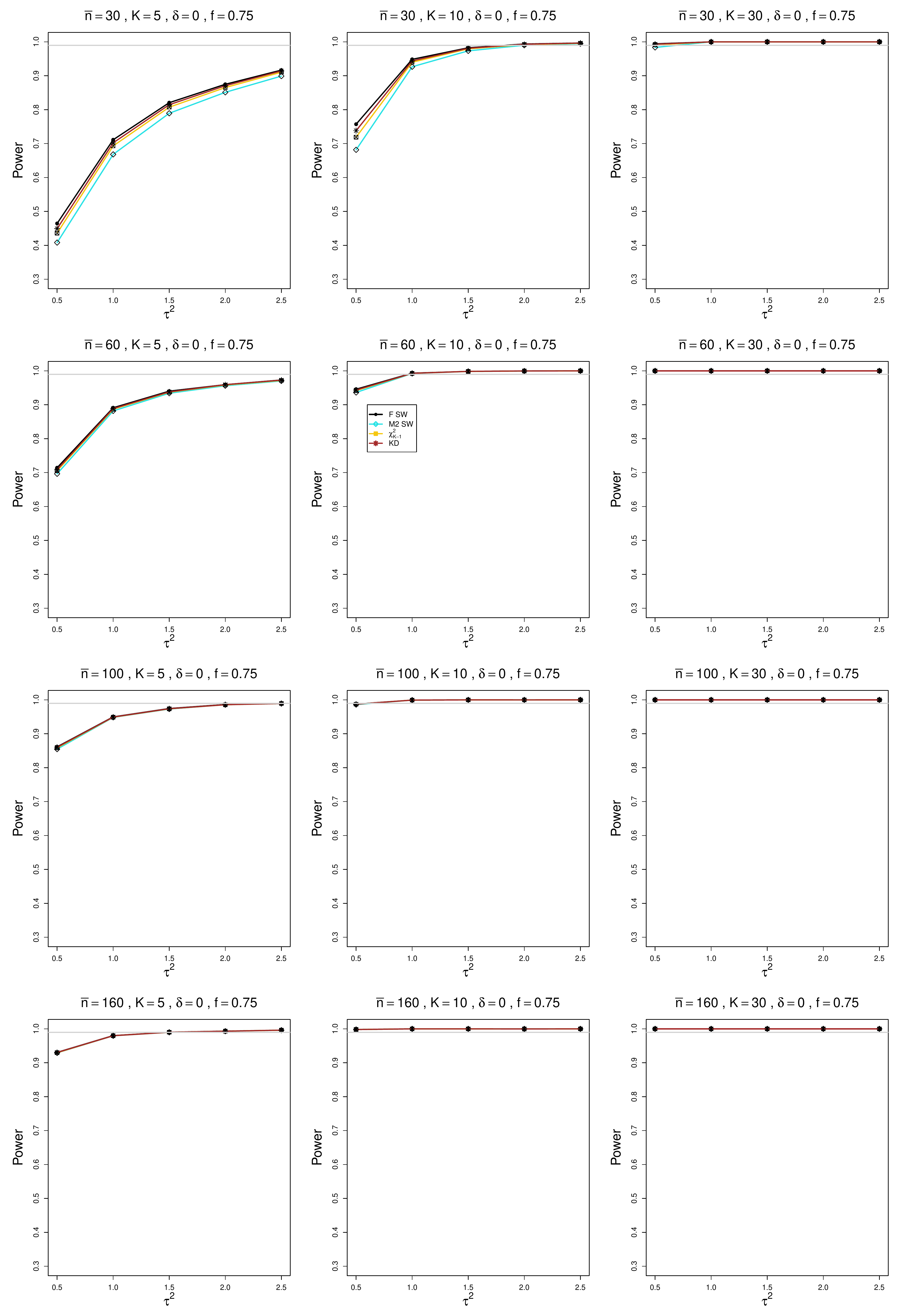}
	\caption{Power for $\delta = 0$, $f = .75$, and unequal sample sizes
		\label{PlotOfPhatAt001delta0andq075SMD_underH0_unequal}}
\end{figure}

\begin{figure}[t]
	\centering
	\includegraphics[scale=0.33]{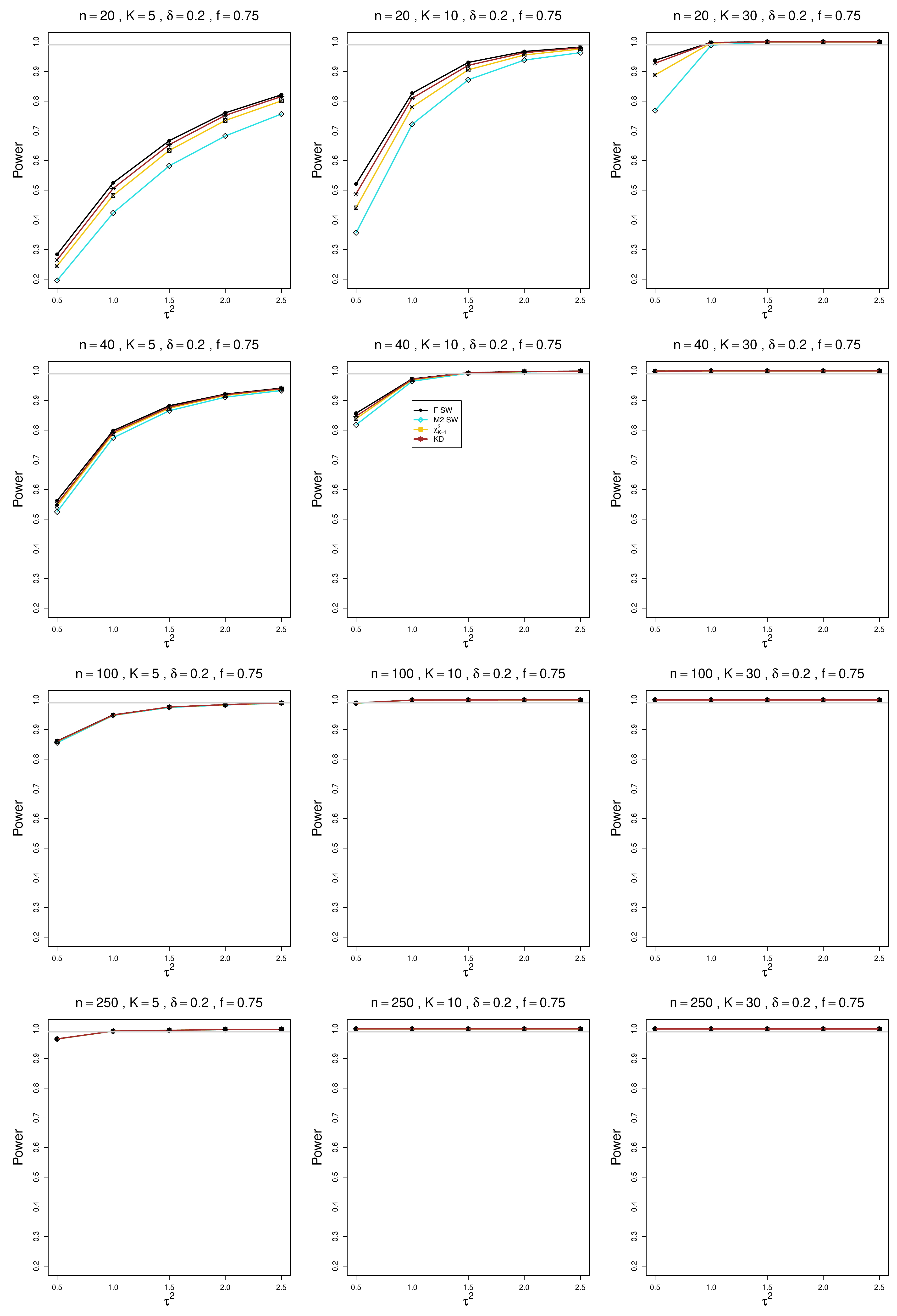}
	\caption{Power for $\delta = 0.2$, $f = .75$, and equal sample sizes
		\label{PlotOfPhatAt001delta02andq075SMD_underH0}}
\end{figure}

\begin{figure}[t]
	\centering
	\includegraphics[scale=0.33]{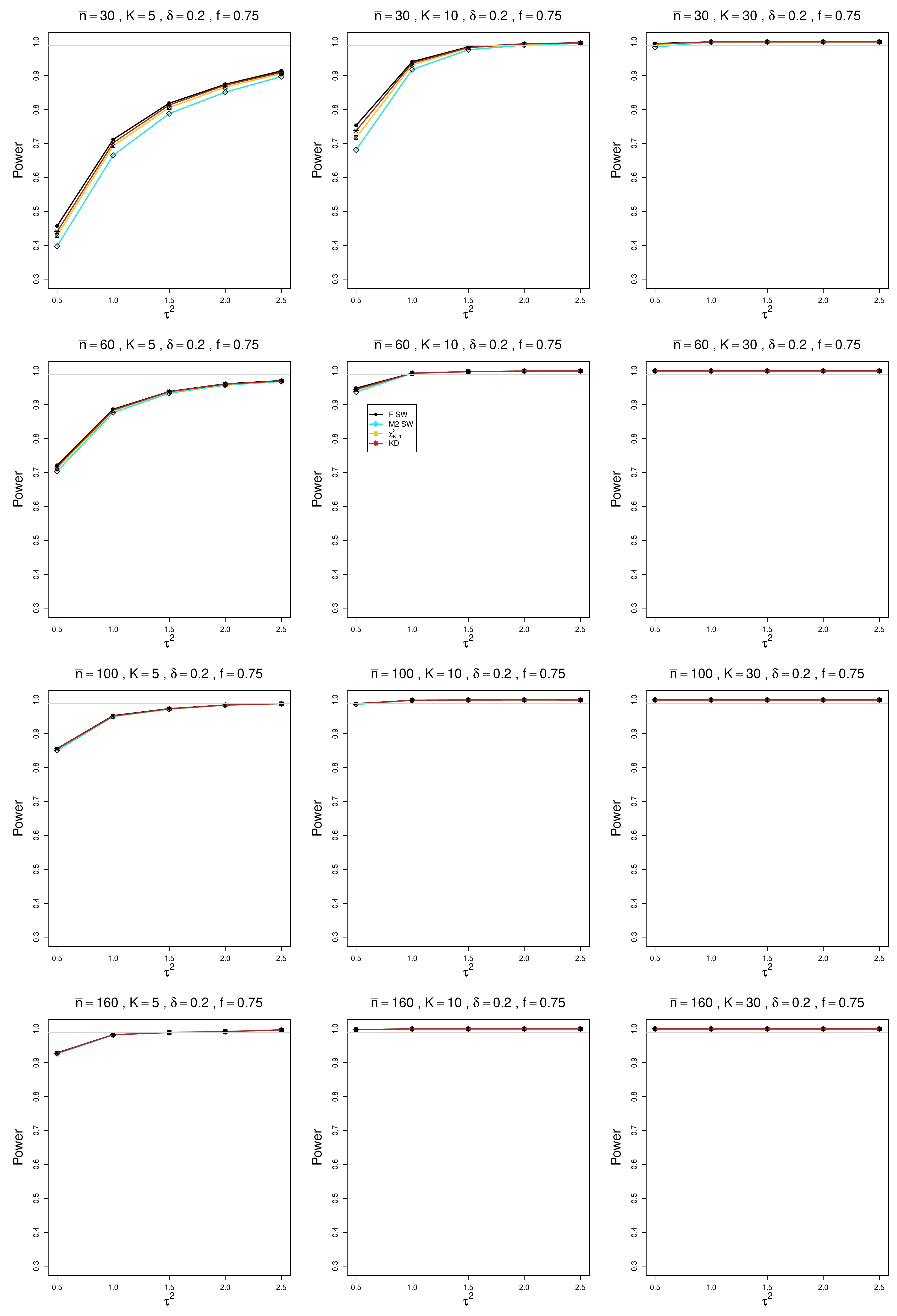}
	\caption{Power for $\delta = 0.2$, $f = .75$, and unequal sample sizes
		\label{PlotOfPhatAt001delta02andq075SMD_underH0_unequal}}
\end{figure}

\begin{figure}[t]
	\centering
	\includegraphics[scale=0.33]{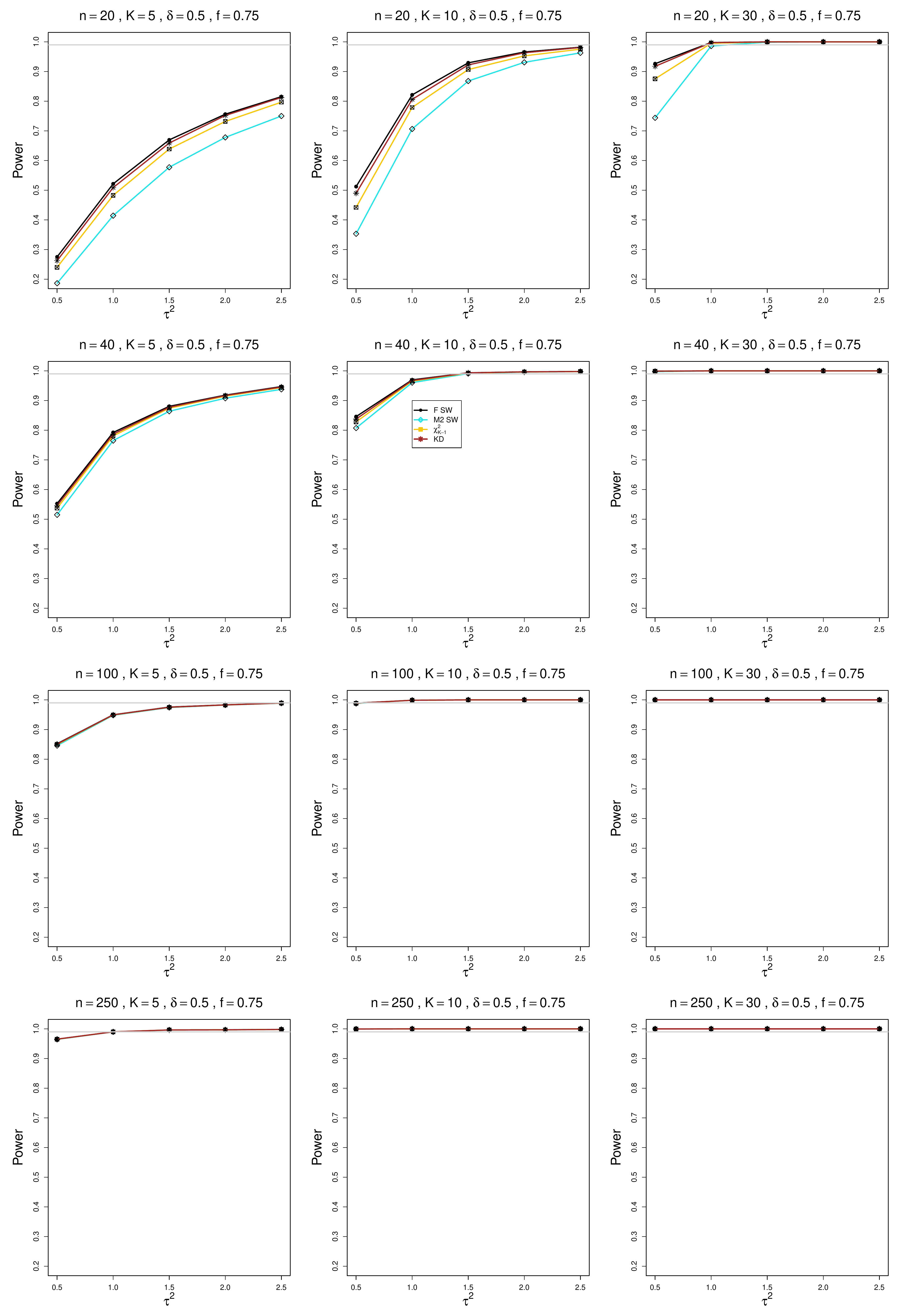}
	\caption{Power for $\delta = 0.5$, $f = .75$, and equal sample sizes
		\label{PlotOfPhatAt001delta05andq075SMD_underH0}}
\end{figure}

\begin{figure}[t]
	\centering
	\includegraphics[scale=0.33]{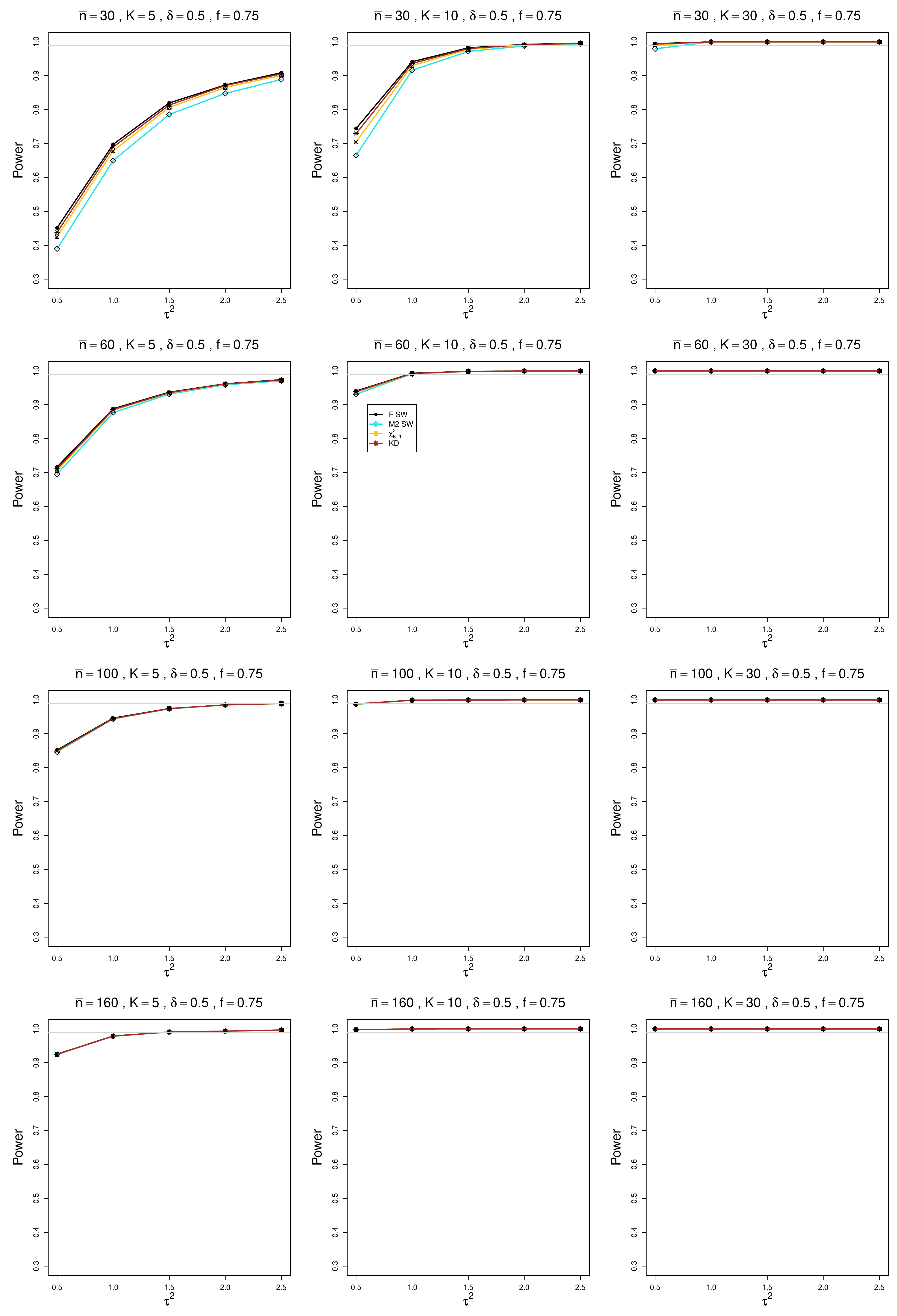}
	\caption{Power for $\delta = 0.5$, $f = .75$, and unequal sample sizes
		\label{PlotOfPhatAt001delta05andq075SMD_underH0_unequal}}
\end{figure}

\begin{figure}[t]
	\centering
	\includegraphics[scale=0.33]{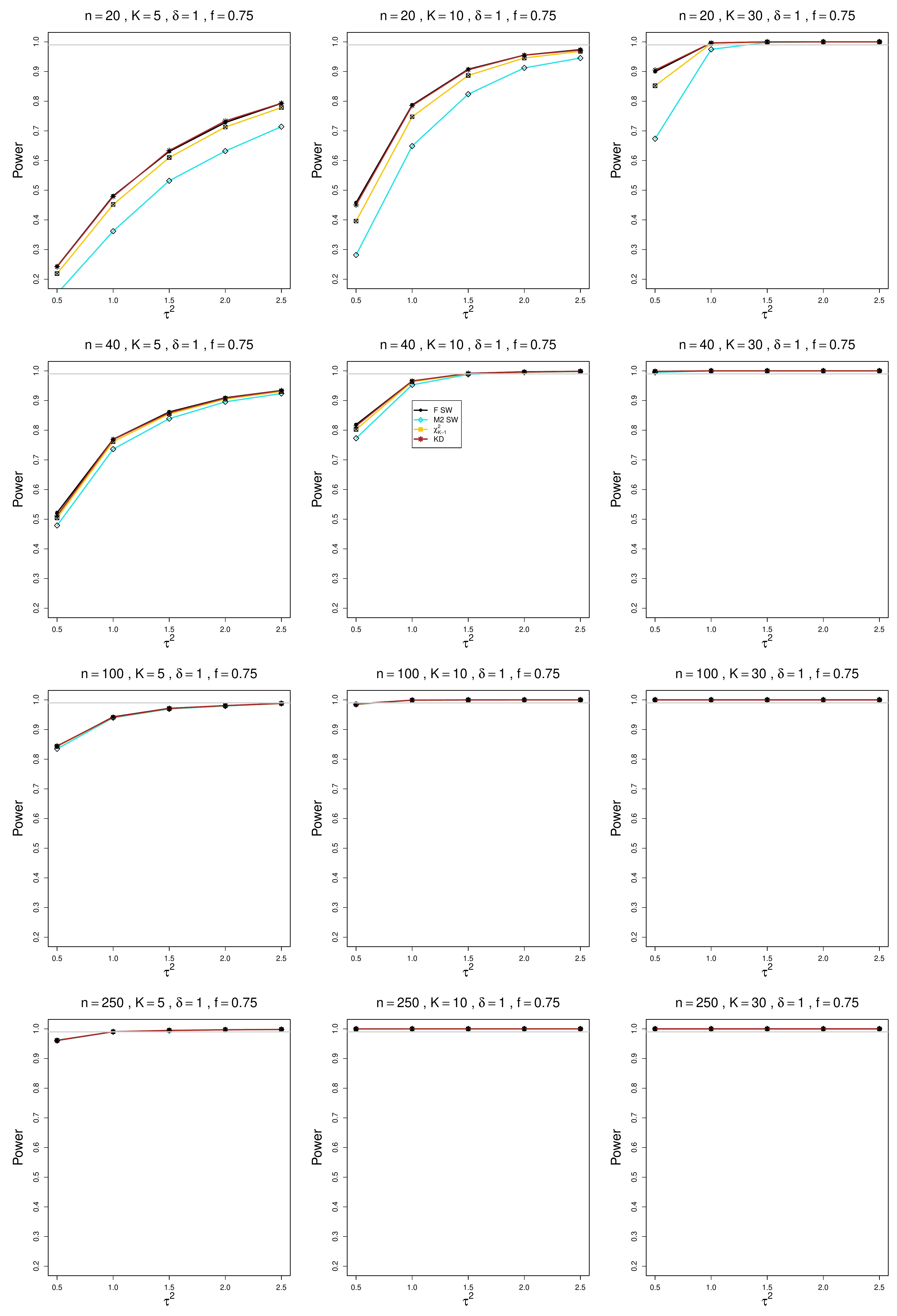}
	\caption{Power for $\delta = 1$, $f = .75$, and equal sample sizes
		\label{PlotOfPhatAt001delta1andq075SMD_underH0}}
\end{figure}

\begin{figure}[t]
	\centering
	\includegraphics[scale=0.33]{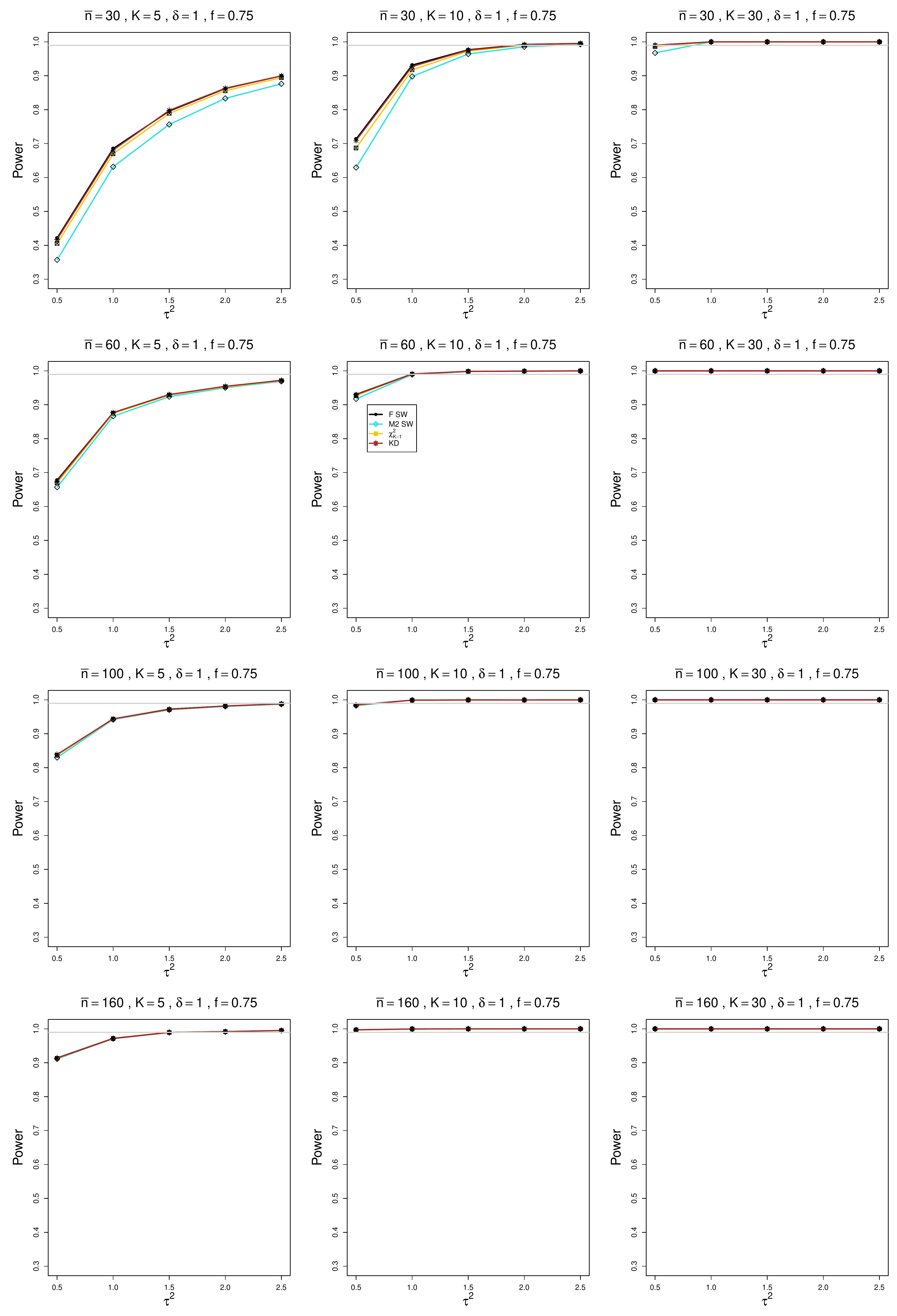}
	\caption{Power for $\delta = 1$, $f = .75$, and unequal sample sizes
		\label{PlotOfPhatAt001delta1andq075SMD_underH0_unequal}}
\end{figure}

\begin{figure}[t]
	\centering
	\includegraphics[scale=0.33]{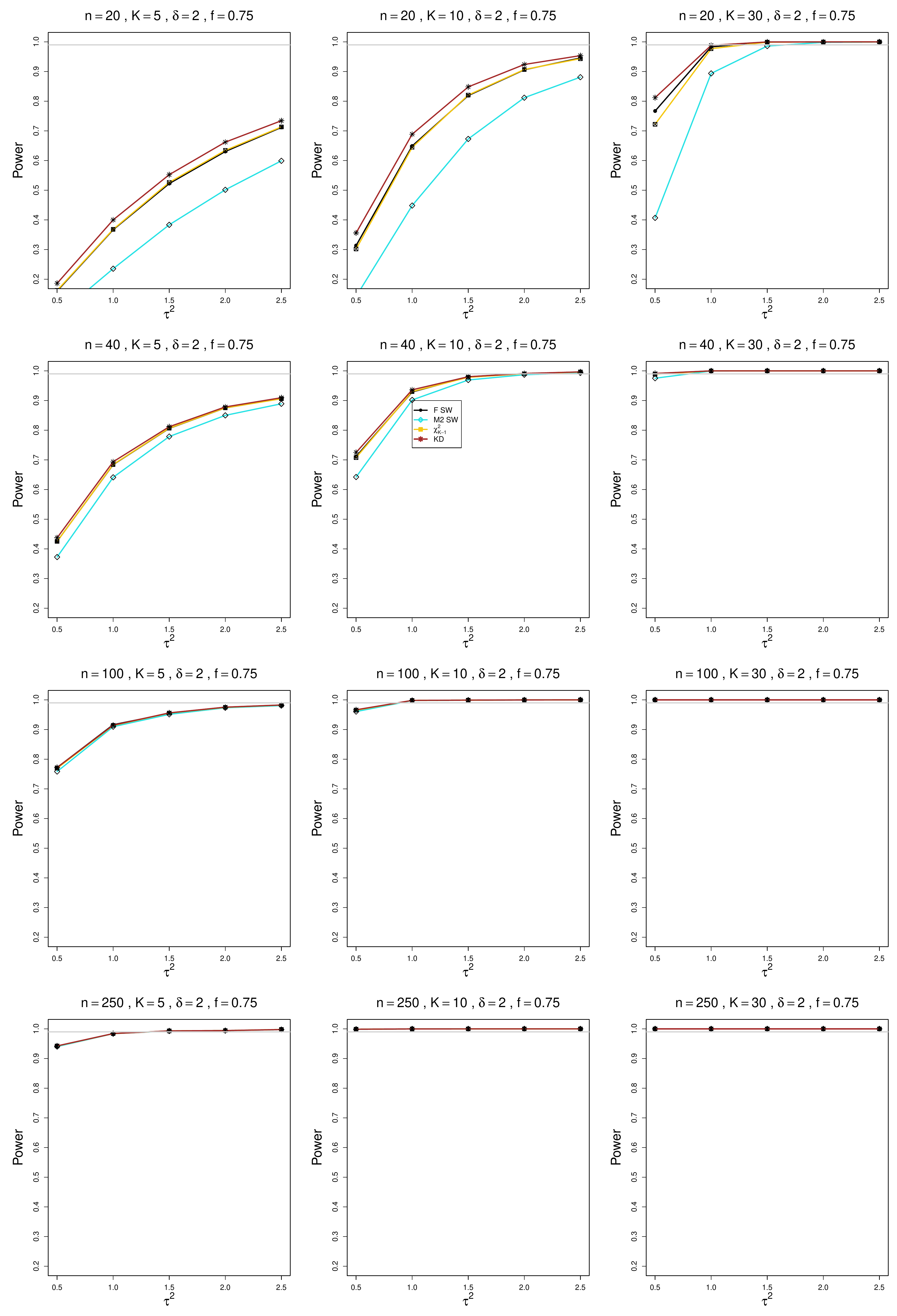}
	\caption{Power for $\delta = 2$, $f = .75$, and equal sample sizes
		\label{PlotOfPhatAt001delta2andq075SMD_underH0}}
\end{figure}

\begin{figure}[t]
	\centering
	\includegraphics[scale=0.33]{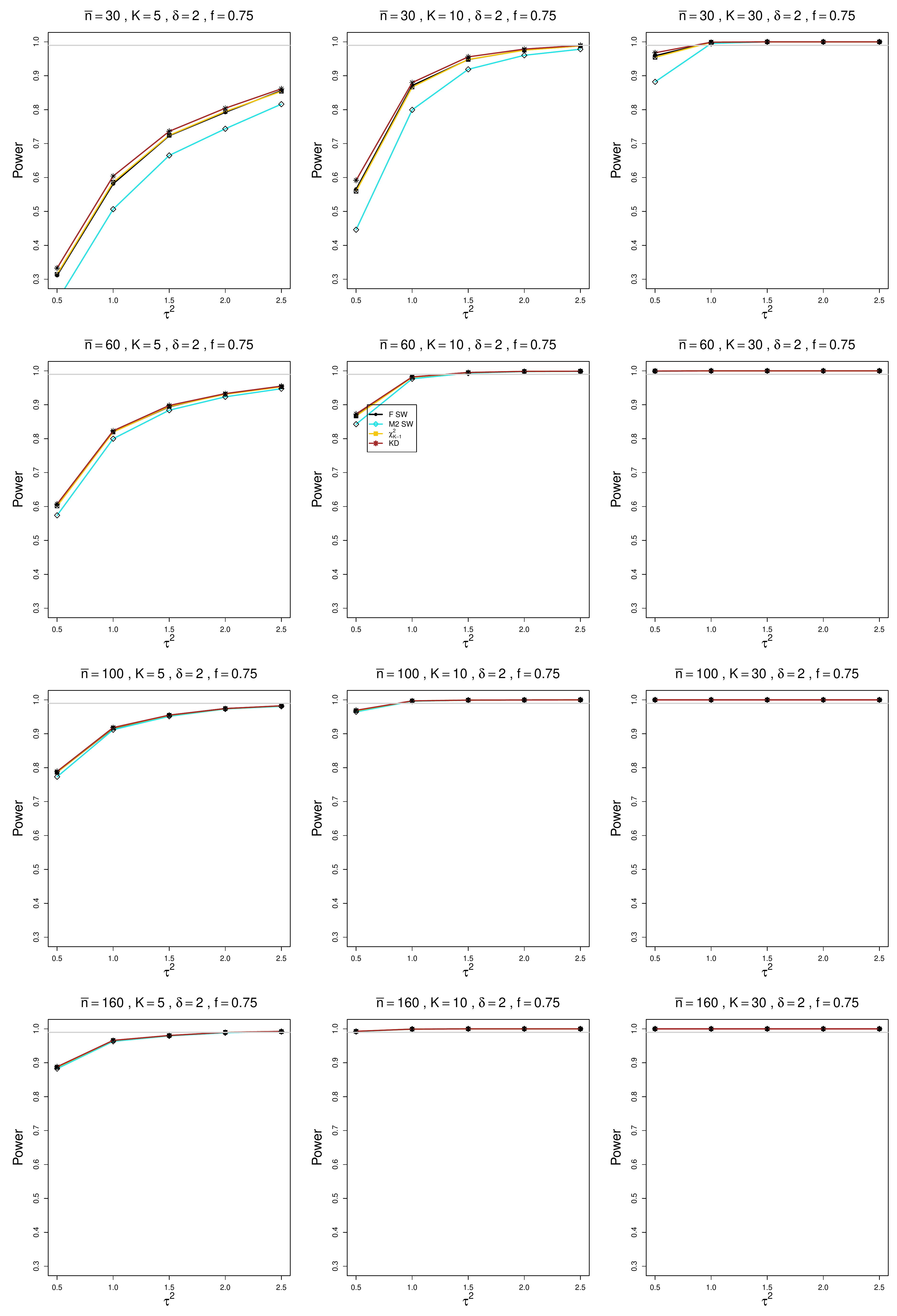}
	\caption{Power for $\delta = 2$, $f = .75$, and unequal sample sizes
		\label{PlotOfPhatAt001delta2andq075SMD_underH0_unequal}}
\end{figure}

\clearpage
\setcounter{figure}{0}
\renewcommand{\thefigure}{E2.\arabic{figure}}
%alpha 0.05 under H0
\subsection*{E2. Power when $\alpha = .05$}

\begin{figure}[t]
	\centering
	\includegraphics[scale=0.33]{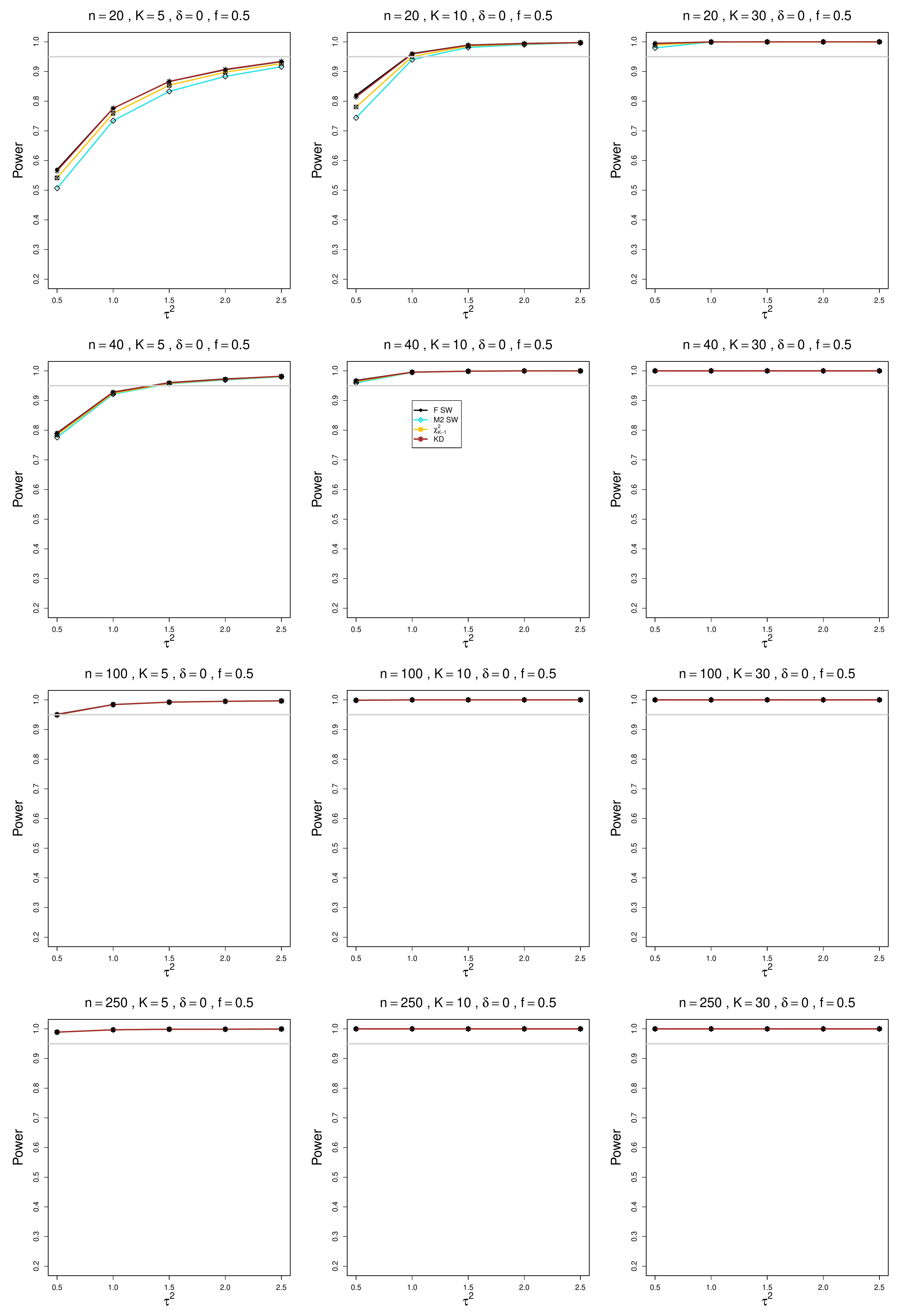}
	\caption{Power for $\delta = 0$, $f = .5$, and equal sample sizes
		\label{PlotOfPhatAt005delta0andq05SMD_underH0}}
\end{figure}

\begin{figure}[t]
	\centering
	\includegraphics[scale=0.33]{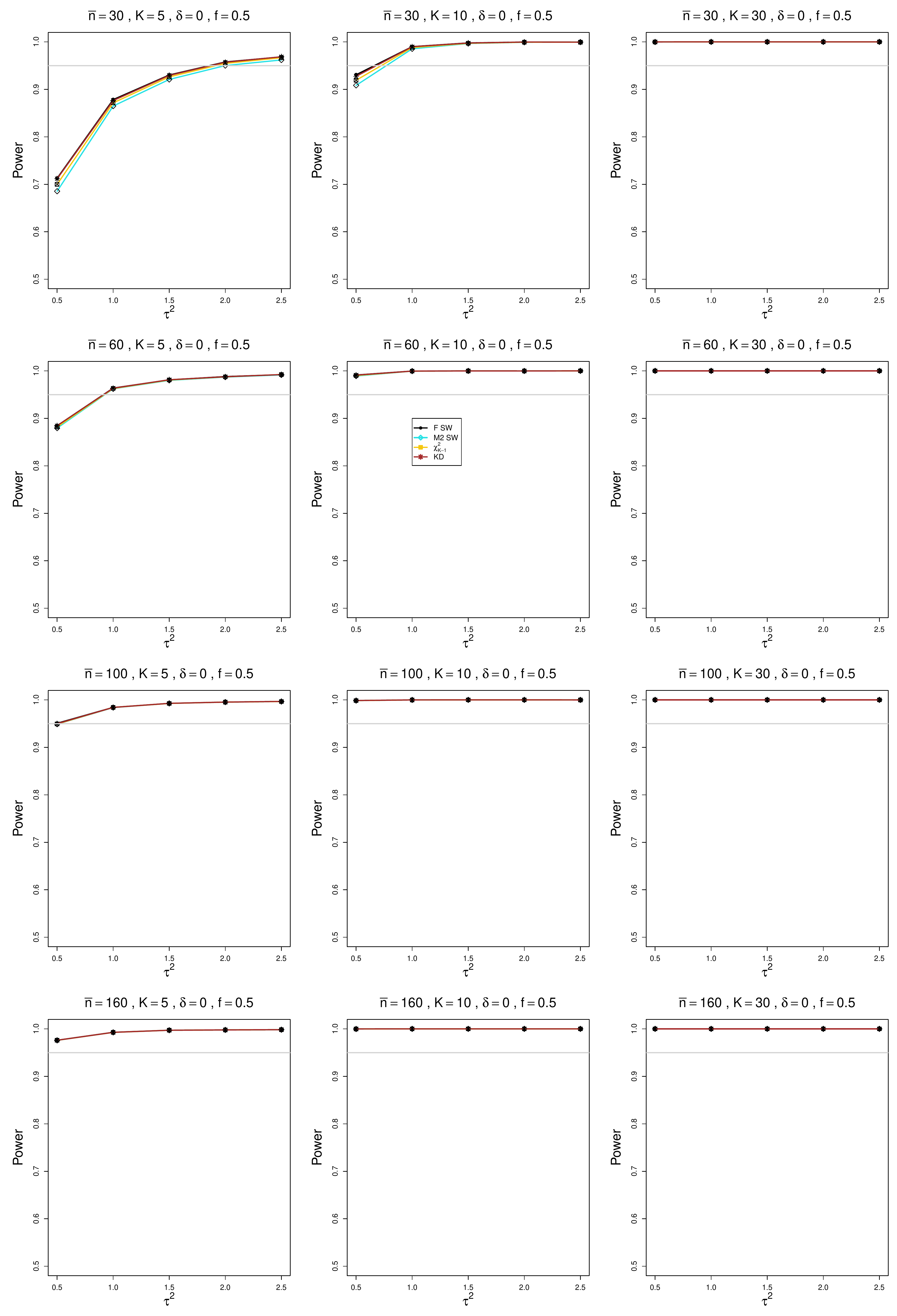}
	\caption{Power for $\delta = 0$, $f = .5$, and unequal sample sizes
		\label{PlotOfPhatAt005delta0andq05SMD_underH0_unequal}}
\end{figure}

\begin{figure}[t]
	\centering
	\includegraphics[scale=0.33]{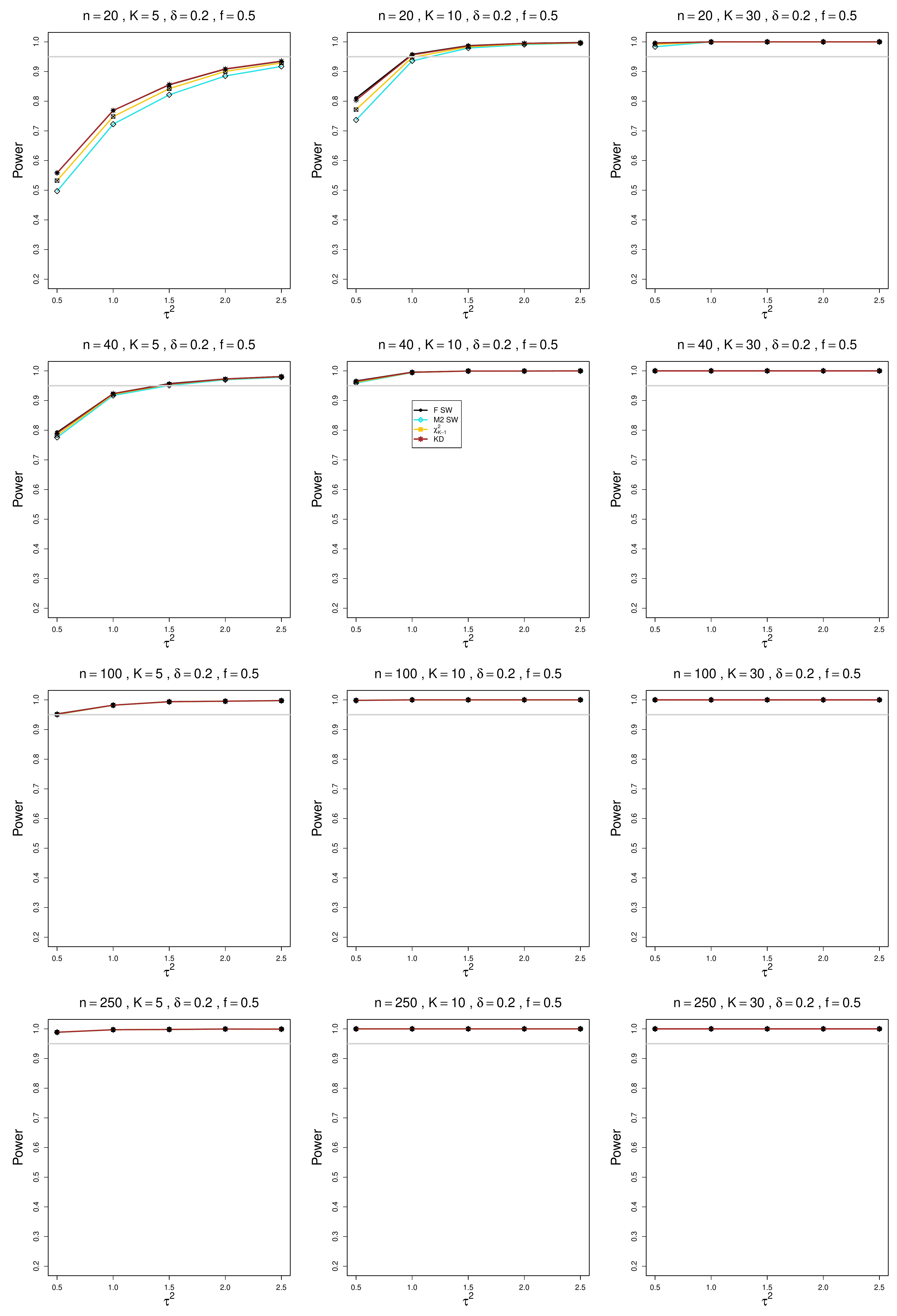}
	\caption{Power for $\delta = 0.2$, $f = .5$, and equal sample sizes
		\label{PlotOfPhatAt005delta02andq05SMD_underH0}}
\end{figure}

\begin{figure}[t]
	\centering
	\includegraphics[scale=0.33]{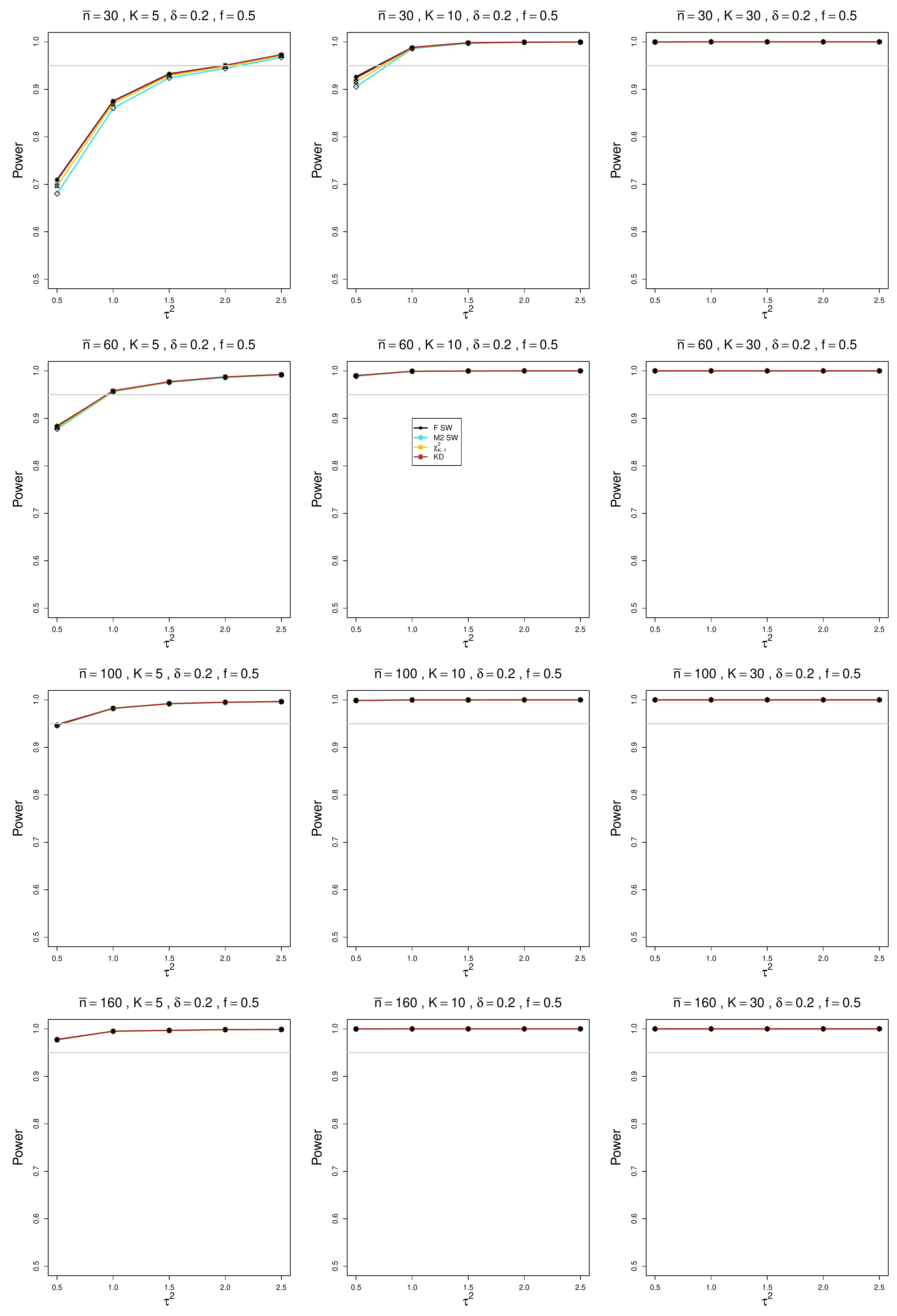}
	\caption{Power for $\delta = 0.2$, $f = .5$, and unequal sample sizes
		\label{PlotOfPhatAt005delta02andq05SMD_underH0_unequal}}
\end{figure}

\begin{figure}[t]
	\centering
	\includegraphics[scale=0.33]{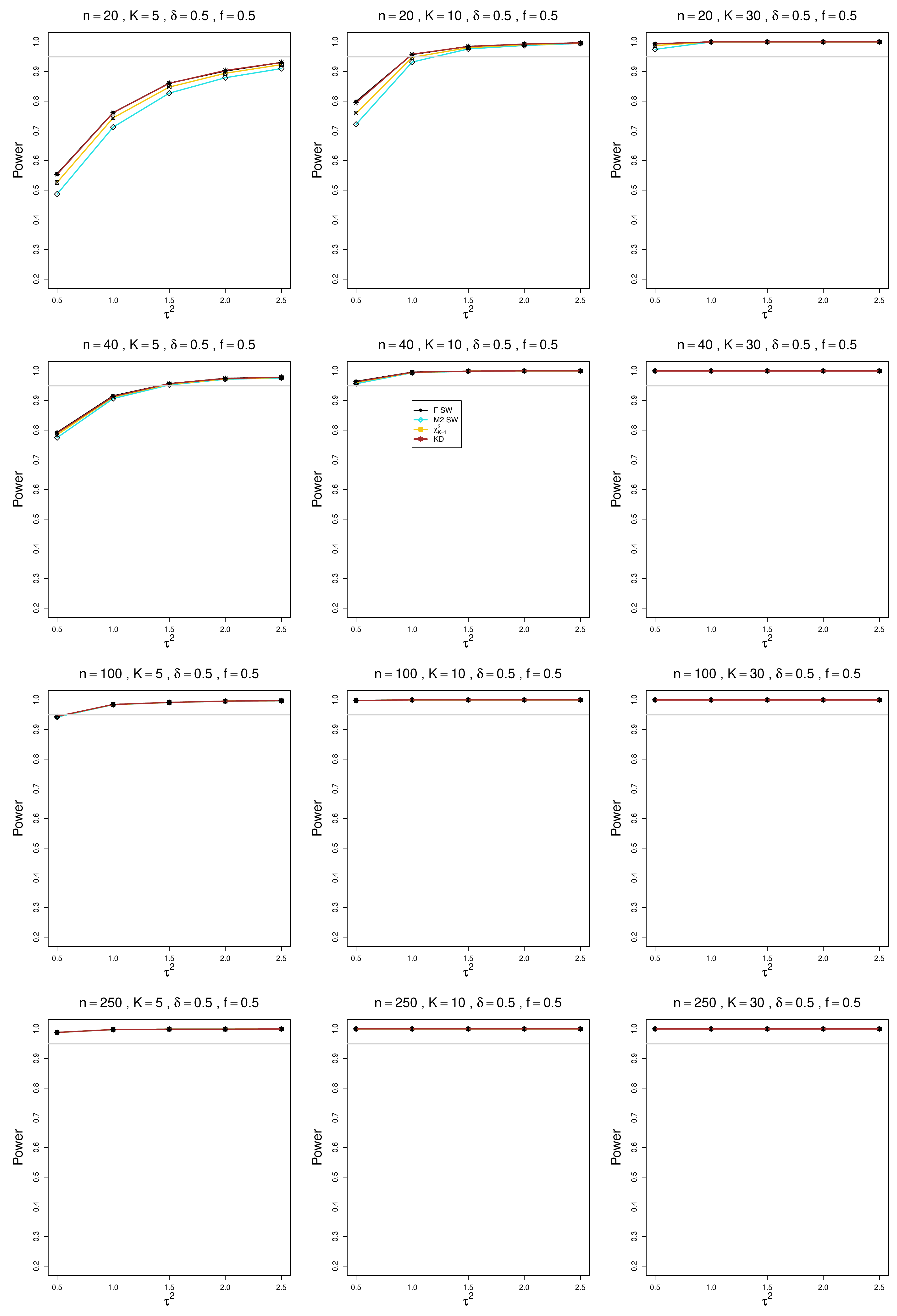}
	\caption{Power for $\delta = 0.5$, $f = .5$, and equal sample sizes
		\label{PlotOfPhatAt005delta05andq05SMD_underH0}}
\end{figure}

\begin{figure}[t]
	\centering
	\includegraphics[scale=0.33]{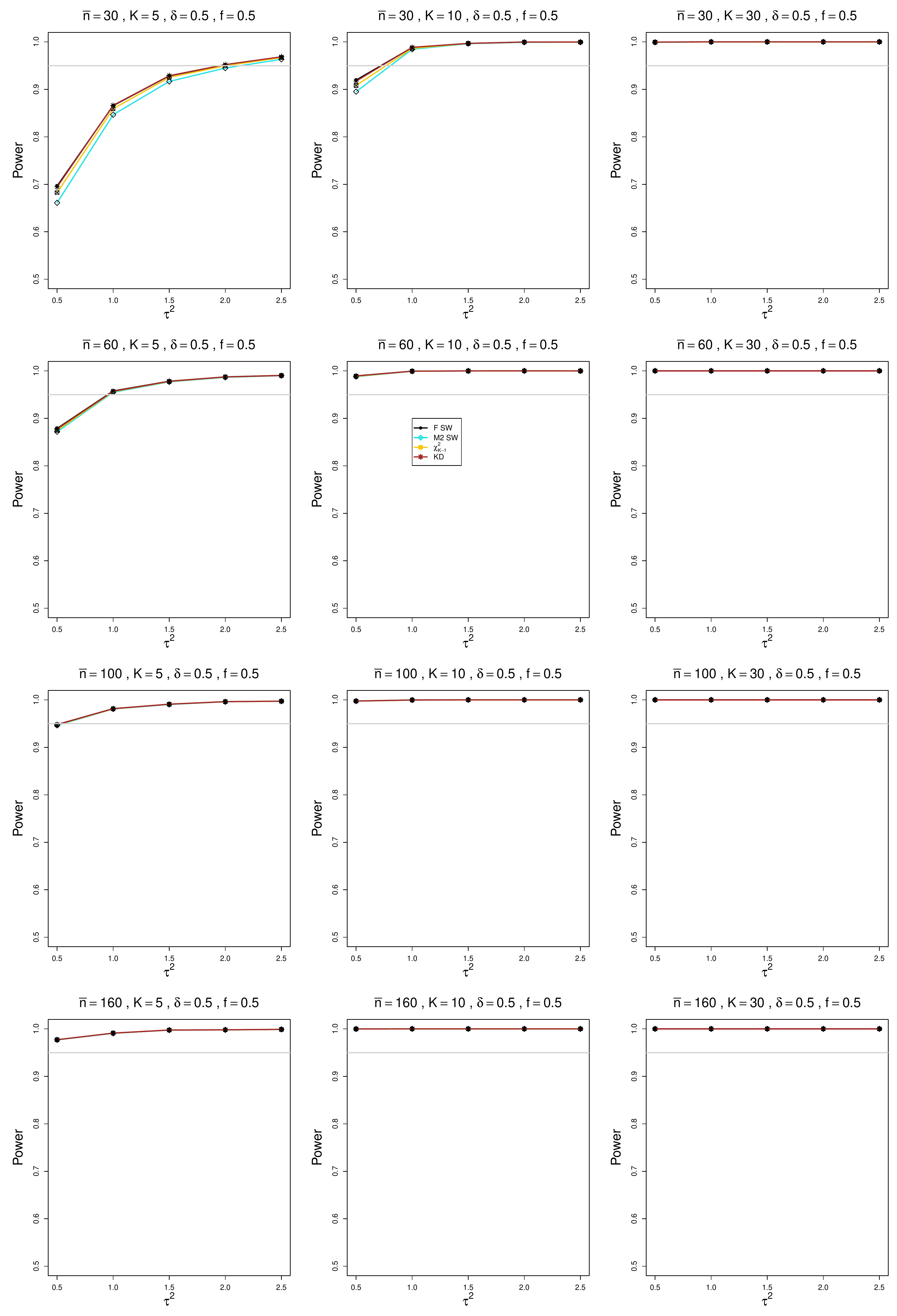}
	\caption{Power for $\delta = 0.5$, $f = .5$, and unequal sample sizes
		\label{PlotOfPhatAt005delta05andq05SMD_underH0_unequal}}
\end{figure}

\begin{figure}[t]
	\centering
	\includegraphics[scale=0.33]{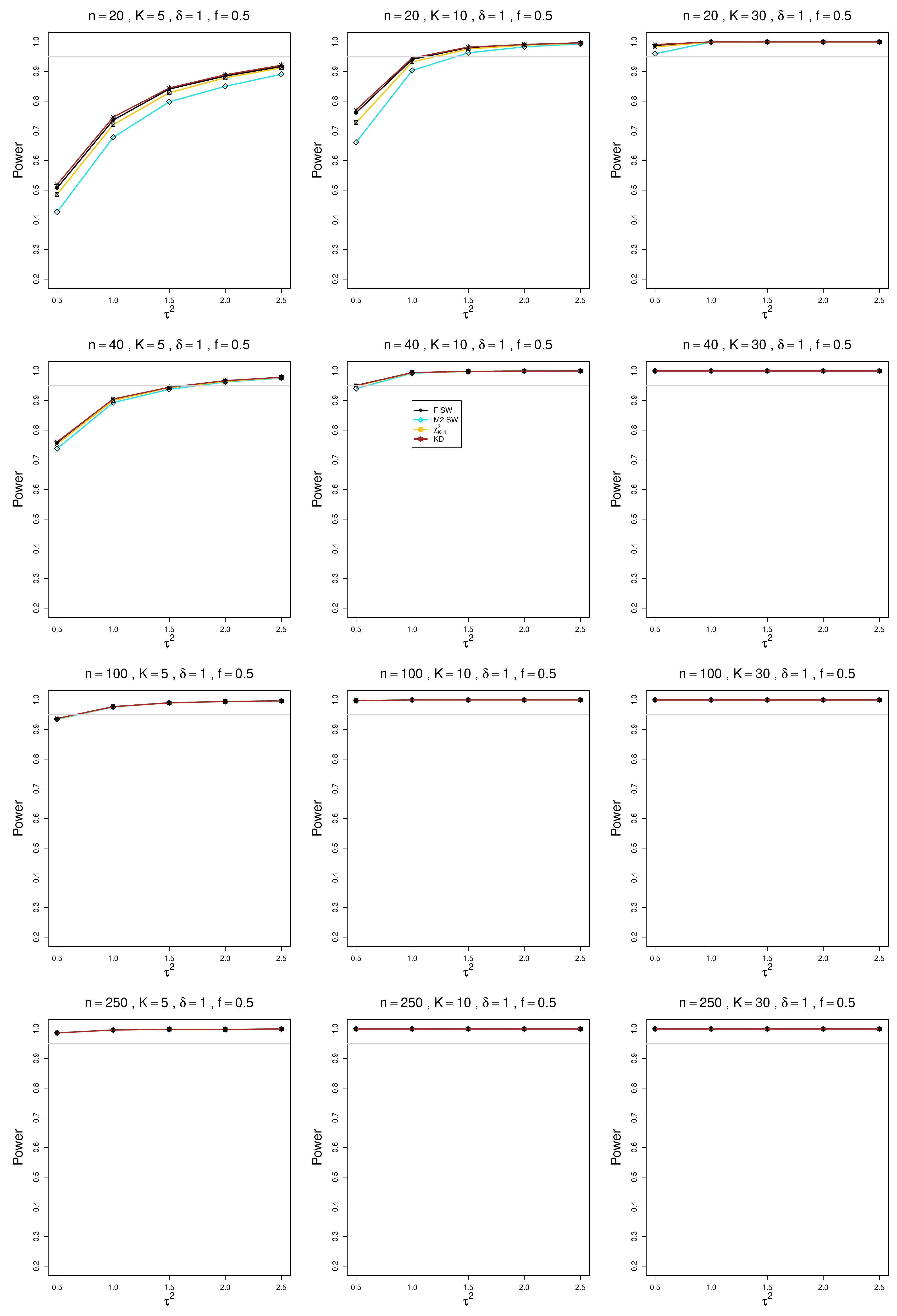}
	\caption{Power for $\delta = 1$, $f = .5$, and equal sample sizes
		\label{PlotOfPhatAt005delta1andq05SMD_underH0}}
\end{figure}

\begin{figure}[t]
	\centering
	\includegraphics[scale=0.33]{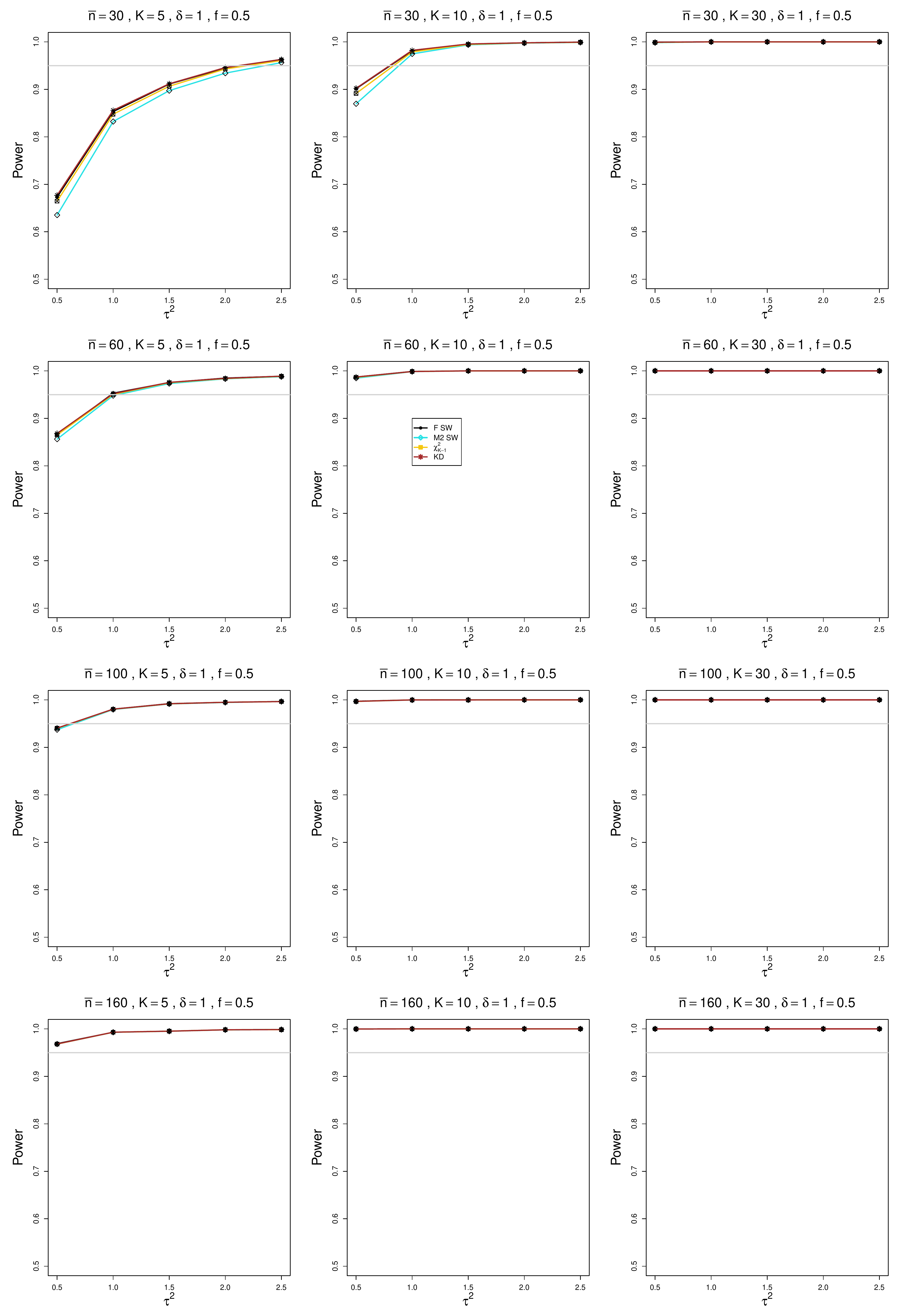}
	\caption{Power for $\delta = 1$, $f = .5$, and unequal sample sizes
		\label{PlotOfPhatAt005delta1andq05SMD_underH0_unequal}}
\end{figure}

\begin{figure}[t]
	\centering
	\includegraphics[scale=0.33]{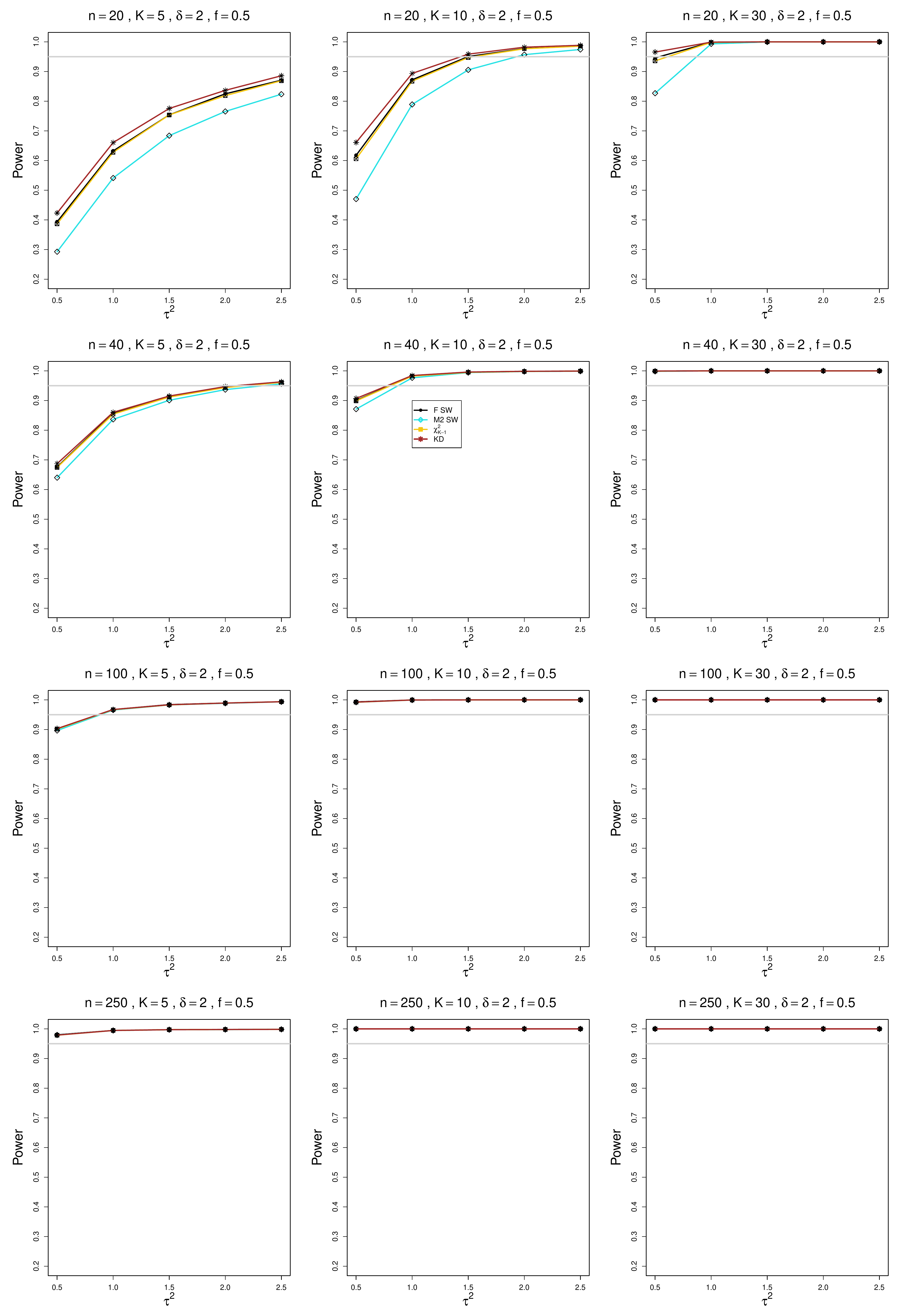}
	\caption{Power for $\delta = 2$, $f = .5$, and equal sample sizes
		\label{PlotOfPhatAt005delta2andq05SMD_underH0}}
\end{figure}

\begin{figure}[t]
	\centering
	\includegraphics[scale=0.33]{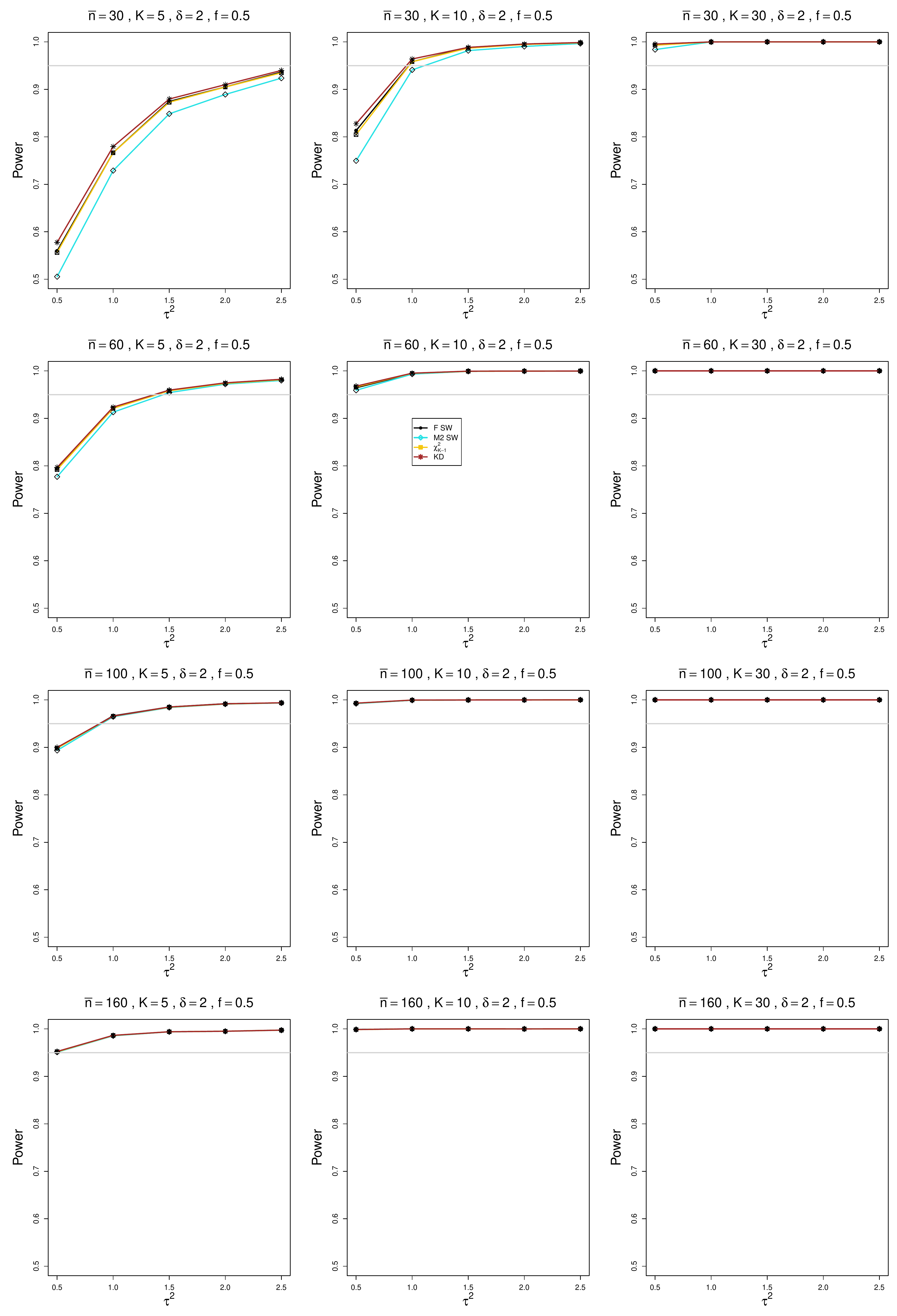}
	\caption{Power for $\delta = 2$, $f = .5$, and unequal sample sizes
		\label{PlotOfPhatAt005delta2andq05SMD_underH0_unequal}}
\end{figure}

%%%%%%%%%%%%%%%%%%%%%%%%%%%%%%%%%%%%%%%%%%%%%%%%%%%%%%%%%%%%%%%%%%%%%%%%%%%%%%%%%%%%%%%%%%%%%%%%%%%%%%%%%%%%%%%%%%%%%%%%%%%%%%
%q=0.75

\begin{figure}[t]
	\centering
	\includegraphics[scale=0.33]{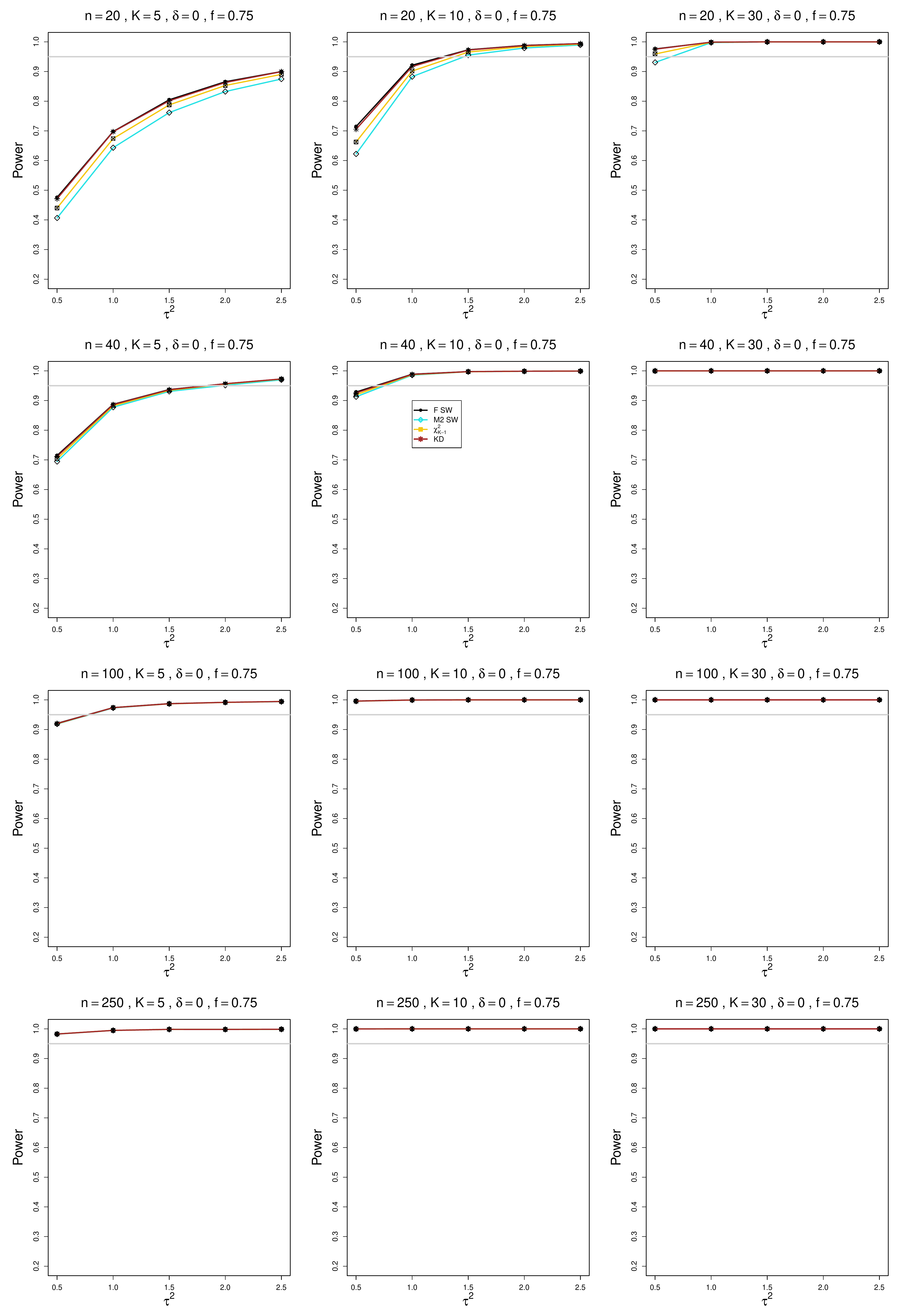}
	\caption{Power for $\delta = 0$, $f = .75$, and equal sample sizes
		\label{PlotOfPhatAt005delta0andq075SMD_underH0}}
\end{figure}

\begin{figure}[t]
	\centering
	\includegraphics[scale=0.33]{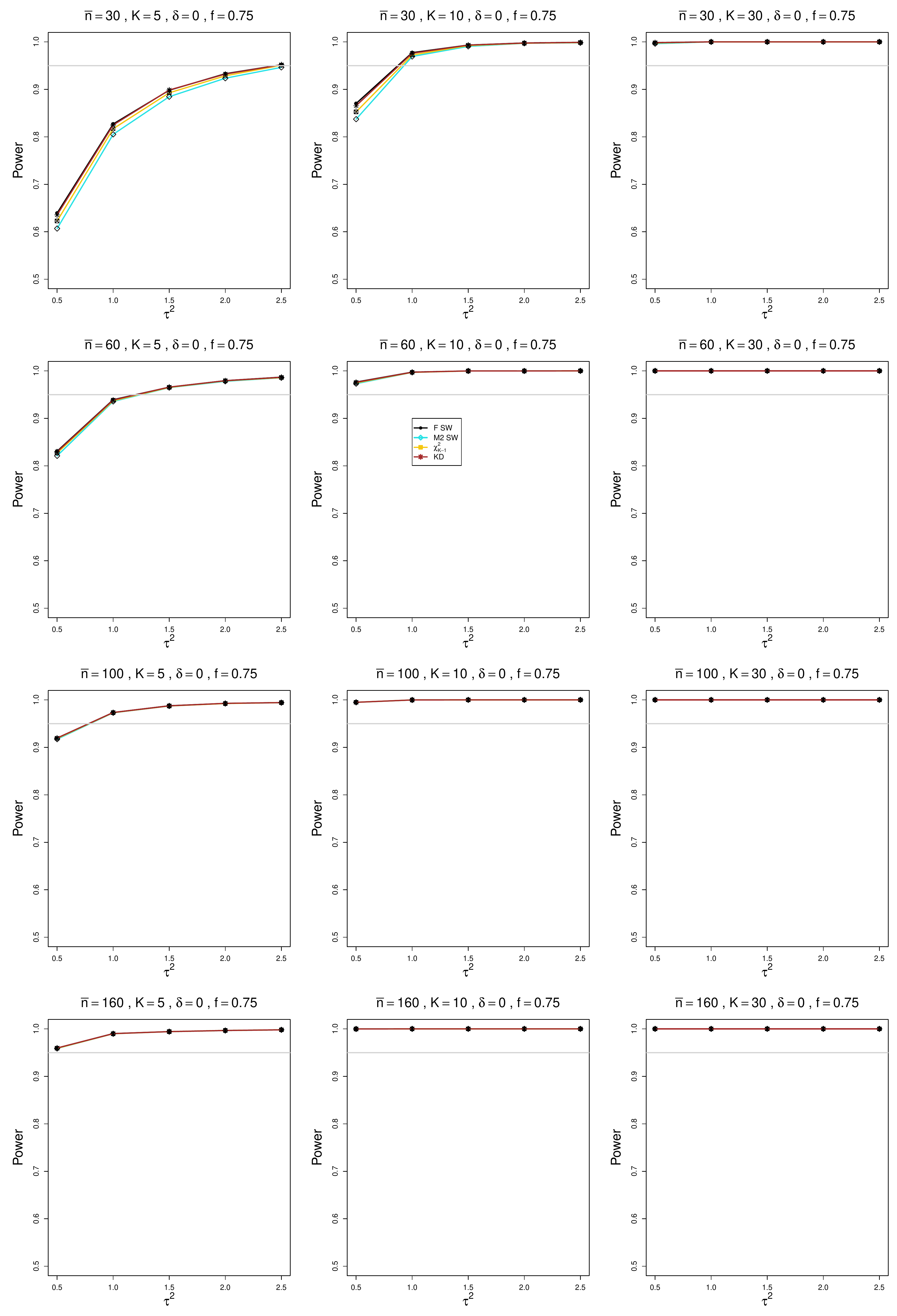}
	\caption{Power for $\delta = 0$, $f = .75$, and unequal sample sizes
		\label{PlotOfPhatAt005delta0andq075SMD_underH0_unequal}}
\end{figure}

\begin{figure}[t]
	\centering
	\includegraphics[scale=0.33]{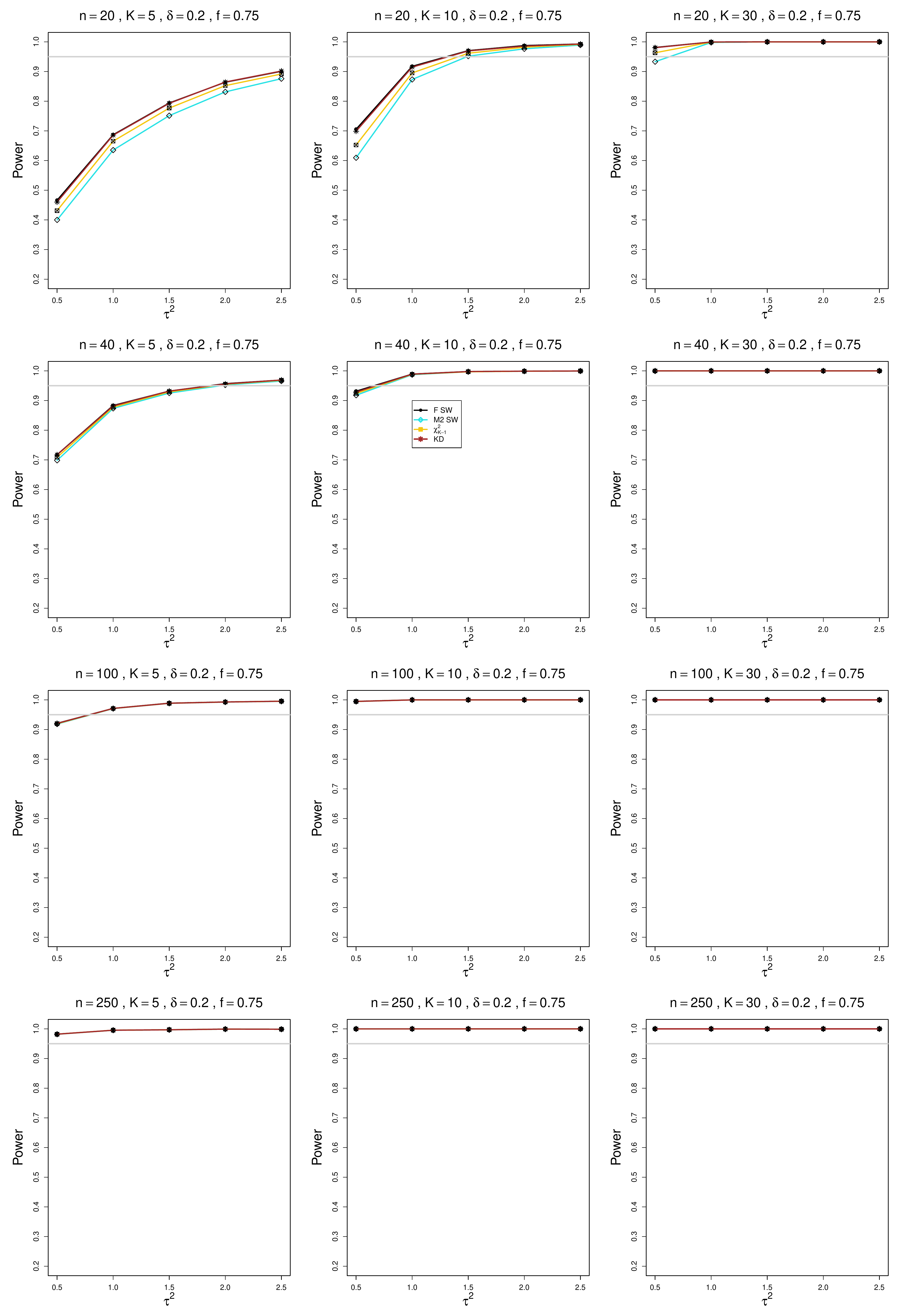}
	\caption{Power for $\delta = 0.2$, $f = .75$, and equal sample sizes
		\label{PlotOfPhatAt005delta02andq075SMD_underH0}}
\end{figure}

\begin{figure}[t]
	\centering
	\includegraphics[scale=0.33]{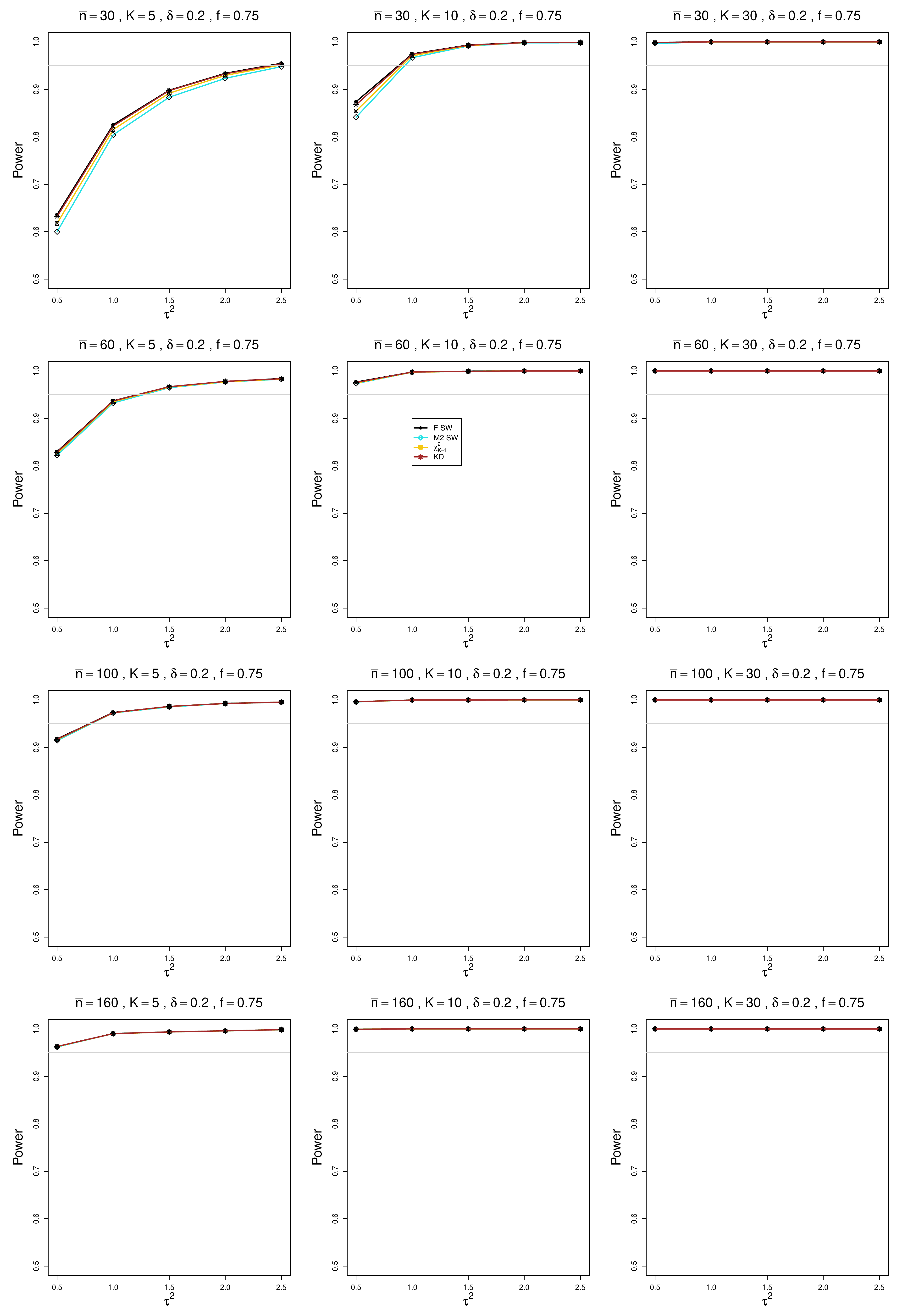}
	\caption{Power for $\delta = 0.2$, $f = .75$, and unequal sample sizes
		\label{PlotOfPhatAt005delta02andq075SMD_underH0_unequal}}
\end{figure}

\begin{figure}[t]
	\centering
	\includegraphics[scale=0.33]{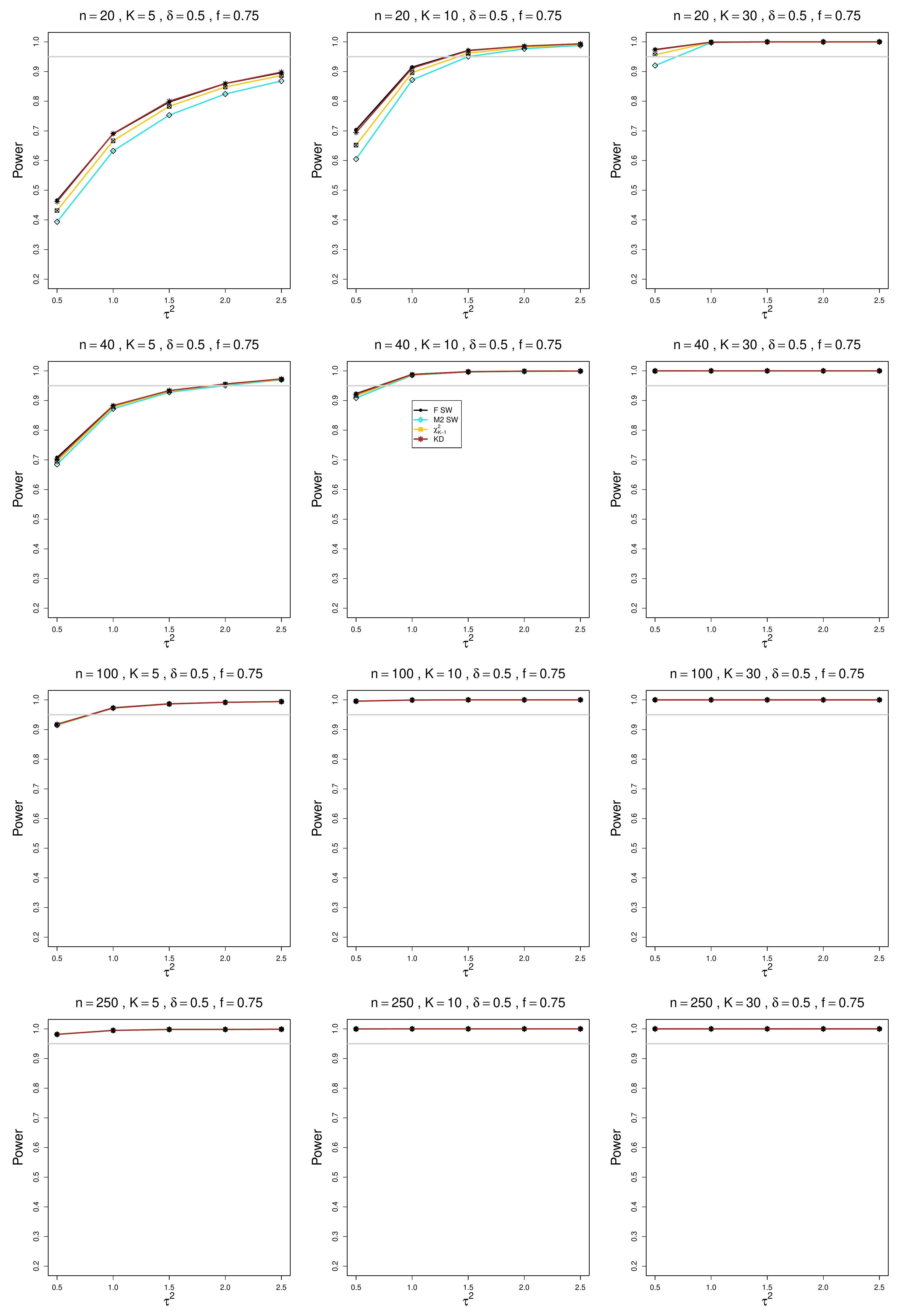}
	\caption{Power for $\delta = 0.5$, $f = .75$, and equal sample sizes
		\label{PlotOfPhatAt005delta05andq075SMD_underH0}}
\end{figure}

\begin{figure}[t]
	\centering
	\includegraphics[scale=0.33]{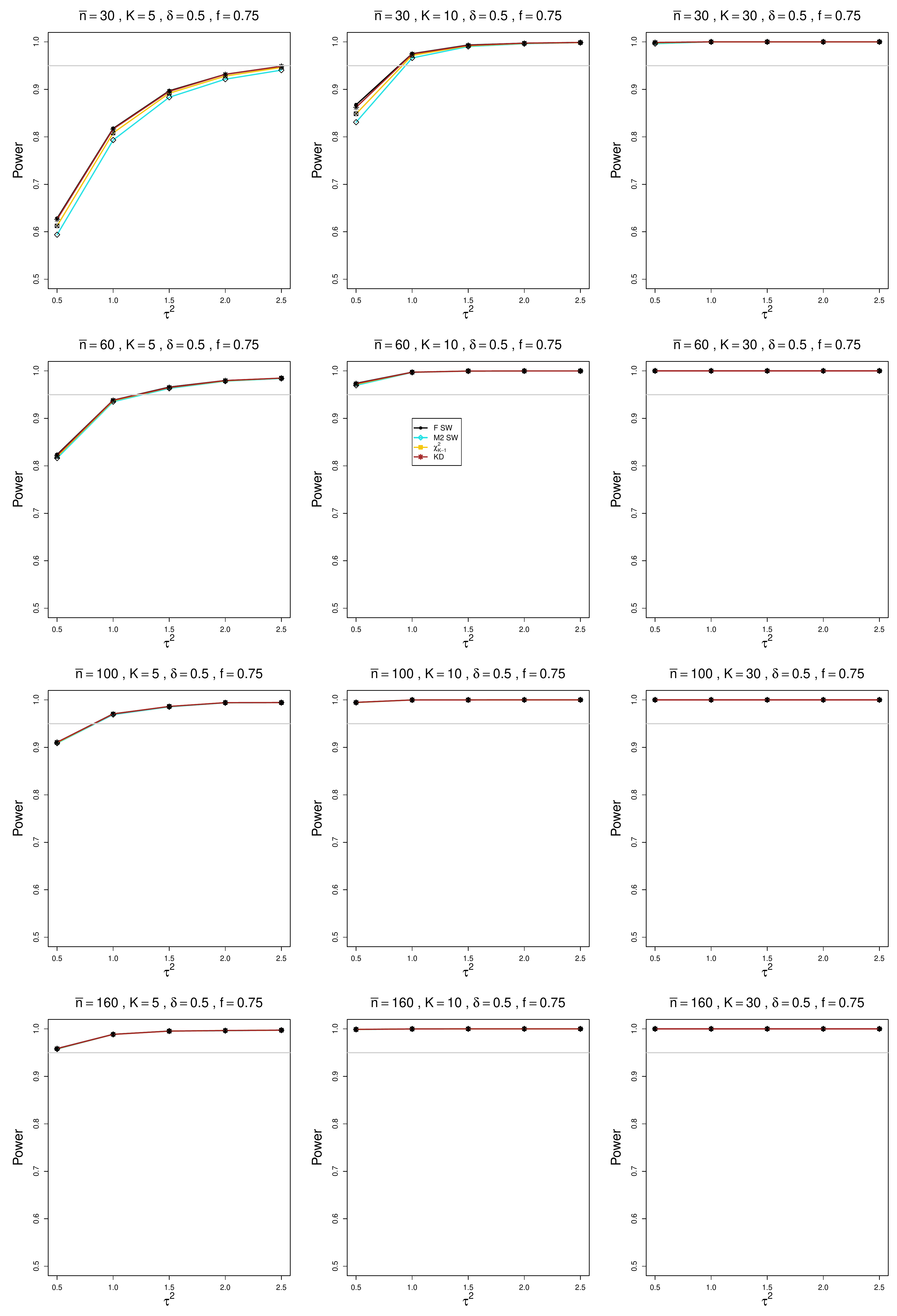}
	\caption{Power for $\delta = 0.5$, $f = .75$, and unequal sample sizes
		\label{PlotOfPhatAt005delta05andq075SMD_underH0_unequal}}
\end{figure}

\begin{figure}[t]
	\centering
	\includegraphics[scale=0.33]{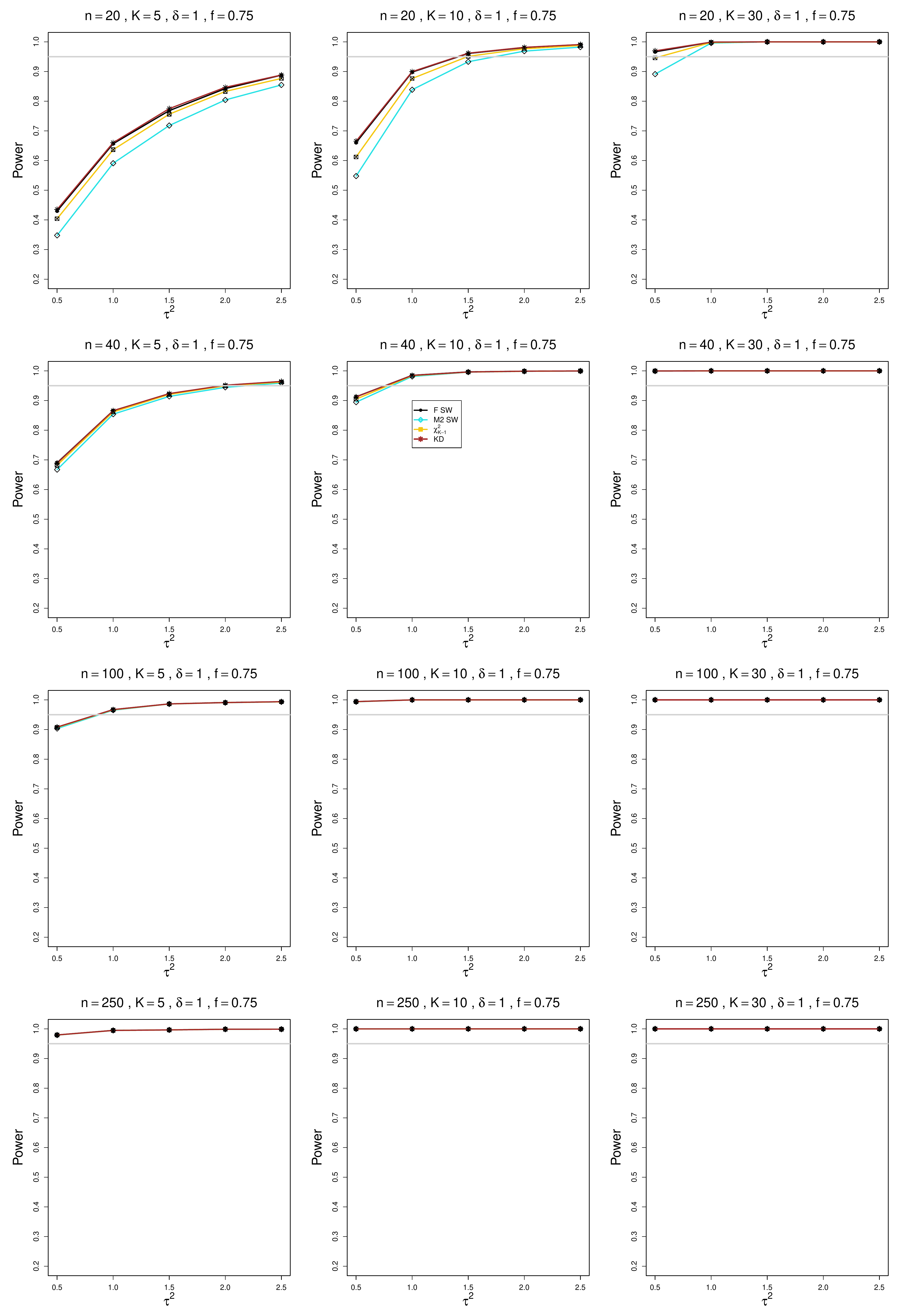}
	\caption{Power for $\delta = 1$, $f = .75$, and equal sample sizes
		\label{PlotOfPhatAt005delta1andq075SMD_underH0}}
\end{figure}

\begin{figure}[t]
	\centering
	\includegraphics[scale=0.33]{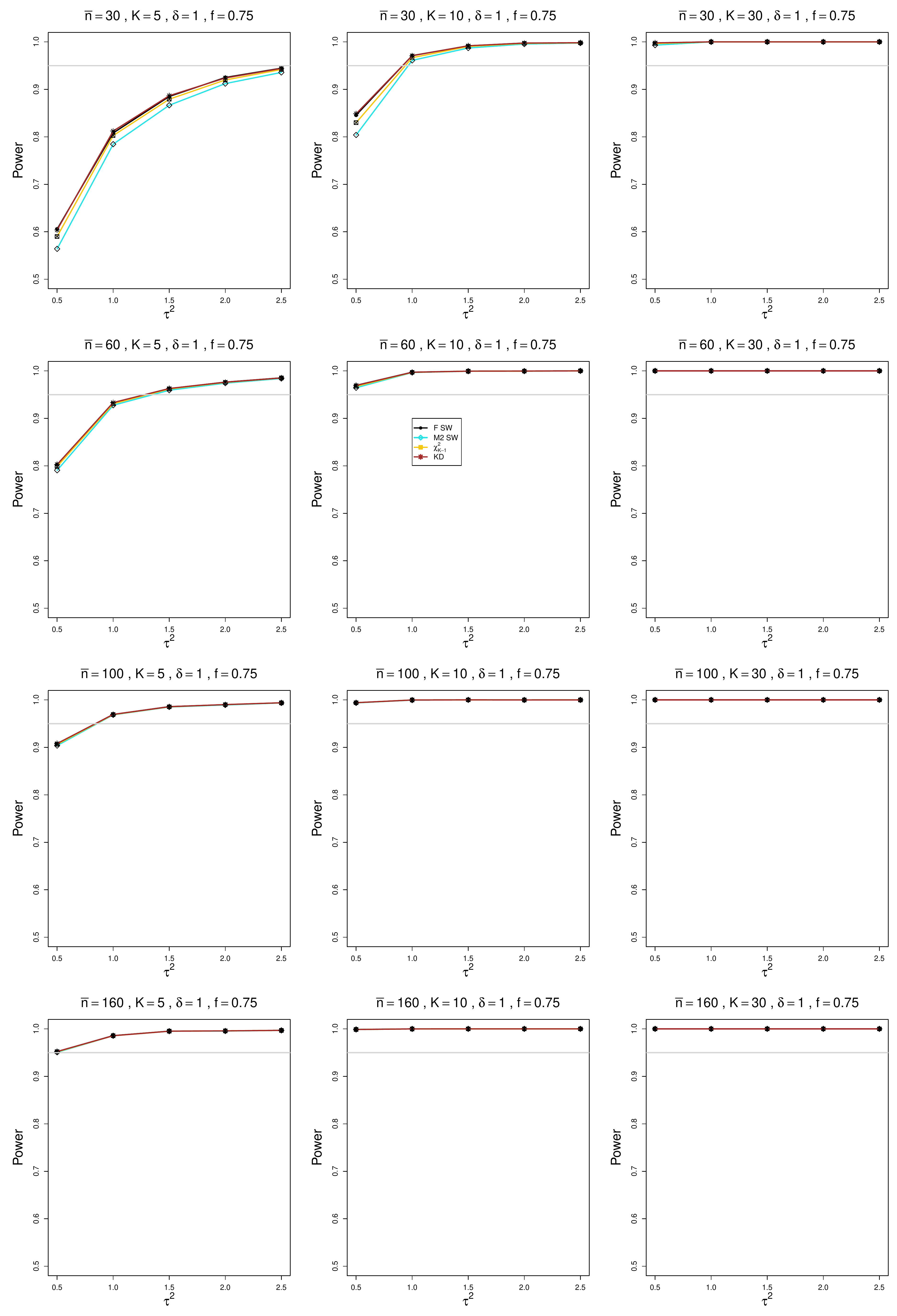}
	\caption{Power for $\delta = 1$, $f = .75$, and unequal sample sizes
		\label{PlotOfPhatAt005delta1andq075SMD_underH0_unequal}}
\end{figure}

\begin{figure}[t]
	\centering
	\includegraphics[scale=0.33]{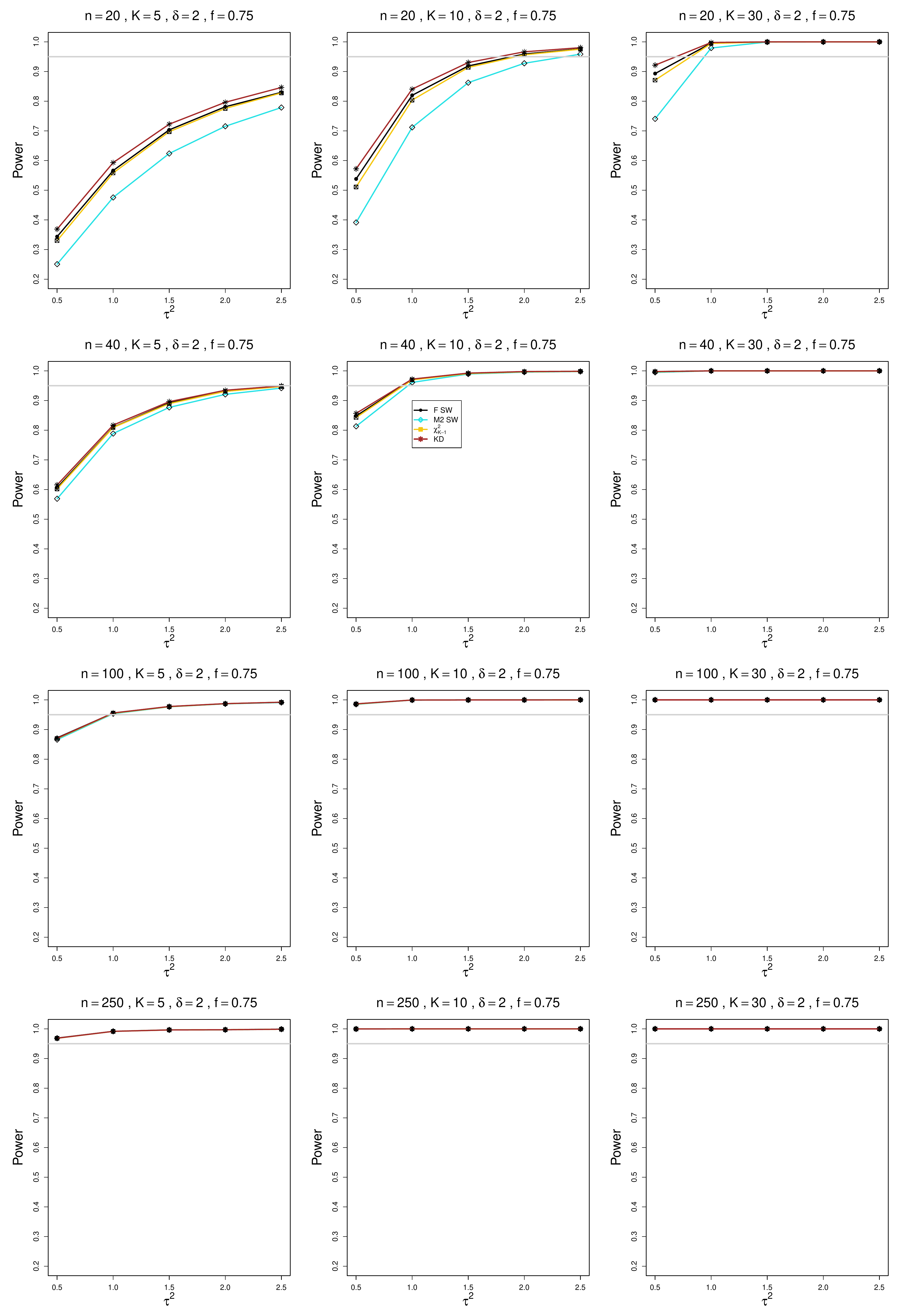}
	\caption{Power for $\delta = 2$, $f = .75$, and equal sample sizes
		\label{PlotOfPhatAt005delta2andq075SMD_underH0}}
\end{figure}

\begin{figure}[t]
	\centering
	\includegraphics[scale=0.33]{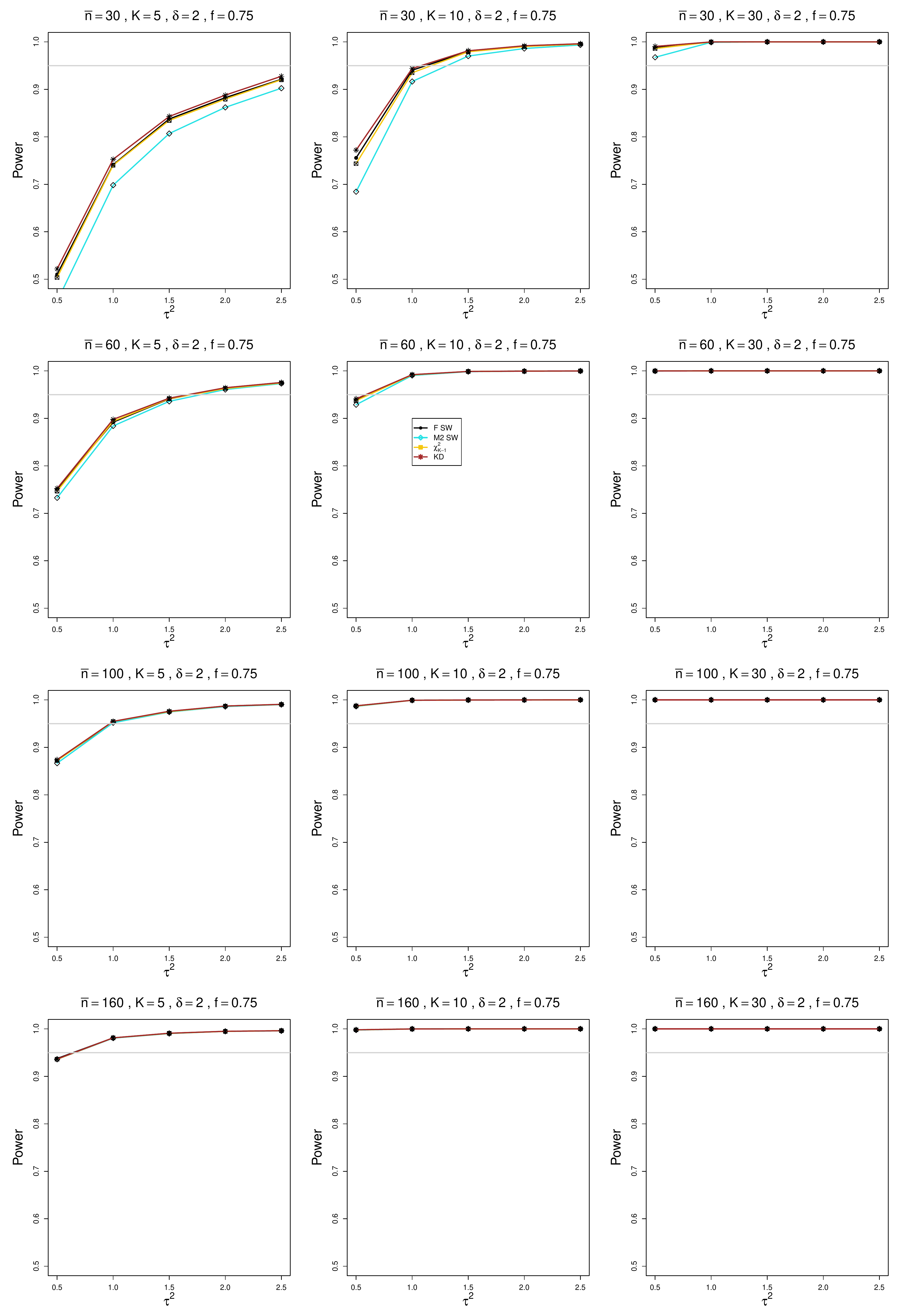}
	\caption{Power for $\delta = 2$, $f = .75$, and unequal sample sizes
		\label{PlotOfPhatAt005delta2andq075SMD_underH0_unequal}}
\end{figure}

\end{document}